\documentclass[11pt]{elsarticle}
\usepackage{epsfig,graphics}
\usepackage{amssymb,amsmath,amsthm,bm, amscd, amsfonts}
\usepackage{booktabs}  
\usepackage{threeparttable}  
\usepackage{multirow}
\usepackage{mathabx}
\usepackage{siunitx}
\usepackage{float}
\biboptions{numbers,sort&compress}
\usepackage{color}

\usepackage{tcolorbox}
\tcbuselibrary{skins}
\usepackage{enumitem}
\usepackage[english]{babel}
\usepackage{algorithm}
\usepackage{algorithmic}

\theoremstyle{plain}
\newtheorem{exam}{Example}[section]

\newtheorem{remark}[exam]{Remark}

\newtheorem{mdef}{Definition}

\def\RR{{\mathbb R}}

\def\ga{\alpha}
\def\gga{\gamma}
\def\gth{\theta}
\def\go{\omega}

\def\gb{\beta}

\def\gl{\lambda}

\def\wt{\widetilde}
\def\n{\noindent}

\def\b0{{\bf 0}}
\def\1{{\bf 1}}

\def\cL{\mathcal L}

\def\wh{\widehat}
\def\wt{\widetilde}

\makeatletter
\def\widebreve{\mathpalette\wide@breve}
\def\wide@breve#1#2{\sbox\z@{$#1#2$}%
     \mathop{\vbox{\m@th\ialign{##\crcr
\kern0.08em\brevefill#1{0.99\wd\z@}\crcr\noalign{\nointerlineskip}%
                    $\hss#1#2\hss$\crcr}}}\limits}
\def\brevefill#1#2{$\m@th\sbox\tw@{$#1($}%
  \hss\resizebox{#2}{\wd\tw@}{\rotatebox[origin=c]{90}{\upshape(}}\hss$}
\makeatletter

\newtcolorbox{nomenclaturebox}{
    enhanced,
    colback=white,
    colframe=black!80,
    arc=0pt,
    boxrule=1pt,
    width=\textwidth,
    title=Nomenclature,
    fonttitle=\bfseries\large,
    boxsep=5pt,
    before upper={\vspace*{5pt}},
    after={\vspace*{15pt}},
    left=10pt,
    right=10pt,
    colbacktitle=white,
    coltitle=black
}

\begin{document}

\begin{frontmatter}




\title{Synchrosqueezed windowed linear canonical transform:\\
A method for mode retrieval from multicomponent signals
with\\ crossing instantaneous frequencies
}
\author[university1]{Shuixin Li}
\ead{lishuixin@zjnu.edu.cn}
\author[university1]{Jiecheng Chen}
\ead{jcchen@zjnu.edu.cn}
\author[university1]{Qingtang Jiang}
\ead{jiangq@zjnu.edu.cn}
\author[university2] {Jian Lu}
\ead{jianlu@szu.edu.cn}


\address[university1]{School of Mathematical Sciences, Zhejiang Normal University, Jinhua 321004, China}
\address[university2]{College of Mathematics and Statistics, Shenzhen University, Shenzhen 518060, China}

\date{}


\vskip -0.5cm



\begin{abstract}
In nature, signals often appear in the form of the superposition of multiple non-stationary signals.
The overlap of signal components in the time-frequency domain poses a significant challenge for signal analysis. One approach to addressing this problem is to introduce an additional chirprate parameter and use the  chirplet transform (CT) to elevate the  two-dimensional time-frequency representation to a three-dimensional  time-frequency-chirprate representation. 
From a certain point of view, the CT  of a signal can be regarded as a windowed special linear canonical transform of that signal, 
undergoing a shift and a modulation.

In this paper, we develop this idea to propose a novel windowed linear canonical transform (WLCT), which provides 
a new time-frequency-chirprate representation. We discuss four  types of WLCTs. 
In addition, we use a 
special X-ray transform to further sharpen the time-frequency-chirprate representation. 
Furthermore, we derive the corresponding three-dimensional synchrosqueezed transform, demonstrating that the WLCTs have great potential for three-dimensional signal separation.
     
\end{abstract}

\end{frontmatter}

\begin{nomenclaturebox}
    \subsection*{Abbreviations}
    \begin{itemize}[leftmargin=*,noitemsep,topsep=3pt]
        \item CT   -- chirplet transform
        \item FrFT -- fractional Fourier transform
        \item IF   -- instantaneous frequency
        \item LCT  -- linear canonical transform
        \item SSO  -- signal separation operator
        \item { SST --synchrosqueezed transform}
        \item SWLCT -- synchrosqueezed windowed linear canonical transform
        \item SXWLCT -- synchrosqueezed X-ray windowed linear canonical transform
        \item WFT  -- windowed Fourier transform
        \item WLCT -- windowed linear canonical transform
        \item XWLCT -- X-ray windowed linear canonical transform
    \end{itemize}

    \subsection*{Symbol Notations }
    \begin{itemize}[leftmargin=*,noitemsep,topsep=3pt]
    \item \( E_{\alpha} \) -- Rényi entropy based on the WLCT defined by \eqref{Ealpha1} or \eqref{Ealpha2}
    \item \( K \) -- number of signal components
    \item \( M_k \) -- the matrix of \( M_k=[\phi_k''(t), 1; -1, 0]\)
    \item \( M_{\lambda, n} \) -- parameter matrix of the WLCT 
    \item \( N \) -- the length of the signal
    \item \( N_c \) -- the size of the chirprate bins
    \item \( N_f \) -- the size of the frequency bins
    \item \( P_n \) -- coefficient factor associated with \(\mathcal{L}_g^{M_{\lambda,n}}(\tau)\) defined by \eqref{def_Pn}
    \item \( \Lambda_x^{g, n}(t,\eta,\lambda) \) -- chirprate reassignment operator defined by \eqref{eq:omega}
    \item \(\mathcal{L}^M_x(u),\ \mathcal{L}^M[x(\cdot)](u),\ \mathcal{L}^M(x)(u) \) -- LCT of \( x(t) \) with matrix \( M \)
    \item \(\mathcal{L}_g^{M_{\lambda,n}}(\tau)\)-- WLCT of Gaussian function \( g(t) \) with matrix \( M_{\lambda, n} \)
    \item \( \Omega_x^{g, n}(t,\eta,\lambda) \) -- frequency reassignment operator defined by \eqref{eq:omega}
    \item \( S_x^{g, n}(t, \eta, \lambda) \) -- SWLCT of \( x(t) \) with matrix \( M_{\lambda, n} \)
    \item \(\mathcal{S}_x^{g, n}(t, \eta, \lambda) \) -- SXWLCT of \( x(t) \) with matrix \( M_{\lambda, n} \)
    \item \( T_x^{g, n}(t, \eta,\lambda) \) -- WLCT of \( x(t) \) with matrix \( M_{\lambda, n} \)
    \item \(\mathcal{T}_x^{g, n}(t, \eta,\lambda) \) -- XWLCT of \( x(t) \) with matrix \( M_{\lambda, n} \)
    \item \( V^h_x(t, \eta) \) -- WFT of \( x(t) \) with window function \( h \)
\end{itemize}
\end{nomenclaturebox}

\section{Introduction}

In real-world environments, signals often consist of multiple components. To accurately describe this, researchers use a simplified mathematical model represented as follows:
\begin{equation}
\label{AHM0}
x(t)=\sum_{k=1}^{K} x_k(t) = \sum_{k=1}^{K} A_k(t) e^{2 \pi i \phi_k(t)},
\end{equation}
where \(A_k(t) > 0\) and \(\phi_k'(t) > 0\) represent the instantaneous amplitude (IA) and instantaneous frequency (IF) of the \(k\)-th component, respectively.
\(A_k(t)\) is a slowly varying function of time.
 Each $x_k(t) = A_k(t) e^{2 \pi i \phi_k(t)}$ represents a constituent component, or alternatively, a mode, of the signal $x(t)$ for $k = 1, 2, \cdots, K$.
 One of the primary objectives in signal processing is the separation of individual modes \(x_k(t)\) from a composite signal. This is pivotal for many applications and represents a fundamental challenge in the field.

Time-frequency analysis is an essential tool for characterizing the time-varying features and restoring the individual signal components of multicomponent non-stationary signals.
 Nevertheless, traditional linear time-frequency methods, such as the  windowed Fourier transform (WFT), also known as the   short time Fourier transform (STFT) \cite{stankovic2013time} 
and the continuous wavelet transform  \cite{daubechies1992ten,mallat1999wavelet}, 
are typically constrained by the uncertainty principle \cite{cohen1995time}, 
making it difficult to achieve high time and frequency 
resolution simultaneously. Recently, Daubechies et al. developed a novel time-frequency analysis technique known as the synchrosqueezed transform (SST) \cite{daubechies2011synchrosqueezed}. 
 Unlike the reassignment method (RM) \cite{auger1995improving}, SST not only sharpens 
   the time-frequency representation but also preserves 
   invertibility. Since the publication of this method, subsequent research and extensions
    have led to a series of interesting and practical results. These include the 2nd-SST \cite{oberlin2015second,oberlin2017second,behera2018theoretical} and 
high-order SST \cite{pham2017high}, the synchroextracting transform (SET) \cite{yu2017synchroextracting}, multiple-SST (MSST) \cite{yu2018multisynchrosqueezing}, demodulation-based SST \cite{wang2013matching,meignen2017demodulation,jiang2017instantaneous,jiang2022instantaneous}, time-reassigned SST \cite{he2019time,li2022theoretical,liu2023local}, and second-order time-reassigned SST \cite{fourer2019second,he2020gaussian}, among others.

\begin{figure}[th]
		\centering
		\begin{tabular}{cccc}
			\resizebox{0.23\textwidth}{!}{\includegraphics{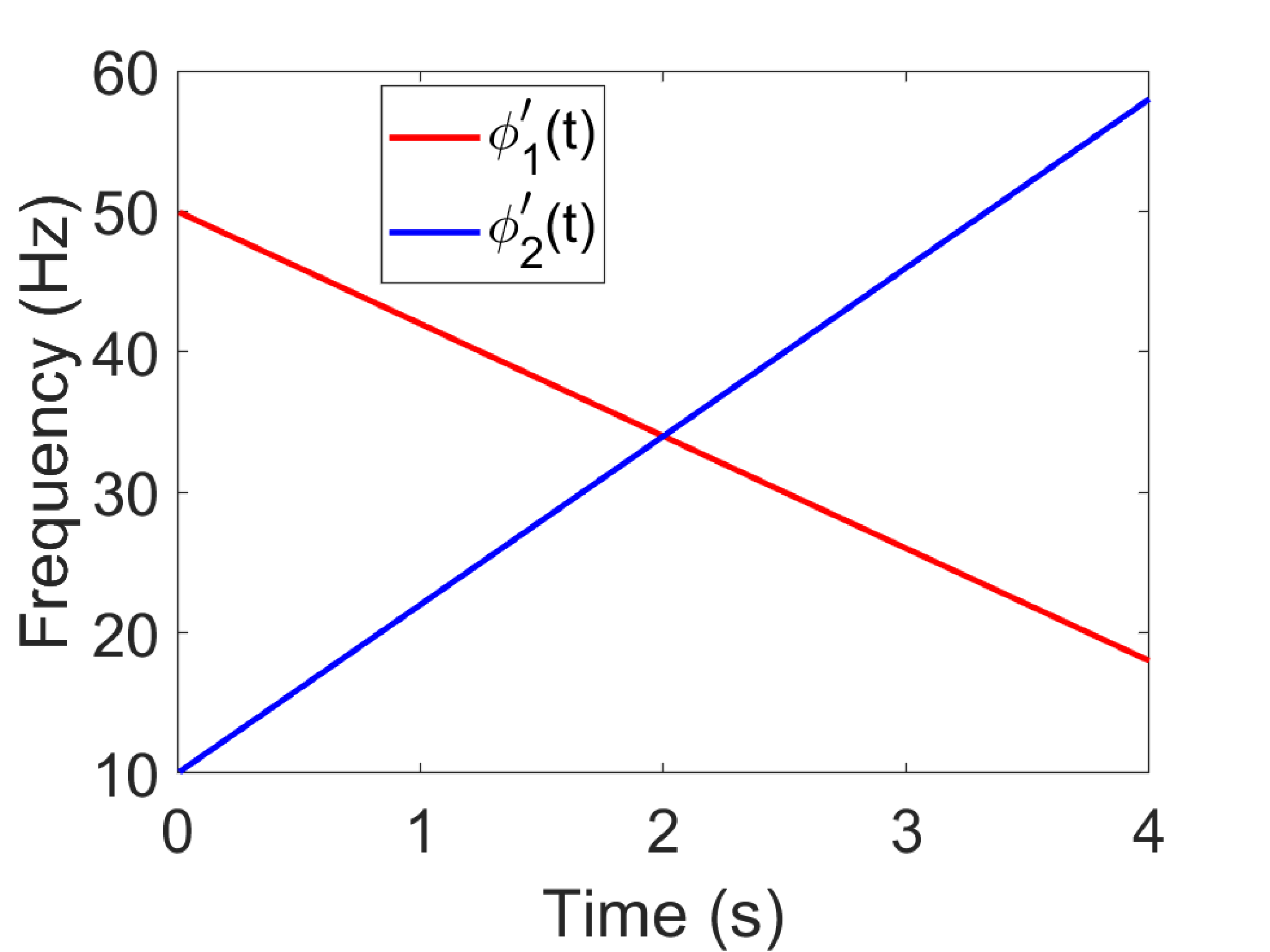}} & 
			\resizebox{0.23\textwidth}{!}{\includegraphics{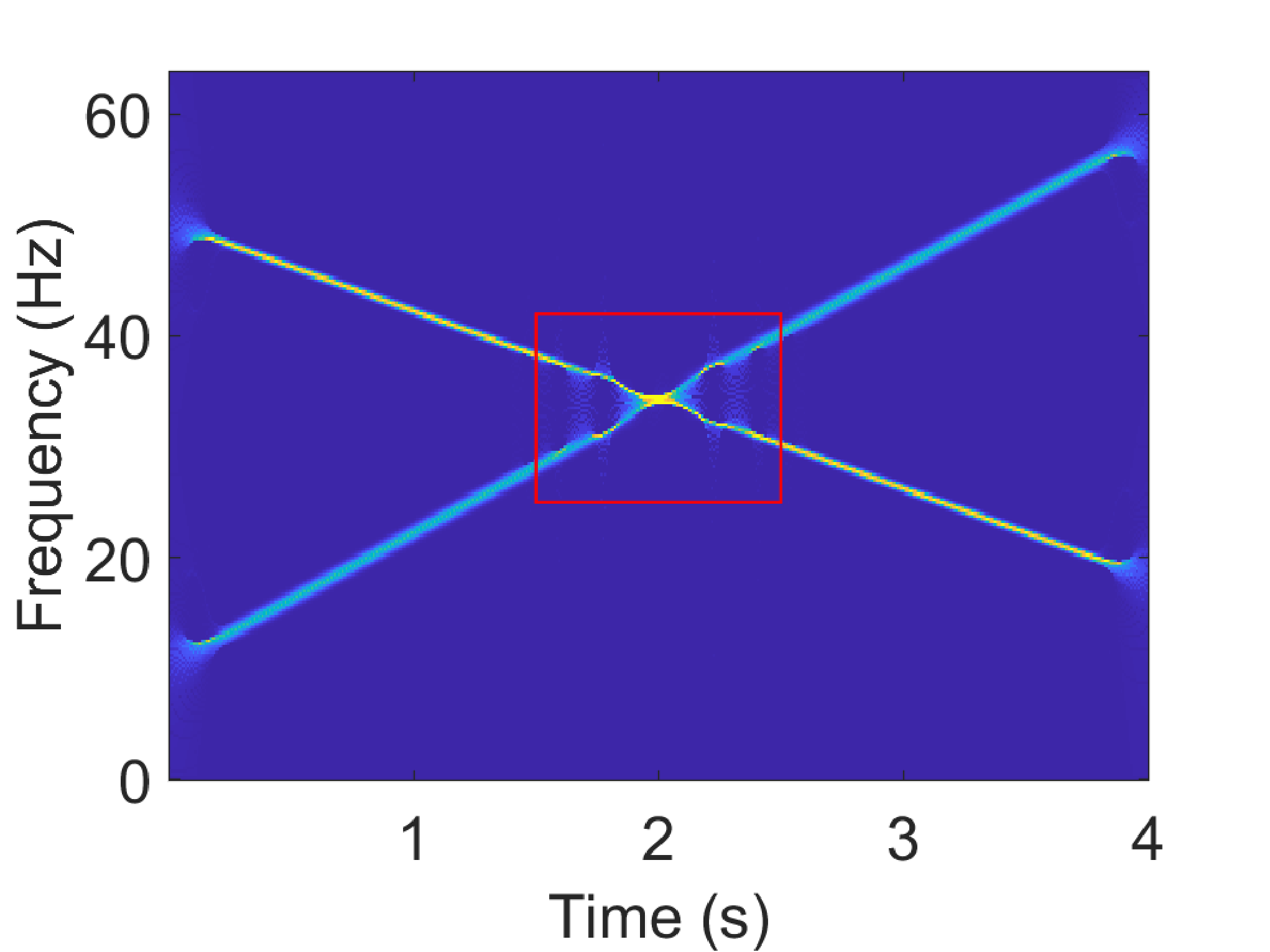}} & 
			\resizebox{0.23\textwidth}{!}{\includegraphics{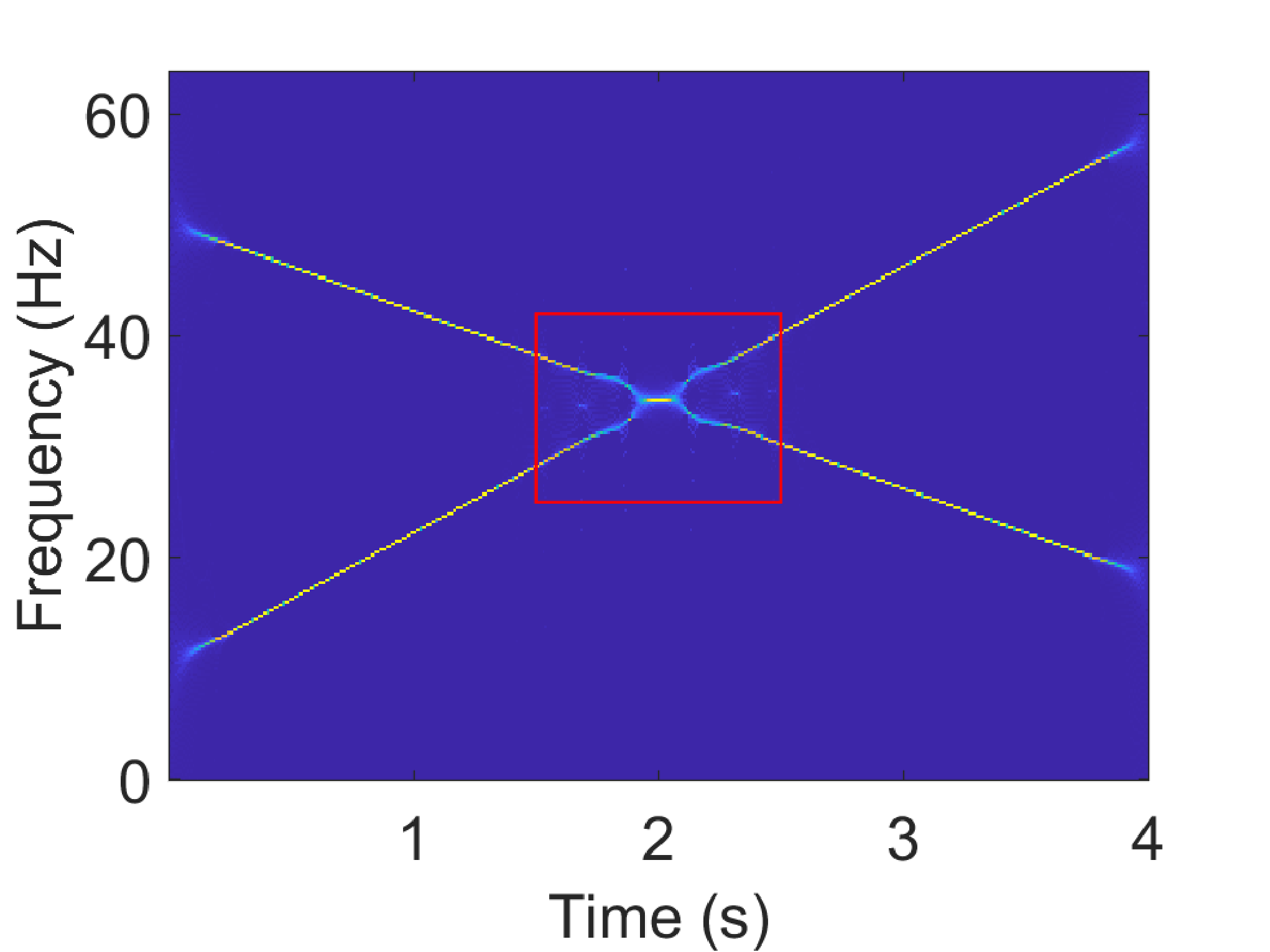}} & 
			\resizebox{0.23\textwidth}{!}{\includegraphics{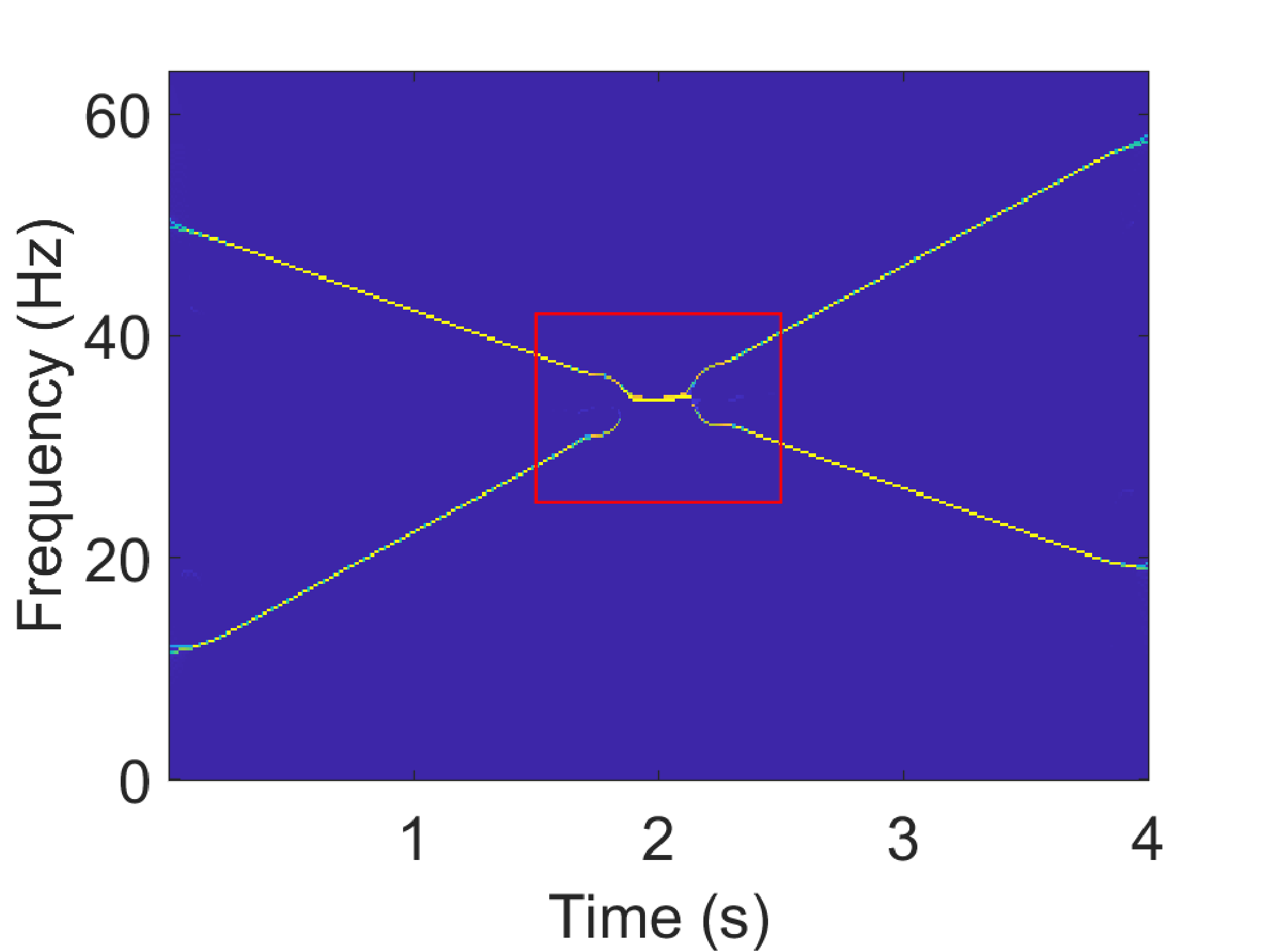}} \\
			(a) & (b) & (c) & (d) \\
			
			\resizebox{0.23\textwidth}{!}{\includegraphics{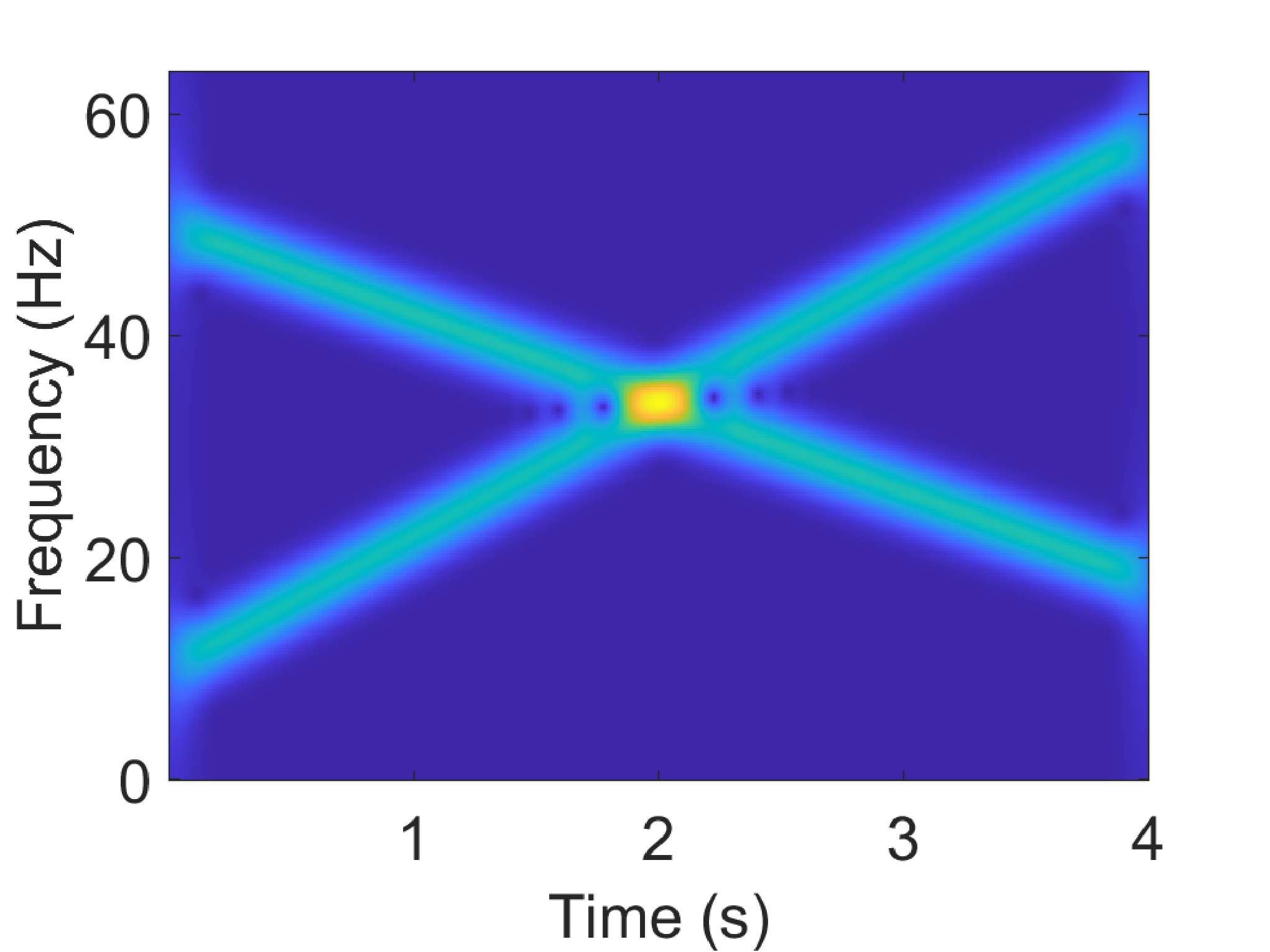}} & 
			\resizebox{0.23\textwidth}{!}{\includegraphics{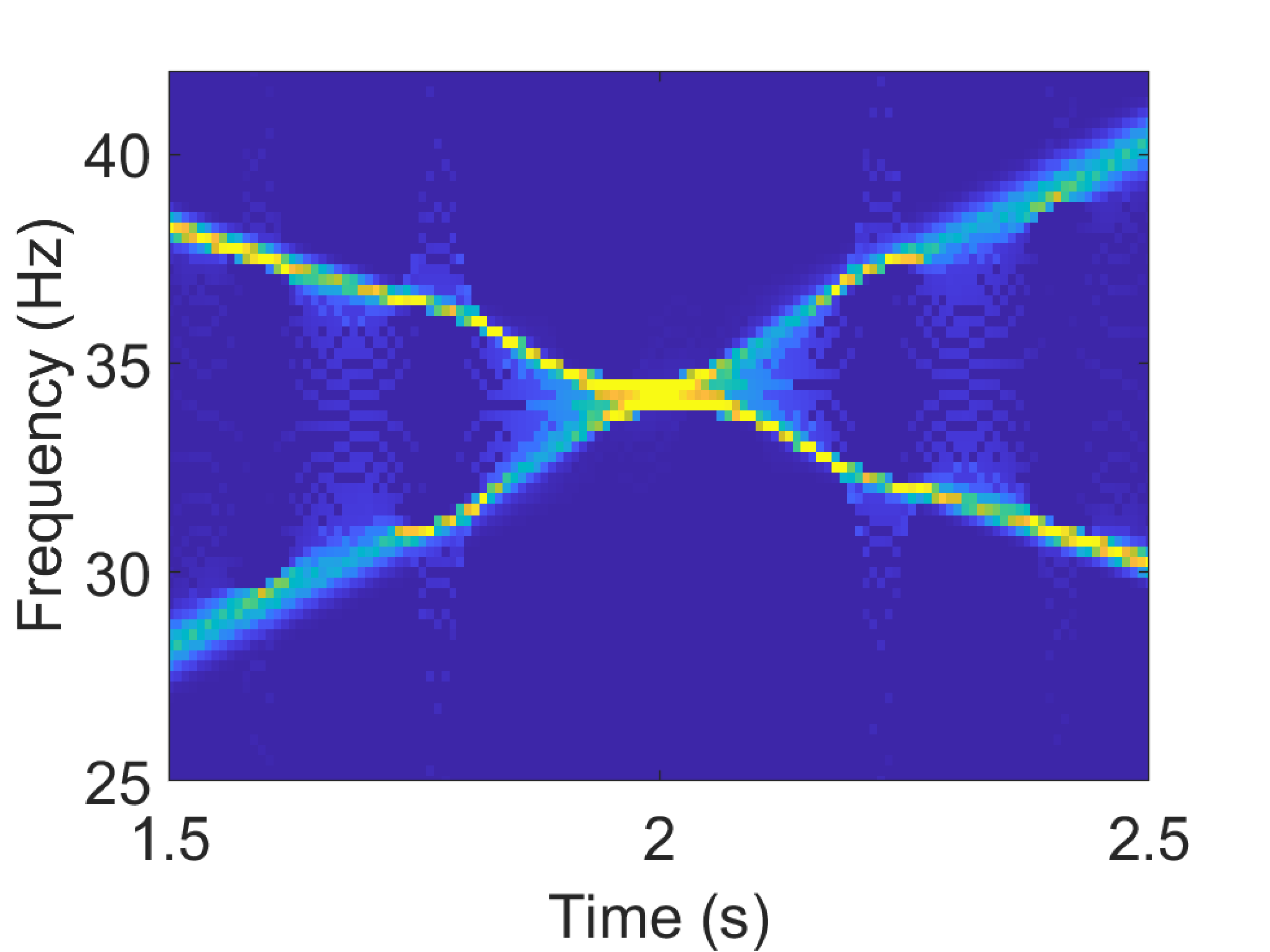}} & 
			\resizebox{0.23\textwidth}{!}{\includegraphics{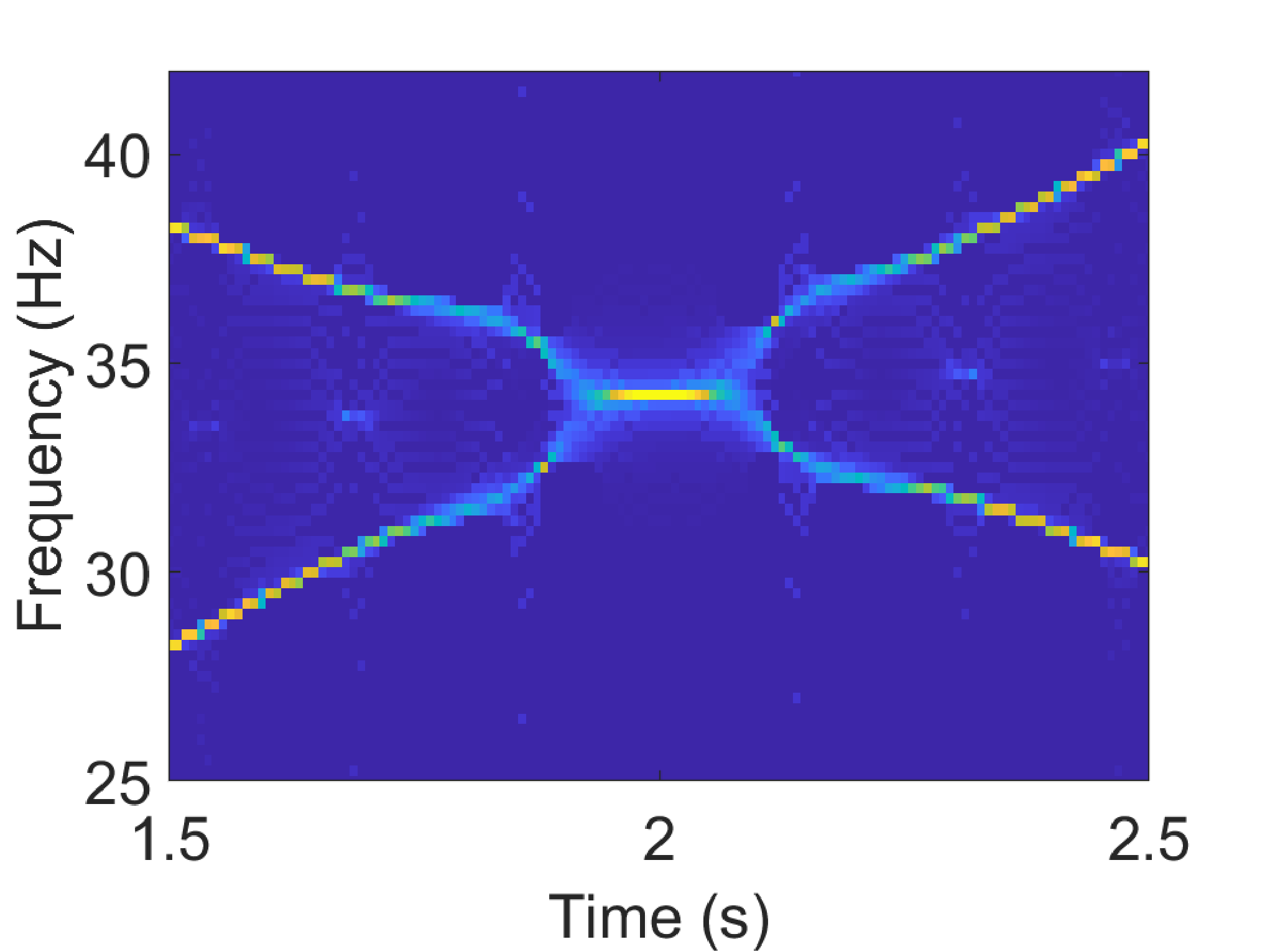}} & 
			\resizebox{0.23\textwidth}{!}{\includegraphics{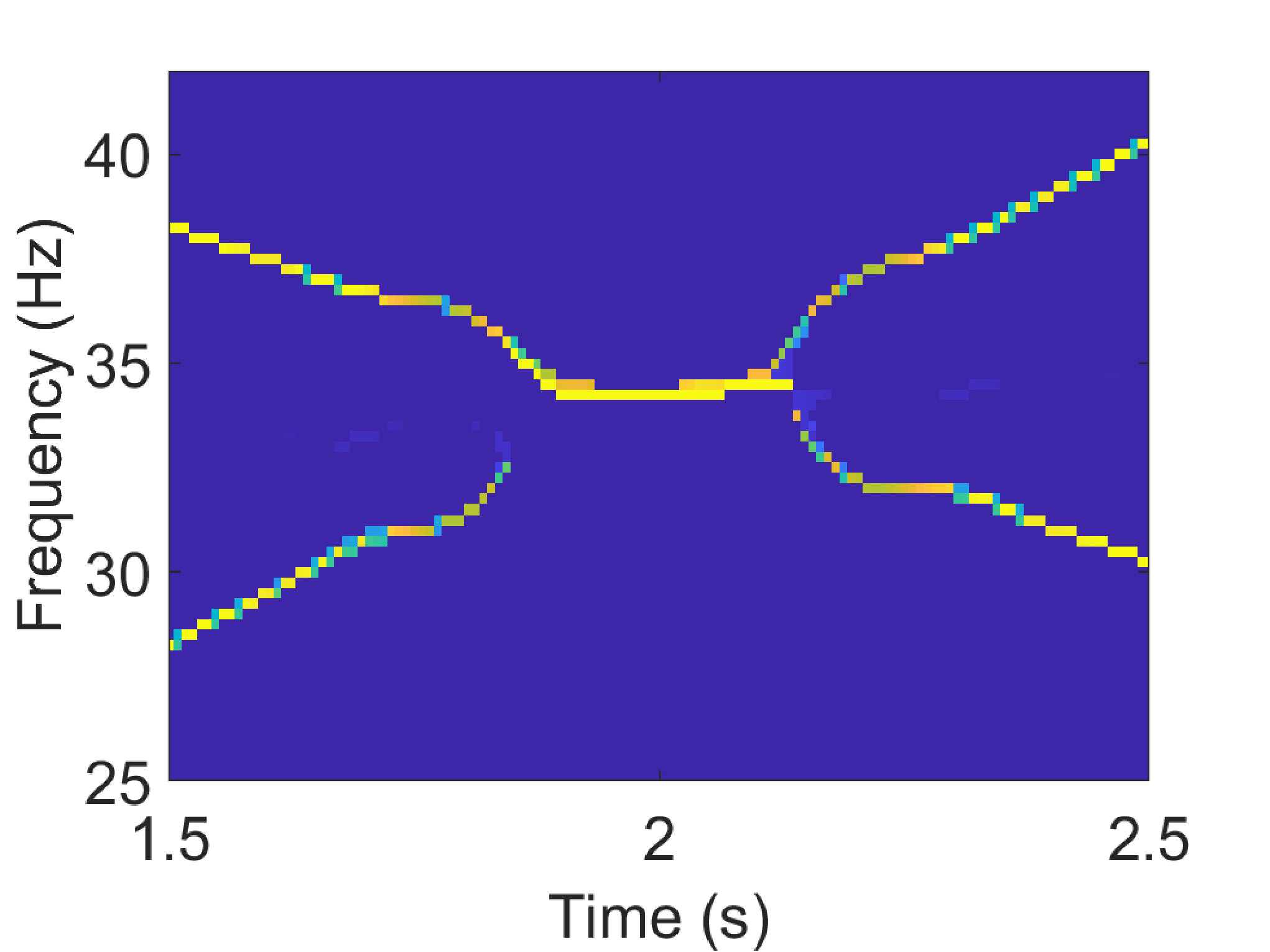}} \\
			(e) & (f) & (g) & (h) \\
		\end{tabular}
		\caption{\small Time-frequency representations of \(x(t)\). (a) IFs of signal; (b) SST; (c) 2nd SST; (d) MSST; (e) STFT; (f)  SST (local zoom); (g) 2nd SST (local zoom); (h) MSST (local zoom).}
		\label{figure:example1}
	\end{figure}

However, when the signal is composed of several components with crossover IF curves, most of the aforementioned methods fail to accurately reflect the signal characteristics in the regions near the intersection points. For example, consider a simple synthetic signal:
\begin{equation}
\label{example1}x(t) = e^{2\pi \mathrm{i}(-4t^2 + 50t)} + e^{2\pi \mathrm{i}(6t^2 + 10t)}, \quad t \in [0,4),
\end{equation}
sampled at \(128\) Hz. 
The IFs of the two components cross at \( t = 2 \) s and \( \eta = 34 \) Hz.
Most time-frequency representations exhibit significant blurring around the intersection point, as illustrated in Fig.~\ref{figure:example1}.

The chirplet transform (CT), proposed by Mann et al. \cite{mann1995chirplet}, offers an effective solution to this issue by extending the two-dimensional time-frequency plane  to a three-dimensional time-frequency-chirprate space. For simplicity of presentation, we use ``chirprate'' to denote ``chirp rate'' in this paper.
 In the time-frequency-chirprate  space, under  certain conditions, the components of non-stationary multicomponent signals can be well separated \cite{li2022chirplet,chui2023analysis}.
Another transformation, known as the wavelet-chirplet transform (WCT), follows a similar concept. It also starts with the wavelet transform and elevates the time-scale plane to a three-dimensional time-scale-chirprate space \cite{chui2021time}.

The linear canonical transform (LCT) is a generalized integral transform that  encompasses the Fourier transform and the fractional Fourier transform (FrFT) \cite{namias1980fractional,almeida1994fractional,tao2009fractional}, serving as a fundamental tool in time-frequency analysis. 
It was independently introduced in the 1970s by Moshinsky, who formulated it within the framework of quantum mechanics \cite{moshinsky1971linear}, and by Collins, who derived it from optical system modeling \cite{collins1970lens}.
Geometrically, the LCT corresponds to an affine transformation in the time-frequency plane. 
Over the past two decades, the combination of the LCT with traditional time-frequency analysis techniques has led to numerous innovative time-frequency analysis methods, such as the short time linear canonical transform  \cite{kou2012windowed,wei2021theory}, the linear canonical wavelet transform \cite{wei2014generalized,guo2018linear}, 
 the linear canonical Stockwell transform \cite{wei2022linear,shah2020linear},
 the Wigner-Ville distribution associated with LCT \cite{bai2012wigner,xin2021new},
 the spectrogram associated with the LCT \cite{zhi2013spectrogram}.
These emerging methods have demonstrated unique strengths in handling non-stationary signals and have yielded a series of rich and meaningful research results.

In this paper, we propose a novel definition of  the windowed linear canonical transform (WLCT), where  the CT is a special case of the WLCT with a  specific parameter matrix. Additionally, we introduce a new parameter through the LCT to elevate the time-frequency plane into a three-dimensional space, which we refer to as the time-frequency-chirprate space (when there is no ambiguity). 

However, significant work remains to be done for accurately estimating the IFs and chirprate.  The signal separation operator (SSO), was first introduced in \cite{chui2016signal} for multicomponent signal separation in the time-frequency plane.
 To recover the signal modes with crossing IFs, the SSO scheme was extended to the time-frequency-chirprate space in \cite{chui2021signal,chui2021analysis}.
 Though the extended SSO scheme shows promise, its dependence on precise IF and chirprate estimations remains a limitation. Similar to the CT, the WLCT exhibits slow decay in the chirprate direction, which further complicates accurate chirprate extraction. To refine the estimations of IFs and chirprates, several methods based on the CT have been developed, including the synchrosqueezed CT \cite{zhu2020frequency,chen2023disentangling}, the 3D extracting transform (TET) \cite{zhu2021three} 
and the multi-synchrosqueezed CT \cite{chen2024multiple,chen2024composite}. 

In this paper, we propose  three-dimensional synchrosqueezed transformations based on the WLCT scheme. Moreover, 
we introduce the composition of the X-ray transform and WLCT, 
 termed X-ray WLCT (XWLCT) to significantly accelerate the decay rate in the chirprate direction. To achieve more accurate estimation of the IFs and chirprates, we further introduce  synchrosqueezed transforms drawing upon the XWLCT. This approach  yields promising results.

The remainder of this paper is organized as follows. Section 2 provides an overview of the definition and fundamental properties of the LCT. In Section 3, we introduce the windowed linear canonical transform (WLCT) 
and derive the associated mode retrieval formulas. The synchrosqueezed WLCT transform is developed in Section 4. Section 5 focuses on the XWLCT and its synchrosqueezed transform, and provides experimental validation. Implementation of the algorithms is described in Section 6. Finally, the paper concludes with a summary in Section 7. 

\section{Preliminary}
 \subsection{Linear canonical transform (LCT)}
 In this paper, the Fourier transform of a signal $x(t)$ is defined as
 \begin{equation}
 \wh x(\xi) := \int_{\mathbb{R}} x(t) e^{-i2\pi \xi t} \, dt.
 \end{equation}

The  LCT has been defined in numerous ways across various applications (e.g., \cite{wolf2013integral,healy2015linear}). 
In this paper, we employ the definition presented in \cite{healy2015linear}. For simplicity, we denote the $2\times 2$  parameter matrix 
$\left[\begin{matrix} a & b \\ c & d \end{matrix}\right]$ as $ [a, b ; c, d]$.

\begin{mdef}\label{de1.1}
Given a matrix $M = [a, b ; c, d]$ with $a, b, c, d \in \mathbb{R}$ and $ad - bc = 1$, the LCT of a function $x(t)$ is defined as
	$$
      \mathcal{L}_x^M(u):= 
      \begin{cases}
      \int_{\mathbb{R}} x(t) K_{M}(u, t) \, dt, & \text{if } b \neq 0; \\
      \sqrt{d} e^{i \pi {cd u^2}} x(du), & \text{if } b = 0.
      \end{cases}
    $$
  where the kernel function $K_M(u, t)$ is given by
  $$
   K_M(u, t):= \frac{1}{\sqrt{ b i}} e^{i \pi \left( \frac{a }{b}{t^2} - \frac{2 }{b} {u t} + \frac{d }{b} {u^2} \right)}.
  $$
\end{mdef}	
Here, the expression \(\frac{1}{\sqrt{b i}}\) is defined as 
		\[
		\frac{1}{\sqrt{b i}} = \frac{\exp\left(-i \frac{\pi}{4} {\rm sign}(b) \right)}{\sqrt{|b|}},
		\]
where \({\rm sign}(b)\) denotes the sign of \(b\), and \(|b|\) is the absolute value of \(b\).
We use the notations \( \mathcal{L}^M_x(u) \),  \( \mathcal{L}^M[x(\cdot)](u) \)  and \( \mathcal{L}^M(x)(u) \)  to denote the LCT of the function \( x(t) \) with respect to matrix $M$. 

The matrix $M =[a, b; c, d]$, subject to the constraint $\det(M) = 1$, possesses three degrees of freedom, allowing for the variation of three independent parameters. This flexibility provides great convenience for analyzing signals.

\subsection{Basic properties}
In this subsection, we present selected fundamental mathematical properties of the LCT  that are essential for subsequent analysis. For detailed proofs and a comprehensive enumeration, we refer the reader to \cite{healy2015linear}. 

Let  \( M = [a, b ; c, d] \) with \(\det(M) = 1\)  be a parameter matrix. 

	\begin{enumerate}
		\item {Additivity:}
		\begin{align}
			\label{additivity property}	\mathcal{L}^{A} \mathcal{L}^{B}[x(t)](u) = \mathcal{L}^{AB}[x(t)](u),
		\end{align}
		where \( A \) and \( B \) are matrices with determinant 1, and \( AB \) denotes their matrix product.
	
		\item {Differentiation Property:}
		Suppose \( y(t) = x'(t) \). Then,
		\begin{align}	
			\label{differential property} \mathcal{L}_y^M(u) = 2\pi \left(aD_u - icu\right) \mathcal{L}_x^M(u),
		\end{align}
		where \( D_u \) denotes the differential operator with respect to \( u \).
	
		\item {Generalized Parseval's Theorem:}
		\begin{equation}
			\label{parseval's property}	\int_{\mathbb{R}} x(t) \overline{y(t)} \, \mathrm{d}t = \int_{\mathbb{R}} \mathcal{L}_x^M(u) \overline{\mathcal{L}_y^M(u)} \, \mathrm{d}u,
		\end{equation}
		where \( \overline {y(t)} \) denotes the complex conjugate of \( y(t) \).
	\end{enumerate}

\section{Windowed linear canonical transform (WLCT): A new definition and properties}

In this section, we propose a novel WLCT.  
We then conduct an in-depth study of its properties, focusing on its application in IF and chirprate estimations as well as mode retrieval.
Our research targets multicomponent signals that exhibit crossover IFs. Notably, the WLCT we propose distinctly differs from the ones explored in the works \cite{kou2012windowed,wei2021theory}.

\subsection{ A new definition  of WLCT} 
Before we introduce our WLCT, let us first recall 
the (modified) windowed Fourier transform (or short-time Fourier transform) of $x(t)$ using a window function $h(t)$, defined by
$$
V^h_x(t, \eta):=\int_\RR x(\tau)e^{-i2\pi \eta (\tau-t)}\overline{h(\tau-t)}d\tau=
\int_\RR x(t+\tau)e^{-i2\pi \eta \tau}\overline{h(\tau)}d\tau
$$ 
By Parseval's theorem, we have 
$$
V^h_x(t, \eta)=
\int_\RR \big(x(t+\cdot)e^{-i2\pi \eta \cdot }\big)^{\wedge}(\go) \; \overline{\wh h(\go)}d\go. 
$$
Therefore, the windowed Fourier transform $V^h_x(t, \eta)$ of $x(t)$ is derived through the following sequence: A time shift of \(t\) units, \(x(\tau) \to x(t+\tau)\); modulation by \(e^{-i2\pi \eta \tau}\), \(x(t+\tau) \to x(t+\tau)e^{-i2\pi \eta \tau}\); a Fourier transform with respect to \(\tau\), \(\big(x(t+\tau)e^{-i2\pi \eta \tau}\big)^{\wedge}(\omega)\); and averaging with a window function  \(g(\omega) =\overline{\widehat h(\omega)}\).

Motivated by the above observation,  we define the  WLCT of \(x(t)\) with a window function \(g(t)\) as follows. 

\begin{mdef}\label{de3.1}
 Given a parameter matrix \(M =[a, b; c,  d]\) with \rm{det}\((M)\) = 1, we define the  WLCT of \(x(t)\) by  
\begin{equation}
\label{def_WLCT}
T_x^{g, M}(t, \eta):=\int_\RR {\mathcal L}^{\wt M}\big(x(t+\cdot)e^{-i2\pi \eta \cdot }\big)(u) \; g(u)du, 
\end{equation}
where for a parameter matrix $M =[a, b; c, d]$,  $\wt M$ denotes  
$$
\wt M:=[d, b; c, a]= \left[\begin{matrix} d & b \\ c & a \end{matrix}\right]. 
$$
\end{mdef}

Note: In  our definition of WLCT, we utilize \(\widetilde{M}\) rather than \(M\) on the right-hand side of \eqref{def_WLCT} solely for the purpose of improving the clarity of the presentation that follows. The rationale for this substitution will be clarified shortly.

By \eqref{additivity property} and  \eqref{parseval's property}, one has 
$$
\int_{\mathbb{R}} {\mathcal L}^{\wt M}([ x(\cdot)])(u) \overline{y(u)} \,du = \int_{\mathbb{R}} x(\tau) { \mathcal{L}^{{\wt M}^{-1}} [\overline{y(\cdot)}](\tau)} d\tau=\int_{\mathbb{R}} x(\tau) { \mathcal{L}^{ M} [y(\cdot)](\tau)} d\tau . 
$$
Hence, the  $T_x^{g, M}(t, \eta)$ can be written as 
\begin{equation}
T_x^{g, M}(t, \eta):=\int_\RR x(t+\tau)e^{-i2\pi \eta \tau } \; {\mathcal L}^{M}(g)(\tau)d\tau.
\end{equation}
Therefore, $T_x^{g, M}(t, \eta)$ can also be considered as a windowed Fourier transform of $x(t)$ with a window function ${\mathcal L}^{M}(g)$ obtained by the LCT of $g(t)$. Such window functions were also considered    by  Pei and Huang in the  adaptive 
short-time Fourier transform  \cite{pei2018adaptive}, where they first applied the FrFT to a real Gaussian window function,  obtaining a  class of complex window functions termed chirp-modulated Gaussian functions.
These chirp-modulated Gaussian functions significantly enhance  energy
concentration in time-frequency representations while maintaining  invertibility \cite{li2022multitaper}. 

\subsection{Special WLCTs}
The choice of the matrix $M$ in the WLCT significantly impacts its properties and applications. Here, we explore the WLCTs with some special choices of $M$. We will consider the case where the window function $g(t)$ is the Gaussian function, defined by:
\begin{equation}
	\label{Gaussian function}	g(t) = e^{-\pi \alpha t^2},  
\end{equation}
where $\ga>0$.  To calculate  $\cL^M_g(u)$, the LCT of  $g(t)$, we need the following formula (refer to formula (23) with $\nu=0$ on p.121 of \cite{bateman1954tables} 
or p.10 of \cite{cohen1995time}) 
\begin{equation}
\label{integral_formula} 
\int_\RR e^{-(\gga+i \gb)t^2 -i \go t}dt =\frac {\sqrt \pi}{\sqrt {\gga+i\gb}} e^{-\frac {\go^2}{4(\gga+i\gb)}}, \qquad \gga>0,   
\end{equation}
where $\sqrt{\gga+i \gb}$ denotes one of the square roots of the complex number  $\gga+i\gb$, with the condition that the real part of \(\sqrt{\gamma + i \beta}\), denoted as Re(\(\sqrt{\gamma + i \beta}\)), is greater than zero.

By \eqref{integral_formula} and direct calculations,  we have that $\cL^M_g(u)$ is given by 
\begin{equation}
\label{LCT_gaussion_bneq0}
\mathcal{L}_g^M(u)= \frac{1}{\sqrt{a+i \alpha b}} e^{- \pi  \frac{ \alpha-i bd \alpha^2 -i a c  }{ b^2 \alpha^2 + a^2} u^2}= C e^{P \pi u^2},  \qquad   b \not=0, 
\end{equation}
where 
\begin{equation}
  C=\frac{1}{\sqrt{a+i \alpha b}}, \quad  P= \frac{i d \alpha + c}{b \alpha - i a}. \label{eq:parameter_P}
\end{equation}
When \( b = 0 \),
\begin{equation*}
   \mathcal{L}_g^M(u) = \sqrt{d} \, e^{-\pi \alpha d^2 u^2} \, e^{i \pi c d u^2},
\end{equation*}
which is a special case of  \eqref{LCT_gaussion_bneq0}  
derived under the conditions \( b = 0 \) and \( ad = 1 \).
Therefore, to encompass all scenarios without distinguishing whether \( b \) is zero or not, we can express \(\mathcal{L}_g^M(u)\) in the unified form  \eqref{LCT_gaussion_bneq0}.
Furthermore, we can derive the derivative of \(\mathcal{L}_g^M(u)\), 
\begin{equation}
    D_u \mathcal{L}_g^M(u) = 2 \pi P u \mathcal{L}_g^M(u). \label{eq:derivative_LCT}
\end{equation}

Consider the analytic signal $x(t)=A(t) e^{i2\pi \phi(t)}$ under the regularity conditions
\begin{equation}
\label{harmonic_cond}
|A'(t)|\le \epsilon_1, \quad |\phi'''(t)|\le \epsilon_2, 
\end{equation}
where $ \epsilon_1, \epsilon_2>0$ are sufficiently small.
Then for a small time-localized interval of \(\tau\), 
\begin{eqnarray*}
&& x(t+\tau)\approx A(t) e^{i2\pi (\phi(t)+\phi'(t)\tau+\frac {\phi''(t)}2 \tau^2)}=x(t) e^{i2\pi  \phi'(t)\tau+i\pi \phi''(t) \tau^2}.  
\end{eqnarray*}
Thus, the WLCT of \(x(t)\) can be expressed as
\begin{align}
    \label{appro_matrix_pre}
T_x^{g, M}(t, \eta)&\approx \int_{\RR} x(t)   e^{i2\pi  (\phi'(t)-\eta)\tau+i\pi \phi''(t) \tau^2}  \cL_g^M(\tau) d\tau  \nonumber\\
	 \qquad &=\frac{1}{\sqrt{-i}} x(t)     \cL^{M_0}(\cL^M(g))(\eta-\phi'(t))=\frac{1}{\sqrt{-i}} x(t)     \cL_g^{M_0 M}(\eta-\phi'(t)), 	
\end{align}
where the matrix \( M_0=[\phi''(t), 1; -1, 0]\), and  the last equation follows the  additivity property \eqref{additivity property} of the LCT.  
 With 
\begin{align}
    \label{eq:matrix_NM}
    M_0M = \left[\begin{matrix}
        \phi''(t) & 1 \\
        -1 & 0
    \end{matrix}\right]
    \left[\begin{matrix}
        a & b \\
        c & d
    \end{matrix}\right]
    = \left[ \begin{matrix}
        a \phi''(t) + c & b \phi''(t) + d \\
        -a & -b
    \end{matrix}
    \right], 
\end{align}
from  \eqref{LCT_gaussion_bneq0}, it follows that
$$
T_x^{g, M}(t, \eta)\approx \frac{x(t)}{\sqrt {-i}} 
\frac{1}{\sqrt{a \phi''(t) +c+ i \alpha ( b \phi''(t) + d)   }} e^{-\pi \frac{- \alpha b + i a}{a \phi''(t) + c +i \alpha
 (b \phi''(t) + d)} (\eta-\phi'(t))^2}.
$$ 
Hence 
\begin{equation}
    \big|  T_x^{g, M}(t, \eta)  \big| \approx  A(t) L^{-\frac{1}{4}} e^{-\frac{\pi \alpha (\eta-\phi'(t))^2}{L}}, 	
    \label{eq:TFMmagnitude}
\end{equation}
where 
$$
L=( a \phi''(t)+c)^2+\alpha^2 ( b \phi''(t)+d )^2.
$$
It follows that the ridge of \(\big| T_x^{g, M}(t,\eta)\big|\) along the  frequency axis \(\eta\)  provides an estimate of the IF \(\phi'(t)\).
In order to estimate the chirprate \(\phi''(t)\), the key step is to investigate the minimization of \(L\), as this process directly contributes to achieving an accurate estimation \cite{chui2021time,chui2023analysis}.

  Next, we examine the minimization of \(L\) by considering several specific choices of the matrix \(M\). We introduce a parameter \(\lambda\) to represent the degree of freedom in selecting the matrix \(M\). 
\begin{enumerate}[label=\Roman*.]
    \item Considering \( M_{\gl, 1}=\left[\lambda, 1; -1,0 \right]\), the expression for \( L \) is
    \begin{equation}
        \label{L_1} L_1(\gl) = \alpha^2 (\phi''(t))^2 + (\lambda \phi''(t) - 1)^2. 
    \end{equation}
 It becomes apparent that the minimization of \( L_1(\gl) \) occurs when \( \lambda= {1}/{\phi''(t)} \).

At this point, 
    \begin{equation}
    \label{gM_a}    \cL^{M_{\gl, 1}}(g)(u) = \frac{1}{\sqrt{\alpha i + \lambda}} e^{-\frac{\pi \alpha u^2}{\lambda^2 + \alpha^2}} e^{-\frac{i \pi \lambda u^2}{\lambda^2 + \alpha^2}}. 
    \end{equation}

    \item With the matrix \( M_{\gl, 2} =\left[1, 0; \lambda,1 \right]\), we derive the following for \( L \):
    \begin{equation}
        \label{L_2} L_{2}(\gl)= \alpha^2 + (\phi''(t) + \lambda)^2.
    \end{equation}
    Notably, \( L_{2}(\gl) \) achieves its minimum value under the condition \( \lambda =- \phi''(t) \).

    For this instance, the associated WLCT of $g(t)$ is
    \begin{equation}
		\label{gM_c}  \cL^{M_{\gl, 2}}(g)(u)  = e^{-\pi \alpha u^2} e^{i \pi \lambda u^2}.
    \end{equation}

    \item Given the matrix \(M_{\gl, 3}=\left[1, -\lambda; 0,1 \right]\), the expression for \( L \) is defined as:
    \begin{equation*}
        L_{3}(\gl) = \alpha^2 (-\lambda \phi''(t) + 1)^2 + (\phi''(t))^2. \label{L_b}
    \end{equation*}
    Interestingly, the value of \( L_{3}(\gl) \) reaches its minimum when \( \lambda \) is set to \( {1}/{\phi''(t)} \).

 In this scenario, 
    \begin{equation}
		\label{gM_b}  \cL^{M_{\gl, 3}}(g)(u)  = \frac{1}{\sqrt{-\alpha \lambda i + 1}} e^{-\frac{\pi \alpha u^2}{\lambda^2 \alpha^2 + 1}} e^{\frac{-i \pi \lambda \alpha^2 u^2}{\lambda^2 \alpha^2 + 1}}.
    \end{equation}

    \item For the matrix \( M_{\gl, 4}=\left[0, 1 ; -1,\lambda \right]\), the quantity \( L \) is given by:
    \begin{equation*}
        L_{4}(\gl)= \alpha^2 (\phi''(t) + \lambda)^2 + 1. \label{L_d}
    \end{equation*}
    It turns out that the minimum of \( L_4(\lambda) \) is attained when \( \lambda \) equals \( -\phi''(t) \).

    Given this context, 
    \begin{equation}
		\label{gM_d}   \cL^{M_{\gl, 4}}(g)(u)= \frac{1}{\sqrt{\alpha i}} e^{-\frac{\pi u^2}{\alpha}} e^{i \pi \lambda u^2}. 
    \end{equation}

    \item When \(M_\gth=[\cos\theta, \sin\theta; -\sin\theta, \cos\theta ]\), $\mathcal L^{M_\gth}(g)(u)$  corresponds to the FrFT of the  Gaussian function \(g\). 
    To ensure that \( \sin\theta \) is not zero, we choose \( \theta \) within the interval \( (0, \pi) \).
    \begin{align*}
        L_5&= \alpha^2 \left(\sin\theta \phi''(t) + \cos\theta\right)^2 + \left(\cos\theta \phi''(t) - \sin\theta\right)^2  \\
         &= \frac{\alpha^2}{1 + \cot^2\theta} \left( \phi''(t) + \cot\theta\right)^2 + \frac{1}{1 + \cot^2\theta} \left(\cot\theta \phi''(t) - 1\right)^2.\nonumber
    \end{align*}
 Setting \( \lambda = \cot\theta \), we write $L_5$ as :
 \begin{equation}
   \label{L_5} L_5(\gl)=\frac{\alpha^2}{1 + \gl^2} \left( \phi''(t) + \gl \right)^2 + \frac{1}{1 + \gl^2} \left(\gl \phi''(t) - 1\right)^2.
 \end{equation}

Taking the derivative of \( L_5(\lambda) \) with respect to \( \lambda \), we have 
    \begin{align*}
        \frac{d L_5}{d\lambda} &= 2(1-\alpha^2 ) \frac{(\lambda \phi''(t))^2 + \lambda^2 \phi''(t) - \lambda - \phi''(t)}{(1 + \lambda^2)^2} \\
        &= 2(1-\alpha^2 ) \frac{(\lambda \phi''(t) - 1)(\lambda + \phi''(t))}{(1 + \lambda^2)^2}.
    \end{align*}

Therefore, the minimum value of \( L_5(\lambda) \) can only  be attained at the points where \( \lambda \) equals \( {1}/{\phi''(t)} \), \( -\phi''(t) \), or \( \lambda \to \pm\infty \).
\begin{align*}
L_5\left({1}/{\phi''(t)}\right) &= \alpha^2 \phi''(t)^2 + \alpha^2, \\
L_5\left(-\phi''(t)\right) &= \phi''(t)^2 + 1, \\
\lim_{\lambda \to \pm\infty} L_5(\lambda) &= \phi''(t)^2 + \alpha^2.
\end{align*}

    It is clear that \( L_5(\lambda) \) get its minimum value when
    \begin{align*}
        \lambda = 
        \begin{cases}
            {1}/{\phi''(t)}, & \text{if } \alpha < 1; \\
            -\phi''(t),          & \text{if } \alpha > 1.
        \end{cases}
    \end{align*}

As we have observed from the above analysis, the two cases where \(\alpha < 1\) and \(\alpha > 1\) exhibit significant disparities. In order to clearly differentiate between these two scenarios and for the sake of more convenient presentation in the subsequent sections, we denote, although the quantities on the left-hand sides of the following two equations are identical:  
    \begin{eqnarray}
		\label{gM_theta5}  && {\mathcal L}^{M_{\lambda, 5}}(g)(u) = \sqrt{\frac{-i - \lambda}{\alpha - i \lambda}} e^{-\frac{\alpha \pi u^2 (1 + \lambda^2)}{\alpha^2 + \lambda^2}} e^{-\frac{i \pi u^2 (1 - \alpha^2) \lambda}{\alpha^2 + \lambda^2}}, \quad \hbox{if $0<\ga<1$},  \\
        \label{gM_theta6}  && {\mathcal L}^{M_{\lambda, 6}}(g)(u) = \sqrt{\frac{-i - \lambda}{\alpha - i \lambda}} e^{-\frac{\alpha \pi u^2 (1 + \lambda^2)}{\alpha^2 + \lambda^2}} e^{-\frac{i \pi u^2 (1 - \alpha^2) \lambda}{\alpha^2 + \lambda^2}}, \quad \hbox{if $\ga>1$}. 
    \end{eqnarray}
\end{enumerate}

Notably, the  functions \eqref{gM_a} and \eqref{gM_b} exhibit a remarkable symmetry: Up to a constant \(\sqrt{\alpha i}\), they are identical under the substitution \(\alpha \leftrightarrow \frac{1}{\alpha}\). This symmetry is also observed between \eqref{gM_c} and \eqref{gM_d}.
Thus, the above first 4 cases can be reduced to the first 2 cases with 
\begin{align*}
M_{\gl, 1}=[\lambda, 1 ; -1, 0], \quad  M_{\gl, 2}=[ 1, 0;  \lambda, 1 ]. 
\end{align*}

To highlight the third variable $\gl$, we write $T_x^{g, M_{\gl, 1}}(t, \eta)$  and  $T_x^{g, M_{\gl, 2}}(t, \eta)$ 
as $T_x^{g, 1}(t, \eta, \gl)$  and  $T_x^{g, 2}(t, \eta, \gl)$ respectively. In addition, write $T_x^{g, M_\gth}(t, \eta)$ as
 $T_x^{g, 5}(t, \eta, \gl)$ or $T_x^{g, 6}(t, \eta, \gl)$ depending on $\ga<1$ or $\ga>1$.   
More precisely, 
\begin{align}
    \label{WLCT definition}
 T_x^{g, n}(t, \eta, \lambda) = \int_{\mathbb{R}} x(t + \tau) \cL^{M_{\gl, n}}(g)(\tau) e^{-2\pi i \eta \tau} \, d\tau,  \; \hbox{for $n=1, 2, 5, 6$}, 
\end{align}
where  $\cL^{M_{\gl, 1}}(g)(\tau)$ and  $\cL^{M_{\gl, 2}}(g)(\tau)$
are defined by \eqref{gM_a} and \eqref{gM_c} respectively, and $\cL^{M_{\gl, 5}}(g)(\tau)$ and  $\cL^{M_{\gl, 6}}(g)(\tau)$
are defined by \eqref{gM_theta5} and \eqref{gM_theta6} respectively.

The chirplet transform (CT) \cite{mann1995chirplet} of $x(t)$ is defined by 
\begin{equation*}
C_{x}(t, \eta, \lambda) := \int_{\mathbb{R}} x(t + \tau) g(\tau) e^{-i 2 \pi \eta \tau} e^{-\pi i \lambda \tau^2} \, d\tau.
\end{equation*}
When \( g(\tau) \) is  the Gaussian function \eqref{Gaussian function}, 
CT is precisely the WLCT  $T_x^{g, 2}(t, \eta, -\lambda)$. 
Consequently, the CT is classified as a specific instance within the broader category of the WLCT.

\subsection{Reconstruction Algorithm }

For a multicomponent signal \( x(t) \) given by \eqref{AHM0}, we assume that \( x(t) \) satisfies the separation conditions:
\begin{align}
    \label{eq:separability1} |\phi_k'(t) - \phi_j'(t)| >2 \Delta_1 \quad \text{or} \quad |\phi_k''(t) - \phi_j''(t)| > 2\Delta_2, \quad t \in \mathbb{R};
\end{align}
or alternatively,
\begin{align}
    \label{eq:separability2} |\phi_k'(t) - \phi_j'(t)| > 2\Delta'_1 \quad \text{or} \quad \frac{|\phi_k''(t) - \phi_j''(t)|}{|\phi_k''(t)| + |\phi_j''(t)|} > \Delta'_2, \quad t \in \mathbb{R},
\end{align}
where \(\Delta_1\), \(\Delta_2\), \(\Delta'_1\), and \(\Delta'_2\) are positive constants.

 \begin{remark} \label{Remark}
Condition \eqref{eq:separability2} implies that the chirprate \(\phi_k''(t)\) of each \(x_k(t)\) must be nonzero. Consequently, we assume \(\phi_k''(t) \neq 0\) for all signal components discussed henceforth in this paper. Additionally, for the purpose of comparing  different types of WLCT $T_x^{g, n}(t, \eta, \lambda)$, we assume that the signals in this paper satisfy both conditions \eqref{eq:separability1} and \eqref{eq:separability2}.
\end{remark}

In \cite{li2022chirplet}, a group  SSO reconstruction algorithm based on the CT was proposed to handle multicomponent signals with crossover IF components.
 Initially, we assume that each signal component 
\(x_k(t) = A_k(t) e^{2\pi i \phi_k(t)}\)
 satisfies condition \eqref{harmonic_cond}. Under this assumption, $x_k(t)$ admits a local linear chirp approximation:
\begin{equation*}
    x_k(t + \tau) \approx x_k(t) e^{2\pi i \phi_k'(t) \tau + \pi i \phi_k''(t) \tau^2}.
\end{equation*}

Building on these concepts and analogous to \eqref{appro_matrix_pre}, we obtain 

 \begin{equation}
	\label{multicomponet approx} T_{x}^{g, n}(t,\eta,\lambda)\approx  \sum_{k = 1}^{K} \frac{1}{\sqrt{-i}} x_k(t) \cL_g^{M_k M_{\lambda,n}}(\eta-\phi_k'(t)).
 \end{equation}
 where \(M_k M_{\lambda,n}\) was defined in \eqref{eq:matrix_NM} by replacing \(M_0\) with the matrix \(M_k=[\phi_k''(t), 1; -1, 0]\) and \(M\) with \(M_{\lambda,n}\) for \(n = 1, 2, 5, 6\).

The synchrosqueezed WLCT (SWLCT, \( S_x^n(t,\xi,\gamma) \)) and synchrosqueezed X-ray WLCT (SXWLCT, \( \mathcal{S}_x^n(t,\xi,\gamma) \)) (both to be defined in subsequent sections) yield highly concentrated time-frequency-chirprate representations. 
The extracted ridge curves \( \check{\xi}_k(t) \) and \( \check{\gamma}_k(t) \), derived from either \( S_x^n(t,\xi,\gamma) \) or \( \mathcal{S}_x^n(t,\xi,\gamma) \), provide  approximations of IFs and chirprates.  This can be represented as:
\[\check{\xi}_k(t) \approx \phi_k'(t) \quad \text{and} \quad \check{\gamma}_k(t) \approx \phi_k''(t), \quad k = 1, 2, \ldots, K.\]

Having established effective approximations for the IFs and chirprates, given the transform \(  T_x^{g, 2}(t, \eta, \lambda)  \)  or \( T_x^{g, 6}(t, \eta, \lambda) \), we can reformulate \eqref{multicomponet approx} into a linear system of equations,
 \begin{align*}
	\begin{bmatrix}
		T_x^{g, n} (t,\check{\xi}_1(t),-\check{\gamma}_1(t)) \\
		T_x^{g, n} (t,\check{\xi}_2(t),-\check{\gamma}_2(t)) \\
	\vdots \\
	   T_x^{g, n} (t,\check{\xi}_K(t),-\check{\gamma}_K(t))
	\end{bmatrix}
	=
	\begin{bmatrix}
	c_{1,1} & c_{1,2} & \cdots & c_{1,K} \\
	c_{2,1} & c_{2,2} & \cdots & c_{2,K} \\
	\vdots & \vdots & \ddots & \vdots \\
	c_{K,1} & c_{K,2} & \cdots & c_{K,K}
	\end{bmatrix}
	\begin{bmatrix}
	x_1(t) \\
	x_2(t) \\
	\vdots \\
	x_K(t)
	\end{bmatrix}.
 \end{align*}
 By solving the linear system, the solution is given by
 \begin{equation}
\label{recover_26}
	 \begin{bmatrix}
		 x_1(t) \\
		 x_2(t) \\
		 \vdots \\
		 x_K(t)
	 \end{bmatrix}
	 =
	 \begin{bmatrix}
		 c_{1,1} & c_{1,2} & \cdots & c_{1,K} \\
		 c_{2,1} & c_{2,2} & \cdots & c_{2,K} \\
		 \vdots & \vdots & \ddots & \vdots \\
		 c_{K,1} & c_{K,2} & \cdots & c_{K,K}
	 \end{bmatrix}
	 ^{-1} 
	 \begin{bmatrix}
		 T_x^{g, n} (t,\check{\xi}_1(t),-\check{\gamma}_1(t)) \\
		 T_x^{ g, n}(t,\check{\xi}_2(t),-\check{\gamma}_2(t)) \\
		 \vdots \\
		 T_x^{ g, n}(t,\check{\xi}_K(t),-\check{\gamma}_K(t))
	 \end{bmatrix},
 \end{equation}
where for \(n=2\),
 \begin{align*}
	 c_{i,j}  = \frac{1}{\sqrt{\alpha - i(\check{\gamma}_j(t) - \check{\gamma}_i(t))}} e^{\frac{-\pi(\check{\xi}_i(t) - \check{\xi}_j(t))^2}{\alpha - i(\check{\gamma}_j(t) - \check{\gamma}_i(t))}}.
 \end{align*}
 and for \(n=6\),
 \begin{align*}
	 {c_{i,j}} &=  \frac{\sqrt{-i+\check{\gamma}_i(t)}}{\sqrt{\alpha + i\check{\gamma}_i(t)}}G_I^{-\frac{1}{2}} e^{\frac{\pi(\check{\xi}_i(t) -\check{\xi}_j(t))^2}{G_I}},\\
	 \text{with} \quad G_I&=\frac{\alpha(1+\check{\gamma}^2_i(t))-i(1-\alpha^2)\check{\gamma}_i(t)}{\alpha^2+\check{\gamma}^2_i(t)}-i\check{\gamma}_j(t).\\
 \end{align*}  
As for \(\cL^{M_{\gl, 1}}(g)(u)\) and \(\cL^{M_{\gl, 5}}(g)(u)\), drawing from similar ideas, we can derive an alternative reconstruction method.
   \begin{equation}
\label{recover_15}
	 \begin{bmatrix}
		 x_1(t) \\
		 x_2(t) \\
		 \vdots \\
		 x_K(t)
	 \end{bmatrix}
	 =
	 \begin{bmatrix}
		 a_{1,1} & a_{1,2} & \cdots & a_{1,K} \\
		 a_{2,1} & a_{2,2} & \cdots & a_{2,K} \\
		 \vdots & \vdots & \ddots & \vdots \\
		 a_{K,1} & a_{K,2} & \cdots & a_{K,K}
	 \end{bmatrix}^{-1}
	 \begin{bmatrix}
		 T_x^{g, n}(t,\check{\xi}_1(t),1/\check{\gamma}_1(t)) \\
		 T_x^{g, n}(t,\check{\xi}_2(t),1/\check{\gamma}_2(t)) \\
		 \vdots \\
		 T_x^{g, n}(t,\check{\xi}_K(t),1/\check{\gamma}_K(t))
	 \end{bmatrix},
 \end{equation}
where for \(n=1\),
  \begin{align*}
	  a_{i,j} = \frac{1}{\sqrt{\alpha\check{\gamma}_j(t) +i(1-\check{\gamma}_j(t) /\check{\gamma}_i(t))}} e^{\frac{-\pi(\alpha-i/\check{\gamma}_i(t))( \check{\xi}_i(t) -  \check{\xi}_j(t))^2}{1-\check{\gamma}_j(t)/\check{\gamma}_i(t)-i\alpha \check{\gamma}_j(t) }}.
  \end{align*}
and for \(n=5\),
  \begin{align*}
	 {a_{i,j}}&= \frac{\sqrt{-i+\check{\gamma}_i(t)}}{\sqrt{\alpha + i \check{\gamma}_i(t)}}{G_{II}}^{-\frac{1}{2}} e^{\frac{\pi(\check{\xi}_i(t) - \check{\xi}_j(t))^2}{G_{II}}},\\
	  \text{with}& \quad G_{II}=\frac{\alpha(1+1/\check{\gamma}^2_i(t))+i(1-\alpha^2)/\check{\gamma}_i(t)}{\alpha^2+1/\check{\gamma}^2_i(t)}-i \check{\gamma}_j(t).\\
  \end{align*}
Note that when $C:=\big[c_{\ell, k}\big]_{1\le \ell, k\le K}$ is singular, $C^{-1}$ in \eqref{recover_26} means the pseudo-inverse of $C$. This  also applies to $A :=\big[a_{\ell, k}\big]_{1\le \ell, k\le K}$ in  \eqref{recover_15}. The modes of a signal can be retrieved by \eqref{recover_26} or \eqref{recover_15} once the
IF and chirprate estimations $\check{\xi}_k(t)$ and $\check{\gamma}_k(t)$ are obtained.  
In the next two sections, we propose the SWLCT and SXWLCT to extract ridge curves \( \check{\xi}_k(t) \) and \( \check{\gamma}_k(t) \) as  IF and chirprate estimations. 

\section{Synchrosqueezed windowed linear canonical transform (SWLCT)}

Using the reassignment technique, the synchrosqueezed algorithm sharpens time-frequency  representations.
In \cite{chen2023disentangling}, Chen and Wu proposed the synchrosqueezed chirplet transform (SCT), which achieves high-resolution representation in the three-dimensional time-frequency-chirprate (TFC) space. As previously discussed, the CT is essentially a subset of the WLCT. This  motivates the development of a synchrosqueezed algorithm for the entire WLCT framework.

  First, we consider the signal \( x(t) \) as a standard linear chirp signal, i.e., 
  \(x(t) = A e^{2 \pi i \phi(t)},\)
  where \( A \) is a positive constant and \( \phi(t) \) is a quadratic function.
  Then
  \begin{align}
T_x^{g, n}(t, \eta,\lambda) :=& \int_{\mathbb{R}} x(t+\tau) e^{-i 2 \pi \eta \tau} \, \mathcal{L}^{M_{\lambda,n}}(g)(\tau) \, d\tau \nonumber \\
=& \int_{\mathbb{R}} x(\tau) e^{-i 2 \pi \eta (\tau-t)} \, \mathcal{L}^{M_{\lambda,n}}(g)(\tau-t) \, d\tau  \label{WLCT_window_shift}\\
 =& \int_{\mathbb{R}} A e^{2 \pi i \left( \phi(t) + \phi'(t) \tau + \frac{1}{2} \phi''(t) \tau^2 \right)} e^{-i 2 \pi \eta \tau} \, \mathcal{L}^{M_{\lambda,n}}(g)(\tau) \, d\tau.\label{WLCT_expansion}
  \end{align}
  Differentiating  \eqref{WLCT_expansion} with respect to \( t \) yields
  \begin{equation}
  \partial_t T_x^{g, n} = 2\pi i \phi'(t) T_x^{g, n} + 2\pi i \phi''(t) T_x^{\tau g, n} \label{eq:partial_t}.
  \end{equation}
 Here, \(T_x^{\tau g, n}(t,\eta,\lambda)\) and \(T_x^{\tau^2 g, n}(t,\eta,\lambda)\) denote the  WLCT of \(x(t)\) as defined in \eqref{WLCT definition}, with  \(\cL^{M_{\gl, n}}(g)(\tau) \)  replaced by \(\tau \cL^{M_{\gl, n}}(g)(\tau) \) and \(\tau^2 \cL^{M_{\gl, n}} (g)(\tau) \).

Under the condition $T_x^{g,n}(t,\eta,\lambda) \neq 0$,   we   can straightforwardly derive 
  \begin{align*}
	  \partial_t\left(\frac{\partial_t T_x^{g, n}}{T_x^{g, n}}\right) &= 2\pi i \phi''(t) \left(1 + \partial_t\left(\frac{T_x^{\tau g, n}}{T_x^{g, n}}\right)\right).
  \end{align*}  
Then, we  define
  \begin{align*}
    \mu_x^{g, n}(t,\eta,\lambda) = \frac{\partial_t\left(\frac{\partial_t T_x^{g, n}}{T_x^{g, n}}\right)}{2\pi i \left(1 + \partial_t\left(\frac{T_x^{\tau g, n}}{T_x^{g, n}}\right)\right)},\quad \omega_x^{g, n}(t,\eta,\lambda) = \frac{\partial_t T_x^{g, n} - 2\pi i   \mu_x^{g, n} T_x^{\tau g, n}}{2\pi i T_x^{g, n}}.
  \end{align*}
 Differentiating \eqref{WLCT_window_shift} directly yields
\begin{align*}
	\partial_t T_x^{g, n}=2\pi i \eta T_x^{g, n}-2 \pi P_n T_x^{\tau g, n }, \label{eq:partial_g}
\end{align*}
where \(P_n\) denotes the value of \(P\) defined in \eqref{eq:parameter_P} with corresponding matrices \(M_{\lambda, n}\):
\begin{equation}
\label{def_Pn}
 P_1 = -\frac{1}{\alpha - i \lambda}, \quad
P_2 = i \lambda - \alpha, \quad P_5=P_6 = \frac{i \lambda \alpha - 1}{\alpha - i \lambda}. 
\end{equation}
Similarly, we have 
\begin{equation*}
	\partial_t T_x^{\tau g, n} =2\pi i \eta T_x^{\tau g, n}-2 \pi P_n T_x^{\tau^2 g, n} -T_x^{g,n}.\label{eq:partial_tg}
\end{equation*}
These expressions yield
\begin{align*}
	\mu_x^{g, n}(t,\eta,\lambda)=  \frac{2\pi P_n(T_x^{g, n}T_x^{\tau^2g, M}-T_f^{\tau g, n}T_f^{\tau g, n})+(T_f^{g, n})^2 }{2\pi i  \left(  (T_x^{\tau g, n})^2- T_f^{g, n} T_f^{\tau^2g, n} \right) }
	= iP_n+\frac{(T_x^{g, n})^2}{{2\pi i  \left(  (T_x^{\tau g, n})^2- T_x^{g, n} T_x^{\tau^2g, n} \right) }},
  \end{align*}
   \begin{align*} 
 \omega_x^{g, n}(t,\eta,\lambda)=\eta+(iP_n-\mu_x^{g, n}) \frac{ T_x^{\tau g, n}}{T_x^{g, n}} 
   = \eta-\frac{T_x^{g, n}T_x^{\tau g, n}}{{2\pi i  \left(  (T_x^{\tau g, n})^2- T_x^{g, n} T_x^{\tau^2g, n} \right) }}.
\end{align*}
The frequency and chirprate reassignment operators are defined as 
\begin{align}
 {\Omega}_x^{g, n}(t,\eta,\lambda) :=\mathrm{Re} \left(\omega_x^{g, n}(t,\eta,\lambda) \right), \quad     {\Lambda}_x^{g, n}(t,\eta,\lambda) :=\mathrm{Re} \left( \mu_x^{g, n}(t,\eta,\lambda)\right). \label{eq:omega}
\end{align}
where \(\mathrm{Re} \) denotes the real part.

\begin{mdef}
Let $T_x^{g, n}(t,\eta,\lambda), n=1, 2, 5, 6$ be the WLCT of a signal $x(t)$  defined by \eqref{WLCT definition}. 
With the  \({\Omega}_x^{g, n}(t,\eta,\lambda)\)   and \({\Lambda}_x^{g, n}(t,\eta,\lambda)\)  defined in \eqref{eq:omega}, then  we define the SWLCTs \(S_x^n(t,\xi,\gamma)\) of \(x(t)\) as follows:	
	
For \(n = 1, 5\), 
	\begin{align}
\label{SWLCT15}		
& S^n_x(t,\xi,\gamma):=\iint_{O_{t, n}}T_x^{ g, n}(t,\eta,\lambda) \delta(\gamma-1/{\Lambda}_x^{g, n}(t,\eta,\lambda)) 
		\delta(\xi-{\Omega}_x^{g, n}(t,\eta,\lambda)) \, d\eta \, d\lambda;   
	\end{align}
and for \(n = 2, 6\), 
	\begin{align}\label{SWLCT26}	
		&S^n_x(t,\xi,\gamma) :=\iint_{O_{t, n}}T_x^{ g, n}(t,\eta,\lambda)  \delta(\gamma+{\Lambda}_x^{g, n}(t,\eta,\lambda)) 
		\delta(\xi- {\Omega}_x^{g, n}(t,\eta,\lambda)) \, d\eta \, d\lambda,
	\end{align}
where 
\begin{equation}
\label{def_Otn}
O_{t, n}:=\left\{(\eta,\lambda):T_x^{g, n}(t, \eta, \gl)\neq 0, (T_x^{\tau g, n}(t, \eta, \gl))^2- T_x^{g, n} (t, \eta, \gl)T_x^{\tau^2g, n}(t, \eta, \gl)\neq 0 \right\}.
\end{equation} 
	\end{mdef}

To achieve effective SWLCTs, the parameter  $\alpha$ must be carefully selected.
An inappropriate selection of \( \alpha \) may lead to an inaccurate representation in the time-frequency-chirprate domain. In practice, employing the R\'enyi entropy to determine the optimal value of \(\alpha\) is an effective approach that has been widely utilized in previous studies, such as \cite{pham2017high,stankovic2001measure}. 

 The R\'enyi entropy based on the WLCT is defined as
\begin{align}
	\label{Ealpha1}E_{\alpha} := \frac{1}{1-\ell} \log_2 \frac{\iint_{\mathbb{R}^2} \int_0^{\infty} \left| T_x^{g_{\alpha}, n}(t,\eta,\lambda) \right|^{2\ell} \frac{d\lambda}{\lambda} dt \, d\eta}{\left( \iint_{\mathbb{R}^2} \int_0^{\infty} \left| T_x^{g_{\alpha}, n}(t,\eta,\gamma) \right|^2 \frac{d\lambda}{\lambda} dt \, d\eta \right)^{\ell}}, \qquad \hbox{if $n=1, 5$}, 
\end{align}
or
\begin{align}
	\label{Ealpha2}E_{\alpha} := \frac{1}{1-\ell} \log_2 \frac{\iiint_{\mathbb{R}^3} \left| T_x^{g_{\alpha}, n}(t,\eta,\lambda) \right|^{2\ell} dt \, d\eta \, d\lambda}{\left( \iiint_{\mathbb{R}^3} \left| T_x^{g_{\alpha}, n}(t,\eta,\lambda) \right|^2 dt \, d\xi \, d\lambda \right)^{\ell}}, \qquad \hbox{if $n=2, 6$},
\end{align}
where $\ell$ is a number greater than 2.
In this paper, we let \( \ell = 2.5 \), a common value used in other studies \cite{sheu2017entropy}.
A decrease in the R\'enyi entropy indicates  increased  concentration of the representation.
 To identify the most concentrated representation, we seek the value of \(\alpha\) that yields the lowest R\'enyi entropy, expressed as follows:
\begin{equation}
\label{entropy_minimization}
\alpha_{\text{opt}}:= \underset{\alpha > 0}{\operatorname{argmin}} \, E_\alpha.
\end{equation}

Once an appropriate value of $\alpha$ is determined, the synchrosqueezing algorithm is employed to obtain the refined time-frequency-chirprate representation $S_x^n(t, \xi, \gamma)$.
The ridge curves $\check{\xi}_k(t)$ and $\check{\gamma}_k(t)$ are extracted from $S_x^n(t,\xi,\gamma)$ using a three-dimensional ridge extractor. This extractor identifies salient features in three-dimensional data through local maxima detection. For implementation details, see \cite{zhang2022local, chen2024multiple}.
Finally, the original signal can be reconstructed from the extracted ridge curves $\check{\xi}_k(t)$ and $\check{\gamma}_k(t)$ through the reconstruction algorithms given in \eqref{recover_26} and \eqref{recover_15}.

\begin{figure}[H]
    \centering
    \begin{tabular}{ccc}
        \resizebox{1.8in}{1.1in}{\includegraphics{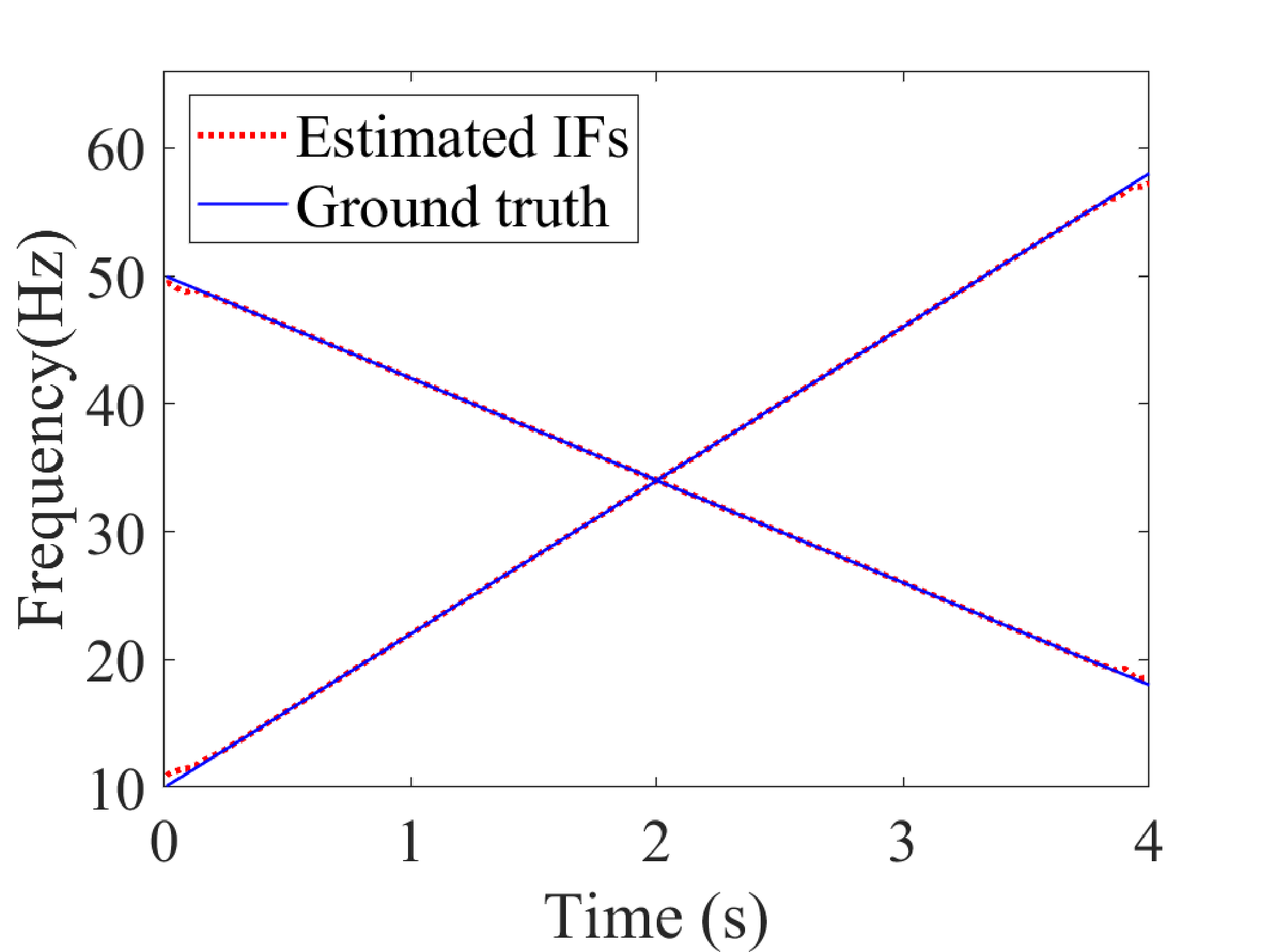}} & 
        \resizebox{1.8in}{1.1in}{\includegraphics{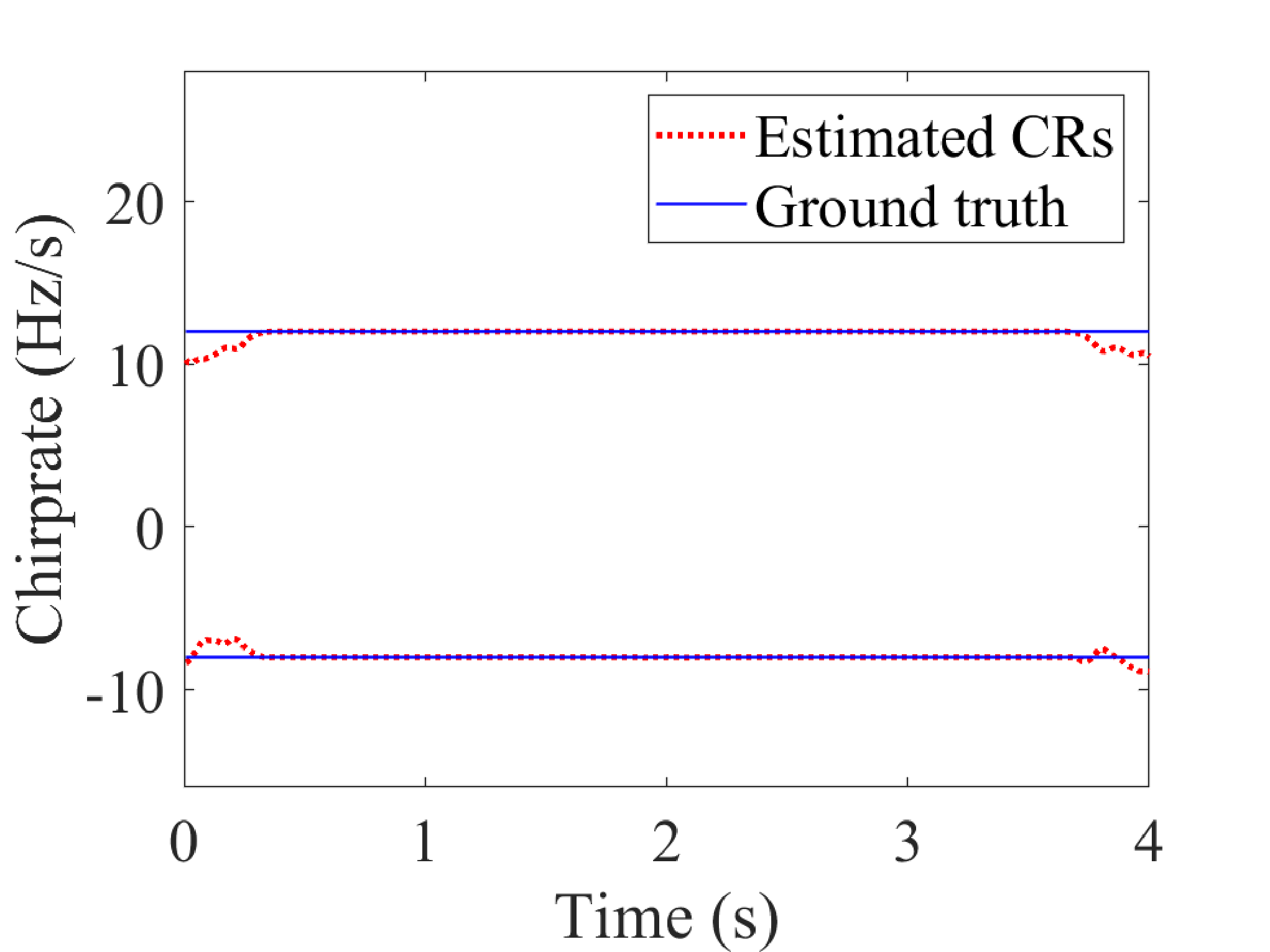}} &
        \resizebox{1.8in}{1.1in}{\includegraphics{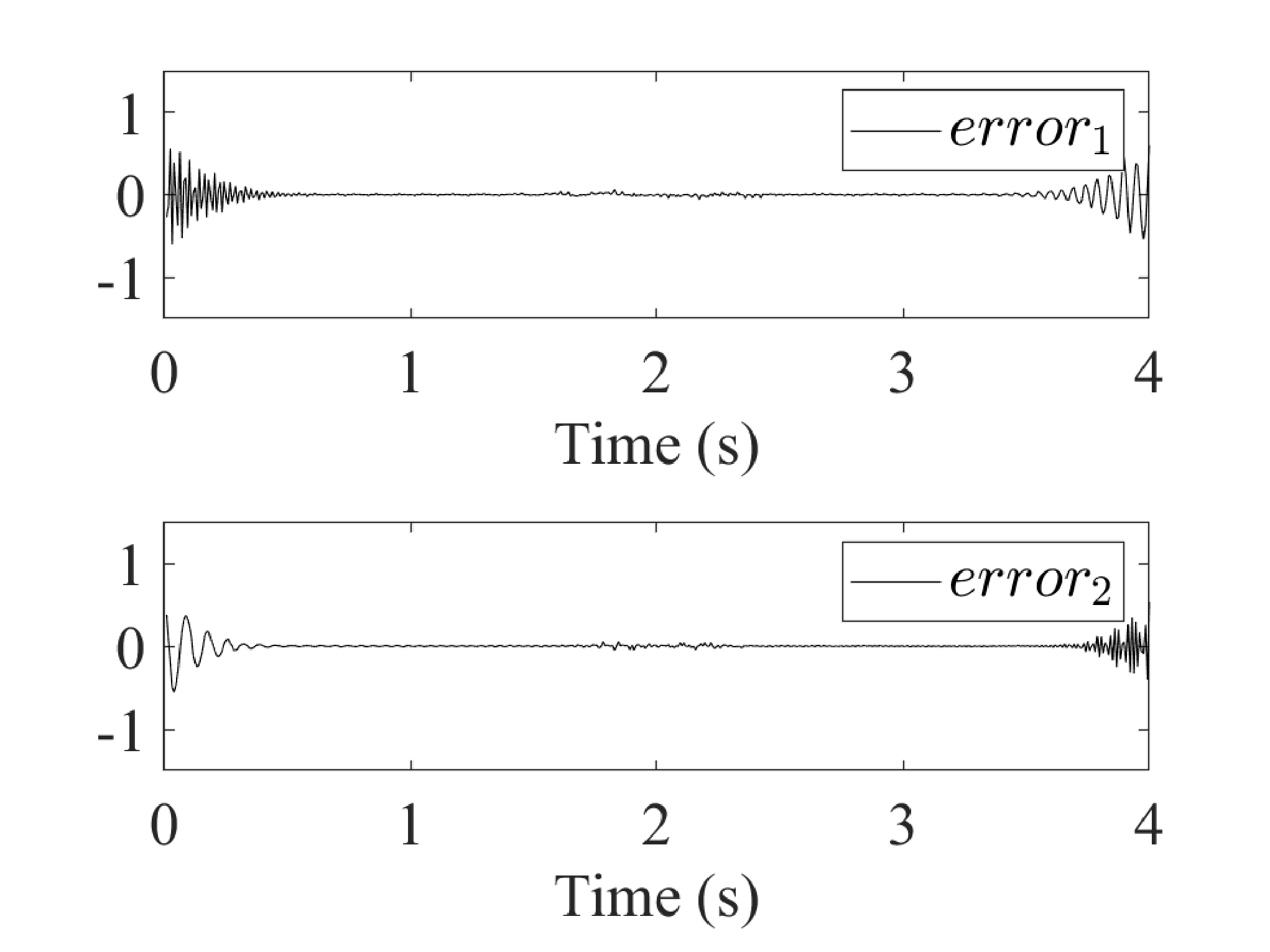}} \\
        \resizebox{1.8in}{1.1in}{\includegraphics{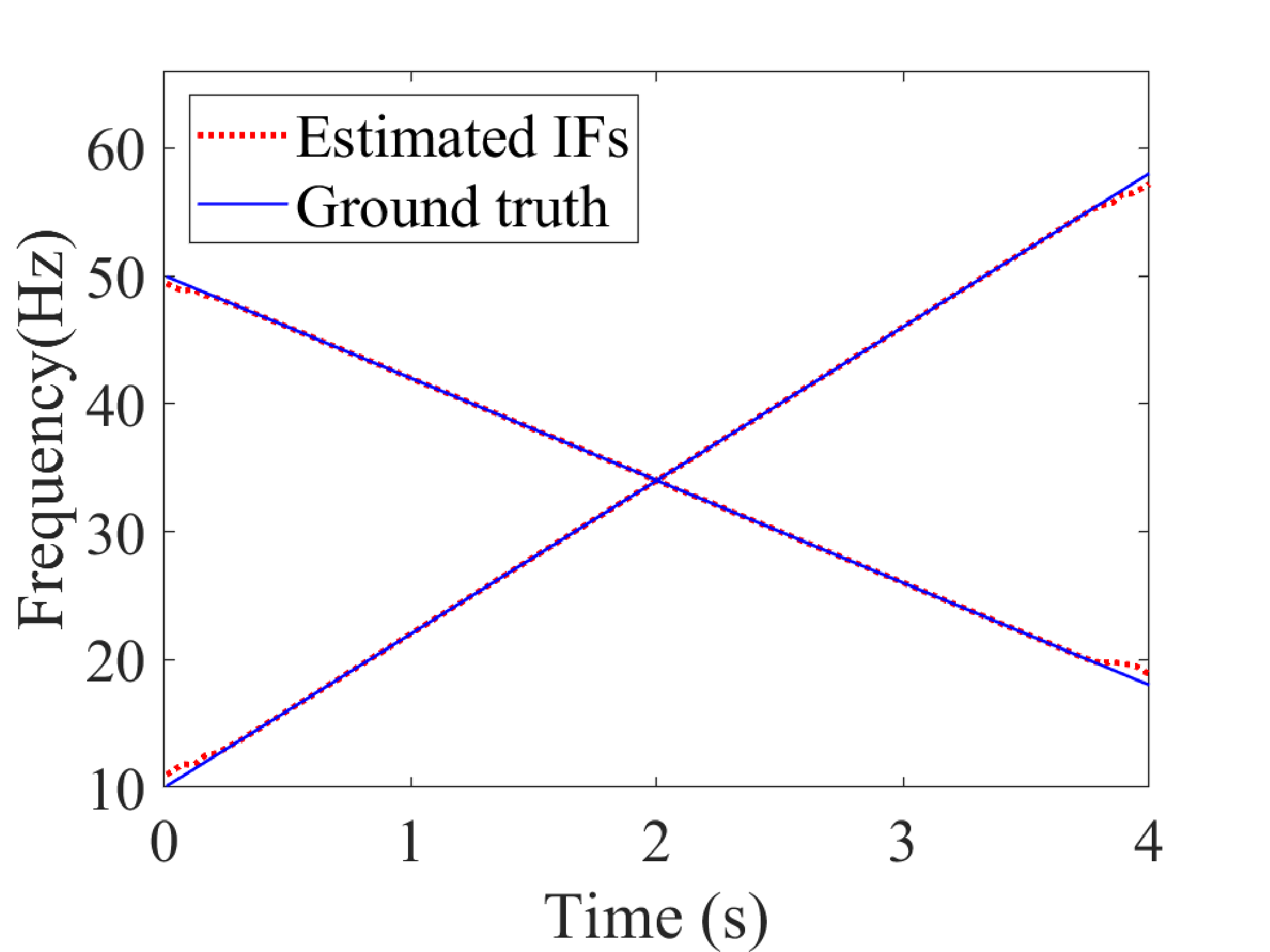}} & 
        \resizebox{1.8in}{1.1in}{\includegraphics{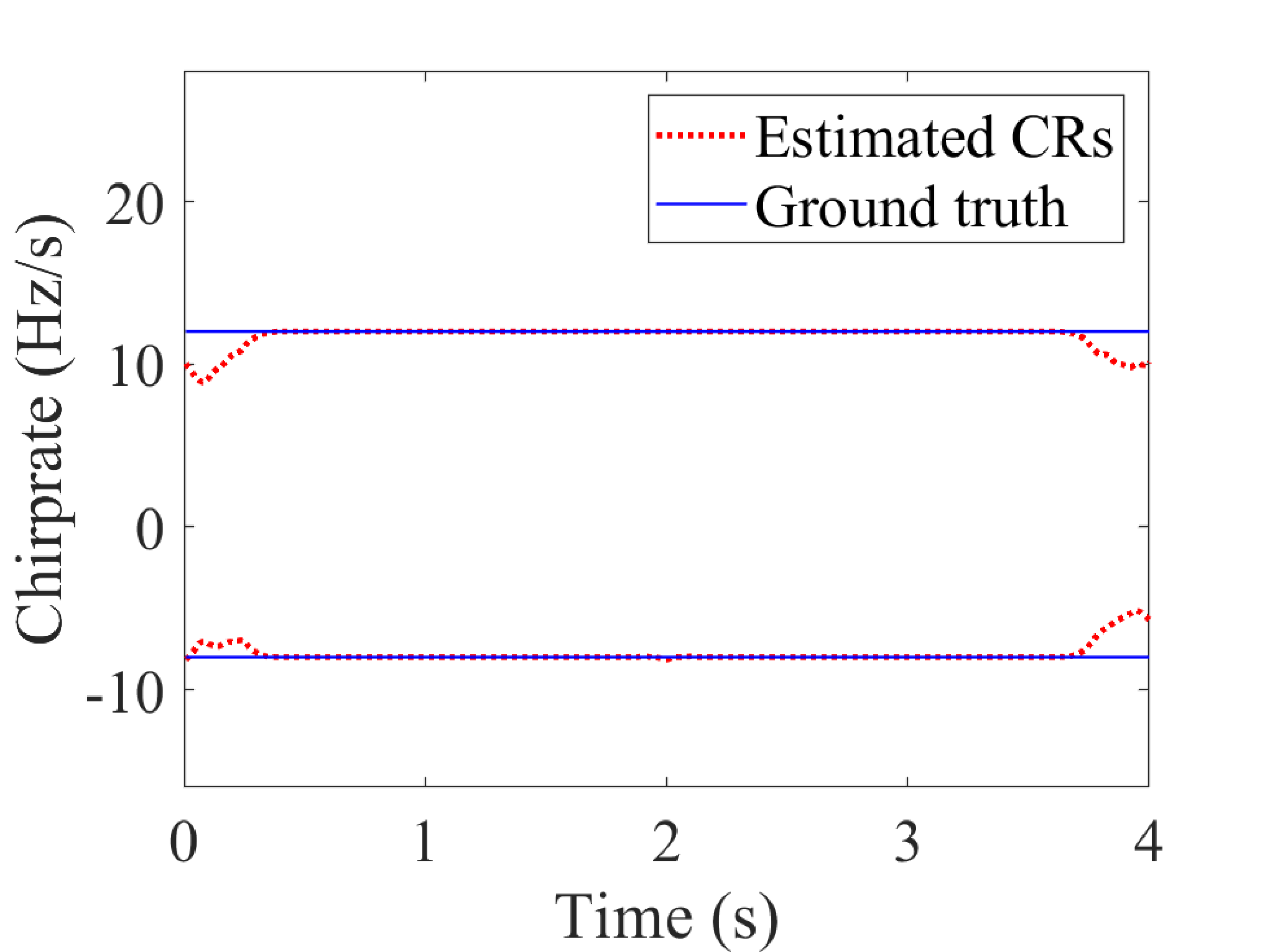}} &
        \resizebox{1.8in}{1.1in}{\includegraphics{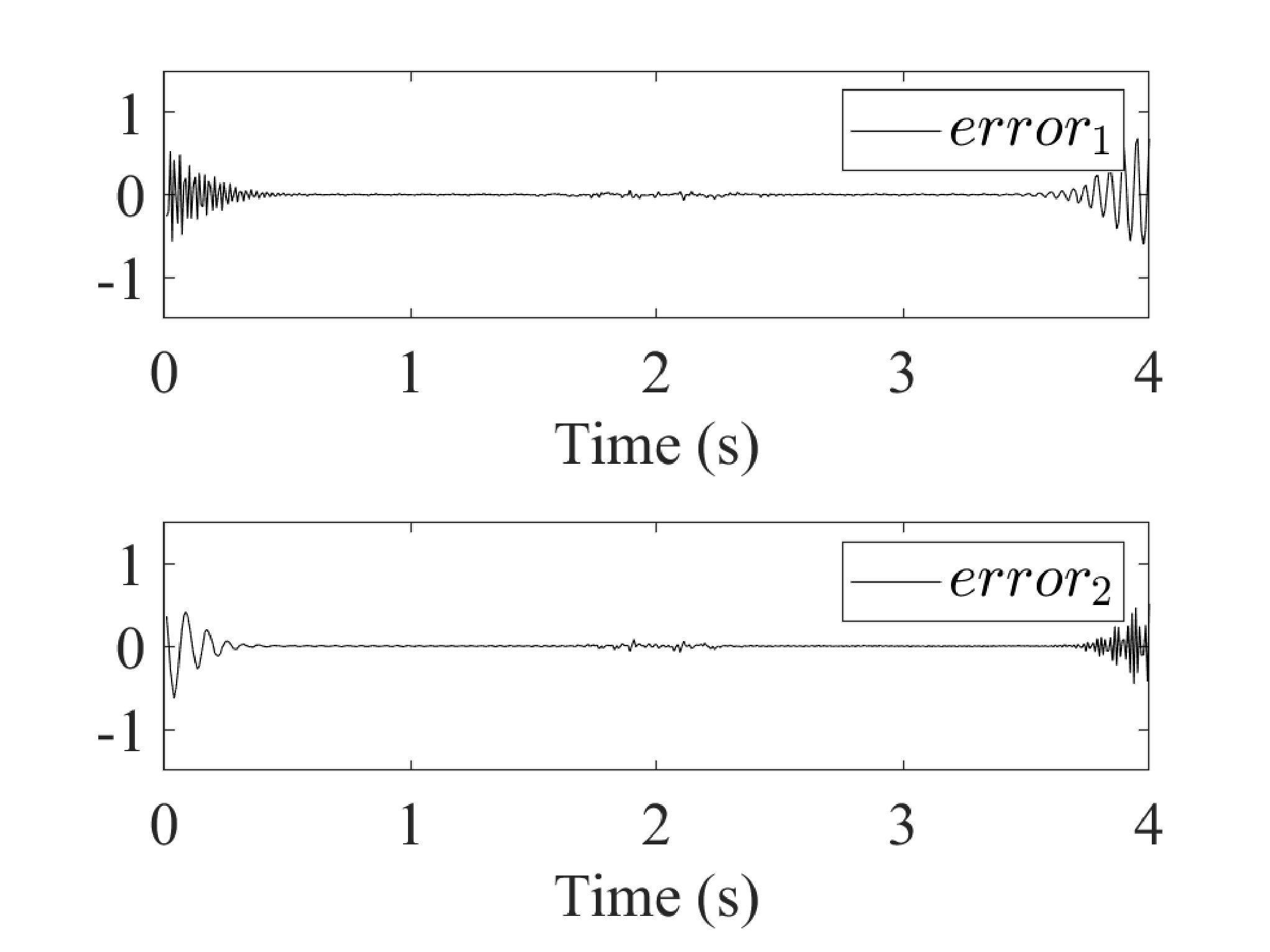}} \\
        \resizebox{1.8in}{1.1in}{\includegraphics{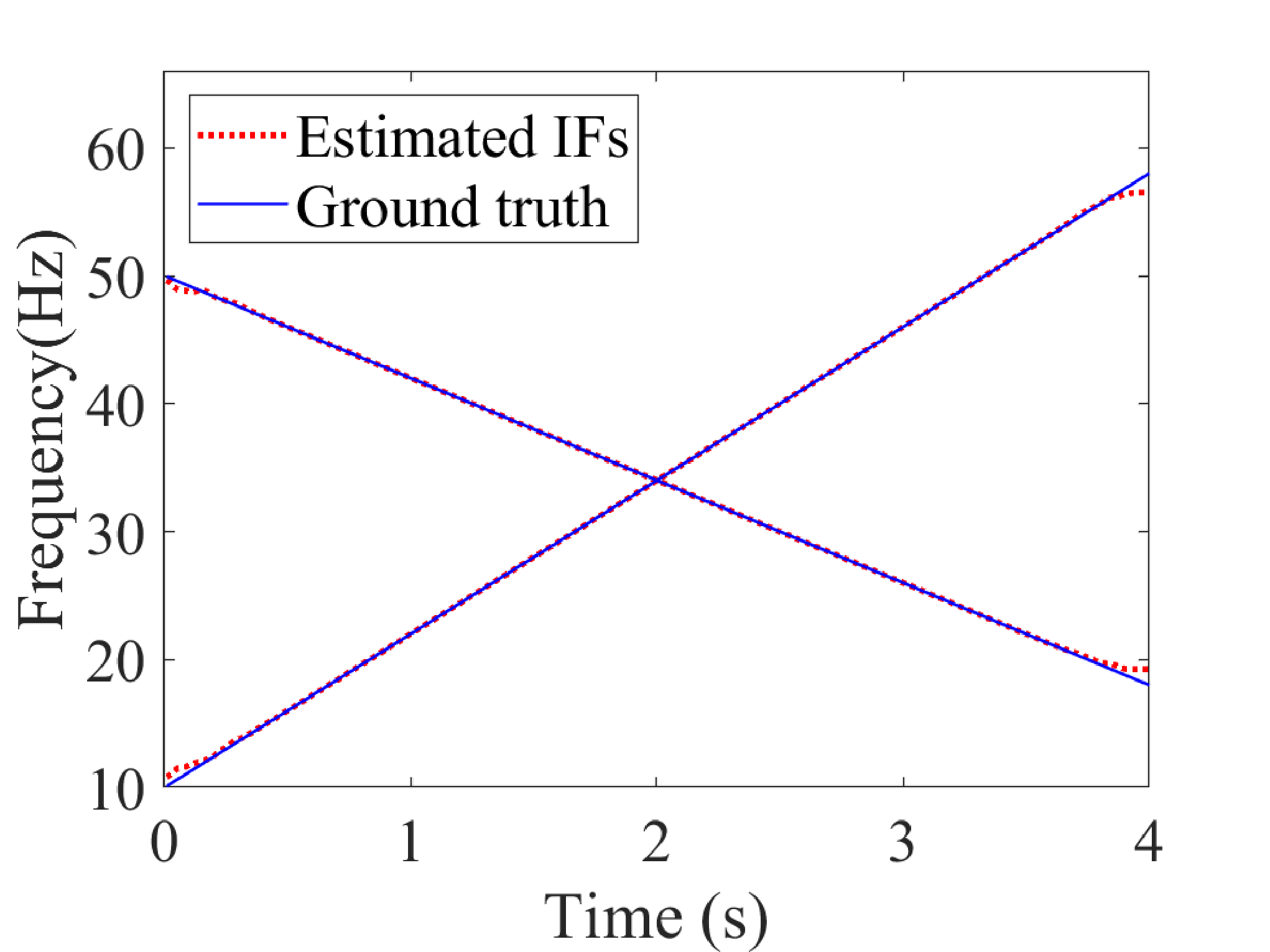}} & 
        \resizebox{1.8in}{1.1in}{\includegraphics{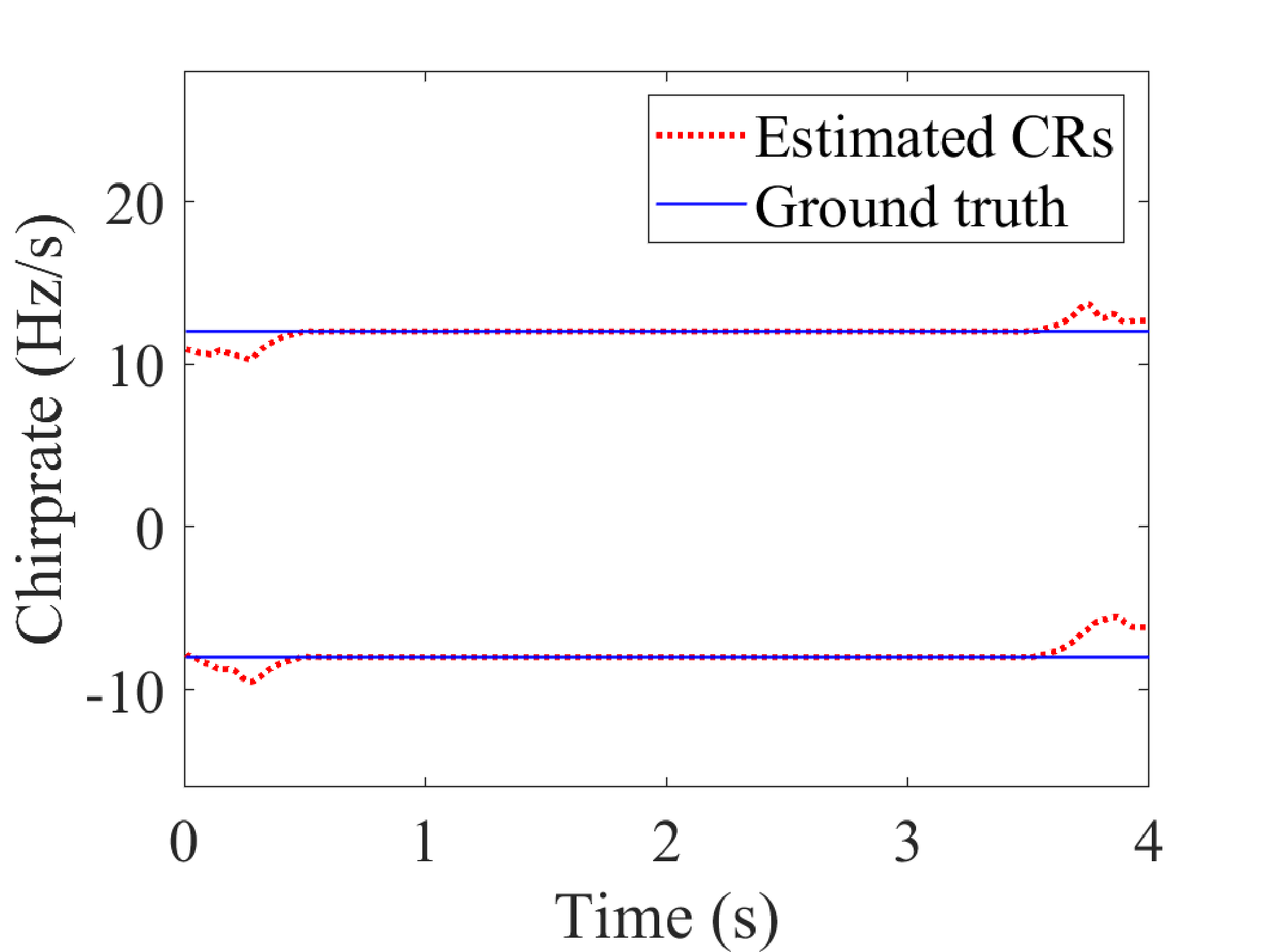}} &
        \resizebox{1.8in}{1.1in}{\includegraphics{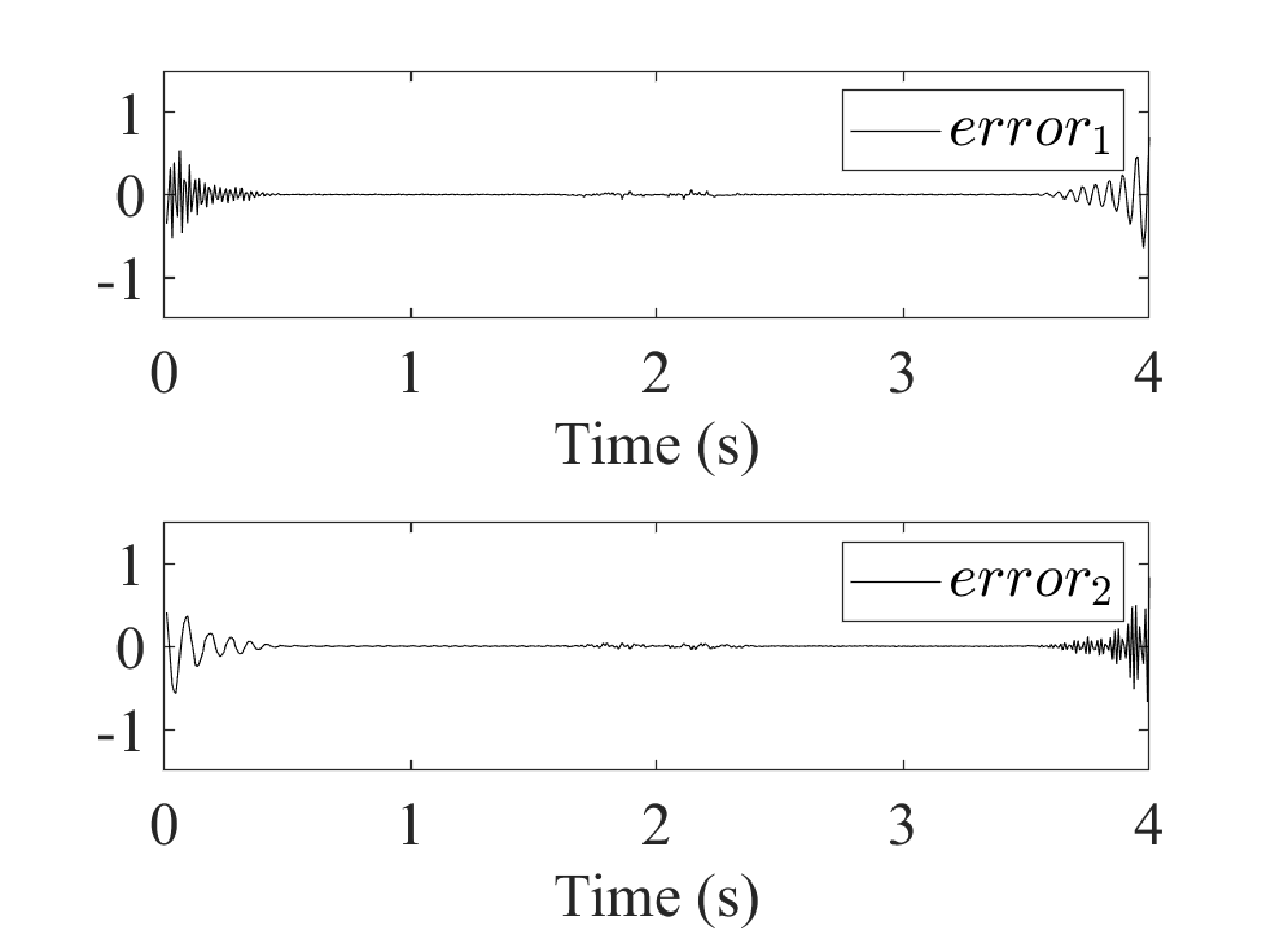}} \\
        \resizebox{1.8in}{1.1in}{\includegraphics{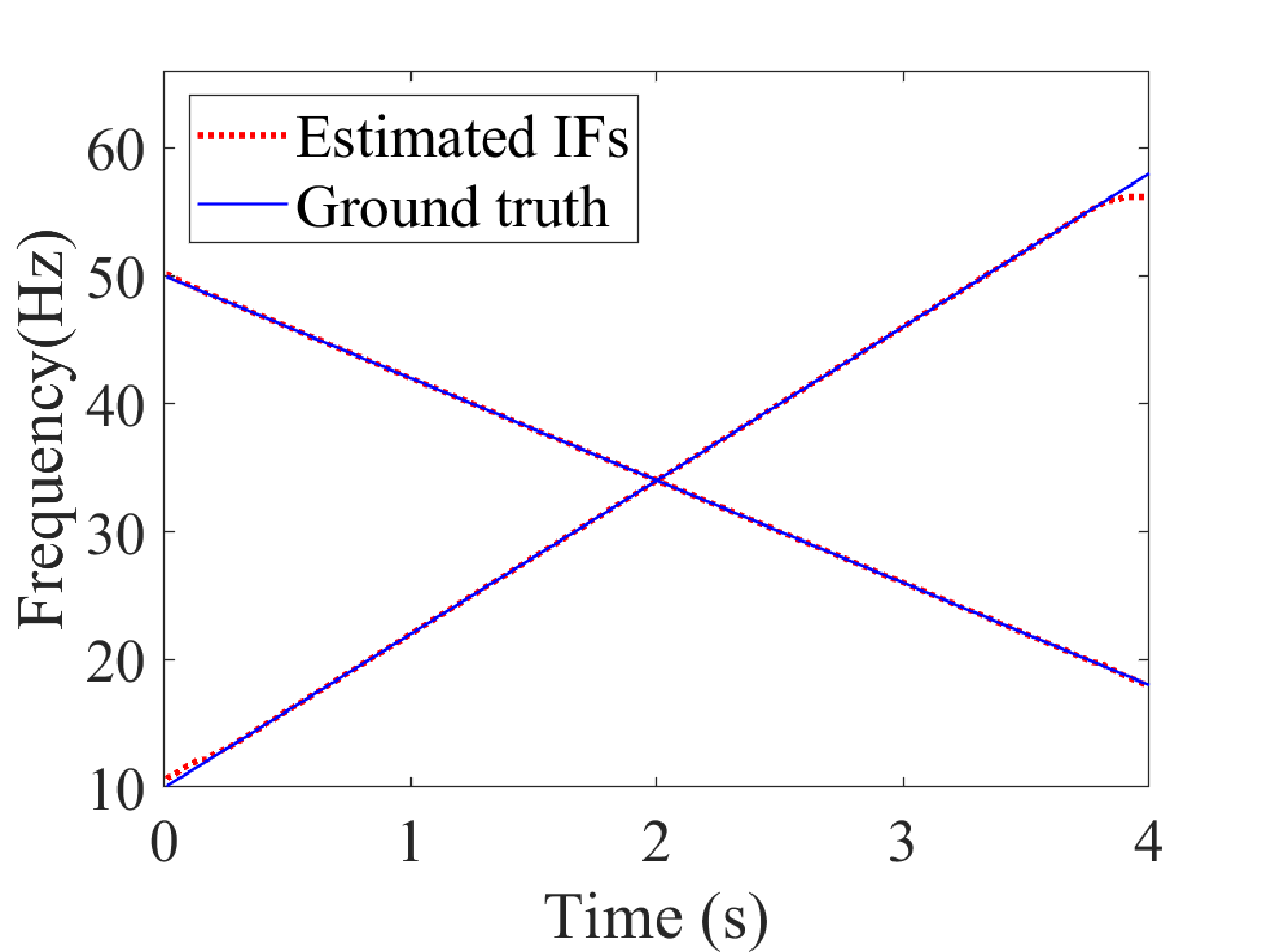}} & 
        \resizebox{1.8in}{1.1in}{\includegraphics{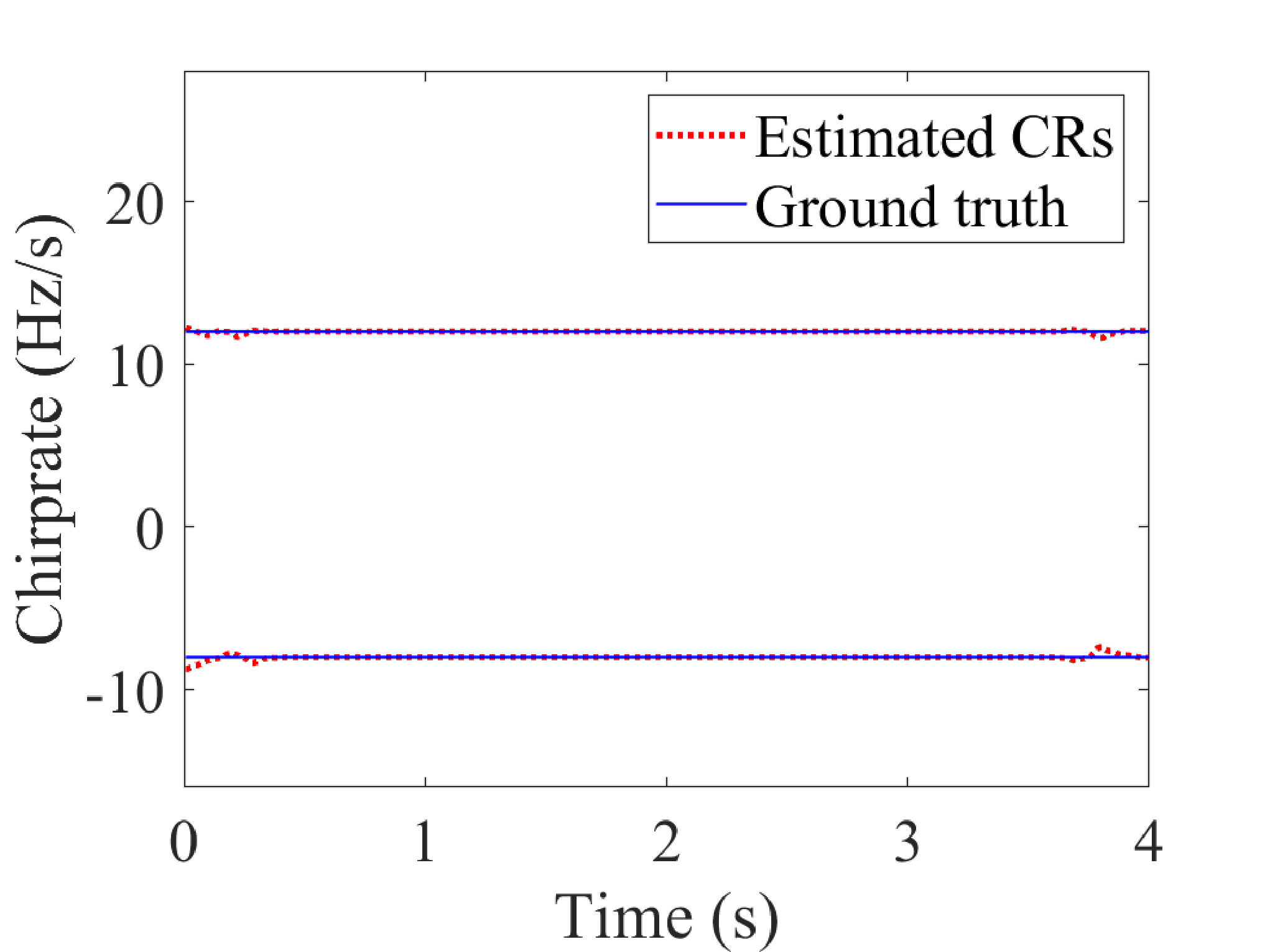}} &
        \resizebox{1.8in}{1.1in}{\includegraphics{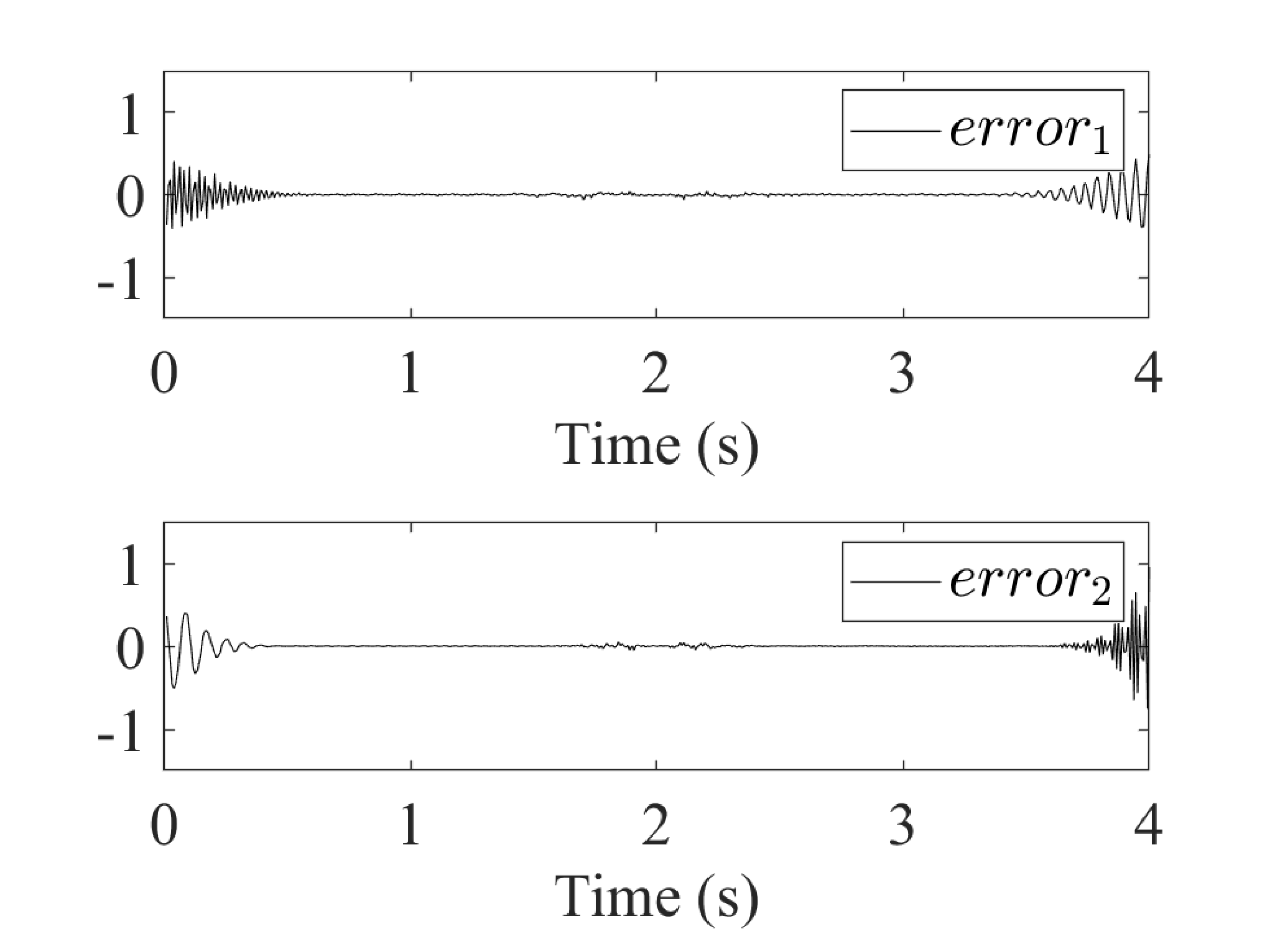}} \\
    \end{tabular}
	\caption{\small IF and chirprate estimations, and real part errors of mode retrieval:  First row by \(S^1_x(t, \xi, \gamma)\), Second row by \(S^5_x(t, \xi, \gamma)\),  Third row by \(S^2_x(t, \xi, \gamma)\) and Fourth row by \(S^6_x(t, \xi, \gamma)\). 
}
\label{figure:recover of X}
\end{figure}

 Next we take the signal \( x(t) \) given by \eqref{example1} as an example to show the effectiveness of SWLCTs. 
The values of \(\alpha\)  obtained from \eqref{entropy_minimization} for \( x(t) \) in \eqref{example1} are  \(\alpha_1=0.050\), \(\alpha_5=0.061\), \(\alpha_2=6.2\)  and \(\alpha_6=23\), corresponding to \(n=1, 5, 2, 6\) respectively.
In Fig.~\ref{figure:recover of X}, we show IF estimation (the 1st column), chirprate (CR) estimation (the 2nd column), and (real part) errors of mode retrieval (3rd column) by SWLCTs. 

To quantify the error of the retrieved mode, we employ the root mean square error (RMSE). Specifically, for a discrete signal \( f= (f_1, f_2, \dots, f_N) \) and its estimation \( \widetilde{f} \) (denoting the reconstructed signal), the RMSE is defined as
\(\text{error} = \frac{1}{\sqrt{N}} \| f - \widetilde{f} \|_2.\)
To mitigate boundary effects, we calculate the error using the central portion of \( f(t) \):
\(
(f_{\left[\frac{N}{8}\right]}, f_{\left[\frac{N}{8}\right]+1}, \dots, f_{\left[\frac{7N}{8}\right]}).
\)
The RMSEs are presented in Table~\ref{tab:example1}, where error1 and error2 denote the RMSEs of two components of \(x(t)\).

\begin{table}[H]
    \centering
    \caption{RMSEs of mode retrieval for \(x(t)\) in \eqref{example1}}
    \label{tab:example1}
    \begin{tabular}{l *{4}{S[table-format=1.4]}}
        \toprule
        {$n$} & \multicolumn{1}{c}{{$n=1$}} & \multicolumn{1}{c}{{$n=5$}} & \multicolumn{1}{c}{{$n=2$}} & \multicolumn{1}{c}{{$n=6$}} \\
        \midrule
        error1 & 0.0117 & 0.0113 & 0.0088 & 0.0118 \\
        error2 & 0.0102 & 0.0112 & 0.0091 & 0.0089 \\
        \bottomrule
    \end{tabular}
\end{table}

Overall, SWLCTs exhibit excellent performance in IF estimation, chirprate estimation, and mode retrieval.
Notably, when \( n = 6 \), the \( S_x^6(t, \xi, \gamma) \) yields the best chirprate estimation near the boundaries compared with the other three methods.

\begin{figure}[H]
    \centering
    \begin{tabular}{cc}
     \resizebox{2.0in}{1.5in}{\includegraphics{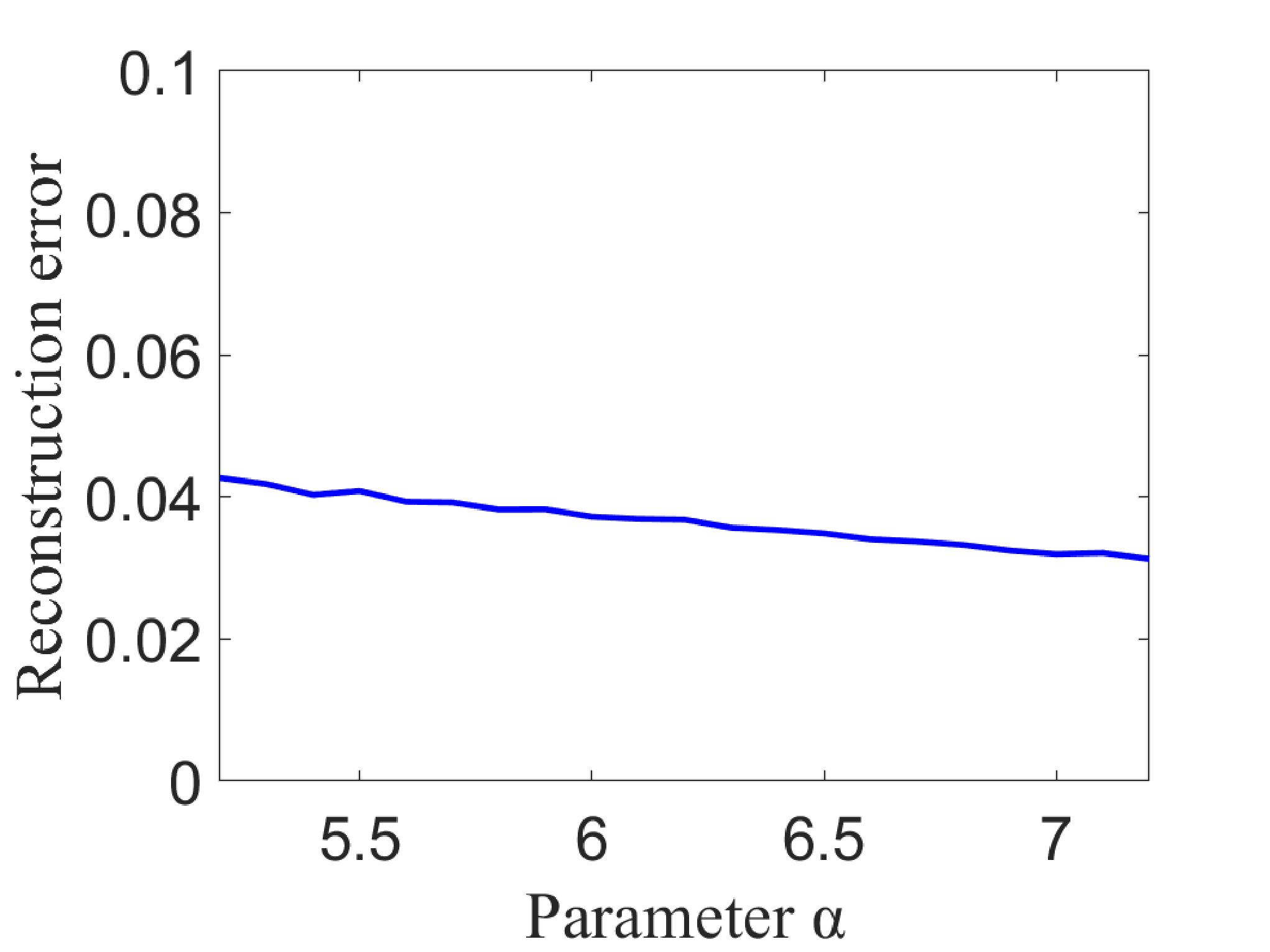}} &
        \resizebox{2.0in}{1.5in}{\includegraphics{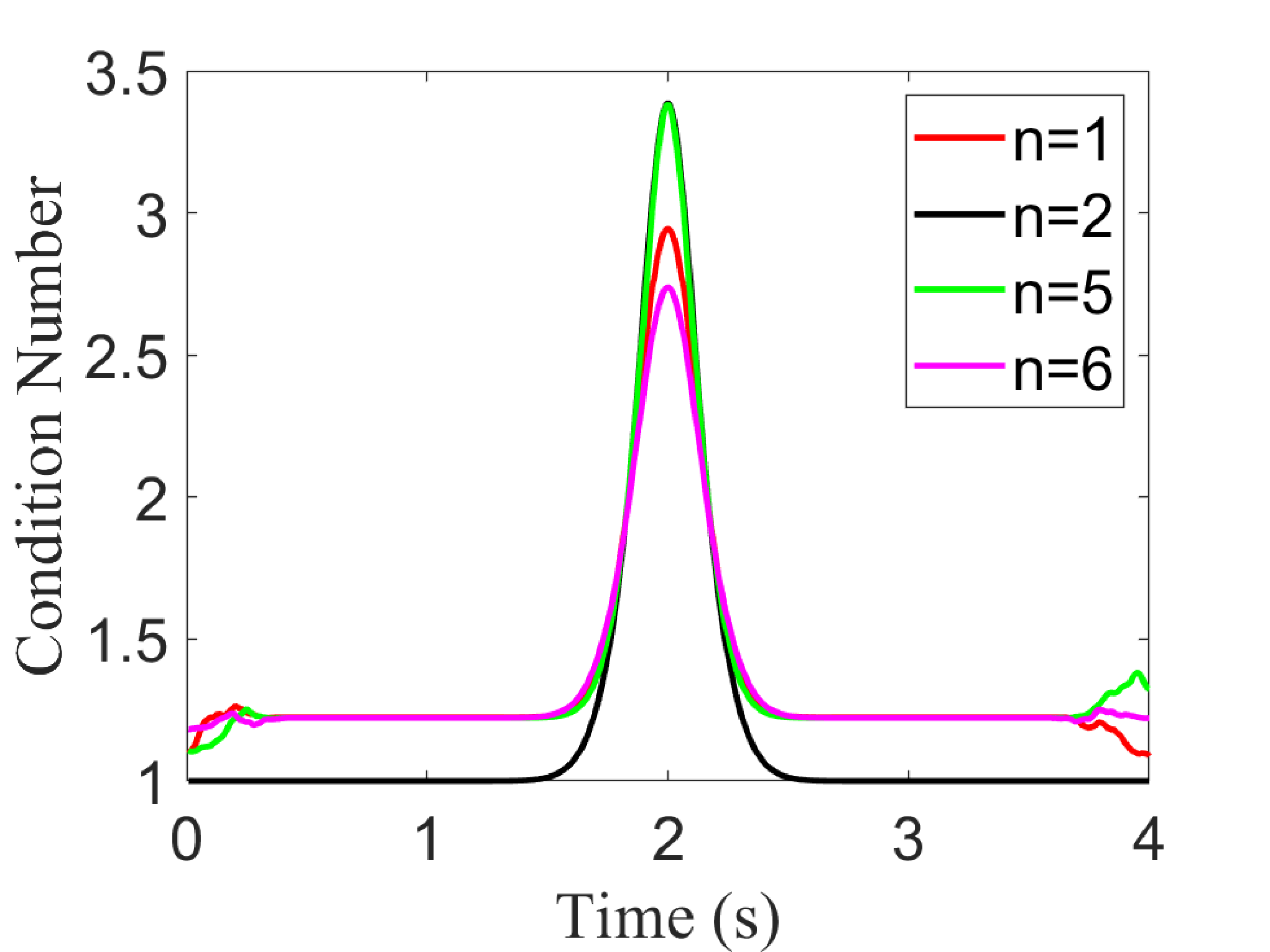}} \\
        (a) & (b) \\
    \end{tabular}
    \caption{\small (a) reconstruction error with estimated IFs and chirprates; 
    (b) condition numbers of inverse coefficient matrices.}
    \label{figure:alpha_sensitivity and condition number}
\end{figure}

Next, let us look at the influence of the window parameter \(\alpha\) on the accuracy of mode retrieval.
Since the optimal $\alpha_n$ values determined by Rényi entropy differ significantly for different $n$, 
we take  \(n=2\) as an example to illustrate the sensitivity of the reconstruction to variations in \(\alpha\).
The \(\alpha\) values are set to \(6.2+0.1j\) for \(j = -10, -9, \cdots, 10\), while the optimal value \(\alpha_2 = 6.2\) is obtained from \eqref{entropy_minimization} for $x(t)$ given by  \eqref{example1}. With each of the parameters $\alpha=6.2+0.1j$,  we extract the IF and chirprate by $S_{x}^{2}(t, \xi, \gamma)$  and then reconstruct the signal components.   
 Panel(a) of  Fig.~\ref{figure:alpha_sensitivity and condition number} shows the 
the reconstruction error of $x(t)$, which is defined as the sum of the RMSEs of the two components of $x(t)$.
As shown in Panel(a) of  Fig.~\ref{figure:alpha_sensitivity and condition number},  the  SWLCT achieves high mode retrieval accuracy when \( \alpha \) is within a reasonable range around \( \alpha_2 \).
This trend generalizes to other types of WLCTs, as they also yield similar accuracy when \( \alpha \) varies within a small region around their corresponding \( \alpha_n \).

As shown in \eqref{recover_26} and \eqref{recover_15}, the reconstruction of the modes relies on the  coefficient matrix \(C\) or \(A\).
The condition number of a matrix serves as an indicator of the sensitivity of a linear system's solution with respect to perturbations in the input data.
For  a non-singular square matrix $M_0$, the condition number $\kappa(M_0)$  is given by
\begin{equation}
\kappa(M_0):= \|M_0\|_p \|M_0^{-1}\|_p,
\label{eq:condition_number}
\end{equation}
where \(\|M_0\|_p\) denotes the \(p\)-norm of the matrix \(M_0\) and \(\|M_0^{-1}\|_p\) denotes the \(p\)-norm of the inverse of \(M_0\).
A singular or non-square matrix \(M\) can also have a condition number (see \cite{golub2013matrix}).
A high condition number indicates that the matrix is ill-conditioned, which can lead to numerical instability in solving linear systems. 

For the signal \(x(t)\) in  \eqref{example1}, we present the  condition numbers (with $p=2$)  of the  coefficient matrices \(C\) and \(A\) in Panel(b) of  Fig.~\ref{figure:alpha_sensitivity and condition number}.
We can observe that the condition numbers  are all relatively small, even near the crossing point of the two components. Specifically, at \(t=2\ \text{s}\), the condition numbers for \(n=2\) and \(n=5\) are slightly larger than those for \(n=1\) and \(n=6\).

\begin{figure}[H]
    \centering
    \begin{tabular}{ccc}
        \resizebox{1.8in}{1.1in}{\includegraphics{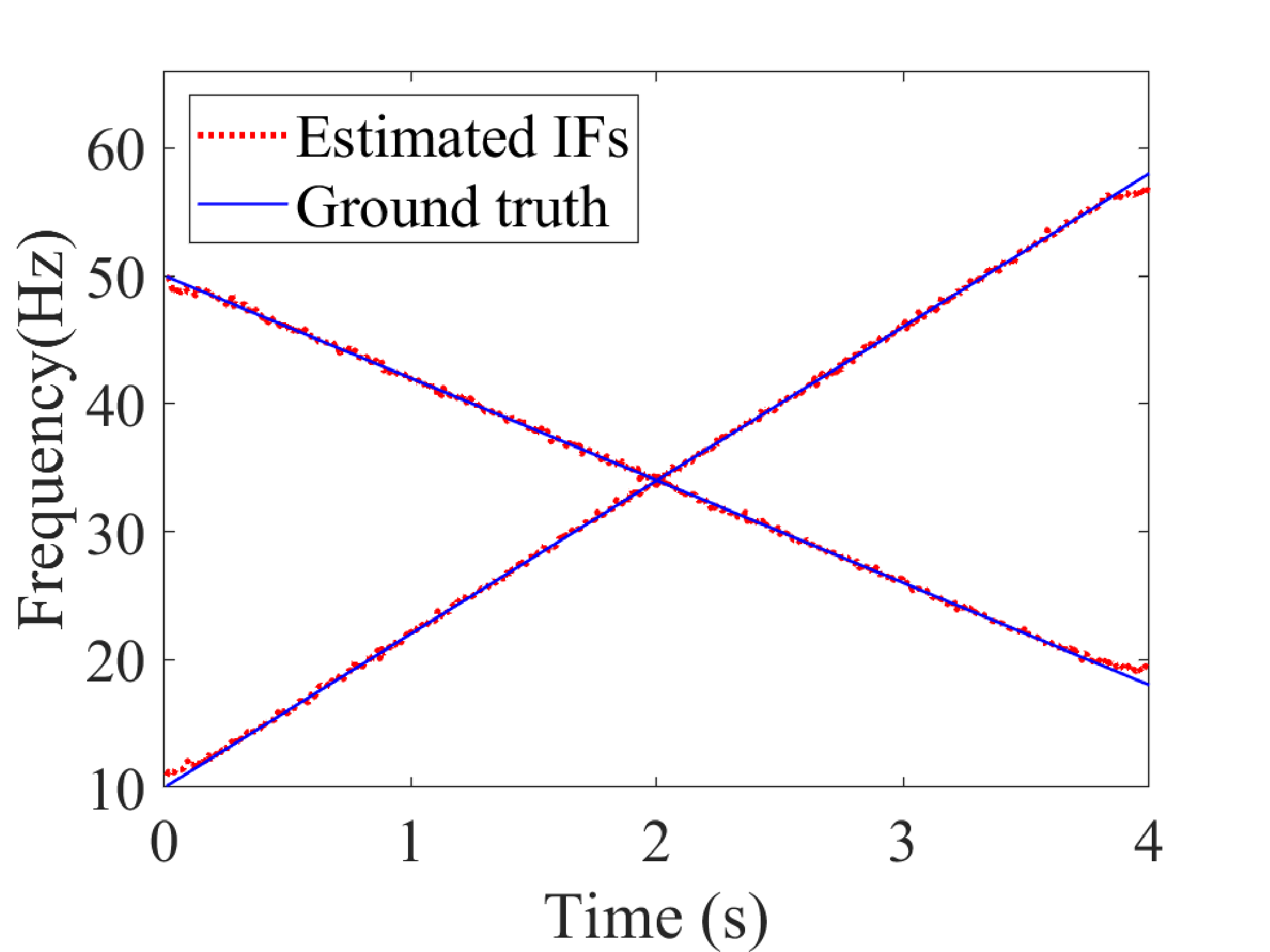}} & 
        \resizebox{1.8in}{1.1in}{\includegraphics{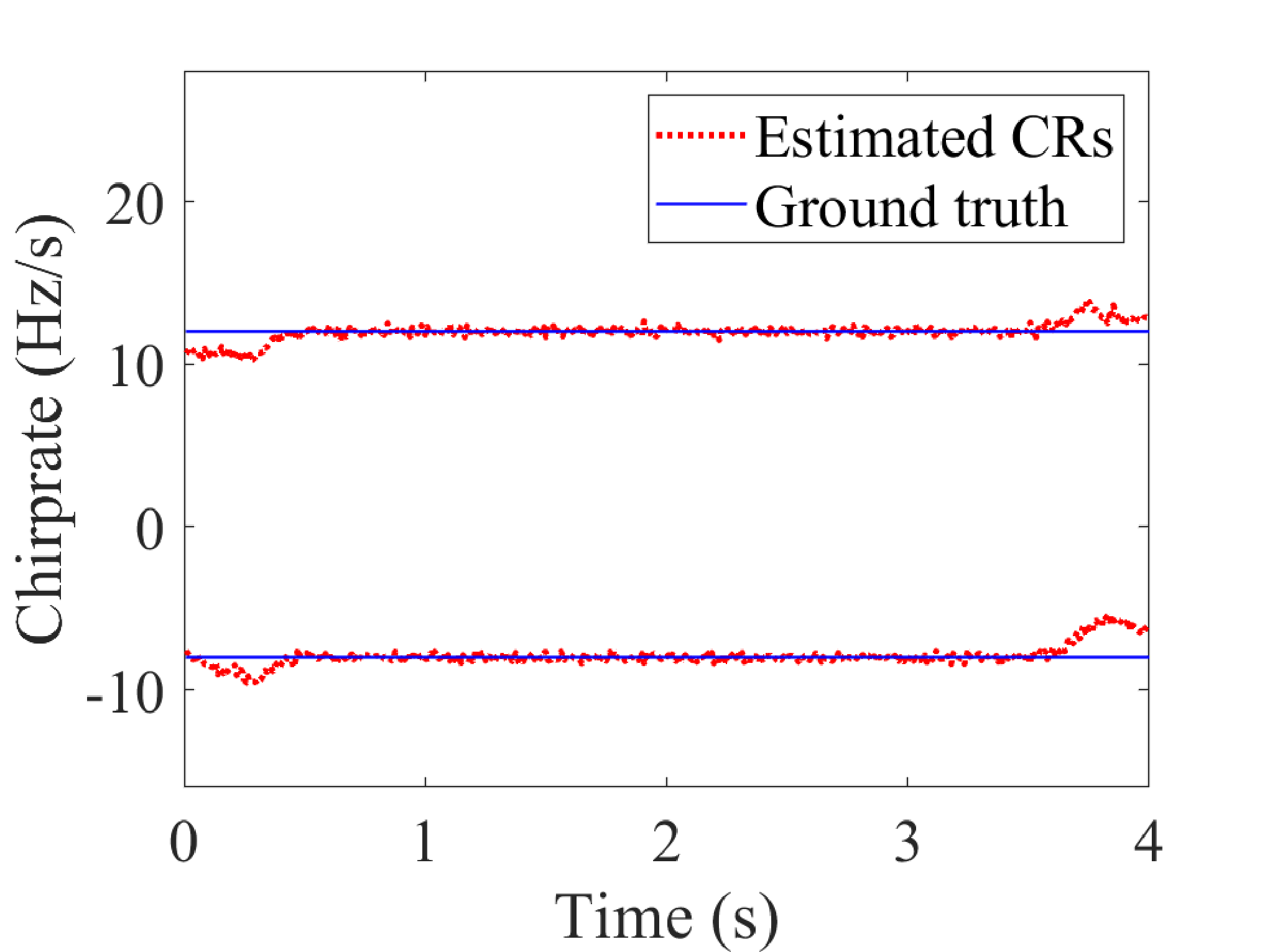}} &
        \resizebox{1.8in}{1.1in}{\includegraphics{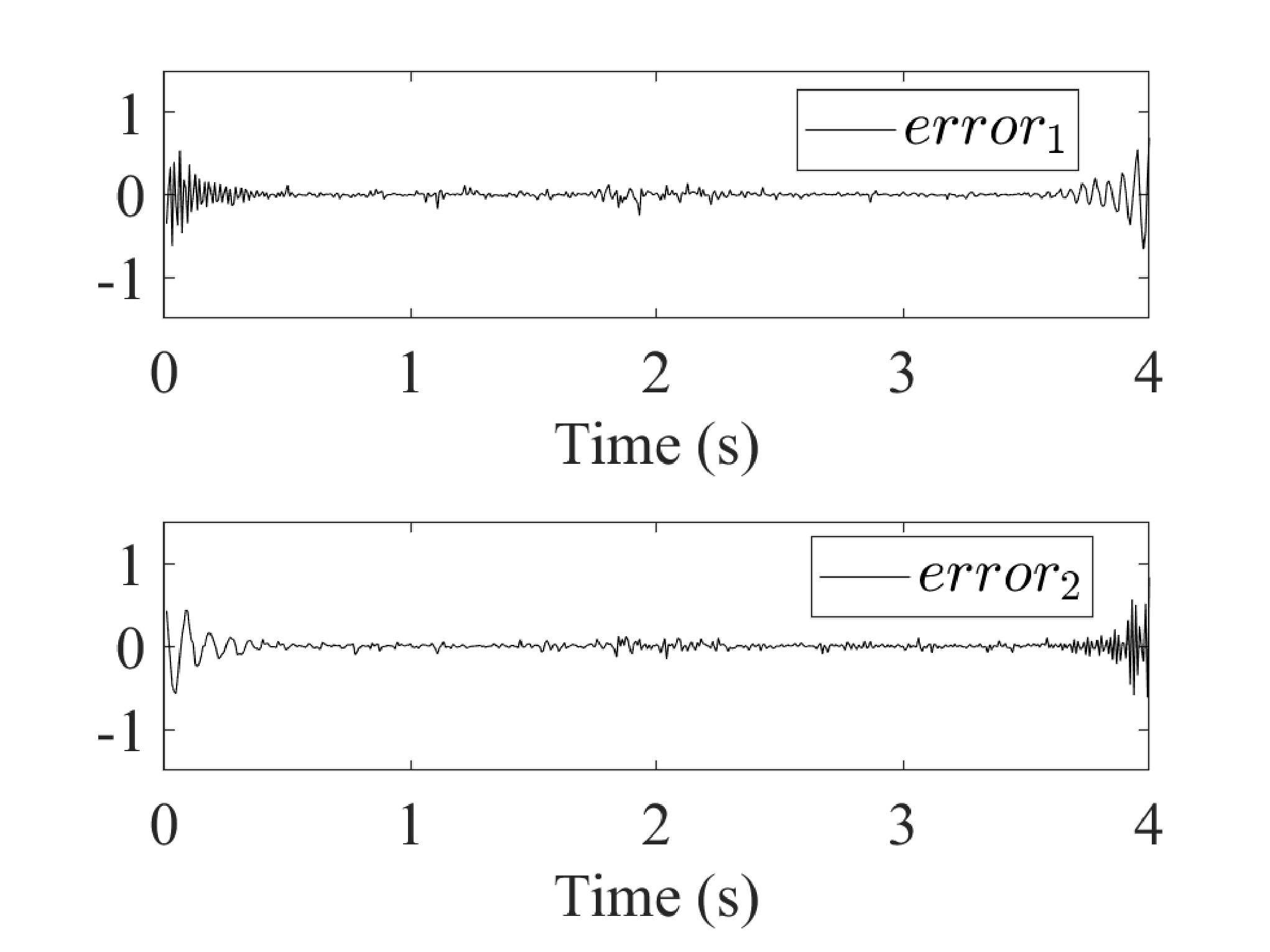}} \\
     (a)  & (b)  & (c)   \\
    \end{tabular}
    \caption{
        (a) IFs with small perturbations; 
        (b) chirprates with small perturbations; 
        (c) Reconstruction errors with small perturbations.
    }
    \label{Figure:Example1_perturbations}
\end{figure}

We take \(n=2\) as an example to show the sensitivity of the reconstruction to  perturbations in IF and chirprate estimations.
First, the individual IF and chirprate estimations are each perturbed by adding independent Gaussian white noise. For each estimation, the noise level is set to achieve a signal-to-noise ratio (SNR) of 15 dB with respect to its own power.
The estimated ridges with small perturbations are shown in panels (a) and (b) of Fig.~\ref{Figure:Example1_perturbations}.
Subsequently, via \eqref{recover_26}, we reconstruct the modes using the perturbed ridges and calculate the reconstruction errors.
The reconstruction errors are shown in Panel(c) of Fig.~\ref{Figure:Example1_perturbations}.
It can be seen that for small perturbations of  IF and chirprate estimations, the reconstruction errors remain small; however, the errors near the crossing point are a little bit larger, which is consistent with the fact that the condition numbers are relatively large near the crossing point.

\section{Synchrosqueezed X-ray windowed linear canonical transform (SXWLCT)}	
Though the SWLCTs can effectively separate the components of certain multicomponent signals, it may not be suitable for all scenarios. For instance, even when the signal components are linear chirp signals, the SWLCTs may not be able to accurately extract the IFs and chirprates.

\begin{figure}[H]
    \centering
    \begin{tabular}{cc}
        \resizebox{2.0in}{1.5in}{\includegraphics{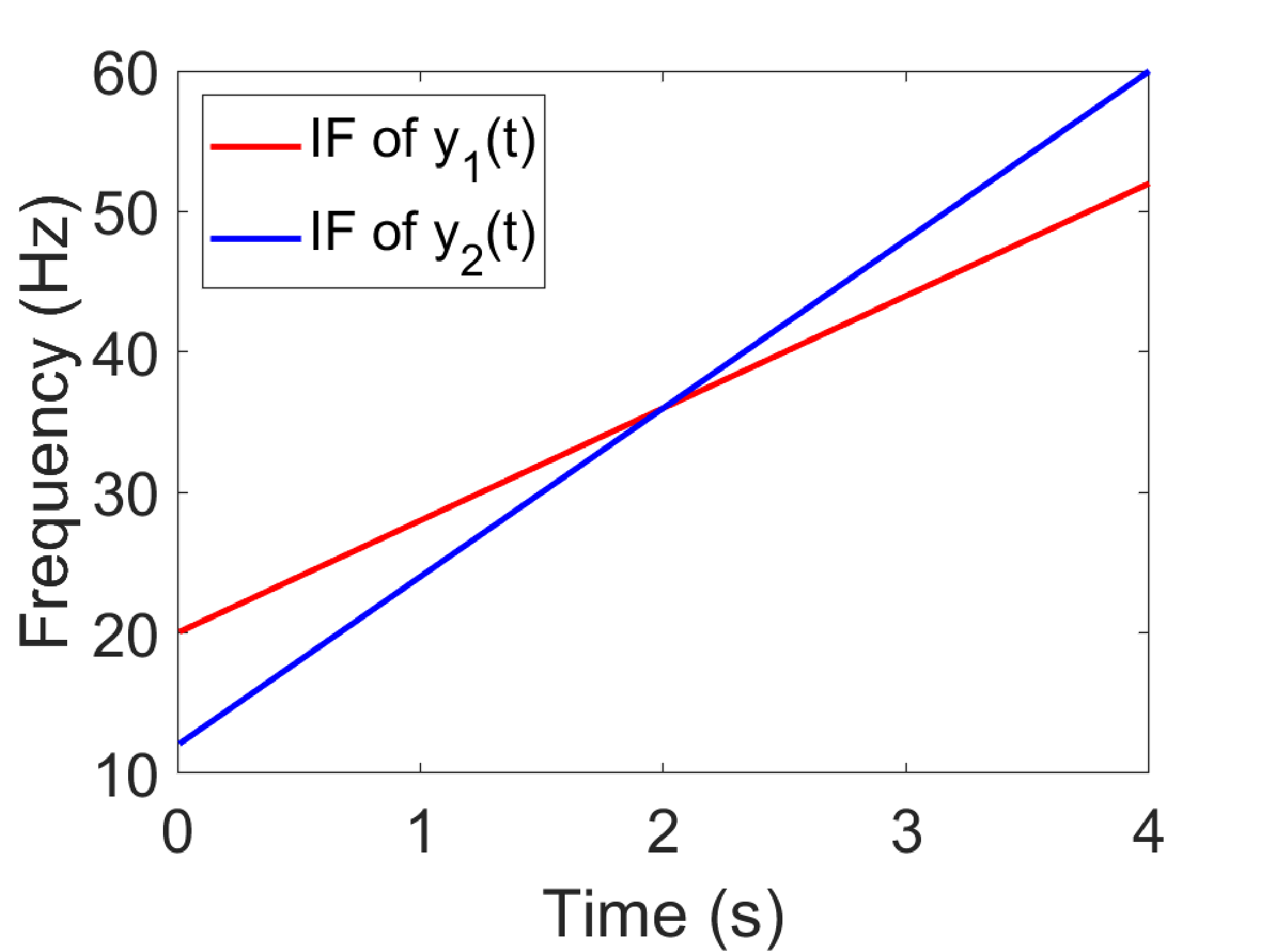}} &
        \resizebox{2.0in}{1.5in}{\includegraphics{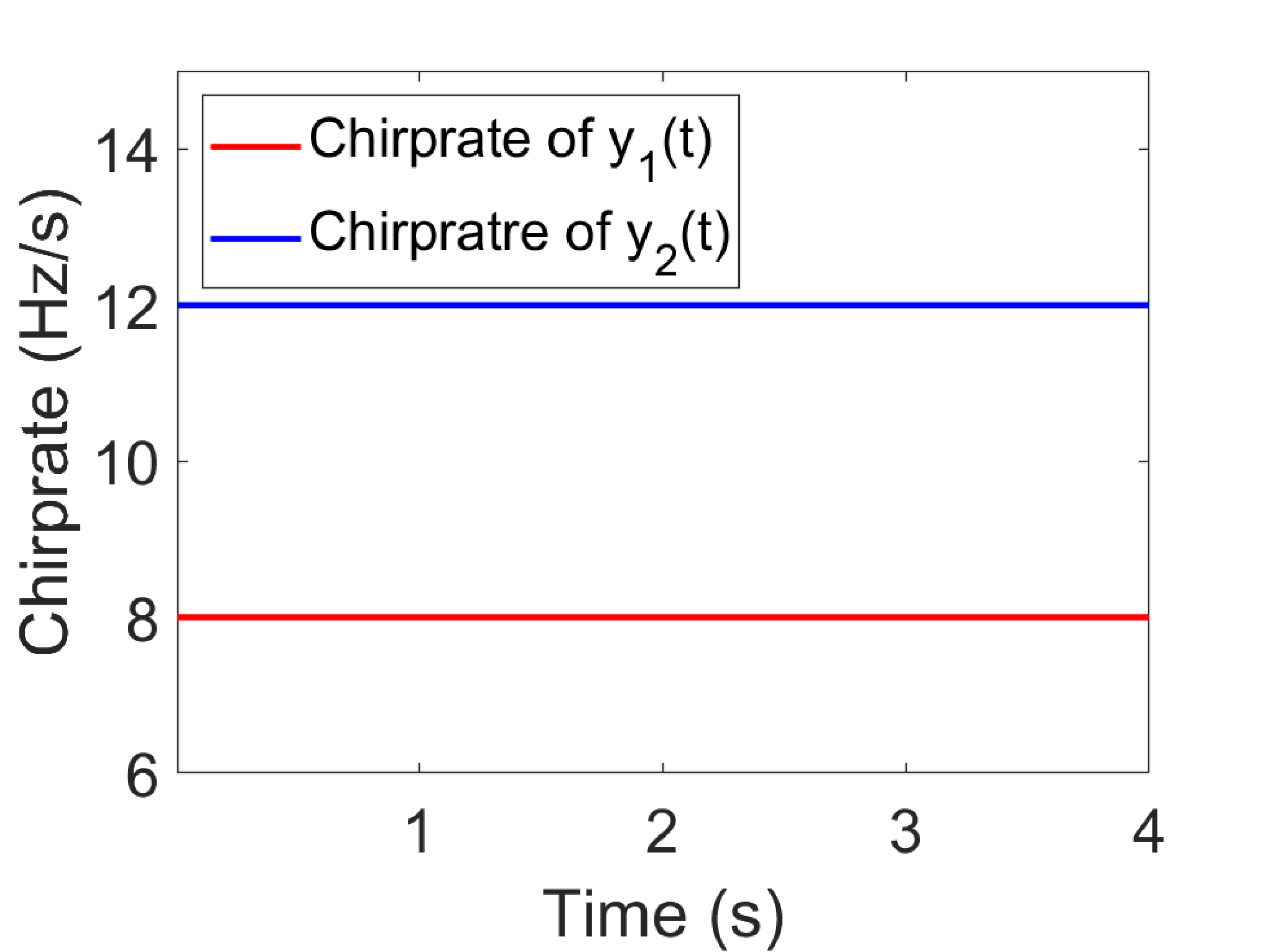}} \\
        (a) & (b) \\
    \end{tabular}
    \caption{\small IFs and chirprates of $y_1, y_2$. (a) IFs; (b) Chirprates.}
    \label{figure:Example2_IFs}
\end{figure}

 Let
\begin{equation}
    \label{example2}
y(t) = y_1(t) + y_2(t), \quad y_1(t) = e^{i 2\pi \phi_1(t)}, \quad y_2(t) = e^{i 2\pi \phi_2(t)},
\end{equation}
where
\begin{equation*}
\phi_1(t) = 4t^2 + 20t, \quad \phi_2(t) = 6t^2 + 12t.
\end{equation*}
The sampling frequency is 128 Hz and the time duration is \([0, 4)\) seconds.
Note that the IFs crossover happens at time \(t = 2\) s and frequency \(\eta = 36\) Hz, and chirprates are 12 and 8 Hz/s respectively.
These  are all depicted in Fig.~\ref{figure:Example2_IFs}.

\begin{figure}[H]
    \centering
    \begin{tabular}{cccc}
        \resizebox{0.22\textwidth}{!}{\includegraphics{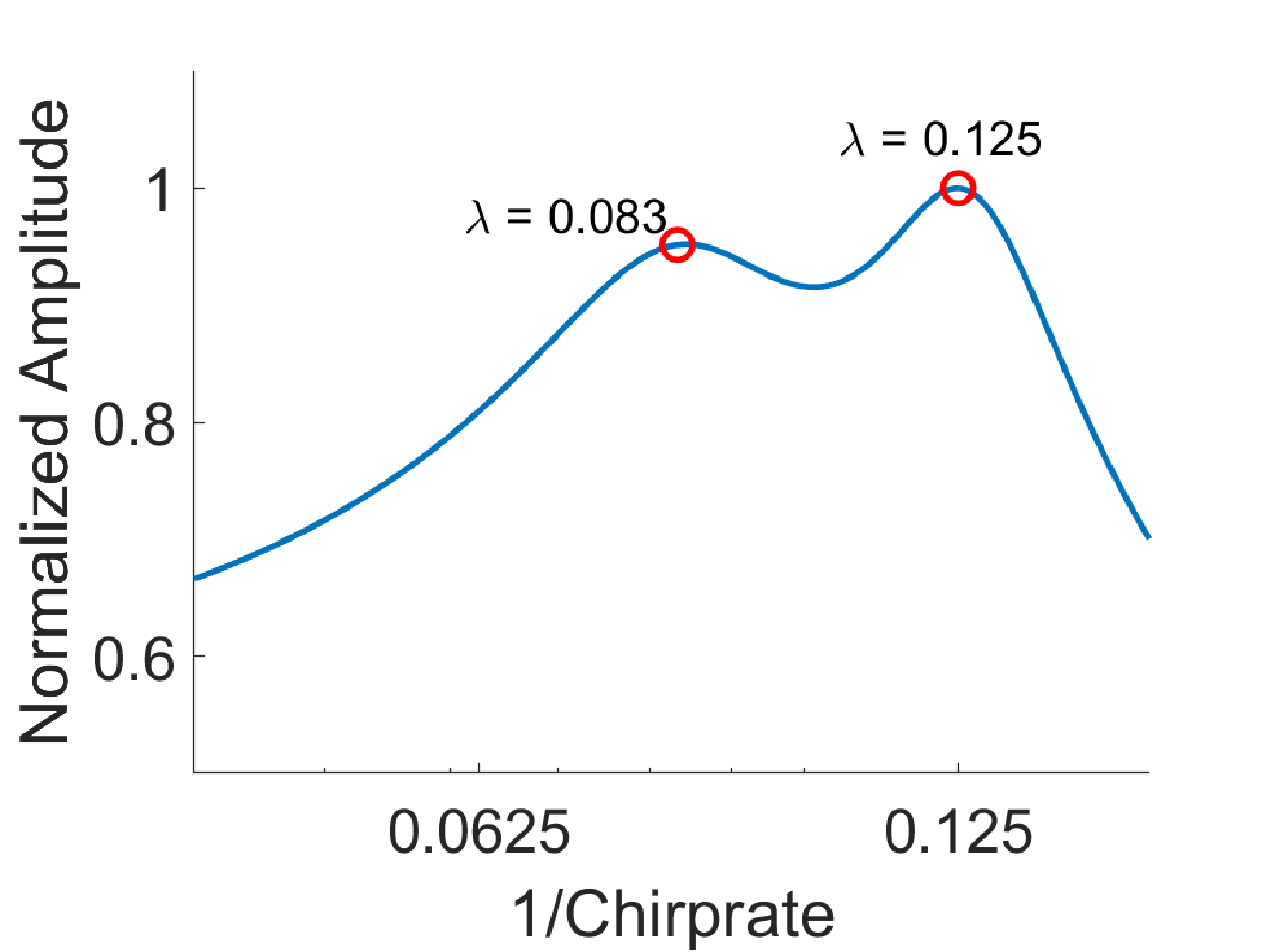}} & 
        \resizebox{0.22\textwidth}{!}{\includegraphics{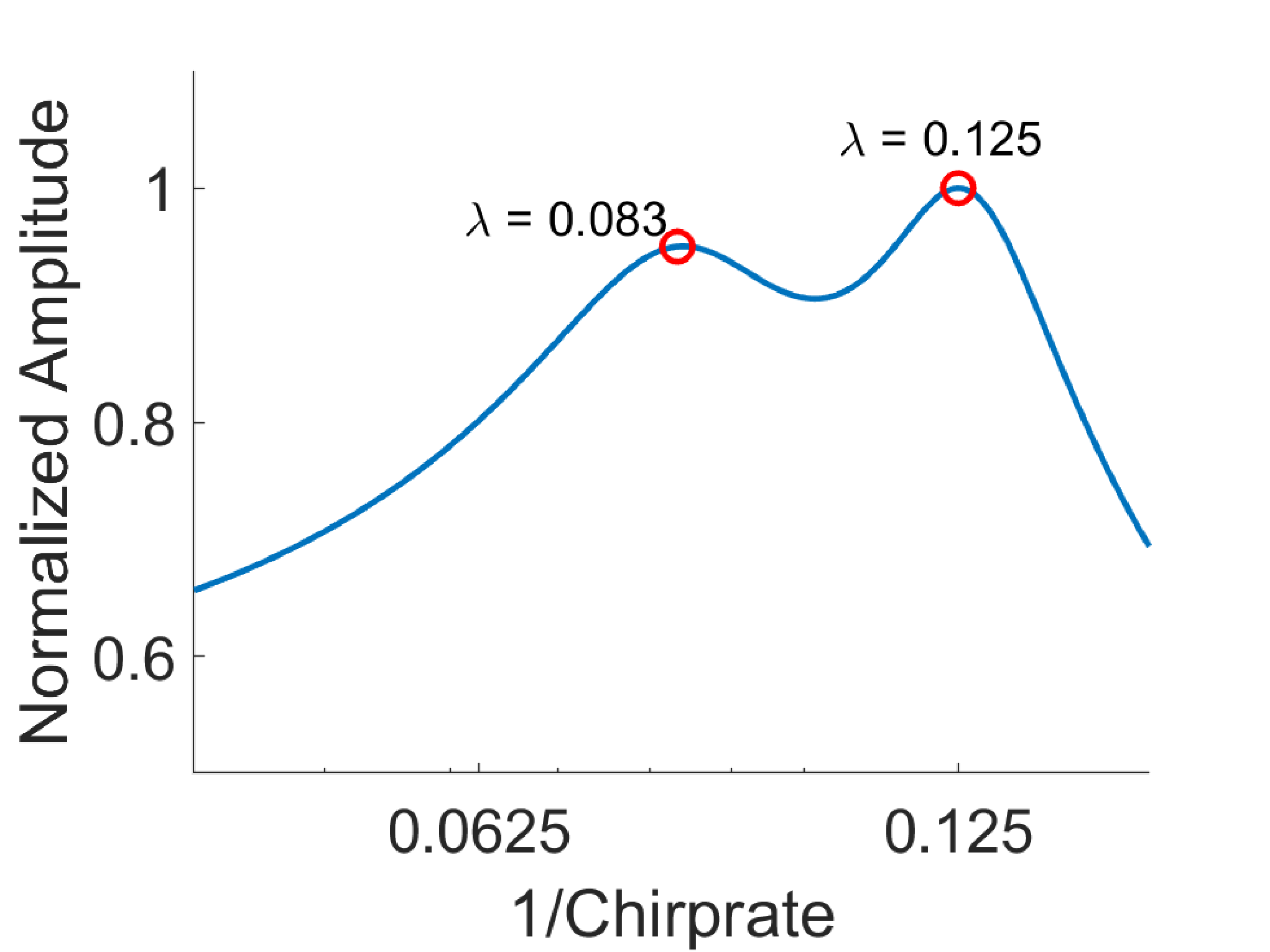}} &
        \resizebox{0.22\textwidth}{!}{\includegraphics{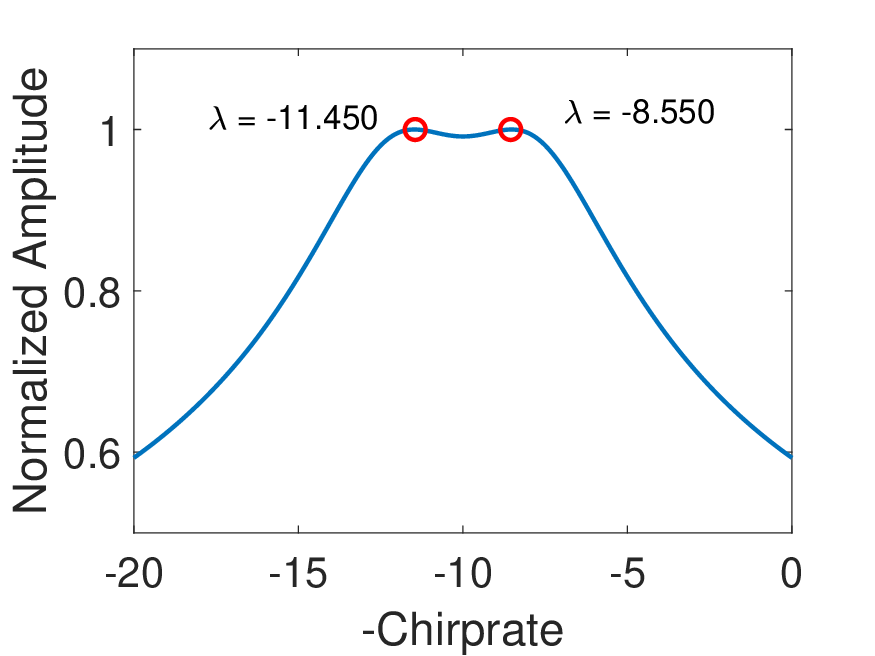}} & 
        \resizebox{0.22\textwidth}{!}{\includegraphics{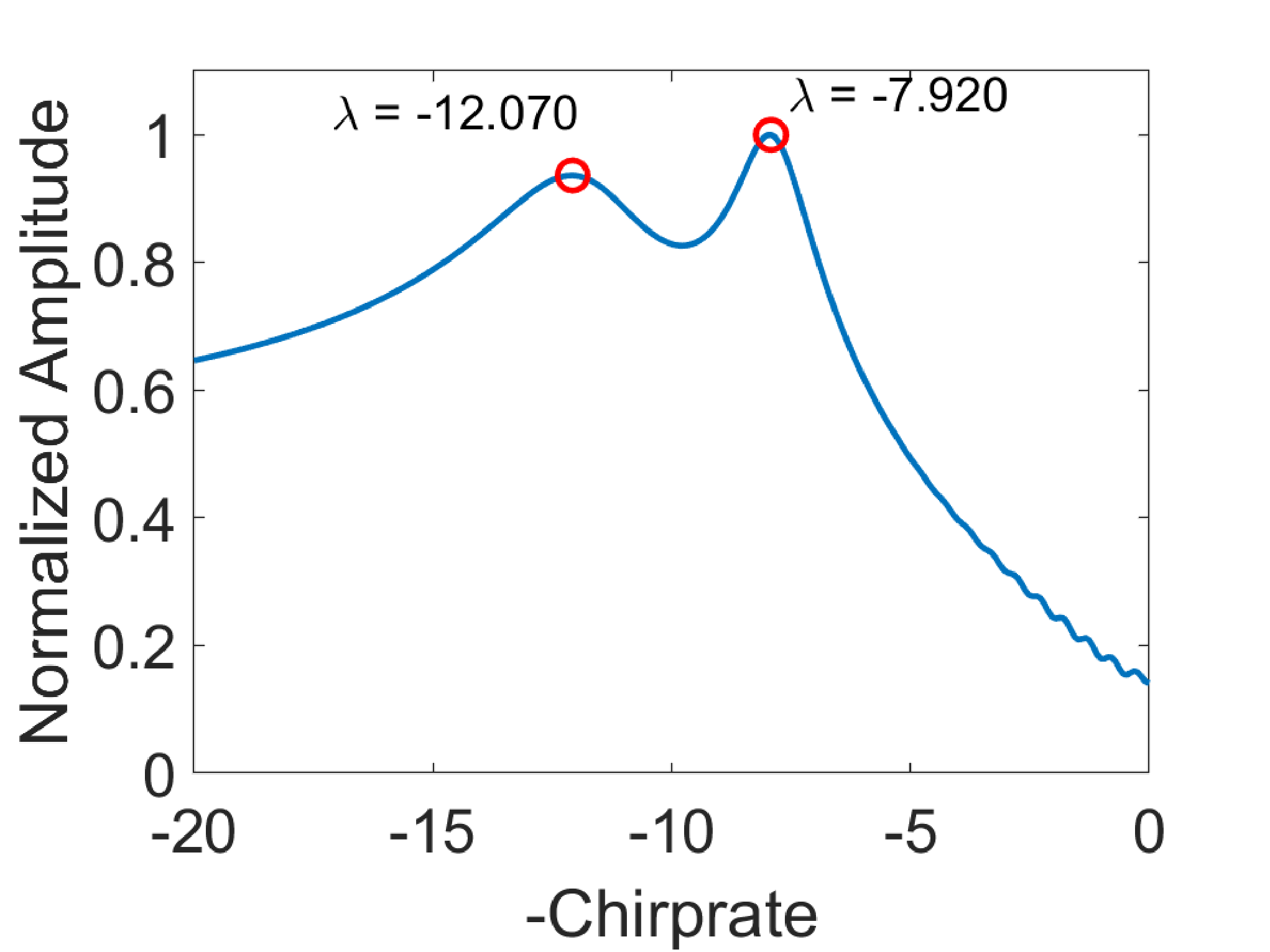}} \\
        (a)  & (b)  & (c) & (d)  \\
    \end{tabular}
    \caption{\small WLCTs of $y(t)$ at $t=2$ s, $\eta=36$ Hz. (a) $T_y^{g,1}(2,36,\lambda)$; (b) $T_y^{g,5}(2,36,\lambda)$; (c) $T_y^{g,2}(2,36,\lambda)$; (d) $T_y^{g,6}(2,36,\lambda)$.}
    \label{figure:WLCT_Y_crossingpoint}
\end{figure}

Through \eqref{entropy_minimization}, we obtain the \(\alpha\) values as follows: For \(\cL^{M_{\gl, 1}}(g)(u) \) and  \(\cL^{M_{\gl, 5}}(g)(u) \), \(\alpha\) is 0.017 and 0.018, respectively; 
for \(\cL^{M_{\gl, 2}}(g)(u) \)  and  \(\cL^{M_{\gl, 6}}(g)(u) \), \(\alpha\) is 2.5 and 80, respectively.
We  examine the slices of \( |T_{y}^{g, n}(t,\eta,\lambda)| \) at the frequency intersection points in Fig.~\ref{figure:WLCT_Y_crossingpoint}.
From panels (a) and (b), distinct local maxima are evident around \(\lambda = 0.125\) and \(\lambda = 0.083\). 
In Panel(d), local maxima also appear at \(\lambda = -7.920\) and \(\lambda = -12.070\), although the latter is less pronounced.
Notably, in Panel(c), peaks are observed at \(\lambda = -11.450\) and \(\lambda = -8.550\), suggesting an obvious deviation from the ground truth.

\begin{figure}[H]
    \centering
    \begin{tabular}{cccc}
        \resizebox{1.5in}{!}{\includegraphics{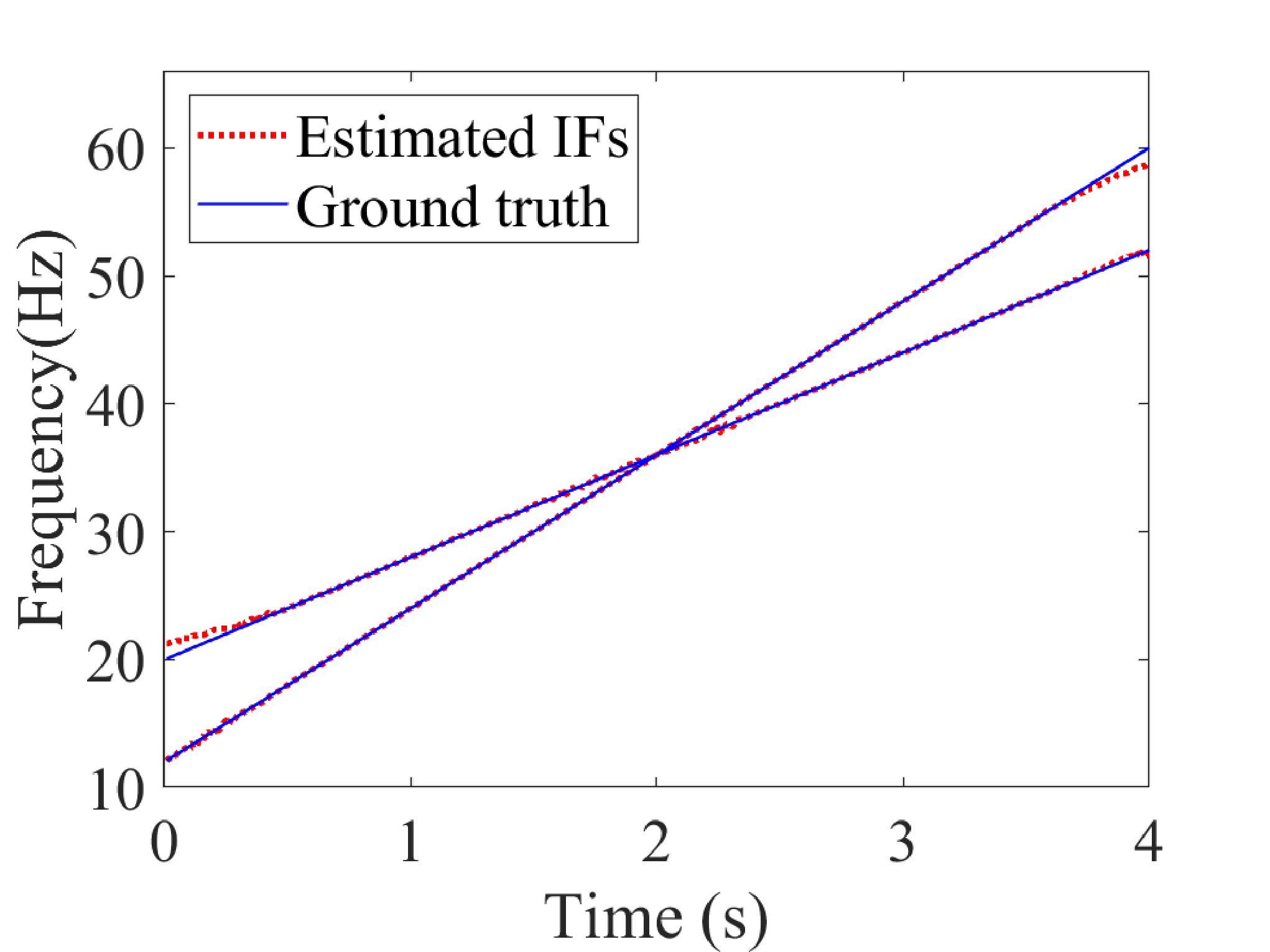}} & 
        \resizebox{1.5in}{!}{\includegraphics{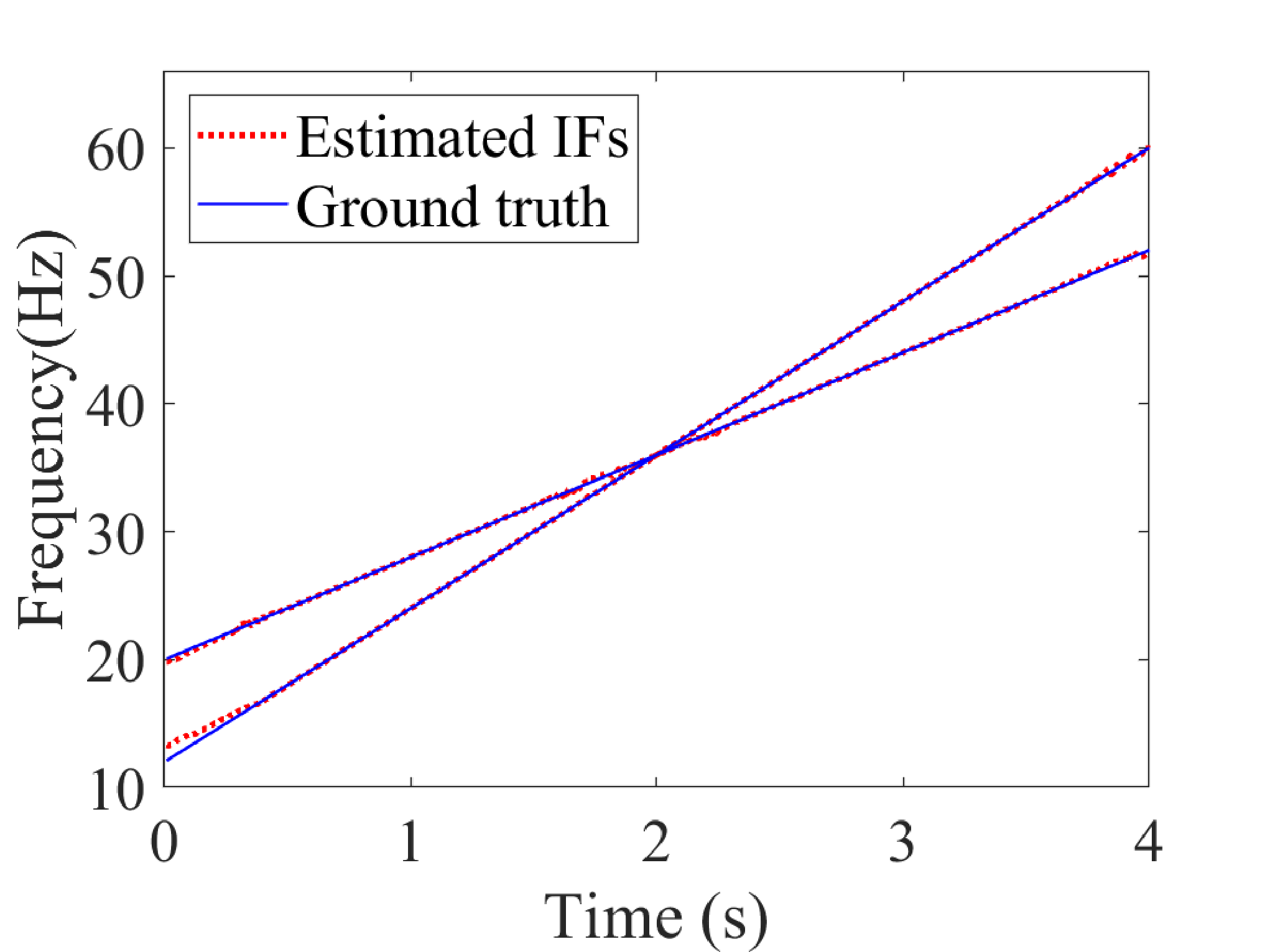}} & 
        \resizebox{1.5in}{!}{\includegraphics{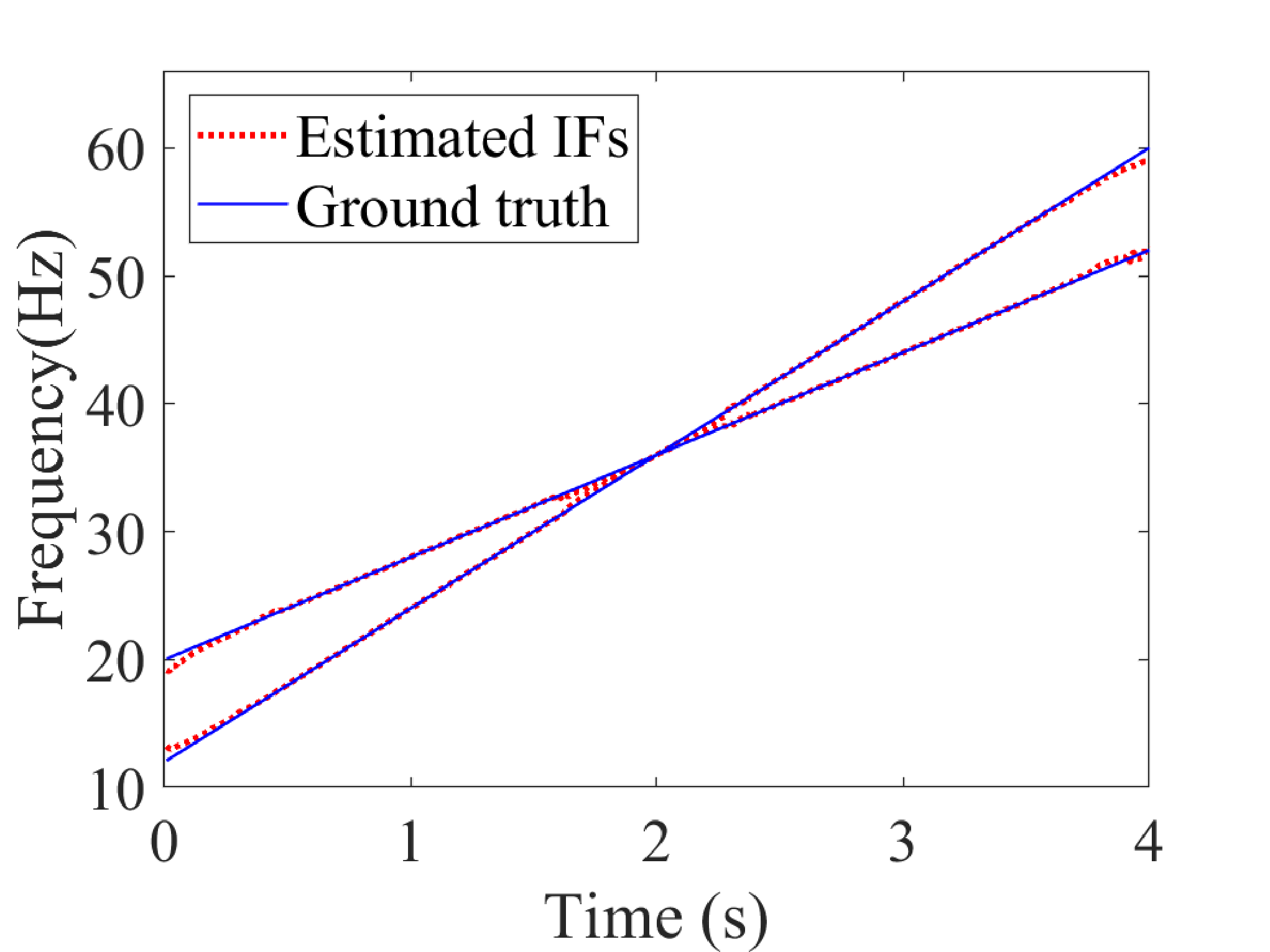}} &
        \resizebox{1.5in}{!}{\includegraphics{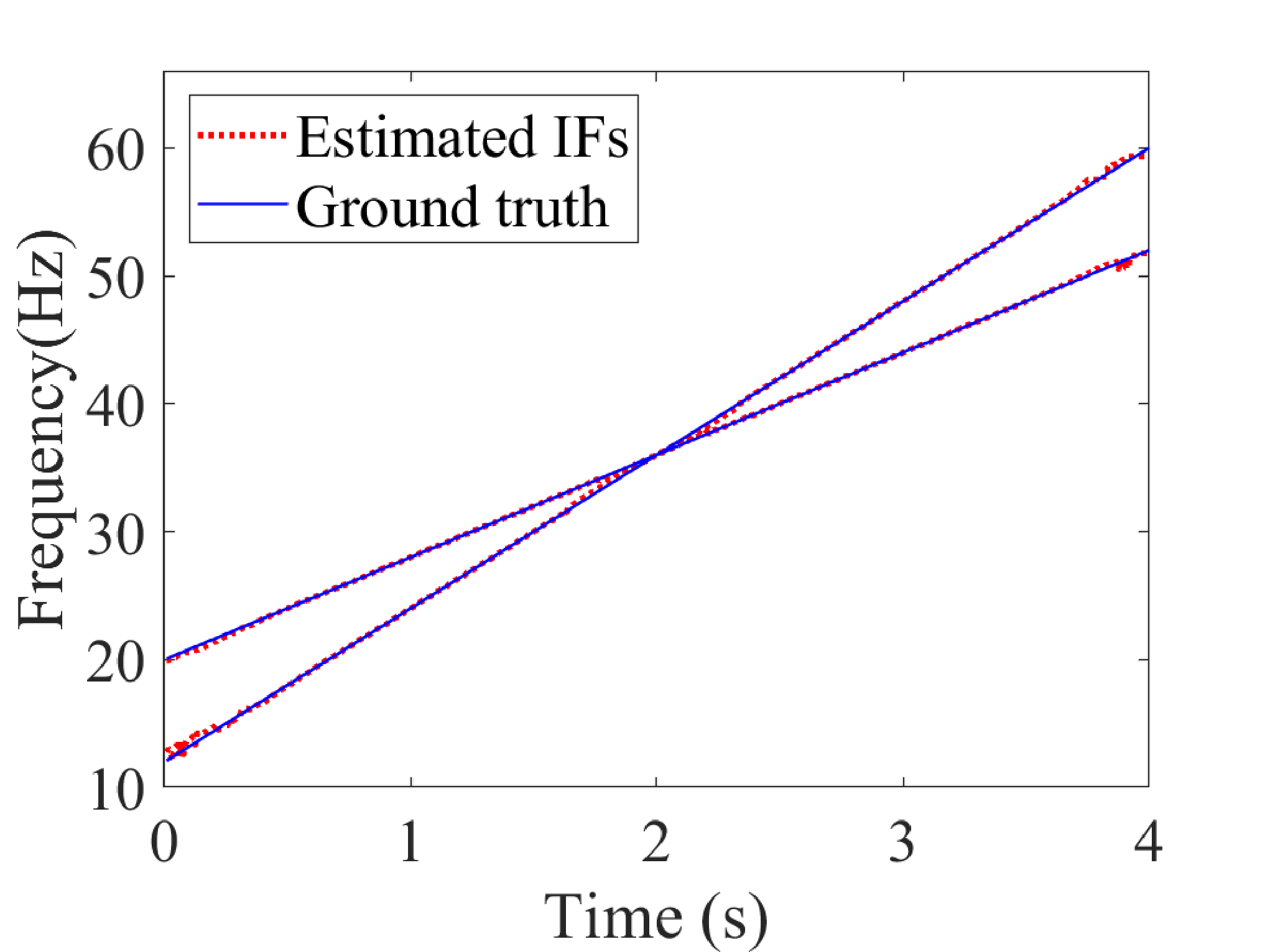}} \\
        \resizebox{1.5in}{!}{\includegraphics{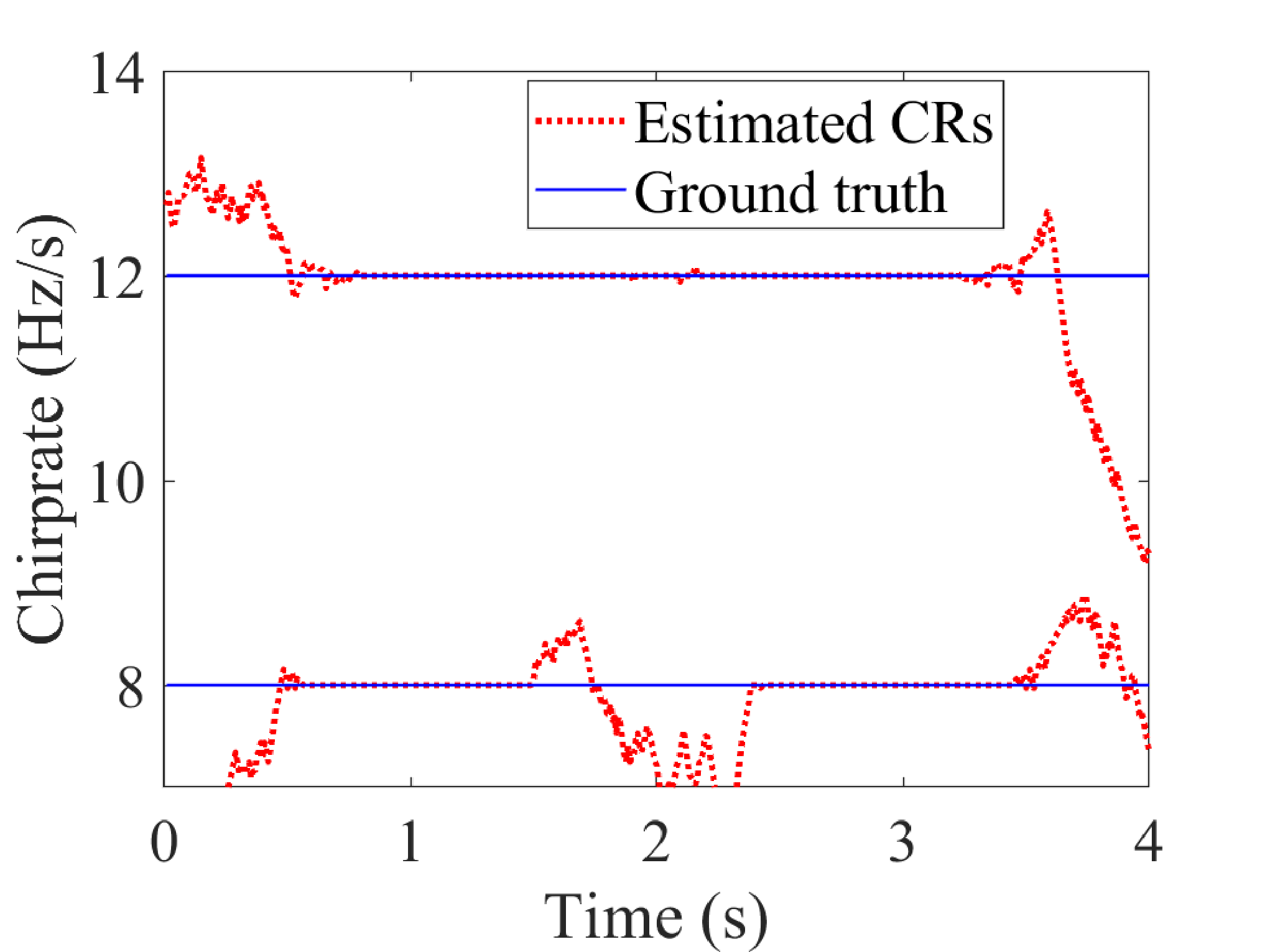}} & 
        \resizebox{1.5in}{!}{\includegraphics{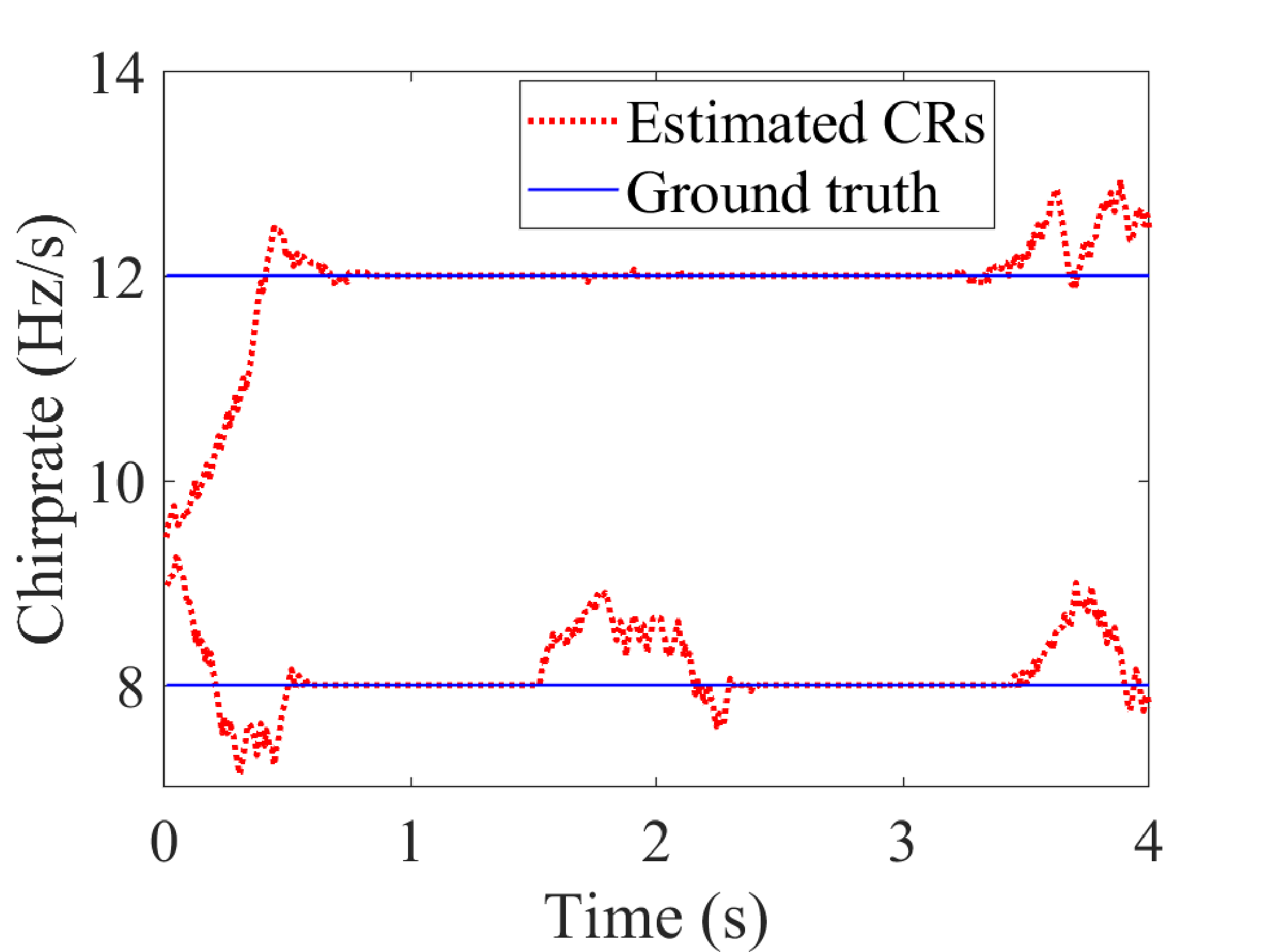}} & 
        \resizebox{1.5in}{!}{\includegraphics{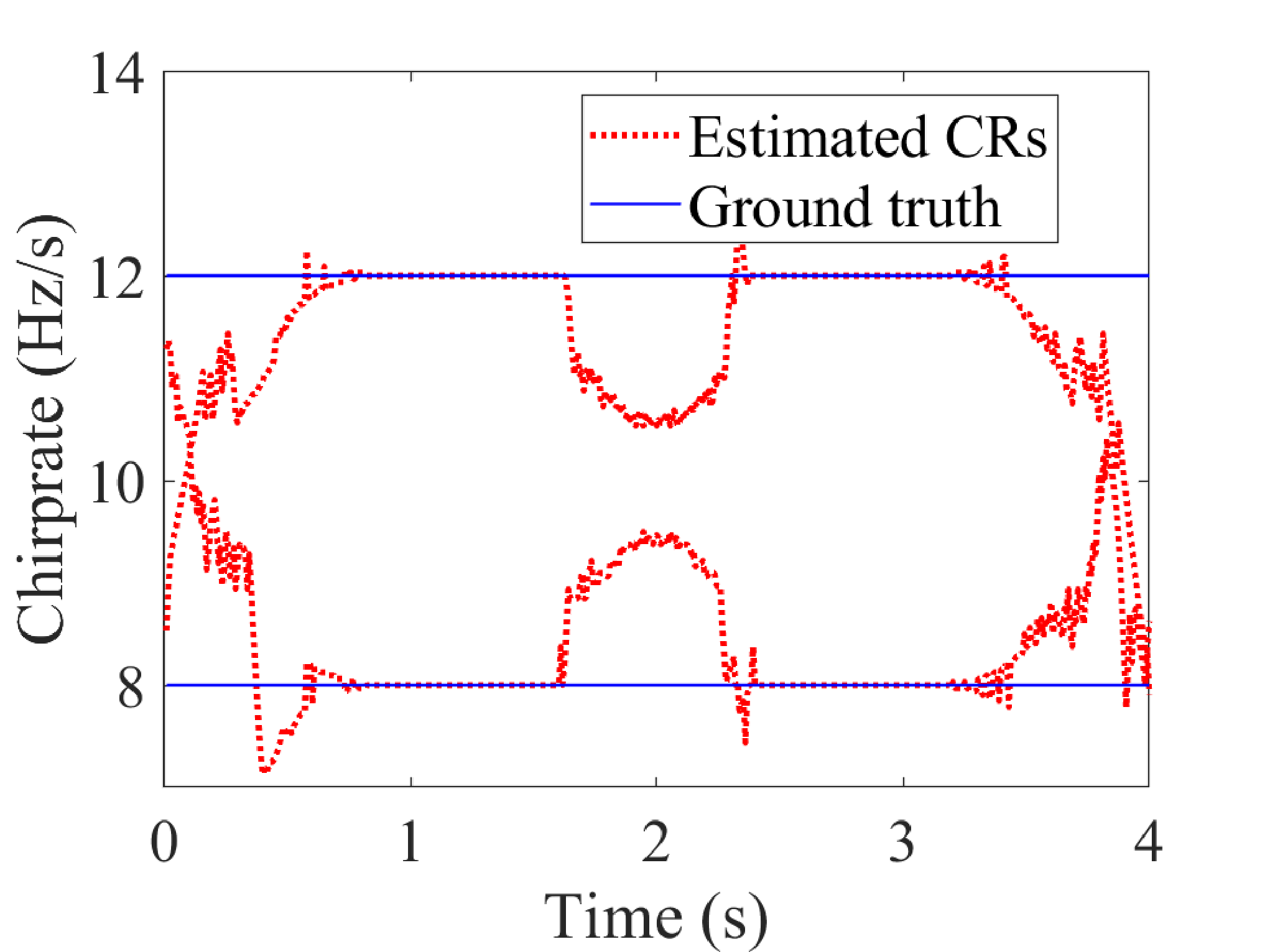}} & 
        \resizebox{1.5in}{!}{\includegraphics{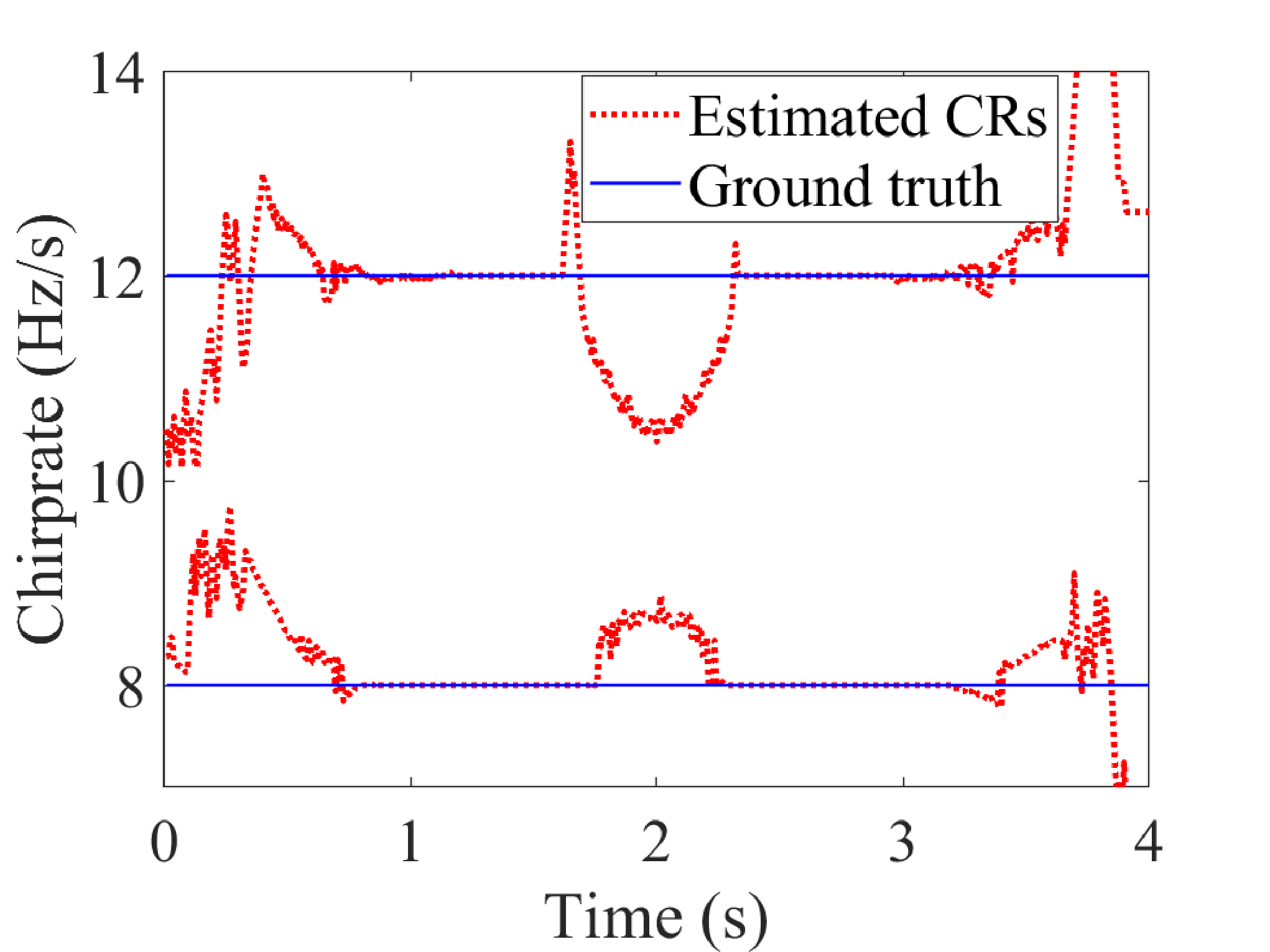}}
    \end{tabular}
	\caption{\small IF and chirprate estimations by SWLCT. 
	First row shows IFs detected by: (1) $S^1_y(t, \xi, \gamma)$; (2) $S^5_y(t, \xi, \gamma)$; 
	(3) $S^2_y(t, \xi, \gamma)$; (4) $S^6_y(t, \xi, \gamma)$. 
	Second row shows corresponding chirprates detected by the same methods.}
    \label{figure:SWLCT of $Y$}
\end{figure}

The IFs and chirprates extracted from SWLCT  are shown in Fig.~\ref{figure:SWLCT of $Y$}.
Near \( t = 2 \) s, \( S^2_y(t,\eta,\lambda)\) and \( S^6_y(t,\eta,\lambda) \) show minor errors in IF extraction and major errors in chirprate extraction. 
In contrast, \( S^1_y(t,\eta,\lambda) \) and \( S^5_y(t,\eta,\lambda)\) have negligible IF errors and more accurate chirprate extraction for high-chirprate components. 

Referring to  \eqref{eq:TFMmagnitude}, the magnitude is given by
\begin{align}
  \label{amplitude} \left| T_x^{g, n}(t,\eta,\lambda) \right| = {L_n(\lambda)}^{-\frac{1}{4}} e^{-\frac{\pi \alpha (\eta-\phi'(t))^2}{L_n(\lambda)}},
\end{align}
where \( L_n(\lambda) \) is defined by \eqref{L_1} for \( n=1 \), \eqref{L_2} for \( n=2 \), \eqref{L_5} for \( n=5 \) or \( n=6 \).
When the frequency \(\eta\) is fixed near \(\phi'(t)\), the decay of \( |T_{y}^{g, n}(t,\eta,\lambda)| \) with respect to \(\lambda\) is determined by the term \( L_n(\lambda)^{-\frac{1}{4}} \). 
This slow decay  gives rise to large errors in chirprate estimation.
To achieve a faster decay rate in the chirprate direction, we propose the X-ray WLCT (XWLCT). 
The X-ray wavelet-chirplet transform was intrduced in \cite{jiang2025synchrosqueezed} to obtain a sharper time-scale-chirprate representation of a signal.

The {continuous X-ray transform} of a 3D function $f(z_1, z_2, z_3)$, where $(z_1, z_2, z_3)\in \mathbb{R}^3$, is denoted by $Pf$. It encompasses the collection of all line integrals of $f$ \cite{averbuch20043d}. 
More precisely, the X-ray transform of $f$ along a given line $s$ parameterized as $s(v) = z + v\theta, v\in \mathbb{R}$,  is defined by 
$$Pf (z, \theta) := \int_{-\infty}^\infty f (z +v\theta) dv.$$
Consider a window function $h(v)$ such that $\int_{-\infty}^\infty h(v) dv=1$. We can then define a localized X-ray transform of $f$ as:
$${\mathcal P} f (z, \theta) := \int_{-\infty}^\infty f (z +v\theta) h(v) dv. $$
This localized variant integrates the function along the line, but critically, it weights the contribution of each point on the line by the window function $h(v)$. This localization proves beneficial in diverse applications by concentrating the transform on a specific segment of the line.

\begin{mdef}
	\label{defineXWLCT} 
	Let \( T_{x}^{g, n}(t,\eta,\lambda) \) be the  WLCTs of \( x(t) \), as defined in  definition \eqref{WLCT definition},  and  \( h(v) \) be a nonnegative window function that 
satisfies \( \int_{-\infty}^\infty h(v) \, dv = 1 \).
	\begin{enumerate}
		\item When \( n = 1 \) or \( 5 \), the XWLCT of \( x(t) \) is defined as
		\begin{equation}
			\label{def_XWLCT_15}
			\mathcal{T}_{x}^{g, n}(t,\eta,\lambda) := \int_{-\infty}^\infty | T_{x}^{g, n}(t+v,\eta+\frac{v}{\lambda},\lambda) | h(v) \, dv,
		\end{equation}
		\item When \( n = 2 \) or \( 6 \), the XWLCT of \( x(t) \) is given by
		\begin{equation}
			\label{def_XWLCT_26}
			\mathcal{T}_{x}^{g, n}(t,\eta,\lambda) := \int_{-\infty}^\infty \left| T_{x}^{g, n}(t+v,\eta-\lambda v,\lambda) \right| h(v) \, dv.
		\end{equation}
	\end{enumerate}
	\end{mdef}

In the definitions presented in  \eqref{def_XWLCT_15} and \eqref{def_XWLCT_26}, the transformations can be interpreted as a (localized) X-ray transform of $| T_{x}^{g, n}(t,\eta,\lambda) |$ along  
the (un-normalized) direction vector \( \beta = \left(1, \frac{1}{\lambda}, 0\right) \) for \eqref{def_XWLCT_15} and \( \beta = \left(1, -\lambda, 0\right) \) for \eqref{def_XWLCT_26} respectively. 
This characteristic is the reason why such transformations are termed as X-ray transforms.

{
Next we explain why \(\mathcal{T}_{x}^{g, n}(t,\eta,\lambda)\) exhibits a faster decay rate than \(T_{x}^{g, n}(t,\xi,\lambda)\) in the \(\lambda\) direction.
When time \( t \) is incremented by a small interval \( v \), we have 
\begin{equation}
    \label{approxtimev} 
    \phi'(t+v) \approx \phi'(t) + \phi''(t)v, \quad \phi''(t+v) \approx \phi''(t).
\end{equation}  
Given the signal \( x(t) = A e^{i2\pi\phi(t)} \), from \eqref{amplitude} and \eqref{approxtimev}, when \(t\) move to \(t+v\), then  \( L_n(\lambda) \) remains essentially unchanged. 
Thus, our focus  shifts to the factor \( (\eta - \phi'(t))^2 \) in the exponent of \( e \). 
Under the condition \eqref{approxtimev}, the factor \((\eta-\phi'(t))^2\) in the XWLCTs assumes two distinct approximate expressions:
\[
(\eta+\frac{v}{\lambda}-\phi'(t+v))^2 \approx (\eta-\phi'(t)+\frac{v}{\lambda}(1-\lambda\phi''(t)))^2,
\]
or
\[
(\eta-\lambda v-\phi'(t+v))^2 \approx (\eta-\phi'(t)-v(\lambda+\phi''(t)))^2.
\]
Thus, for  \(n=1,5\), we have:
\begin{equation}
  \label{XWLCTmagnitude_1} \mathcal{T}_{x}^{g, n}(t,\eta,\lambda) = \left| T_{x}^{g, n}(t+v,\eta+\frac{v}{\lambda},\lambda) \right| \approx{L_n}^{-\frac{1}{4}} e^{-\frac{\pi \alpha \left(\eta-\phi'(t)+\frac{v}{\lambda}(1-\lambda\phi''(t))\right)^2}{L_n}},
\end{equation}
and for \(n=2,6\),
\begin{equation}
	\label{XWLCTmagnitude_2} \mathcal{T}_{x}^{g, n}(t,\eta,\lambda) = \left| T_{x}^{g, n}(t+v,\eta-\lambda v,\lambda) \right| \approx {L_n}^{-\frac{1}{4}} e^{-\frac{\pi \alpha \left(\eta-\phi'(t)-v(\lambda+\phi''(t))\right)^2}{L_n}}.
\end{equation}
Both \eqref{XWLCTmagnitude_1} and \eqref{XWLCTmagnitude_2} show that a larger \( v \) leads to faster decay of the XWLCTs magnitude with respect to \(\lambda\). However, a good approximation requires a small \( v \) (see \eqref{approxtimev}). To address this contradiction, we use a local integration with a window function \( h(v) \) in \eqref{def_XWLCT_15} and \eqref{def_XWLCT_26}. 
  }

\begin{figure}[H]
    \centering
    \begin{tabular}{cccc}
        \resizebox{0.22\textwidth}{!}{\includegraphics{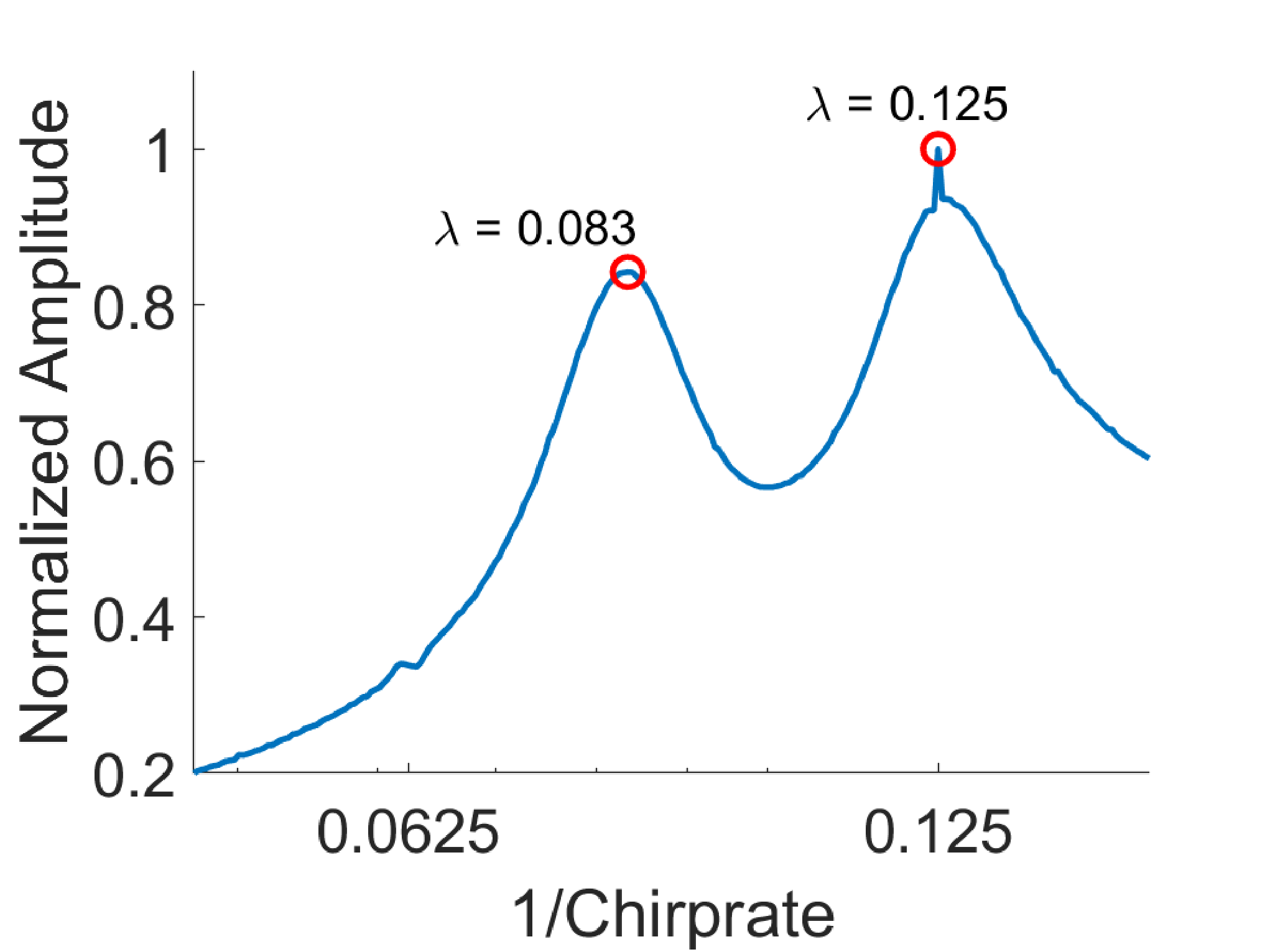}} & 
        \resizebox{0.22\textwidth}{!}{\includegraphics{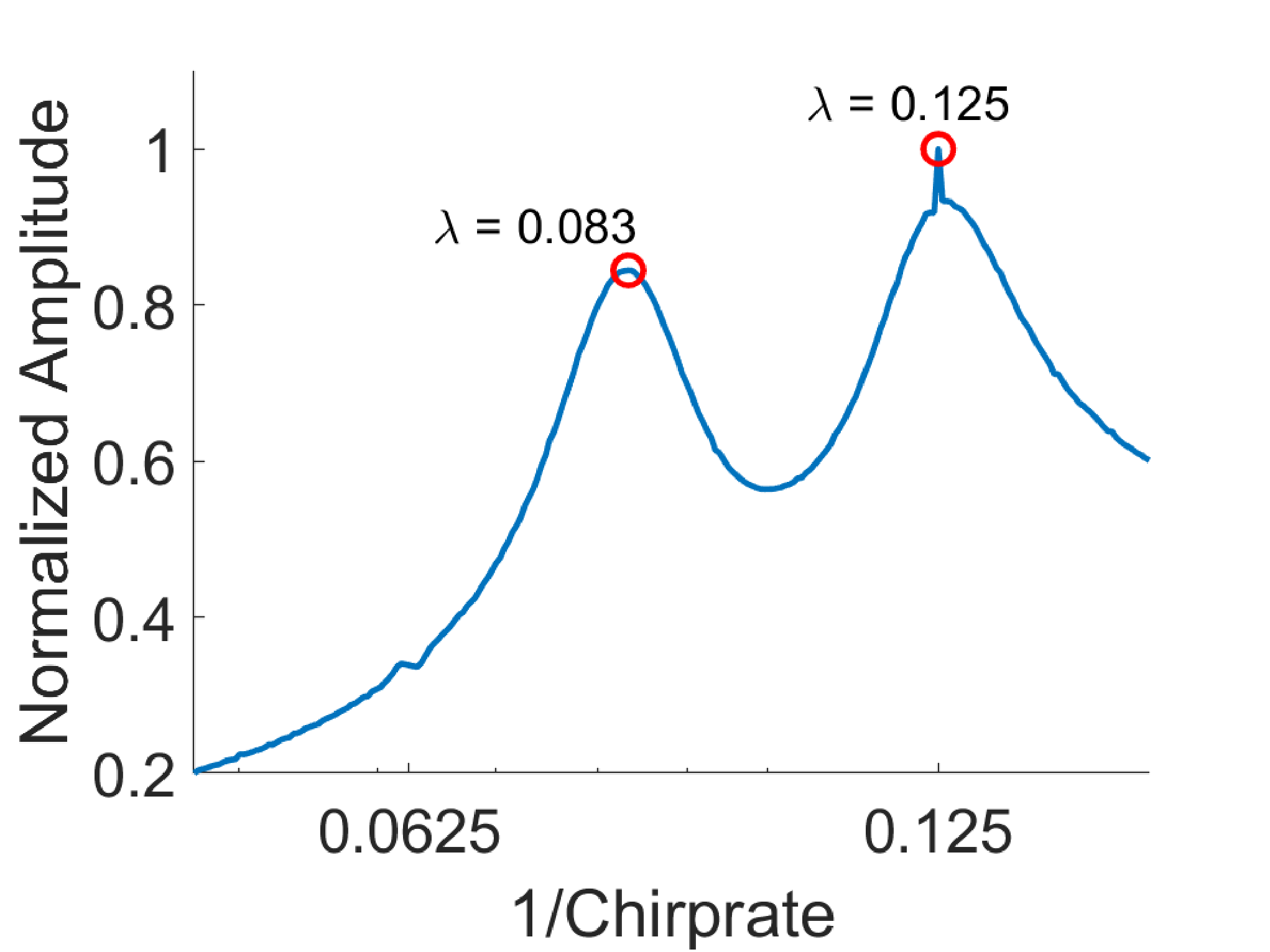}} &
        \resizebox{0.22\textwidth}{!}{\includegraphics{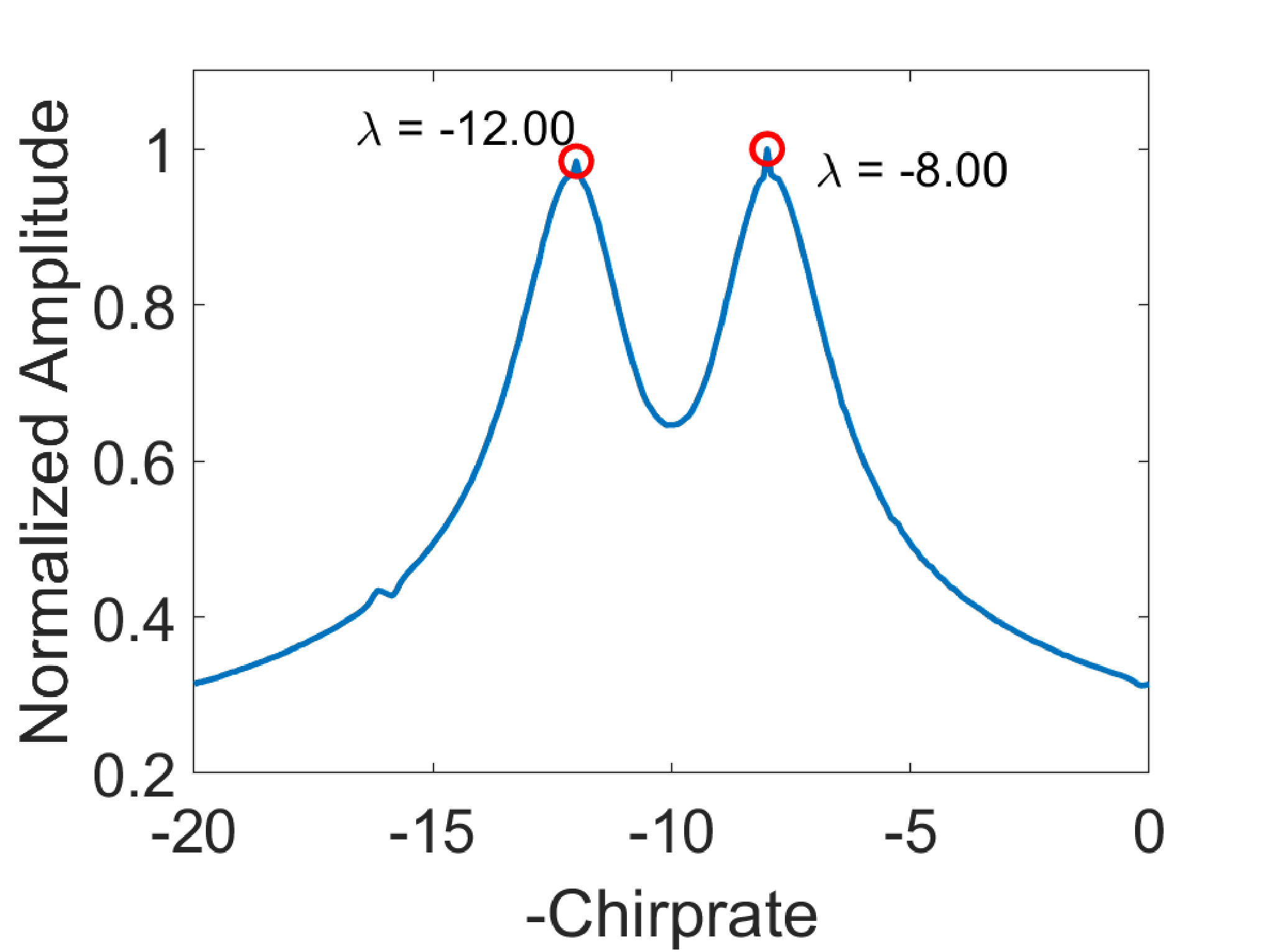}} & 
        \resizebox{0.22\textwidth}{!}{\includegraphics{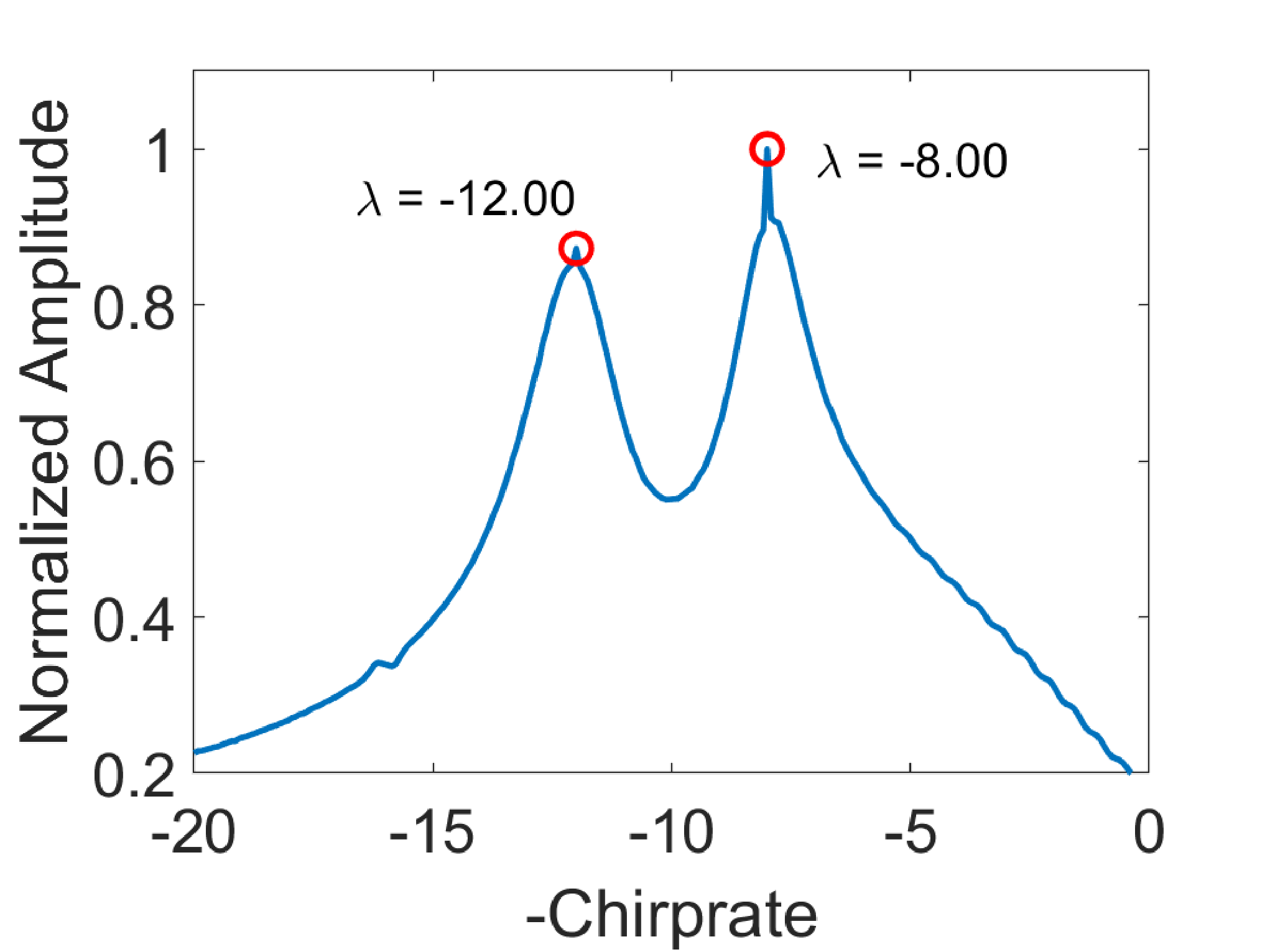}} \\
        (a)  & (b)  & (c) & (d)  \\
    \end{tabular}
    \caption{\small XWLCTs of $y(t)$ at $t=2$ s, $\eta=36$ Hz. (a) $\mathcal{T}_y^{g,1}(2,36,\lambda)$; (b) $\mathcal{T}_y^{g,5}(2,36,\lambda)$; (c) $\mathcal{T}_y^{g,2}(2,36,\lambda)$; (d) $\mathcal{T}_y^{g,6}(2,36,\lambda)$.}
    \label{figure:XWLCT_Y_crossingpoint}
\end{figure}

To illustrate the capability of the XWLCTs, we present  the slice of \(\mathcal{T}_{y}^{g, n}(t,\eta,\lambda)\) at the intersection point  in Fig.~\ref{figure:XWLCT_Y_crossingpoint}. 
As shown in panels (c) and (d) of Fig.~\ref{figure:XWLCT_Y_crossingpoint}, two distinct peaks are clearly observed when \(\lambda\) is set to \(-12\) and \(-8\), whereas panels (a) and (b) show the peaks occur at \(\lambda=0.125\) and \(\lambda=0.083\) respectively.
By comparing with Fig.~\ref{figure:WLCT_Y_crossingpoint}, it can be observed that the XWLCTs $\mathcal{T}_{y}^{g, n}(t,\eta,\lambda)$ significantly increase the decay rate in the \(\lambda\) direction, thereby forming a more concentrated time-frequency-chirprate representation. 

To further enhance the XWLCTs, we introduce the synchrosqueezed XWLCTs (SXWLCTs) based on the XWLCTs.

\begin{mdef}
Let  $\mathcal{T}_x^{ g, n}(t,\eta,\lambda)$ be the  XWLCTs of a signal $x(t)$ defined by  \eqref{def_XWLCT_15} or \eqref{def_XWLCT_26}.
 The SXWLCTs of $x(t)$ are defined as follows:
	\begin{align}
&\hbox{For $n=1, 5$}, \quad 		\mathcal{S}^n_x(t,\xi,\gamma):=	\iint_{O_{t, n}} \mathcal{T}_x^{ g, n}(t,\eta,\lambda) \delta(\gamma-1/{\Lambda}_x^{g, n}(t,\eta,\lambda)) 
		\delta(\xi- {\Omega}_x^{g, n}(t,\eta,\lambda)) \, d\eta \, d\lambda;
	\end{align}
	\begin{align}
&\hbox{and for $n=2, 6$}, \quad 	\mathcal{S}^n_x(t,\xi,\gamma):=	\iint_{O_{t, n}} \mathcal{T}_x^{ g, n}(t,\eta,\lambda)
		\delta(\gamma+{\Lambda}_x^{g, n}(t,\eta,\lambda)) 
		\delta(\xi- {\Omega}_x^{g, n}(t,\eta,\lambda)) \, d\eta \, d\lambda, 
	\end{align}
where \({\Omega}_x^{g, n}(t,\eta,\lambda)\), \({\Lambda}_x^{g, n}(t,\eta,\lambda)\), and $O_{t, n}$  are defined by  \eqref{eq:omega} and \eqref{def_Otn} respectively.
	\end{mdef}

\begin{figure}[H]
    \centering
    \begin{tabular}{ccc}
        \resizebox{1.8in}{1.1in}{\includegraphics{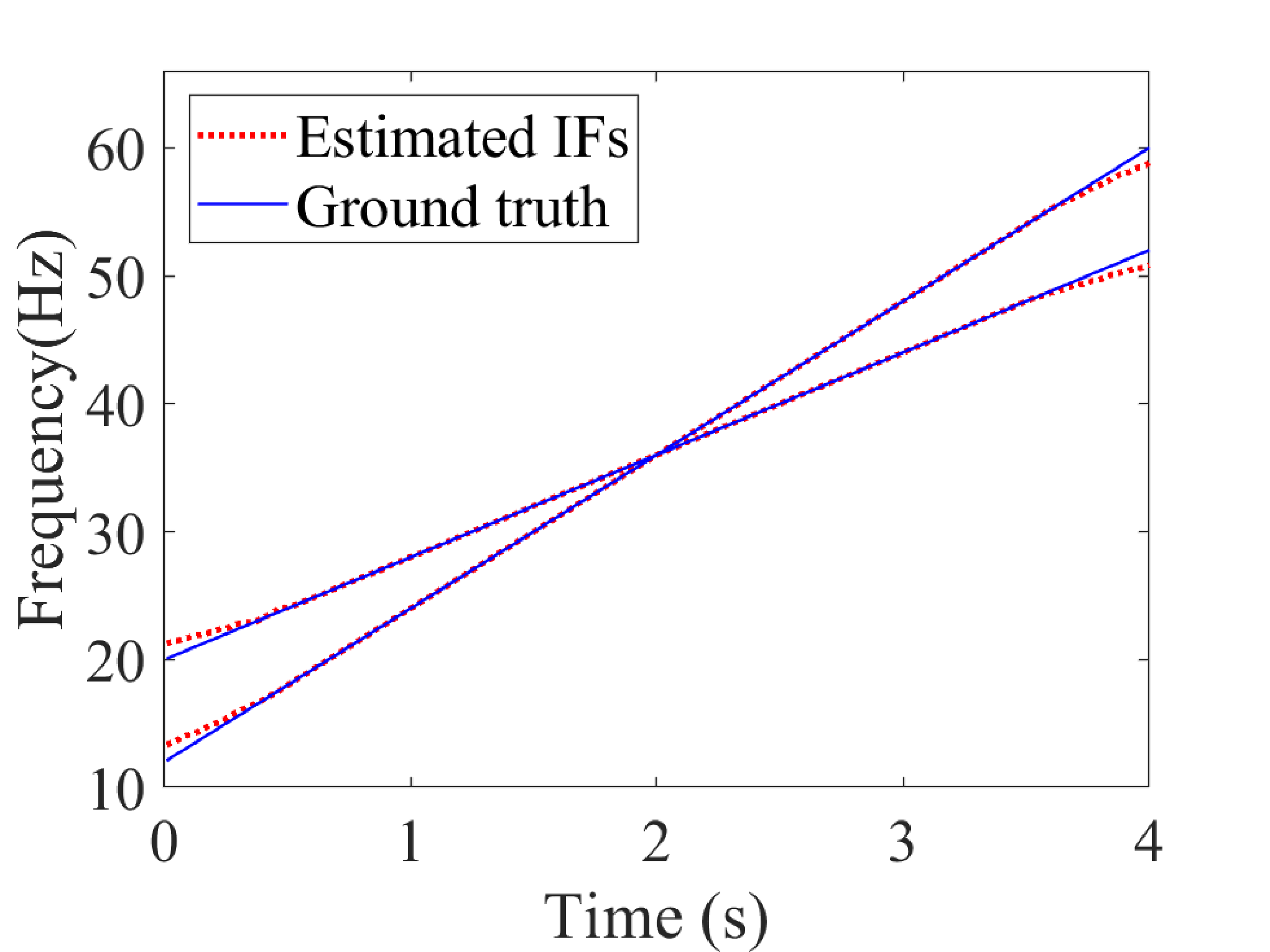}} & 
        \resizebox{1.8in}{1.1in}{\includegraphics{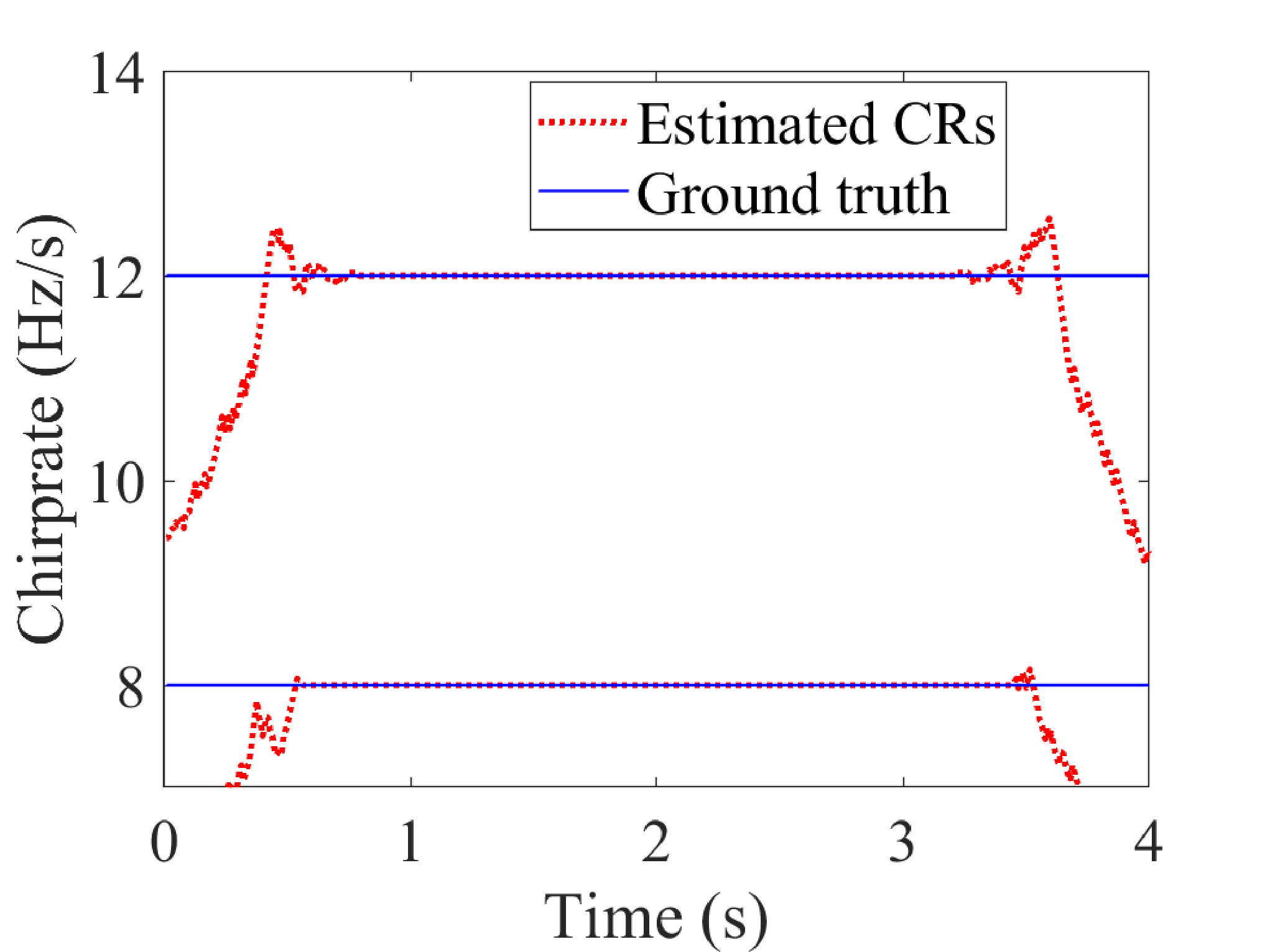}} &
        \resizebox{1.8in}{1.1in}{\includegraphics{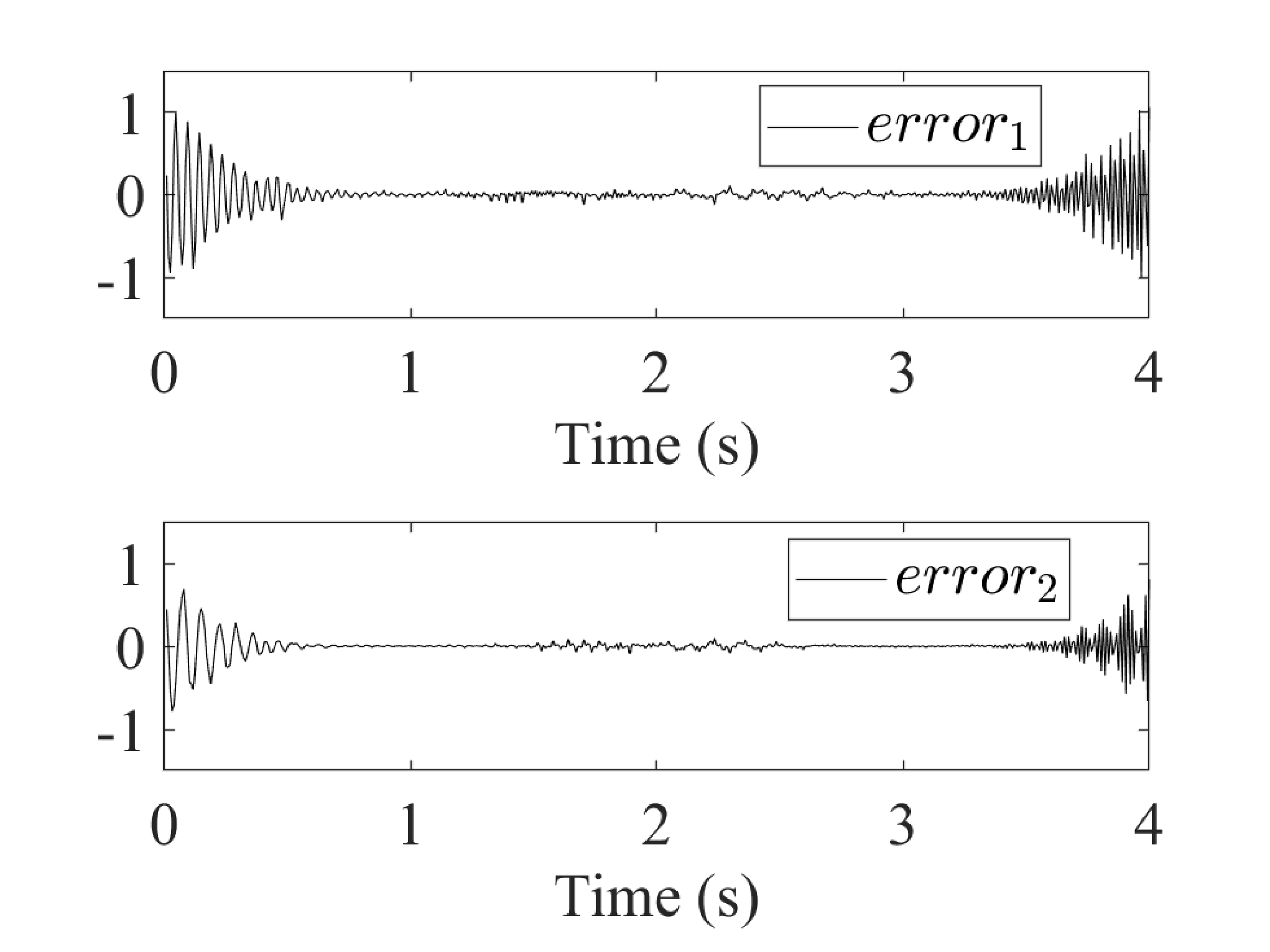}} \\
        \resizebox{1.8in}{1.1in}{\includegraphics{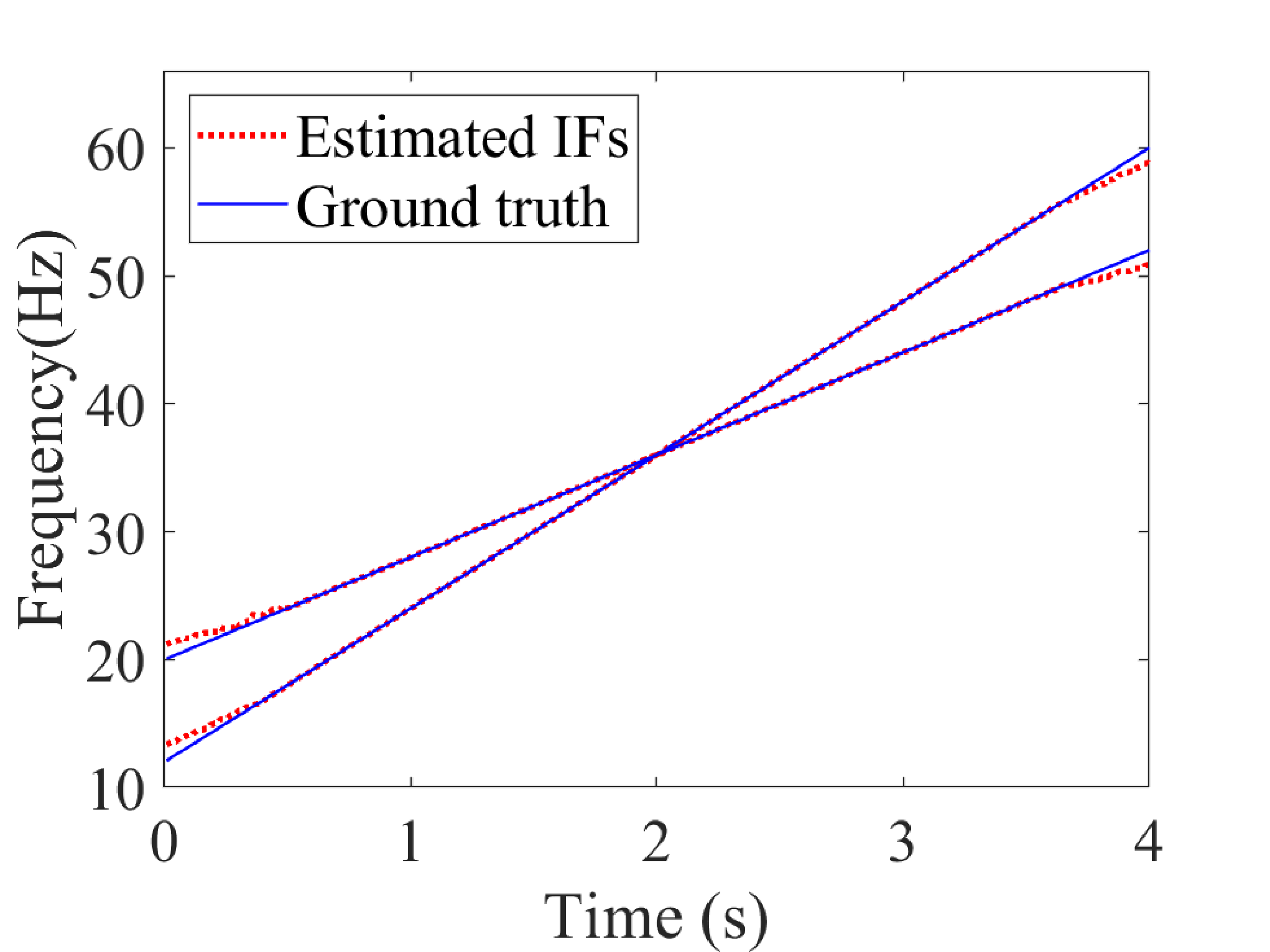}} & 
        \resizebox{1.8in}{1.1in}{\includegraphics{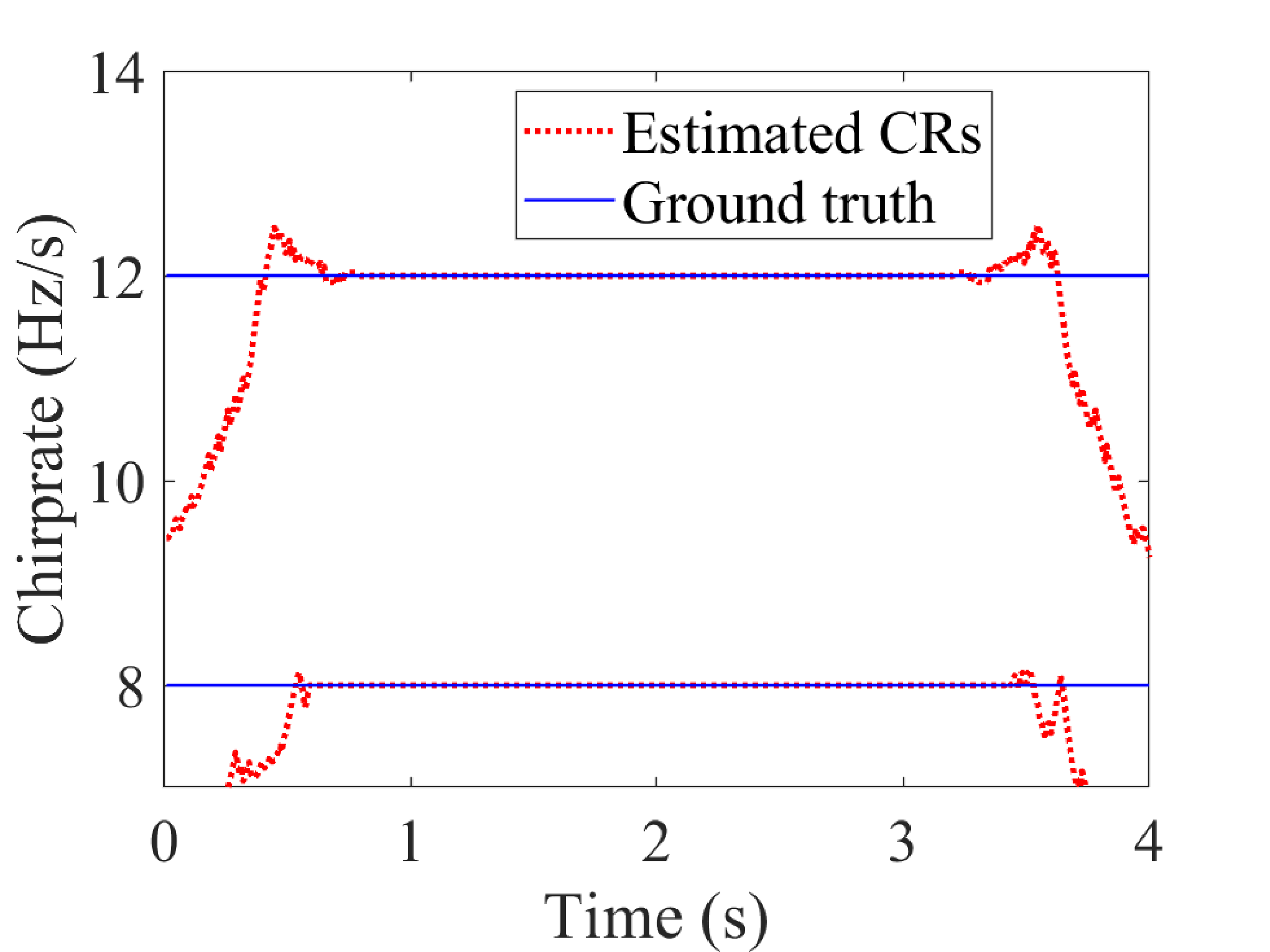}} &
        \resizebox{1.8in}{1.1in}{\includegraphics{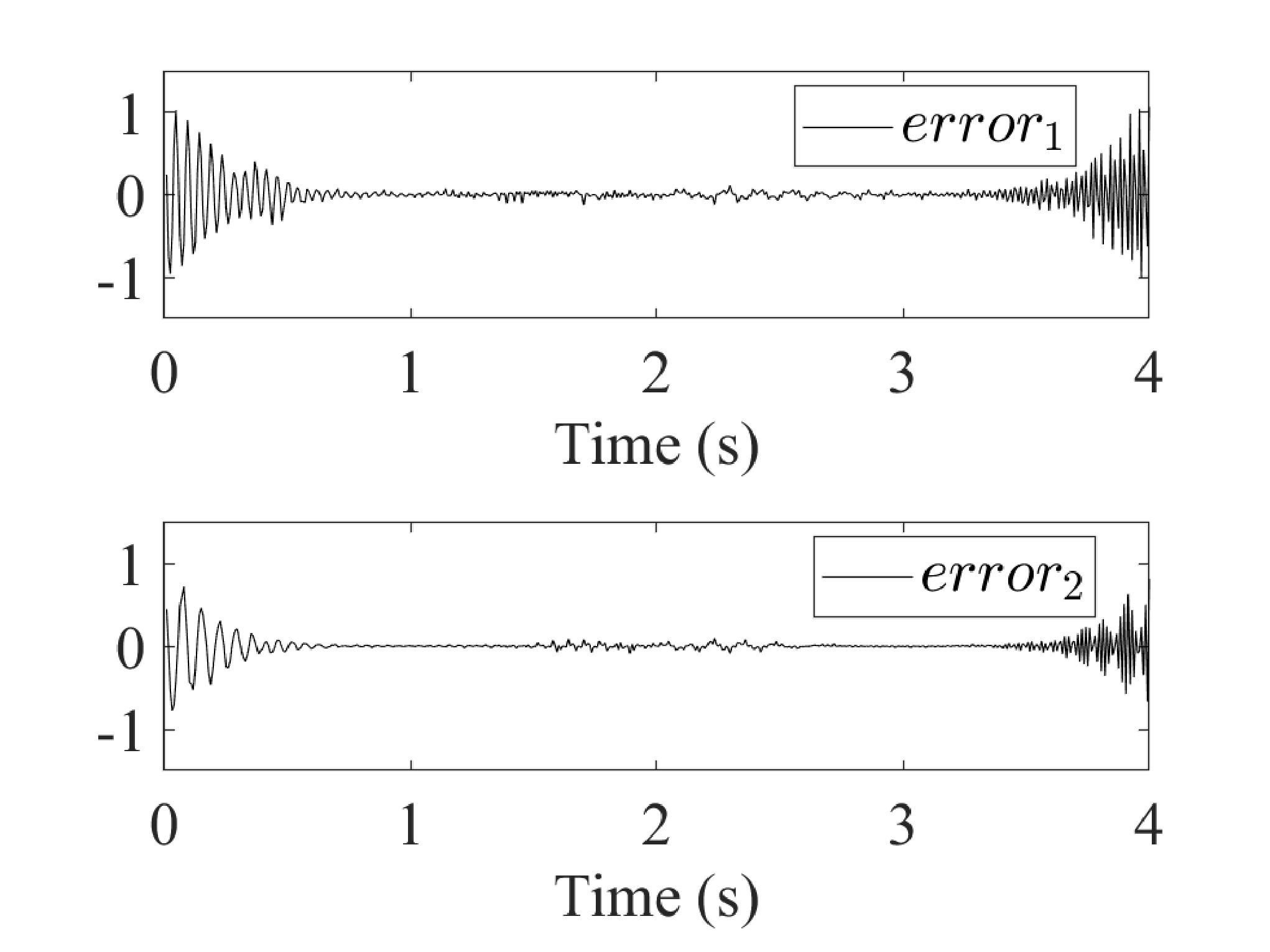}} \\
        \resizebox{1.8in}{1.1in}{\includegraphics{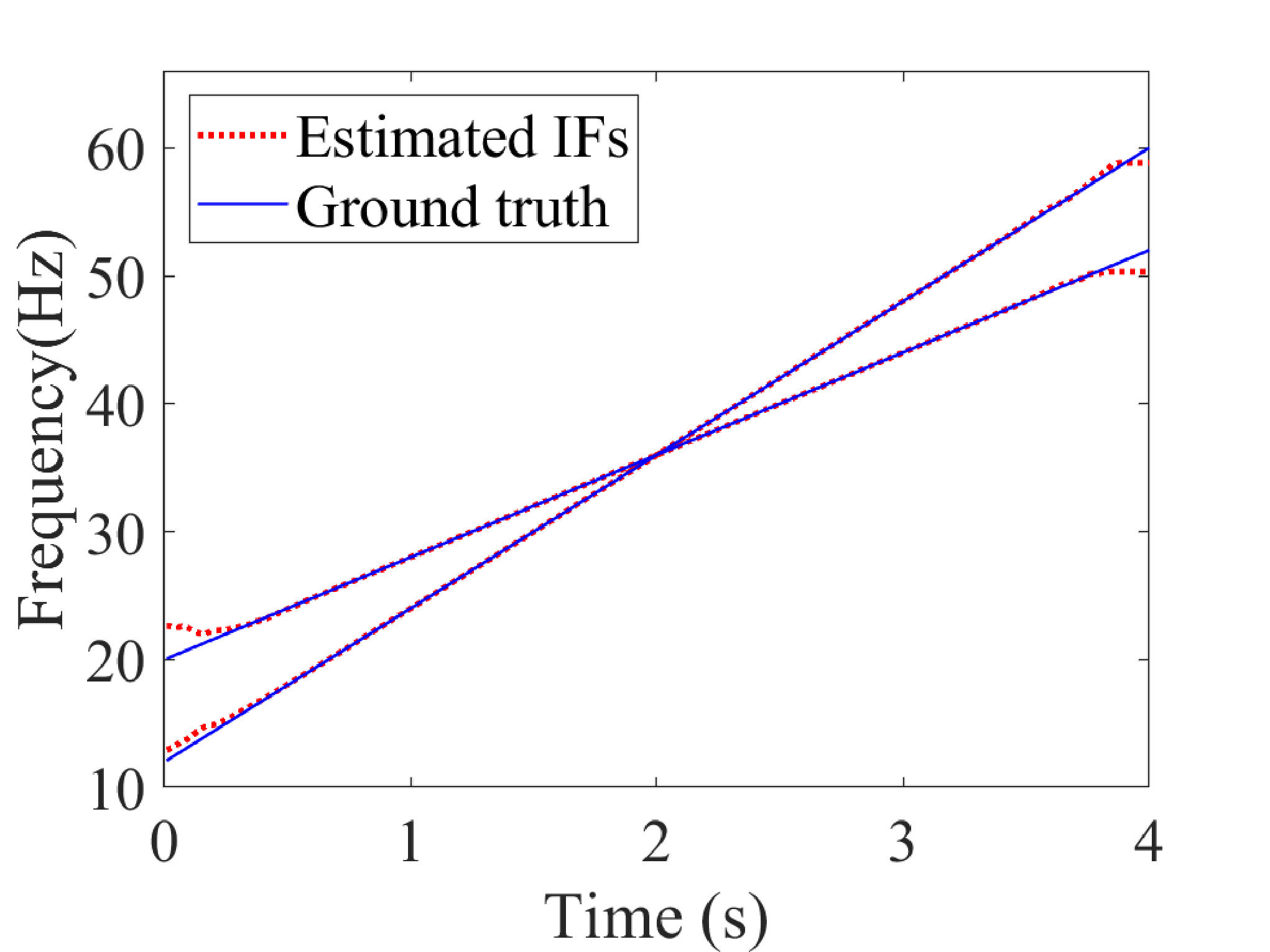}} & 
        \resizebox{1.8in}{1.1in}{\includegraphics{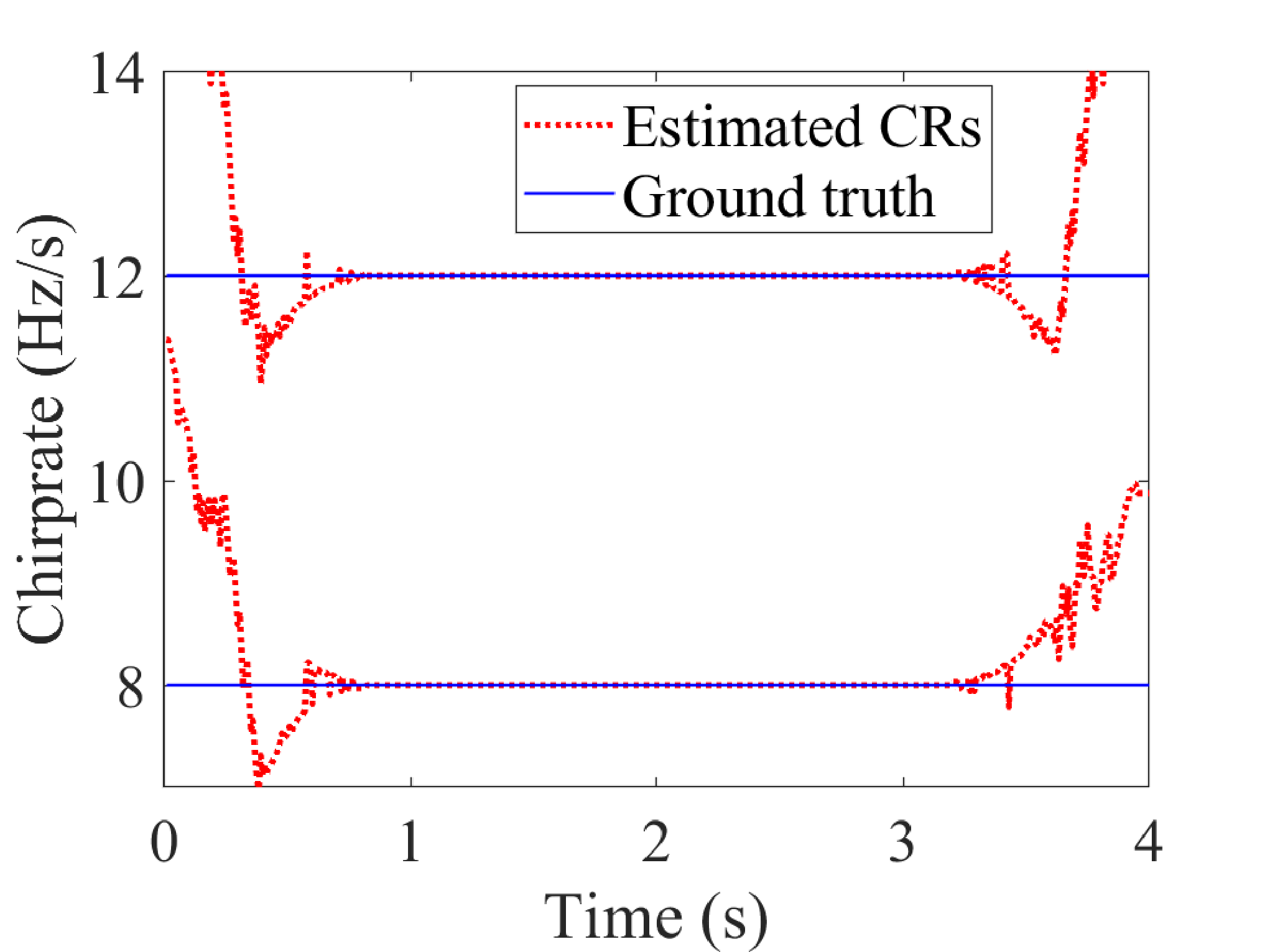}} &
        \resizebox{1.8in}{1.1in}{\includegraphics{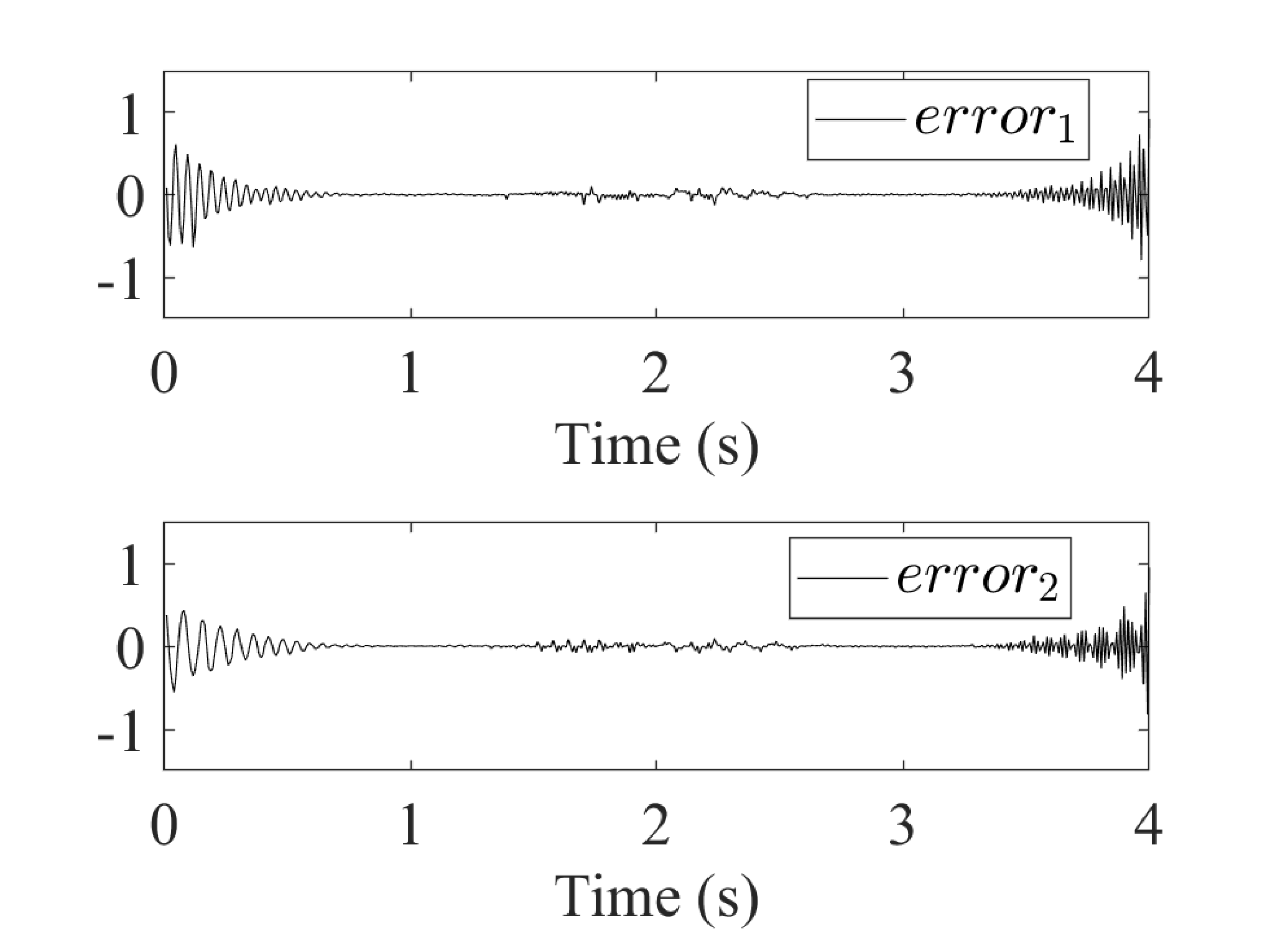}} \\
        \resizebox{1.8in}{1.1in}{\includegraphics{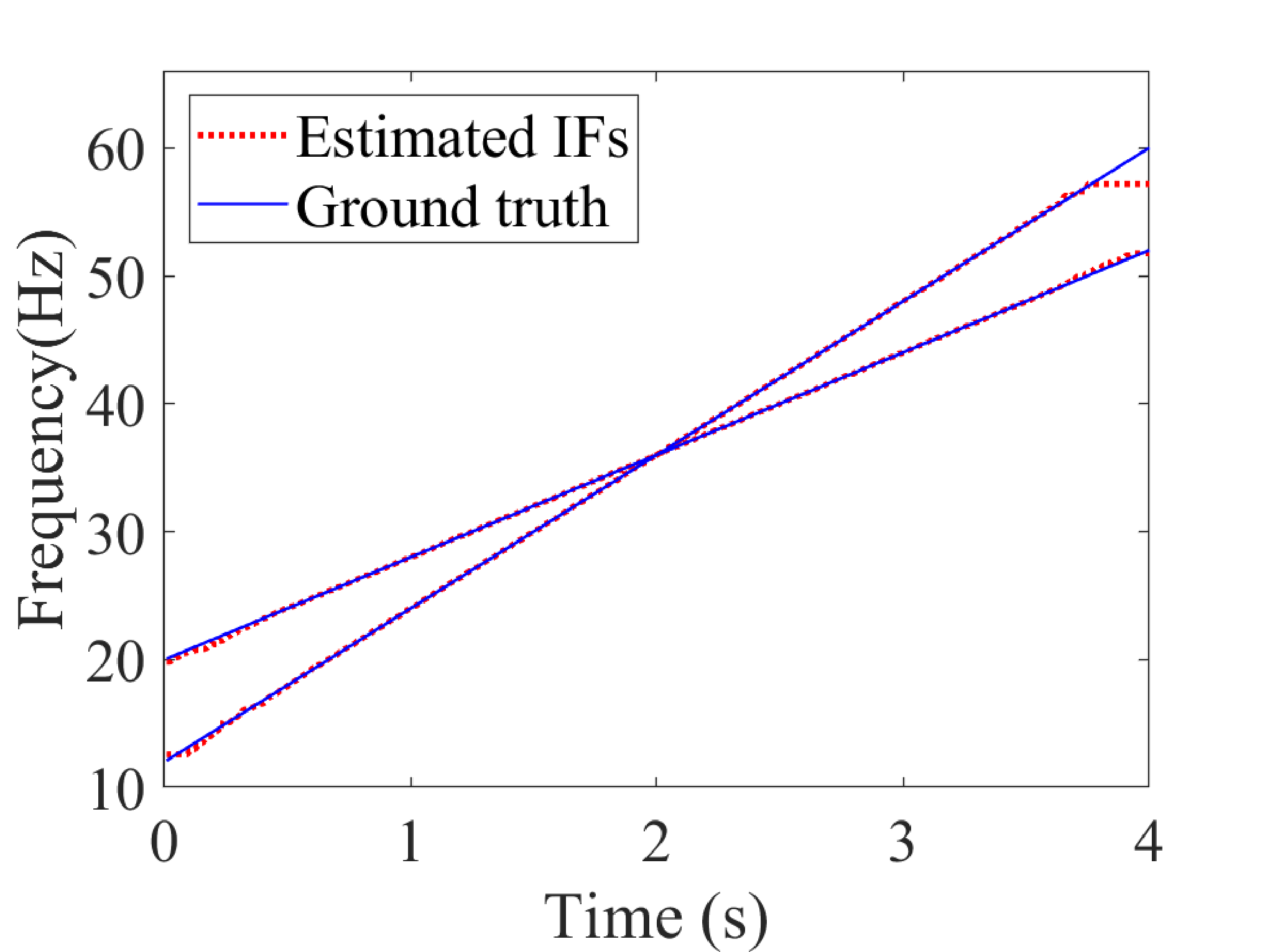}} & 
        \resizebox{1.8in}{1.1in}{\includegraphics{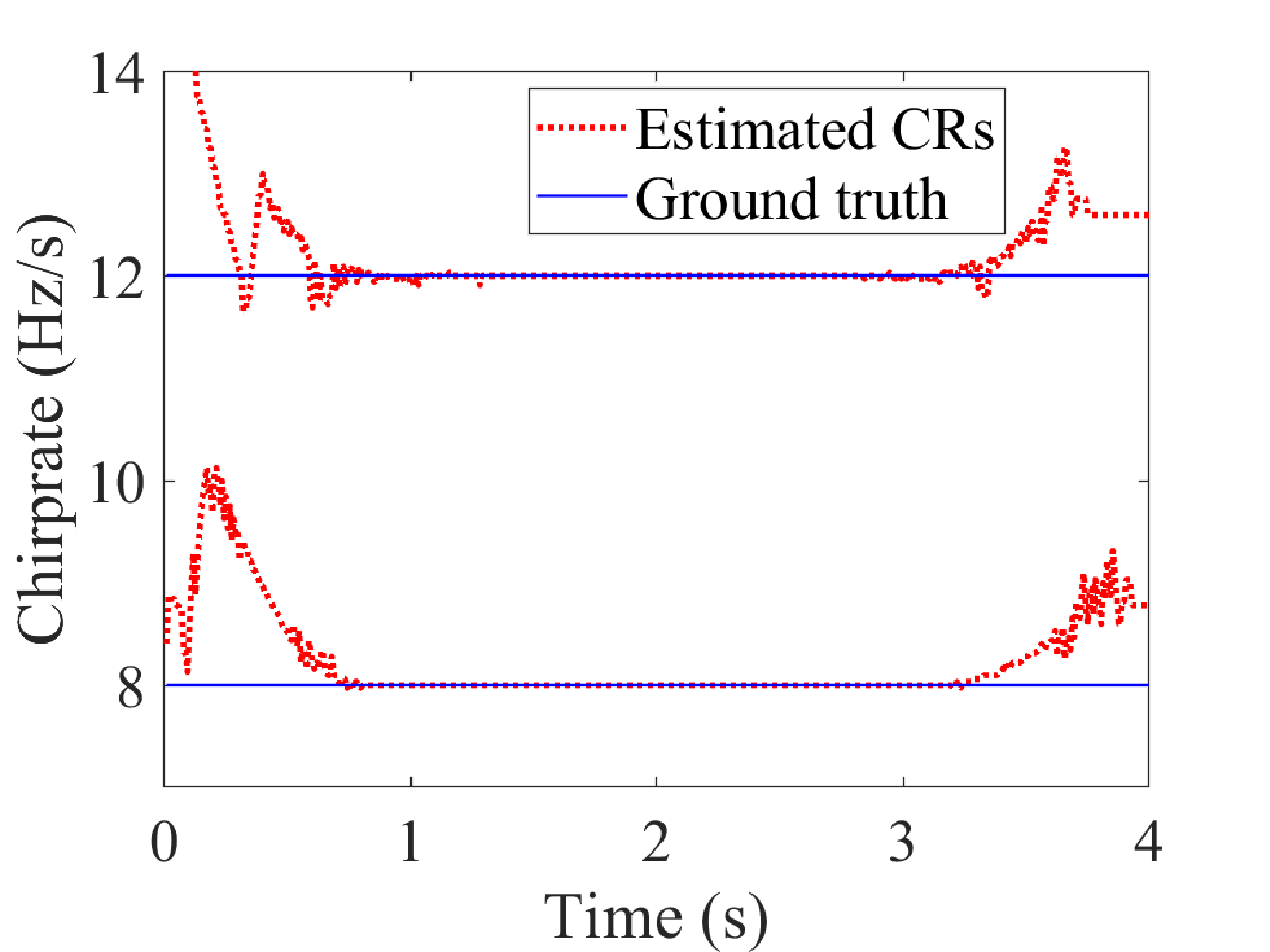}} &
        \resizebox{1.8in}{1.1in}{\includegraphics{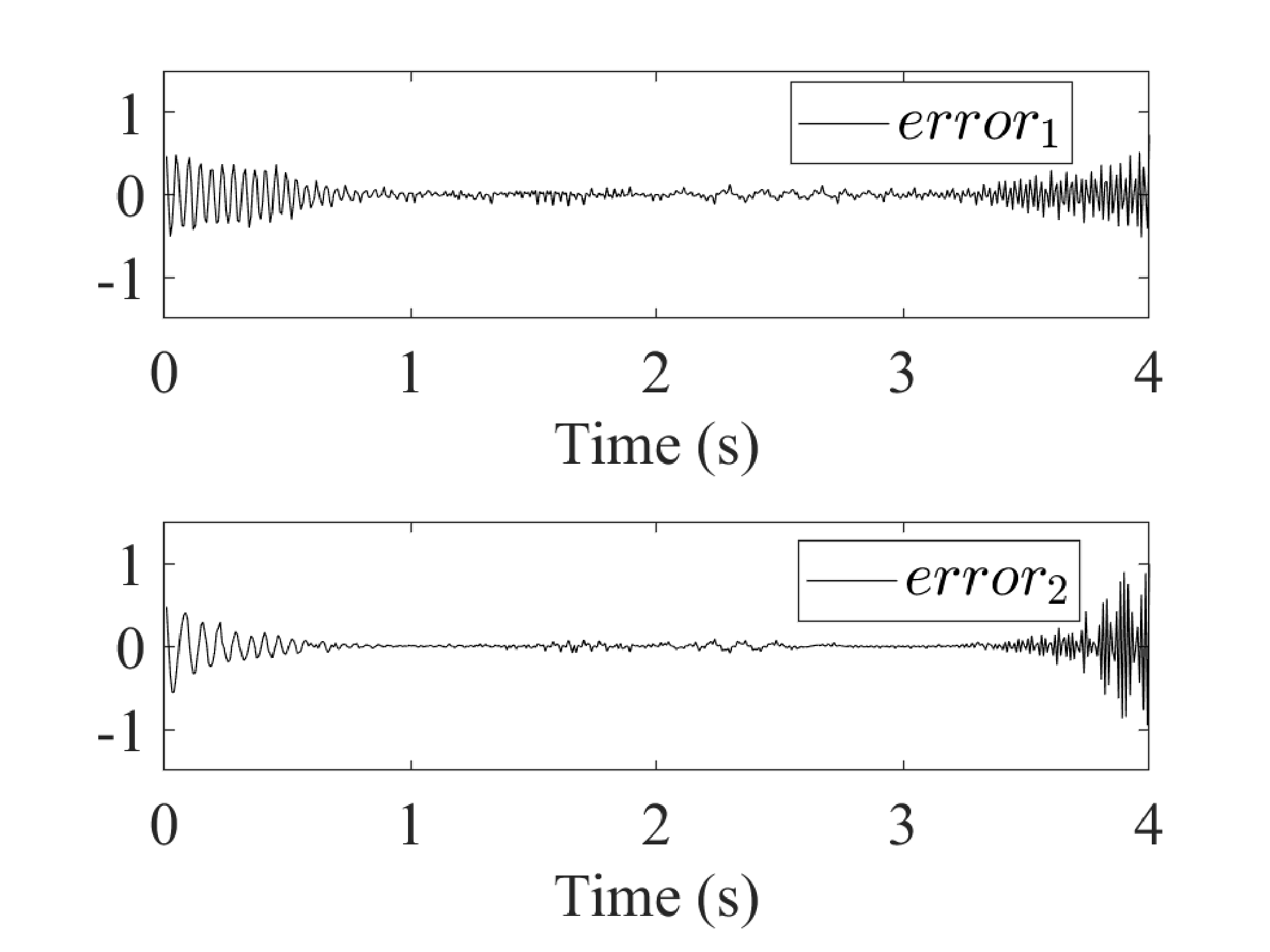}} \\
    \end{tabular}
    	\caption{\small  IF and chirprate estimations, and real part errors of mode retrieval by SXWLCT.
      First row by \(\mathcal{S}^1_y(t, \xi, \gamma)\), Second row by \(\mathcal{S}^5_y(t, \xi, \gamma)\),  Third row by \(\mathcal{S}^2_y(t, \xi, \gamma)\) and Fourth row by \(\mathcal{S}^6_y(t, \xi, \gamma)\).} 
     \label{figure:SXWLCT of $Y$}
\end{figure}

The results of the IF estimations (first column), chirprate estimations (second column), and the real-part errors of mode retrieval (third column) obtained via SXWLCTs are presented in Fig.~\ref{figure:SXWLCT of $Y$}. 
Compared with SWLCTs (see Fig.~\ref{figure:SWLCT of $Y$}), the SXWLCTs show higher accuracy in both IF and chirprate estimation, with improved mode retrieval capability.
Notably, in the mode retrieval process, the  $\mathcal{S}^1_x(t, \xi, \gamma)$, $\mathcal{S}^5_x(t, \xi, \gamma)$, and $\mathcal{S}^6_x(t, \xi, \gamma)$ exhibit significantly smaller errors when recovering high-chirprate components.

\begin{figure}[H]
    \centering
    \begin{tabular}{cccc}
        \resizebox{1.5in}{!}{\includegraphics{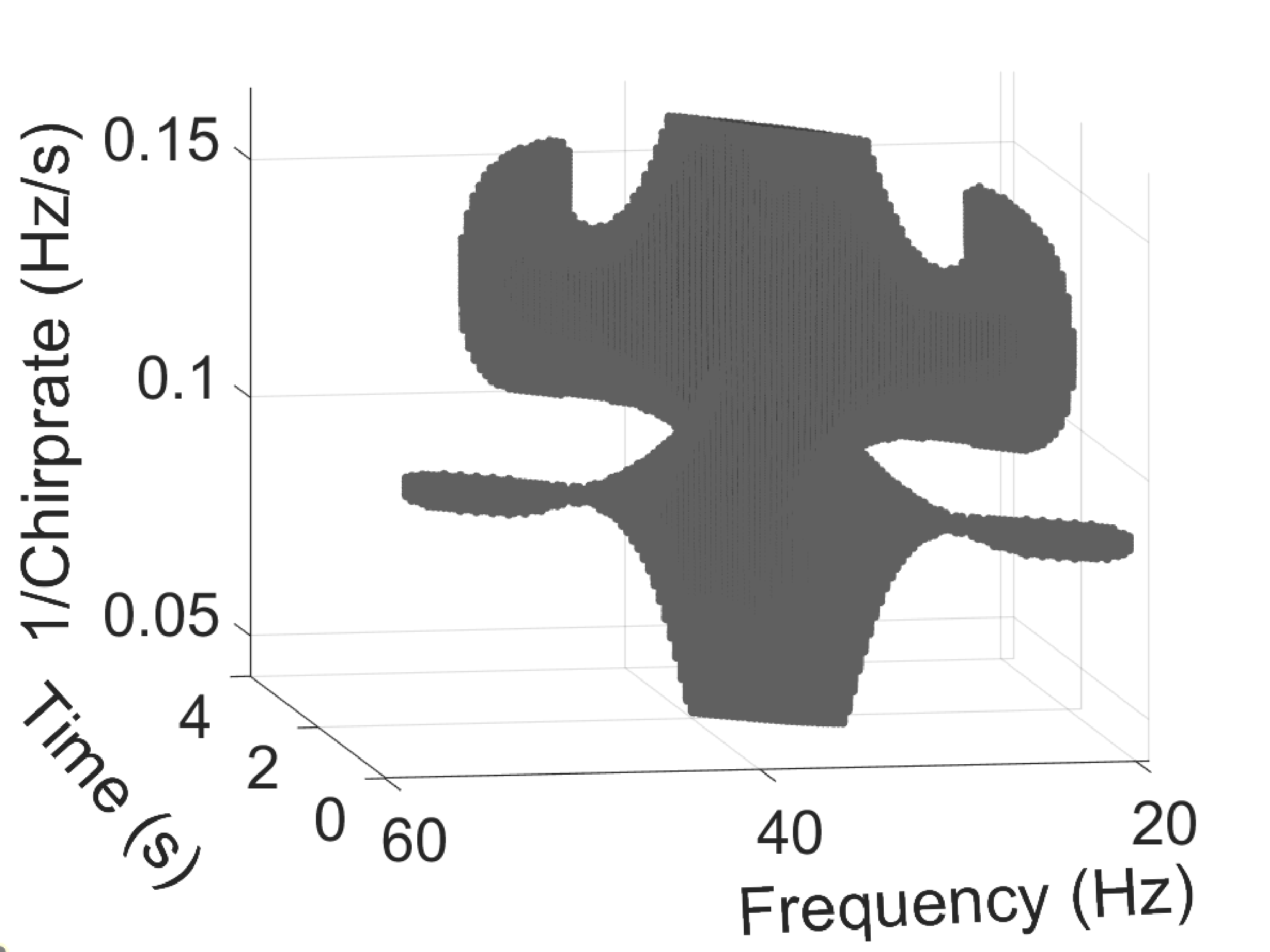}} & 
        \resizebox{1.5in}{!}{\includegraphics{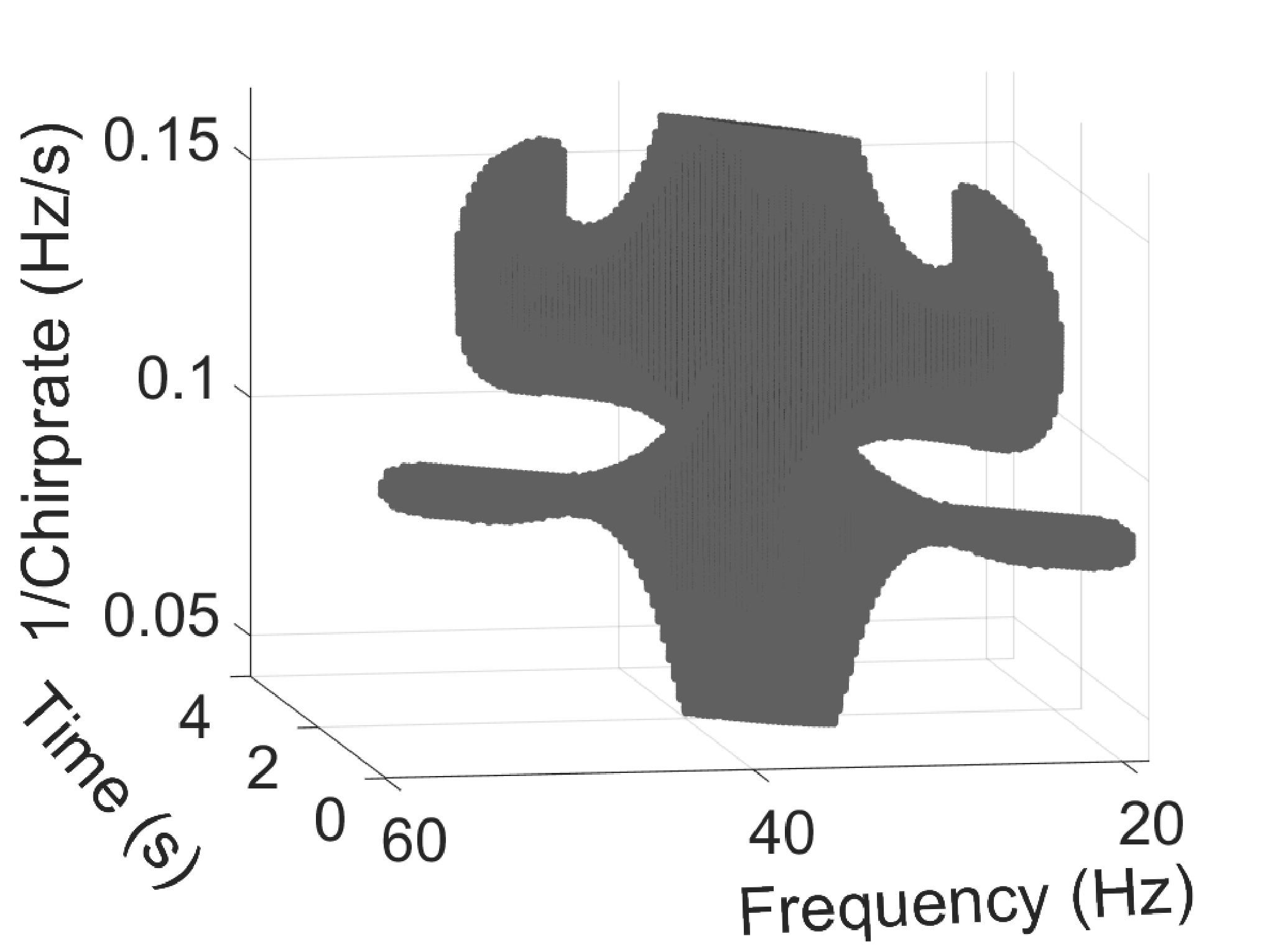}} & 
        \resizebox{1.5in}{!}{\includegraphics{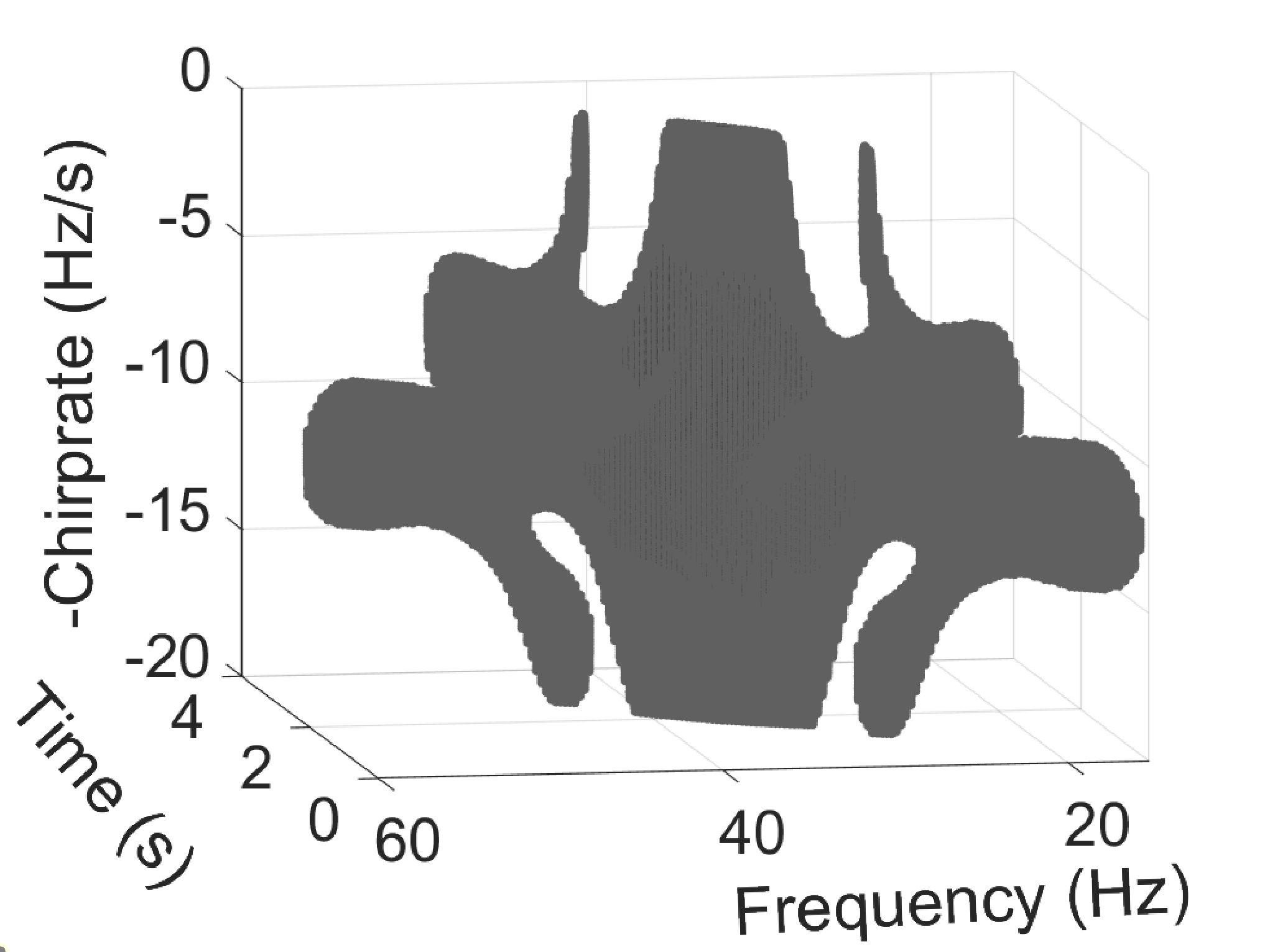}} &
        \resizebox{1.5in}{!}{\includegraphics{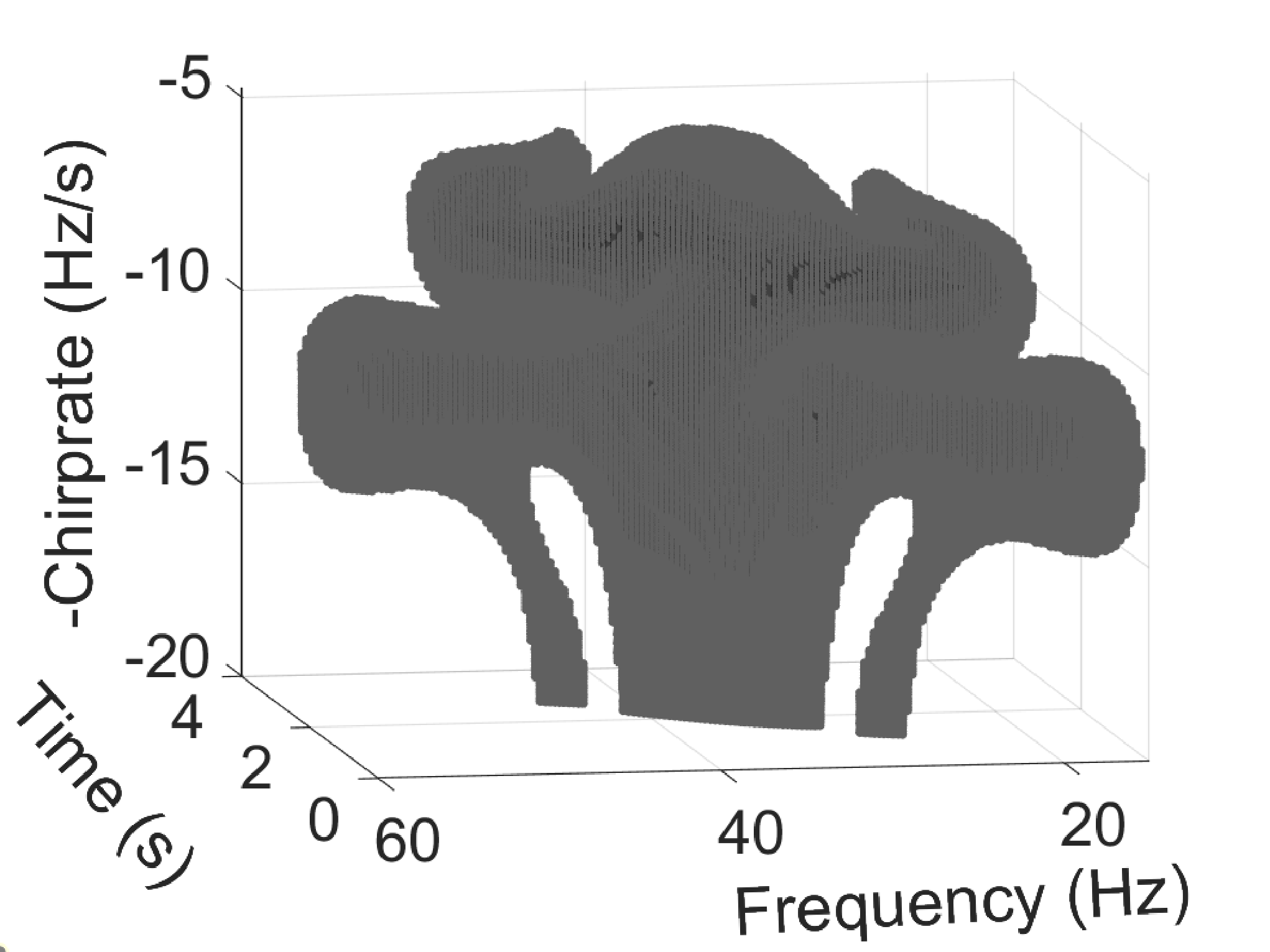}} \\
        \resizebox{1.5in}{!}{\includegraphics{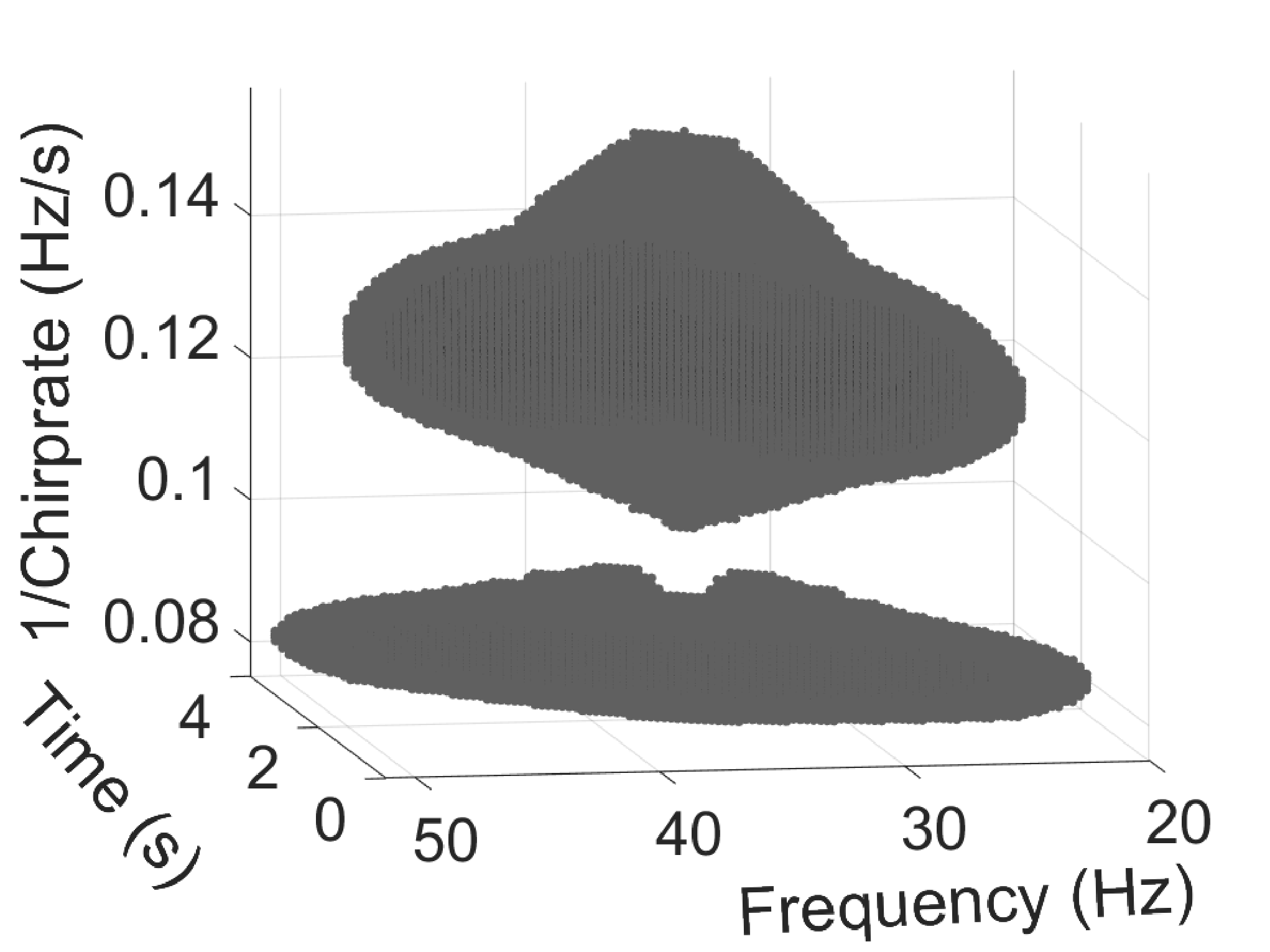}} & 
        \resizebox{1.5in}{!}{\includegraphics{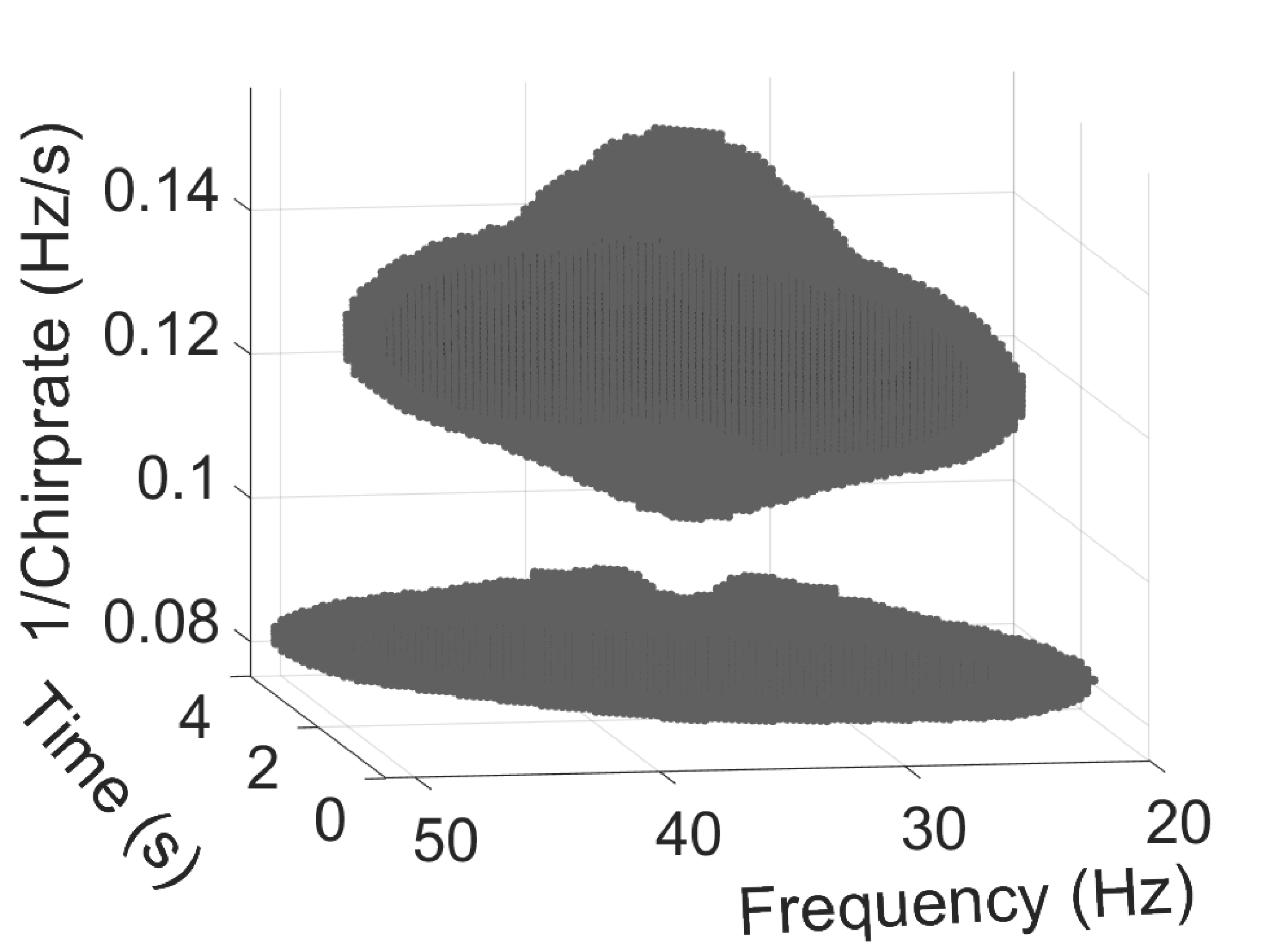}} & 
        \resizebox{1.5in}{!}{\includegraphics{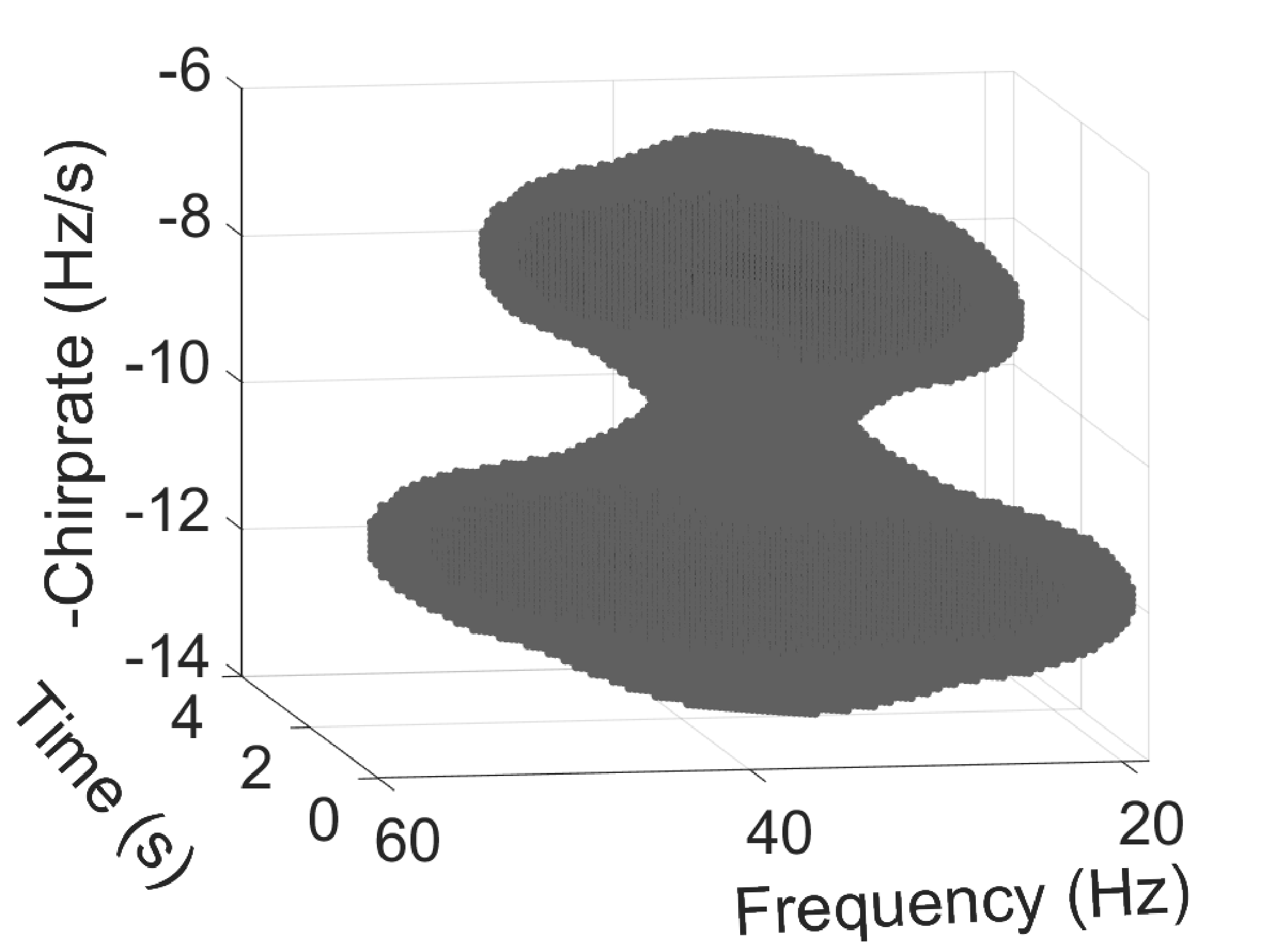}} & 
        \resizebox{1.5in}{!}{\includegraphics{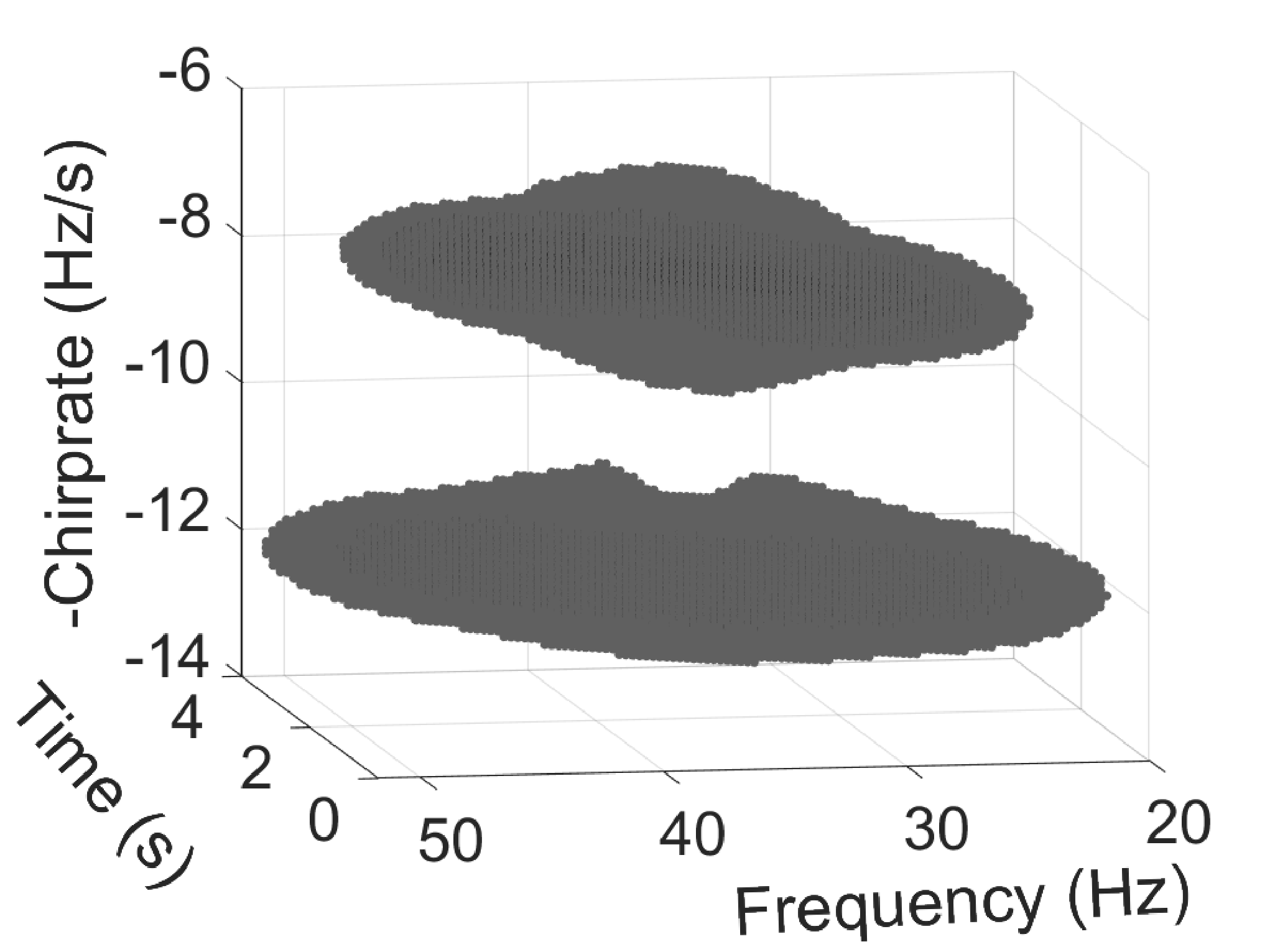}}
    \end{tabular}
	\caption{Three-dimensional representations of the WLCTs and XWLCTs. First row (left to right): WLCTs with $n=1, 5, 2, 6$; Second row: Corresponding XWLCTs.}
    \label{figure:XWLCT_Y_3D}
\end{figure}

Furthermore, for a more intuitive understanding, we present the three-dimensional distributions of the WLCTs and XWLCTs in Fig.~\ref{figure:XWLCT_Y_3D}. The XWLCT is designed to enhance concentration specifically along the chirprate dimension by accelerating its decay, thereby isolating the signal from cross-component interference. 
This refined representation in the chirprate domain is the key factor that allows the SXWLCTs to provide more accurate estimations of both the chirprate and the IF than the SWLCTs.

\begin{figure}[H]
    \centering
    \begin{tabular}{ccc}
        \resizebox{1.8in}{1.1in}{\includegraphics{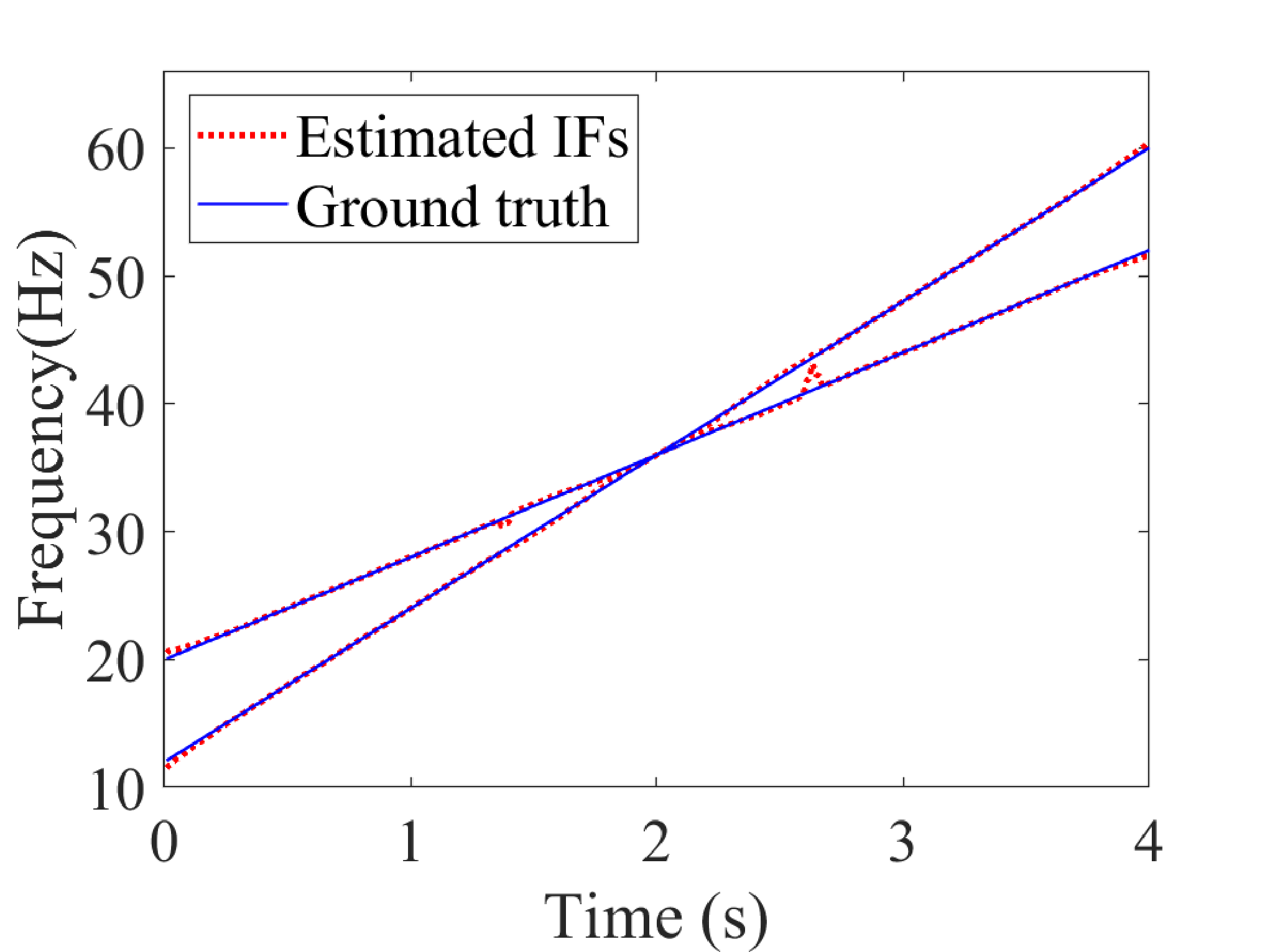}} & 
        \resizebox{1.8in}{1.1in}{\includegraphics{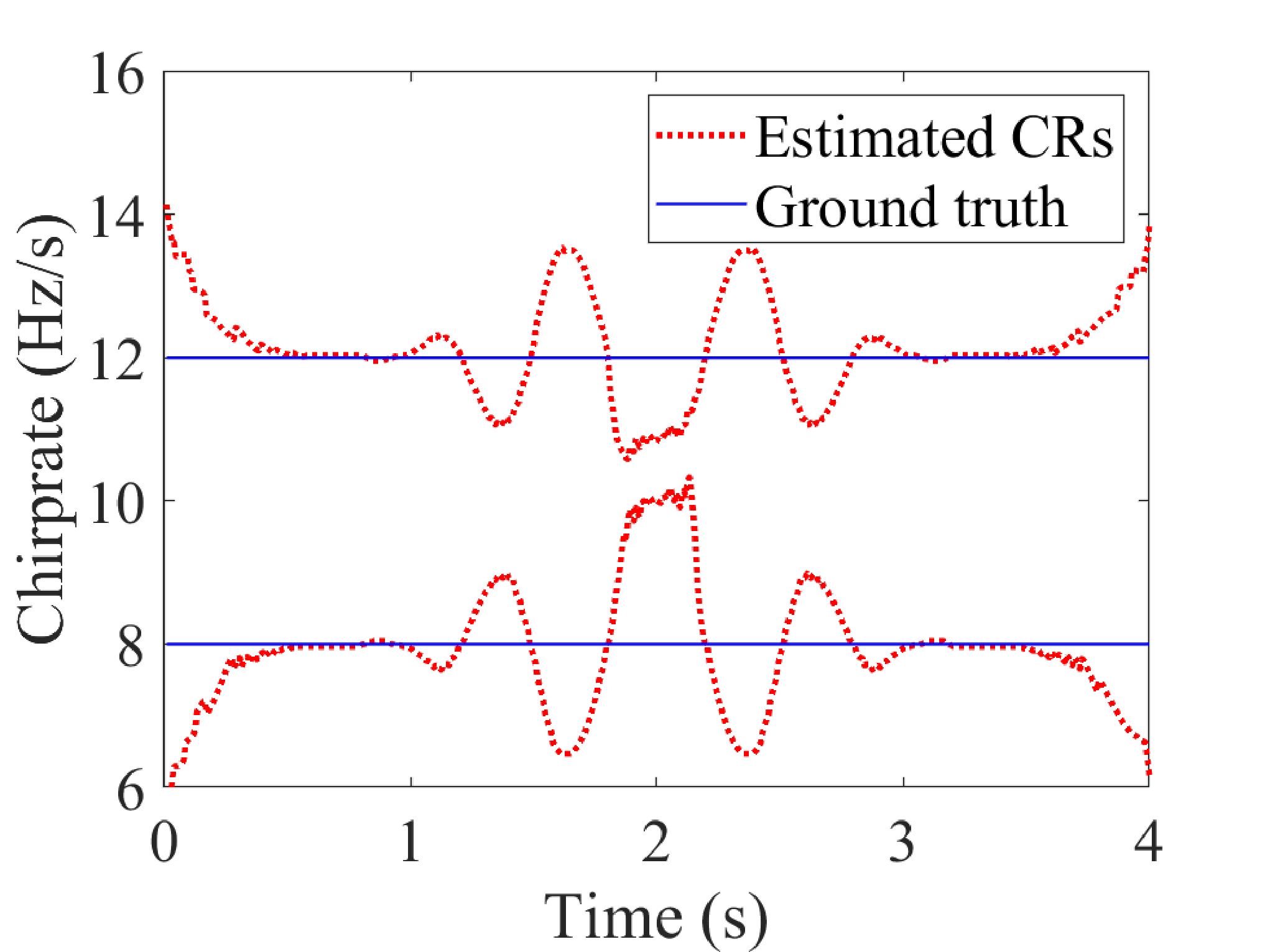}} &
        \resizebox{1.8in}{1.1in}{\includegraphics{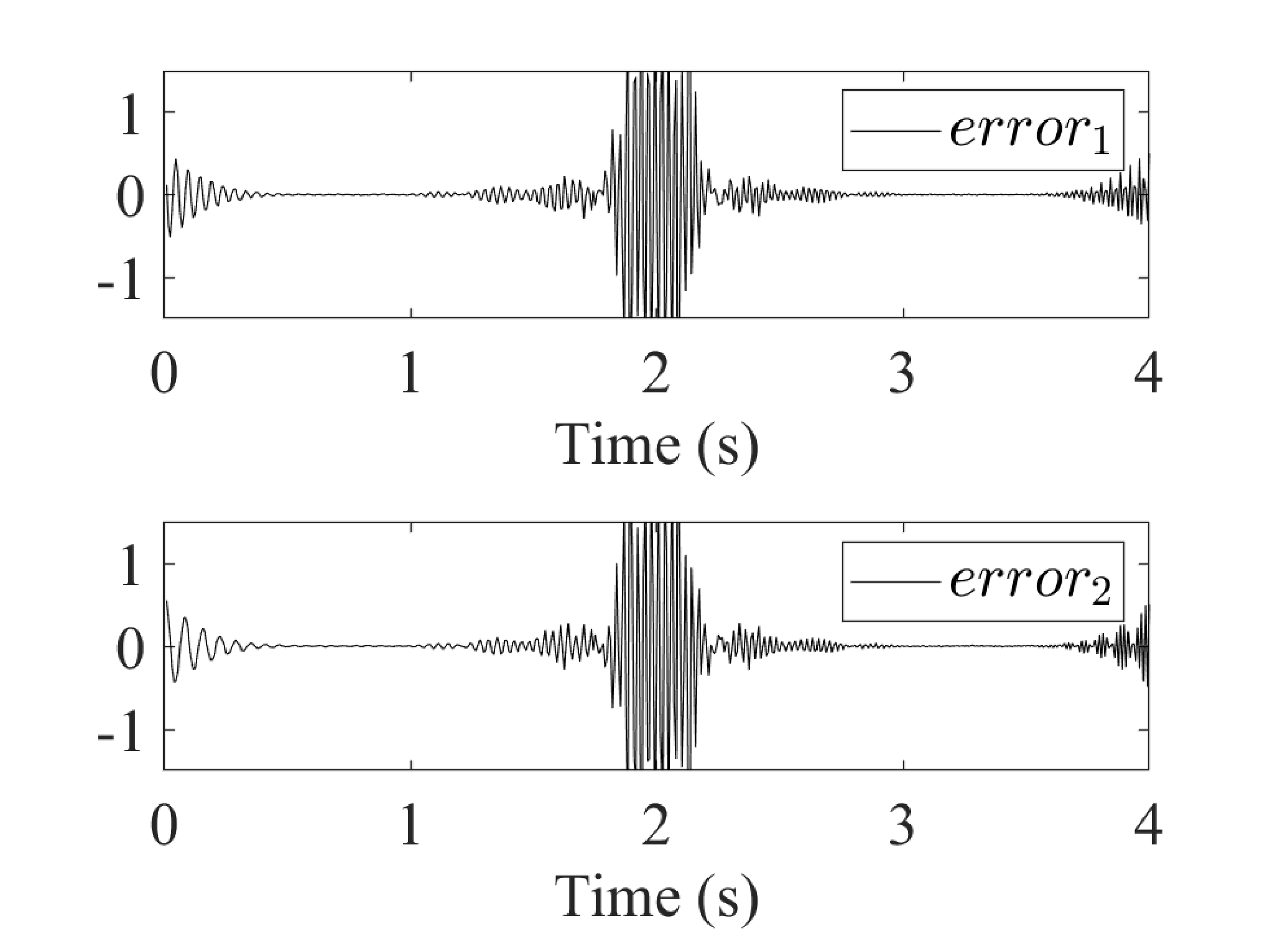}} \\
        \resizebox{1.8in}{1.1in}{\includegraphics{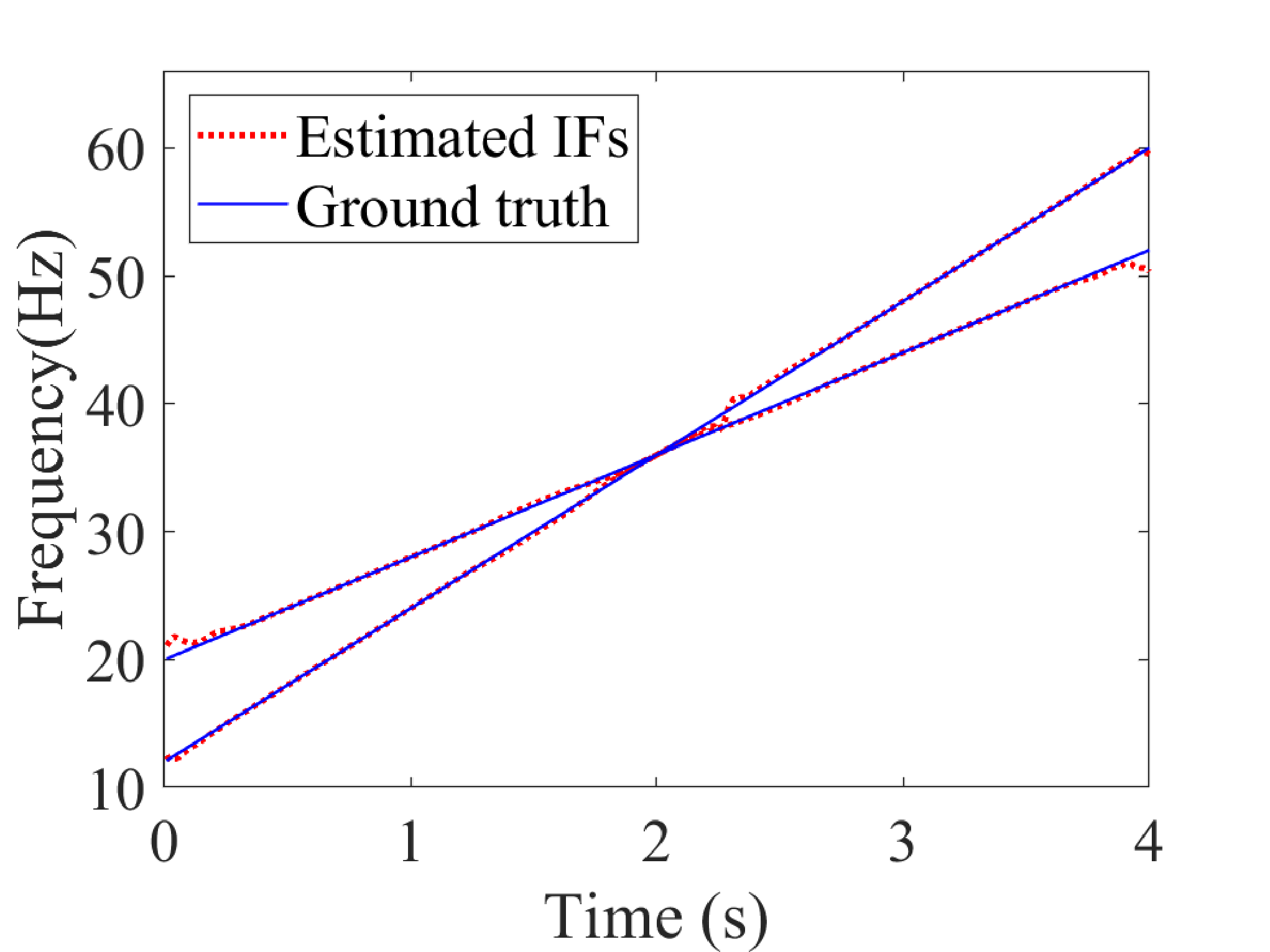}} & 
        \resizebox{1.8in}{1.1in}{\includegraphics{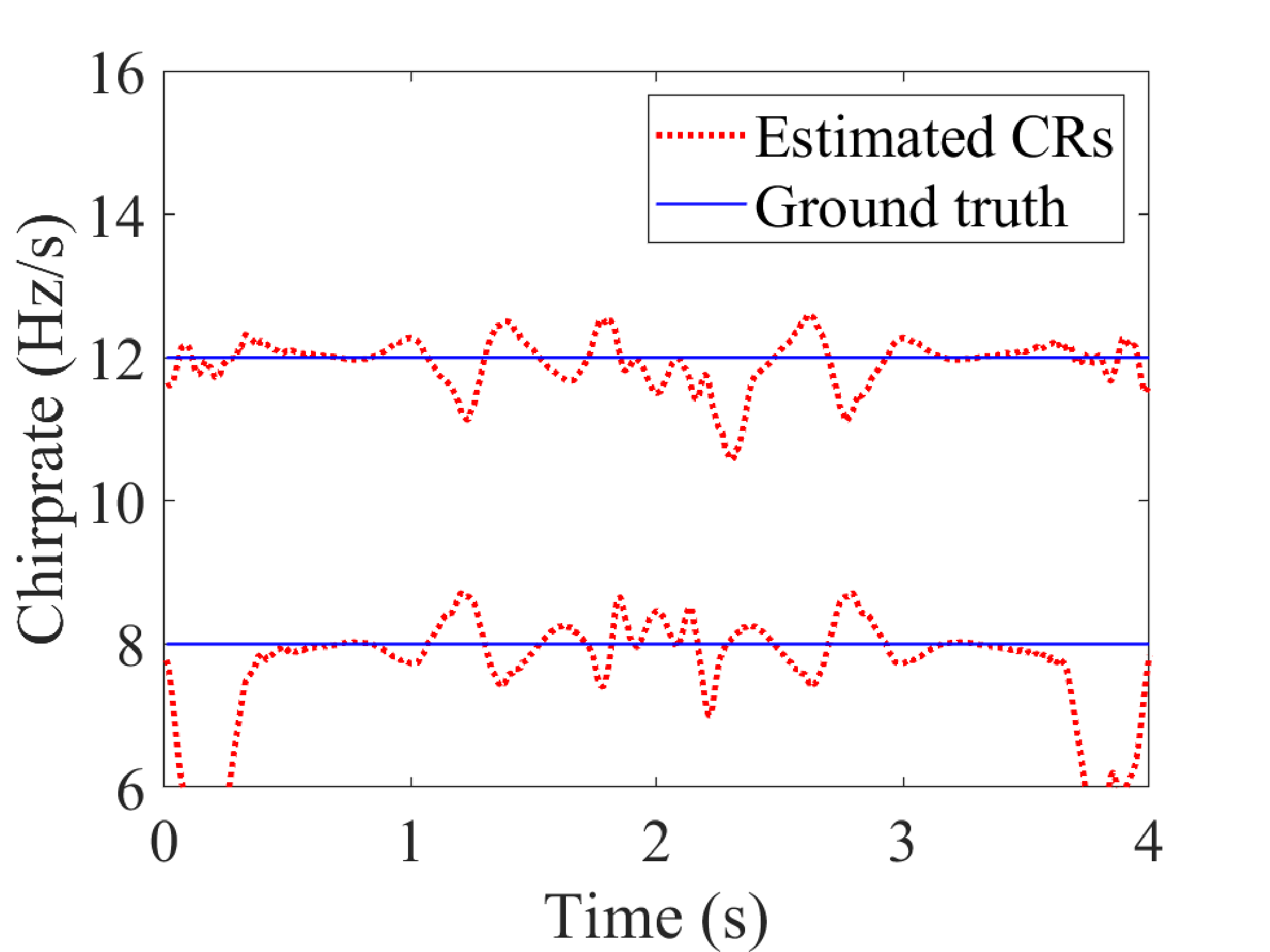}} &
        \resizebox{1.8in}{1.1in}{\includegraphics{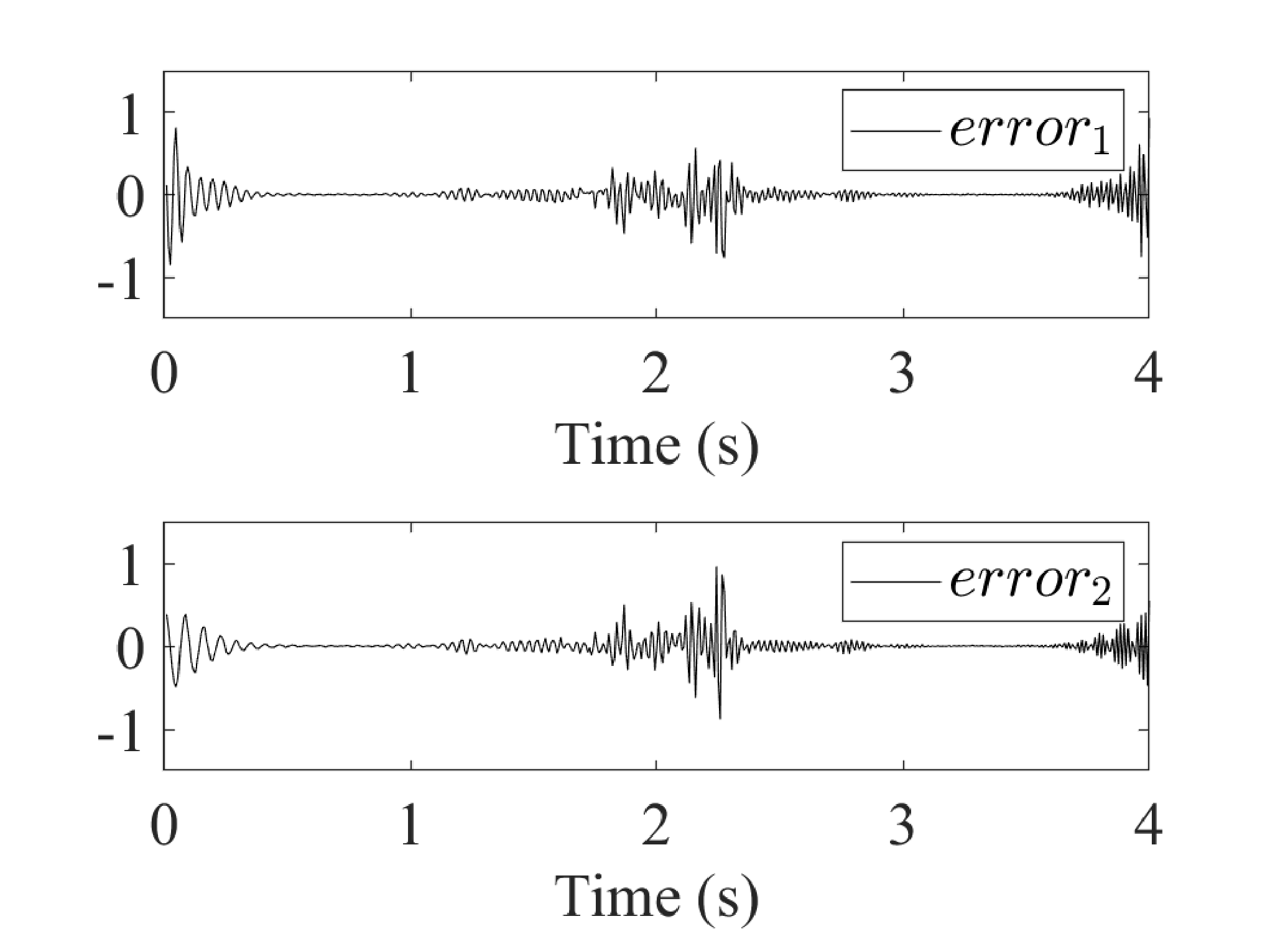}} \\
    \end{tabular}
    	\caption{\small IF and chirprate estimations, along with real part errors of mode retrieval using TET and MESCT. First row: Results from TET. Second row: Results from MESCT (with three squeezes).}
     \label{figure:TET and MESCT of y}
\end{figure}

For a comparative analysis, Fig.~\ref{figure:TET and MESCT of y} presents the results obtained by the TET \cite{zhu2021three} and MESCT \cite{chen2024multiple} methods for the signal in \eqref{example2}. 
Visually, the TET and MESCT   exhibit significant deviations in estimating the IFs and chirprates, which leads to considerable errors in the subsequent mode retrieval. 
The performance is evaluated using the RMSE of the retrieved modes, as summarized in Table \ref{tab:example2}. The consistently lower RMSE values of our method confirm its superior capability in processing such signals.

\begin{table}[H]
    \centering
    \caption{RMSEs of mode retrieval for \(y(t)\) in \eqref{example2}}
    \label{tab:example2}
    \begin{tabular}{l S[table-format=1.4] S[table-format=1.4] S[table-format=1.4] c S[table-format=1.4] c}
        \toprule
 {\(\mathcal{S}^n_y(t, \xi, \gamma)\) }      & {$n=1$} & {$n=5$} & {$n=2$} & {$n=6$} & {TET} & {MESCT} \\
        \midrule
        error1 &  0.0359 &0.0256 & 0.0376 & 0.0585 & 0.5785 &  0.0812 \\
        error2 &0.0225 &0.0253 & 0.0215 & 0.0273 & 0.5725 & 0.0788 \\
        \bottomrule
    \end{tabular}
\end{table}

\begin{figure}[H]
    \centering
    \begin{tabular}{cc}
        \resizebox{2.0in}{1.5in}{\includegraphics{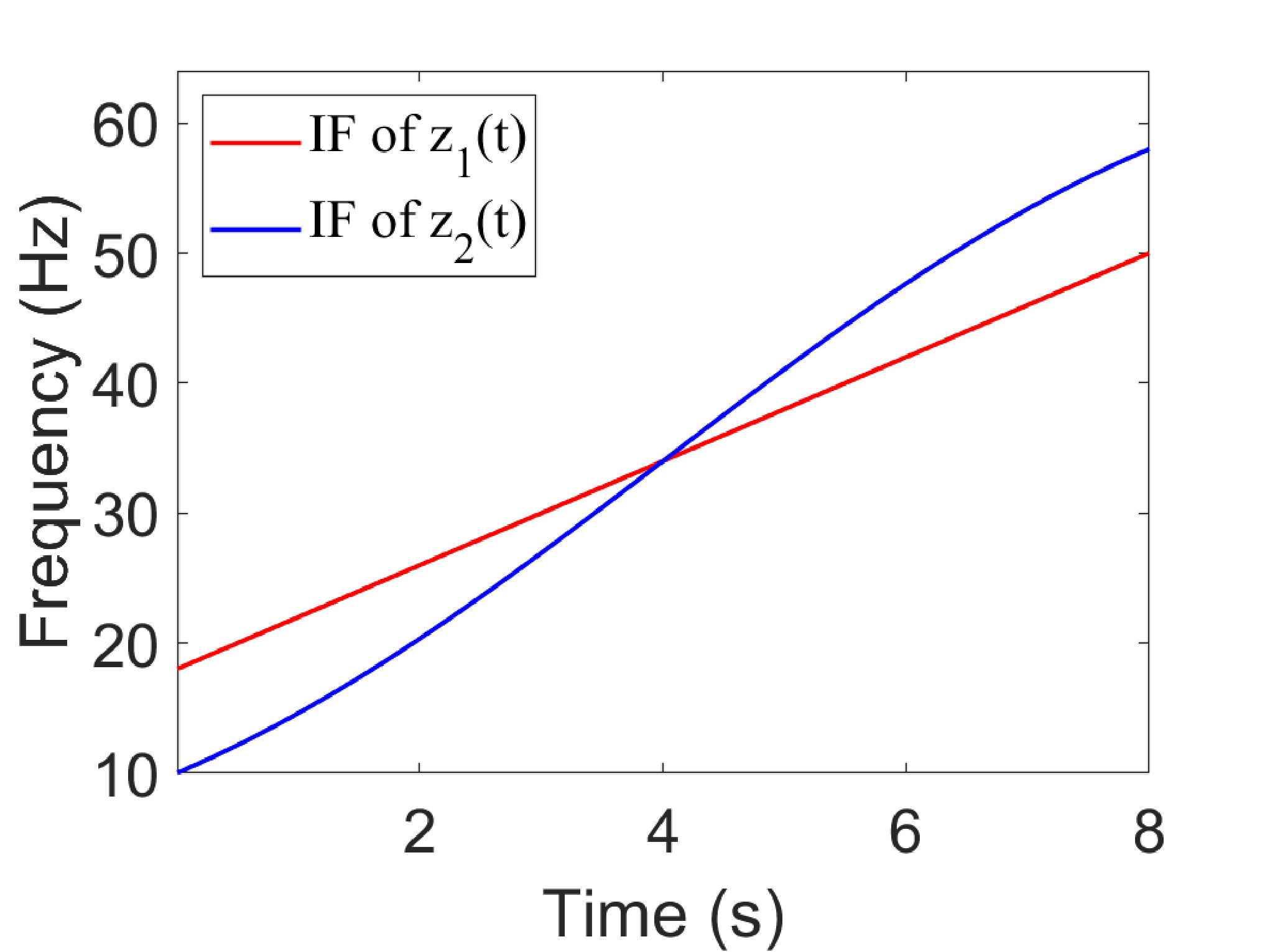}} &
        \resizebox{2.0in}{1.5in}{\includegraphics{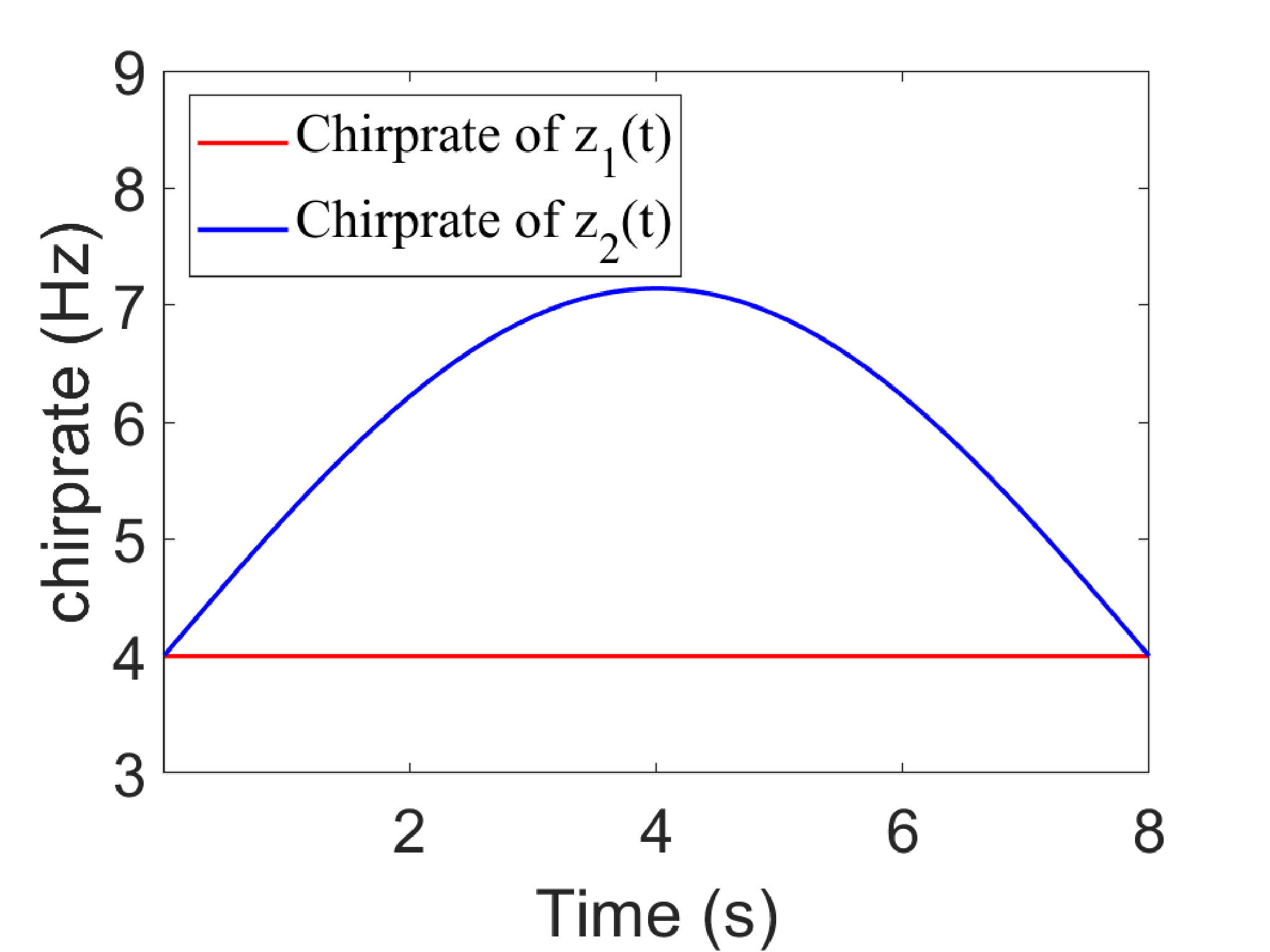}} \\
        (a) & (b) \\
    \end{tabular}
    \caption{\small IFs and chirprates of $z_1, z_2$. (a) IFs; (b) Chirprates.}
    \label{figure:Example3_IFs}
\end{figure}

Next, we consider 
\begin{equation}
	\label{example3}z(t) = z_1(t) + z_2(t), \quad z_1(t) = e^{i 2\pi \phi_1(t)}, \quad z_2(t) = e^{i 2\pi \phi_2(t)}, \quad t \in [0, 8),
\end{equation}
where \( \phi_1(t) = 2t^2 + 18t \) and \( \phi_2(t) = 2t^2 + 18t + \frac{64}{\pi} \cos\left(\frac{\pi t}{8} + \frac{\pi}{2}\right) \). 
The sampling frequency is 128 Hz and  corresponding window parameters are 0.062, 0.06, 1, and 29 for \( n = 1, 5, 2, 6 \).
The IFs are expressed as \(\phi_1'(t) = 4t + 18\) and \(\phi_2'(t) = 4t + 18 - 8 \sin\left(\frac{\pi t}{8} + \frac{\pi}{2}\right)\), respectively, and they intersect at the point \(t = 4\) s, \(\eta = 34\) Hz. The chirprates of the two modes are defined as \(\phi_1''(t) = 4\) and \(\phi_2''(t) = 4 - \pi \cos\left(\frac{\pi t}{8} + \frac{\pi}{2}\right)\).
The IFs and chirprates of \(z_1(t), z_2(t)\) are shown in Fig.~\ref{figure:Example3_IFs}.

\begin{figure}[H]
    \centering
    \begin{tabular}{ccc}
        \resizebox{1.8in}{1.1in}{\includegraphics{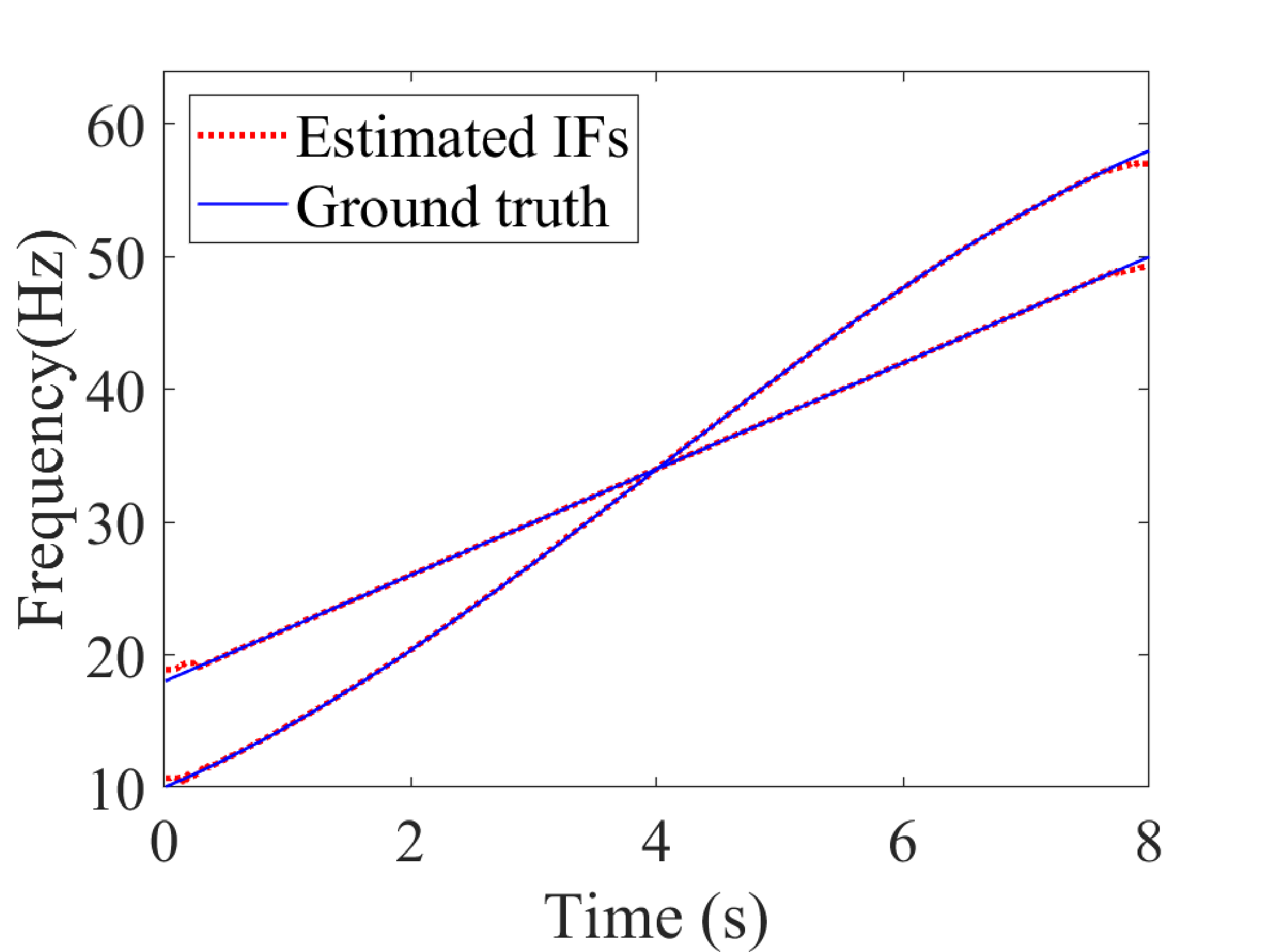}} & 
        \resizebox{1.8in}{1.1in}{\includegraphics{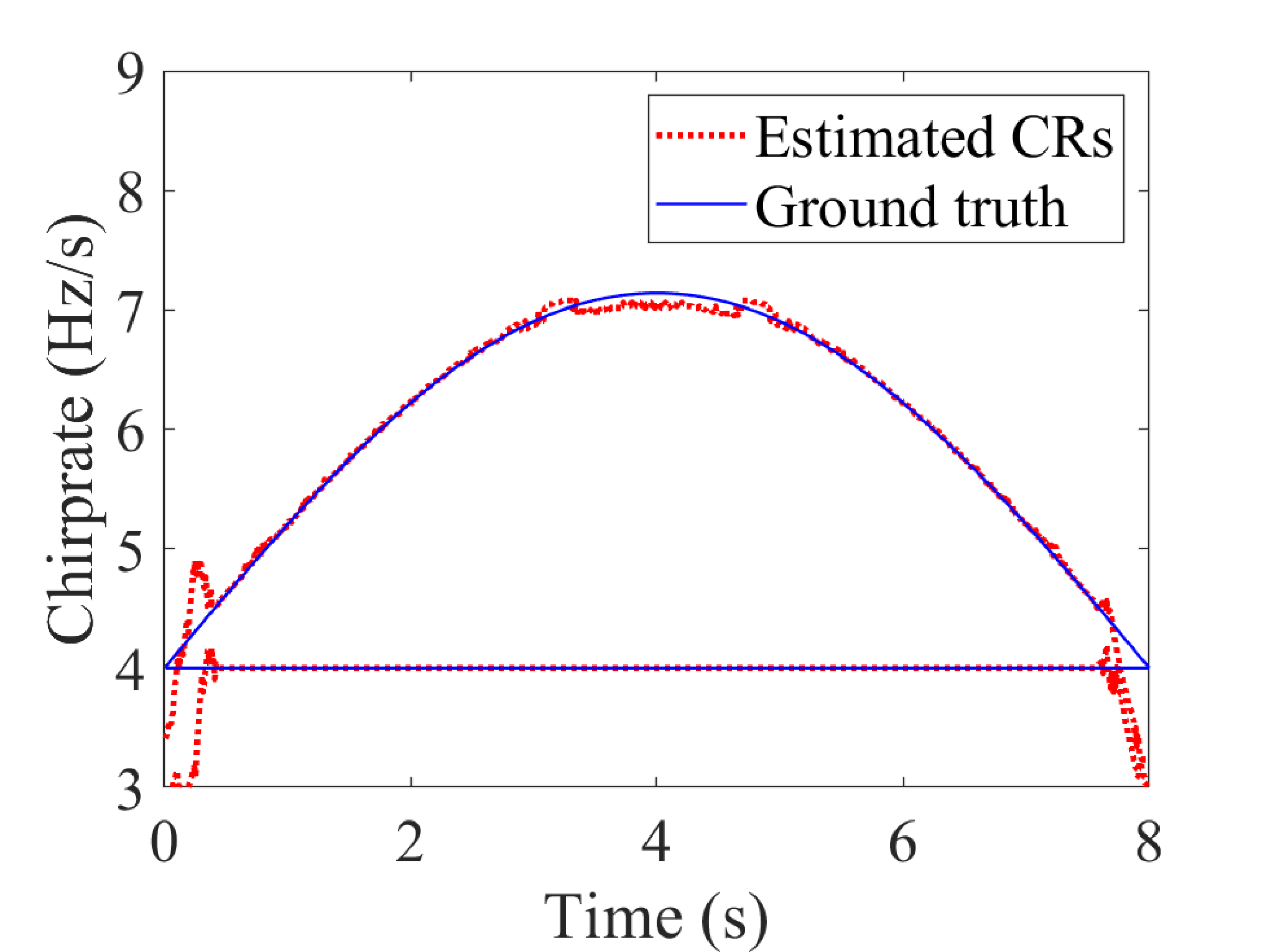}} &
        \resizebox{1.8in}{1.1in}{\includegraphics{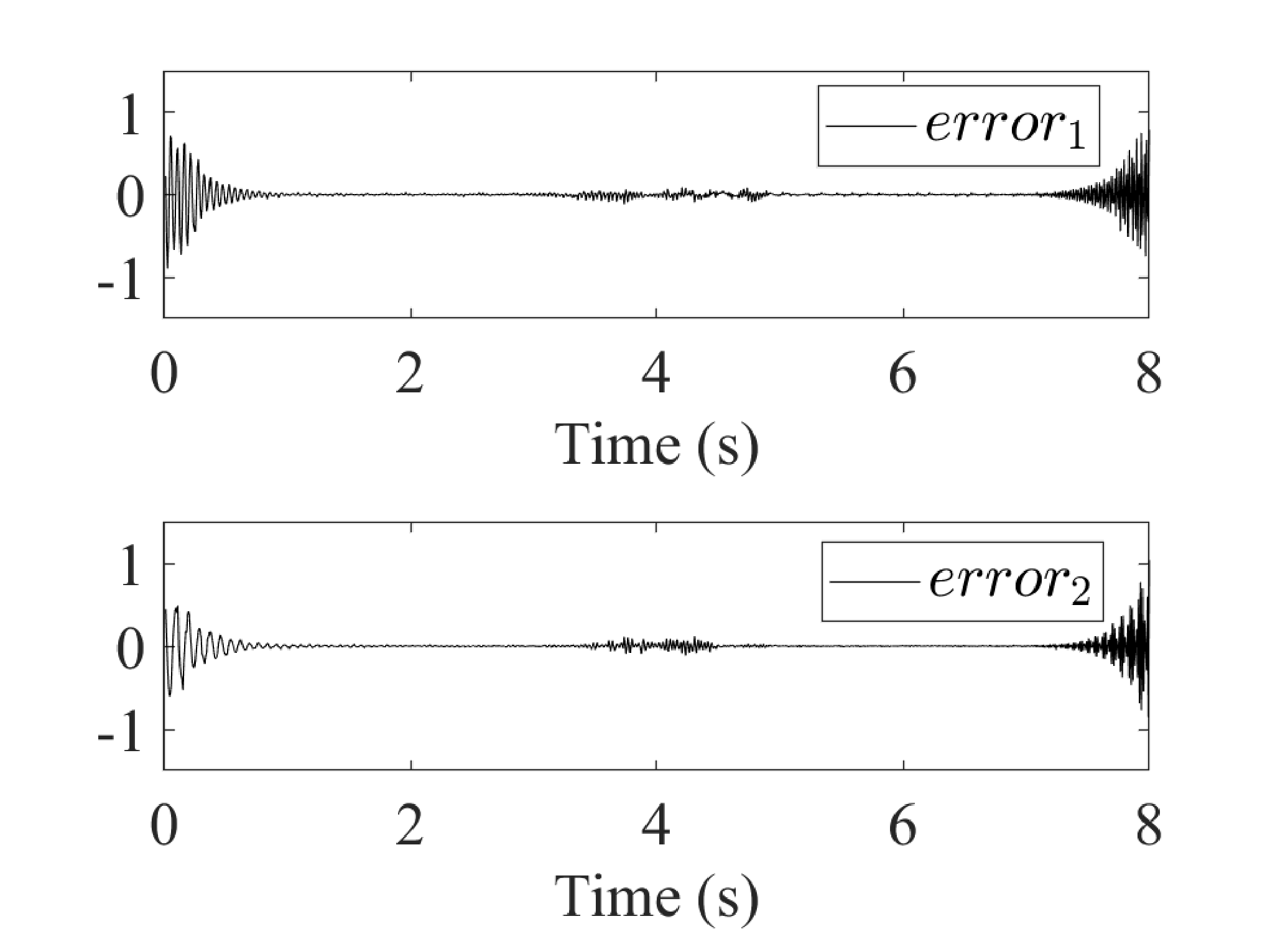}} \\
        \resizebox{1.8in}{1.1in}{\includegraphics{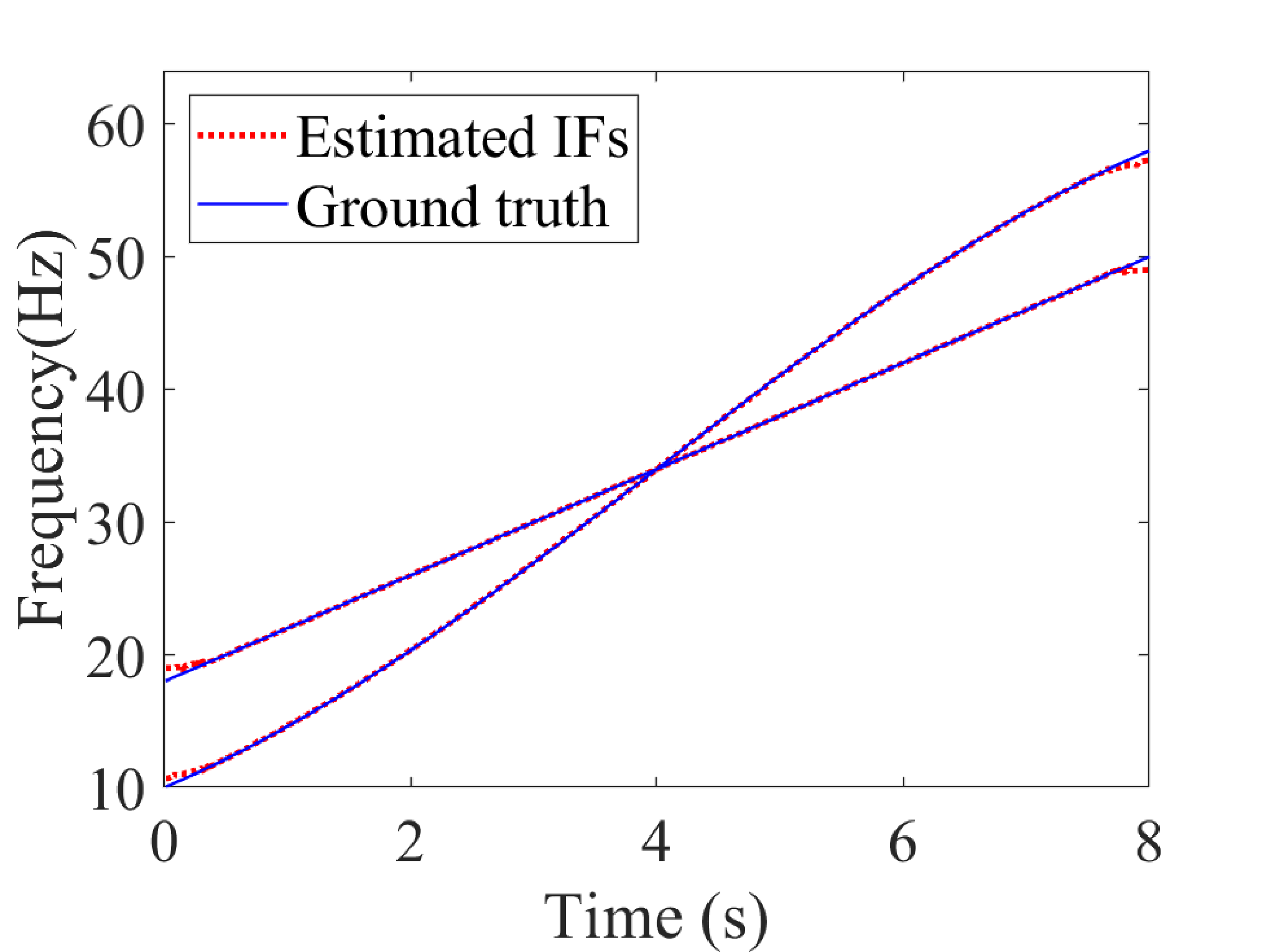}} & 
        \resizebox{1.8in}{1.1in}{\includegraphics{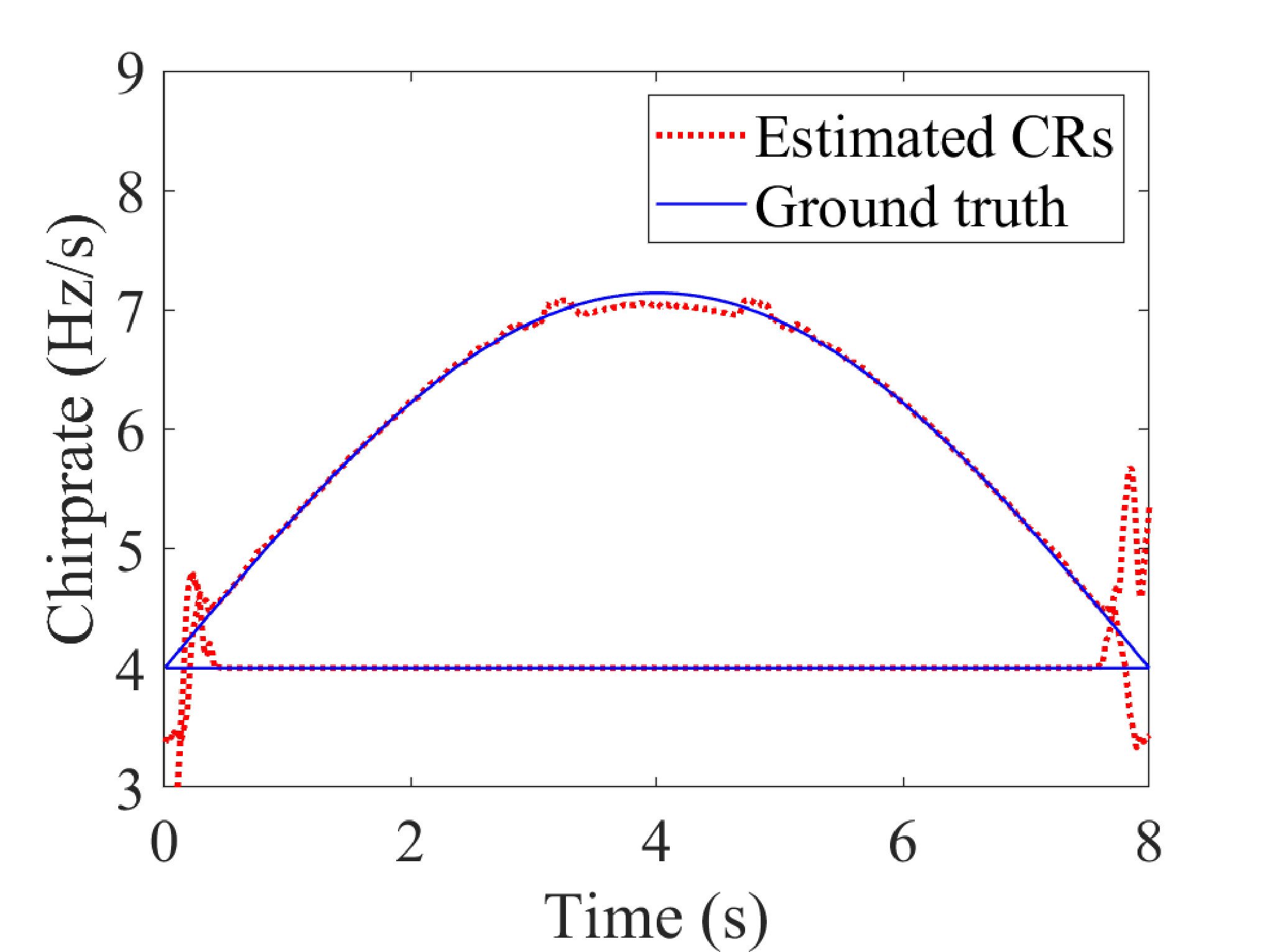}} &
        \resizebox{1.8in}{1.1in}{\includegraphics{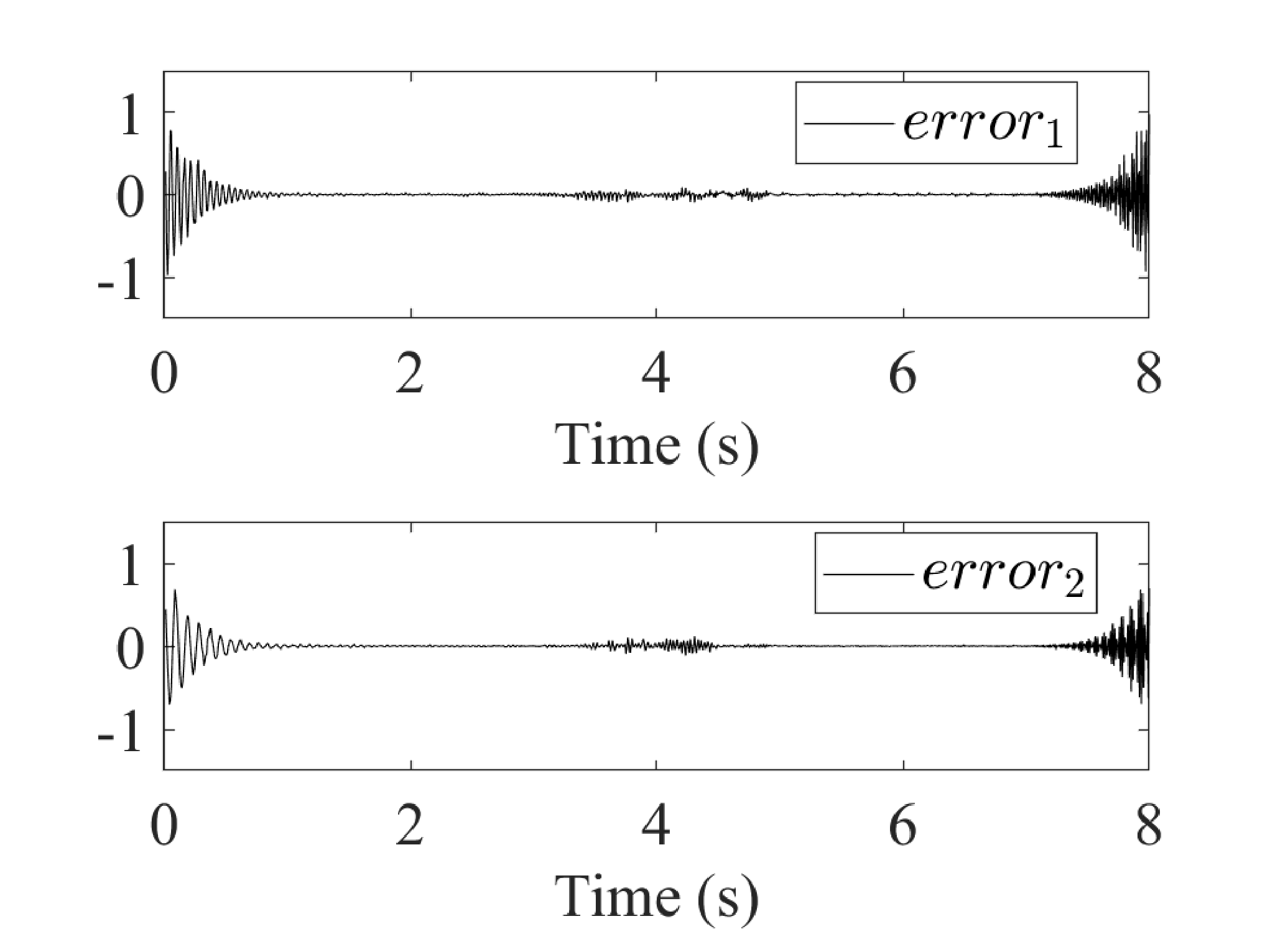}} \\
        \resizebox{1.8in}{1.1in}{\includegraphics{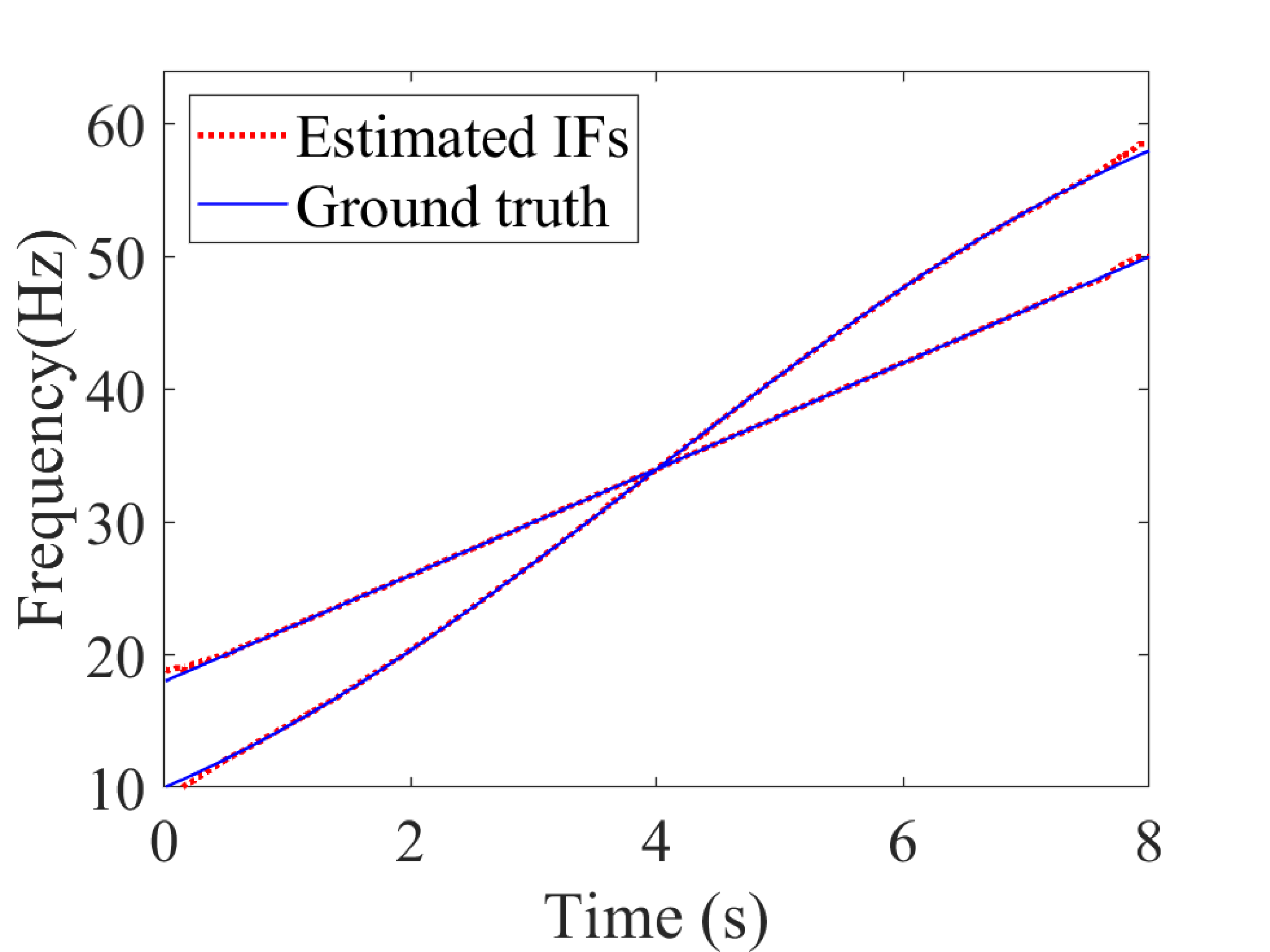}} & 
        \resizebox{1.8in}{1.1in}{\includegraphics{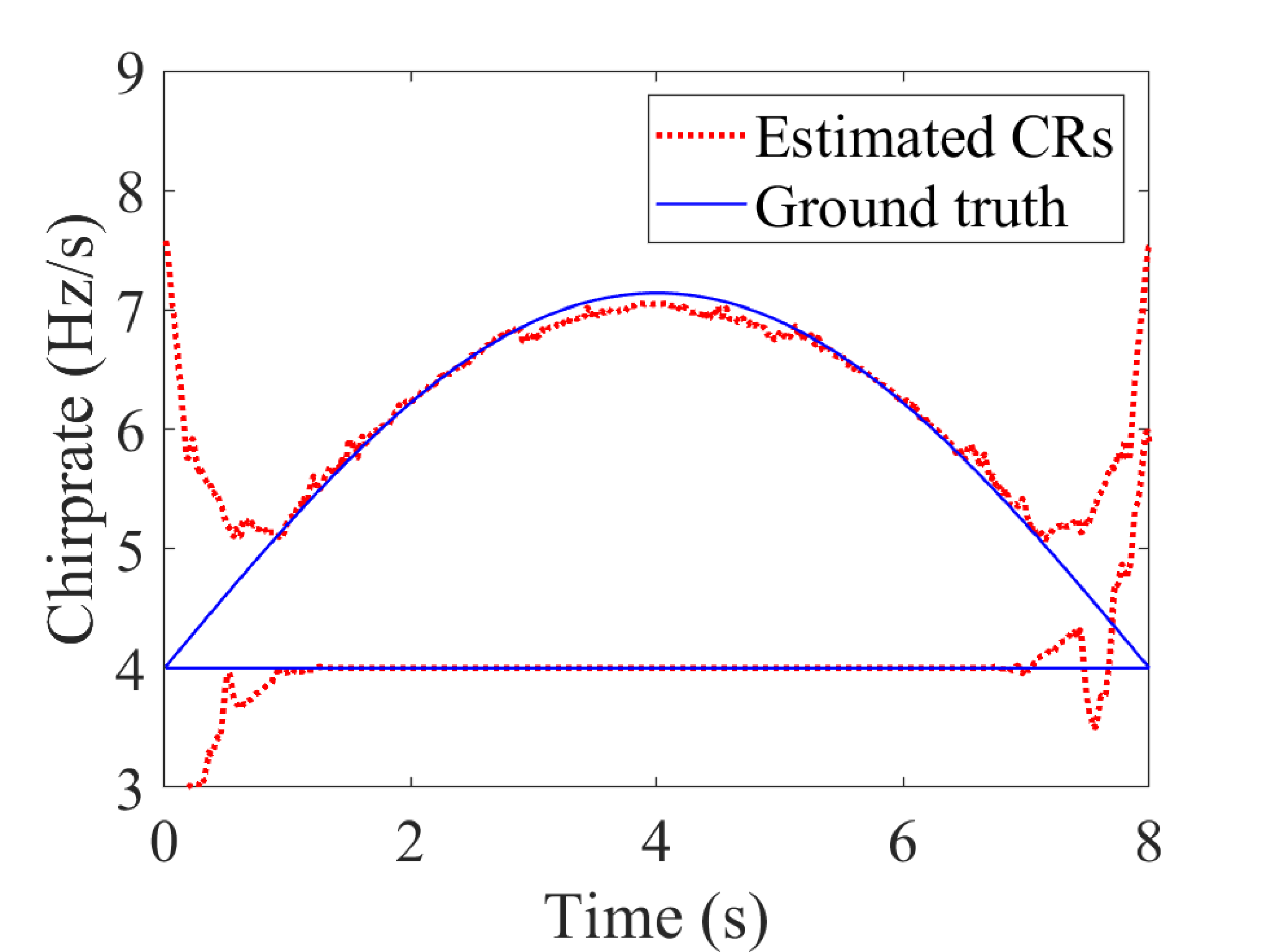}} &
        \resizebox{1.8in}{1.1in}{\includegraphics{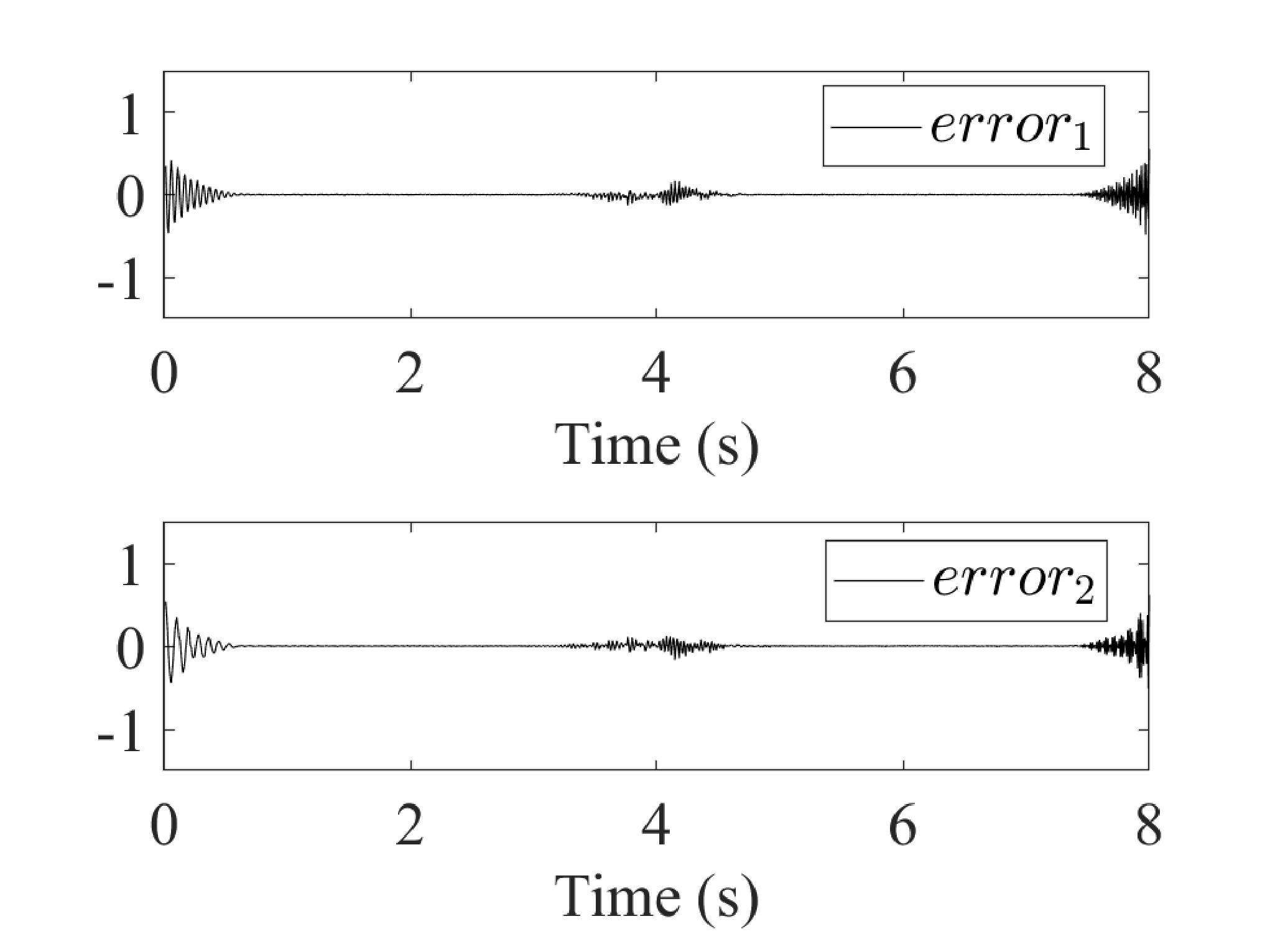}} \\
        \resizebox{1.8in}{1.1in}{\includegraphics{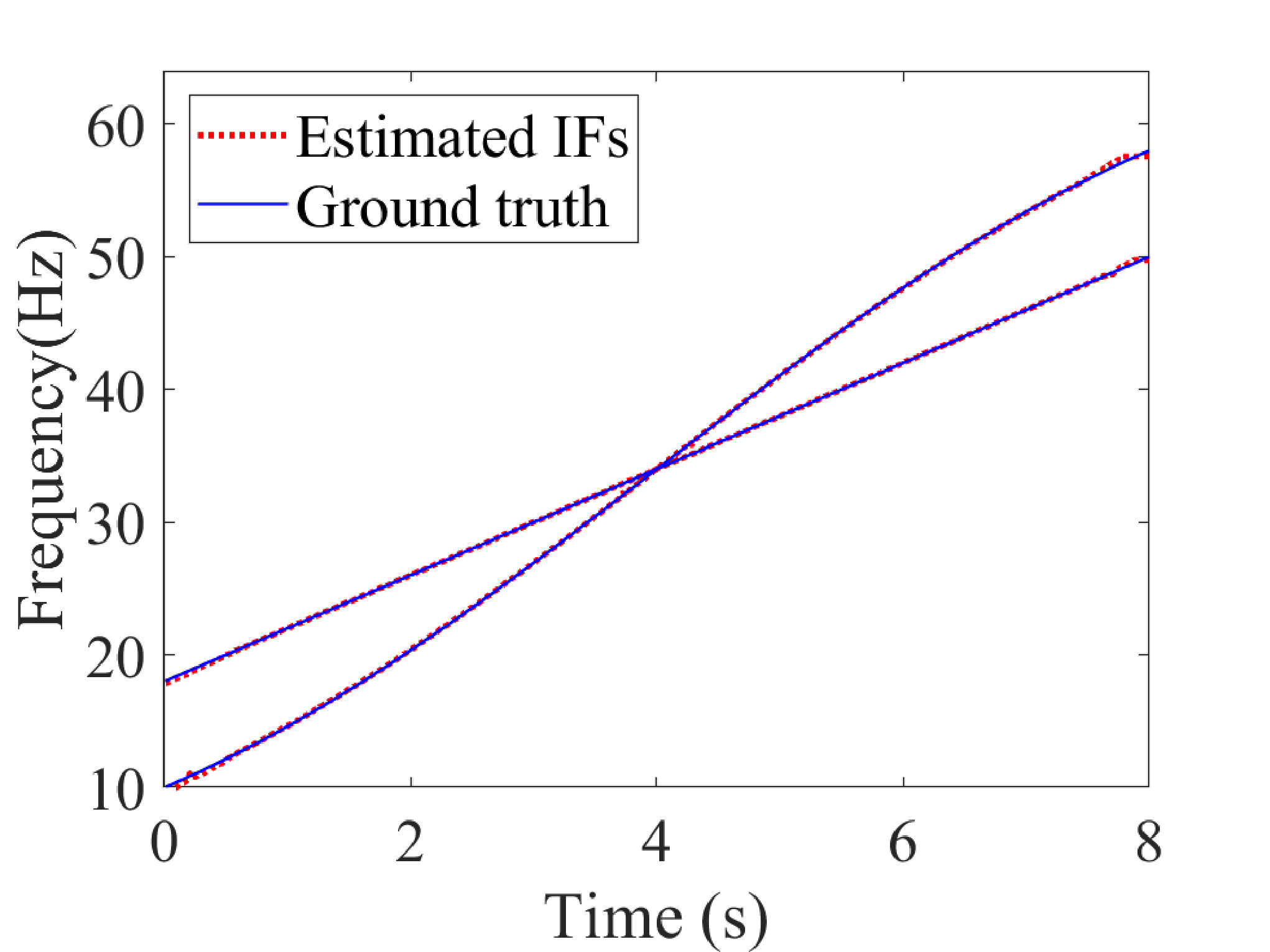}} & 
        \resizebox{1.8in}{1.1in}{\includegraphics{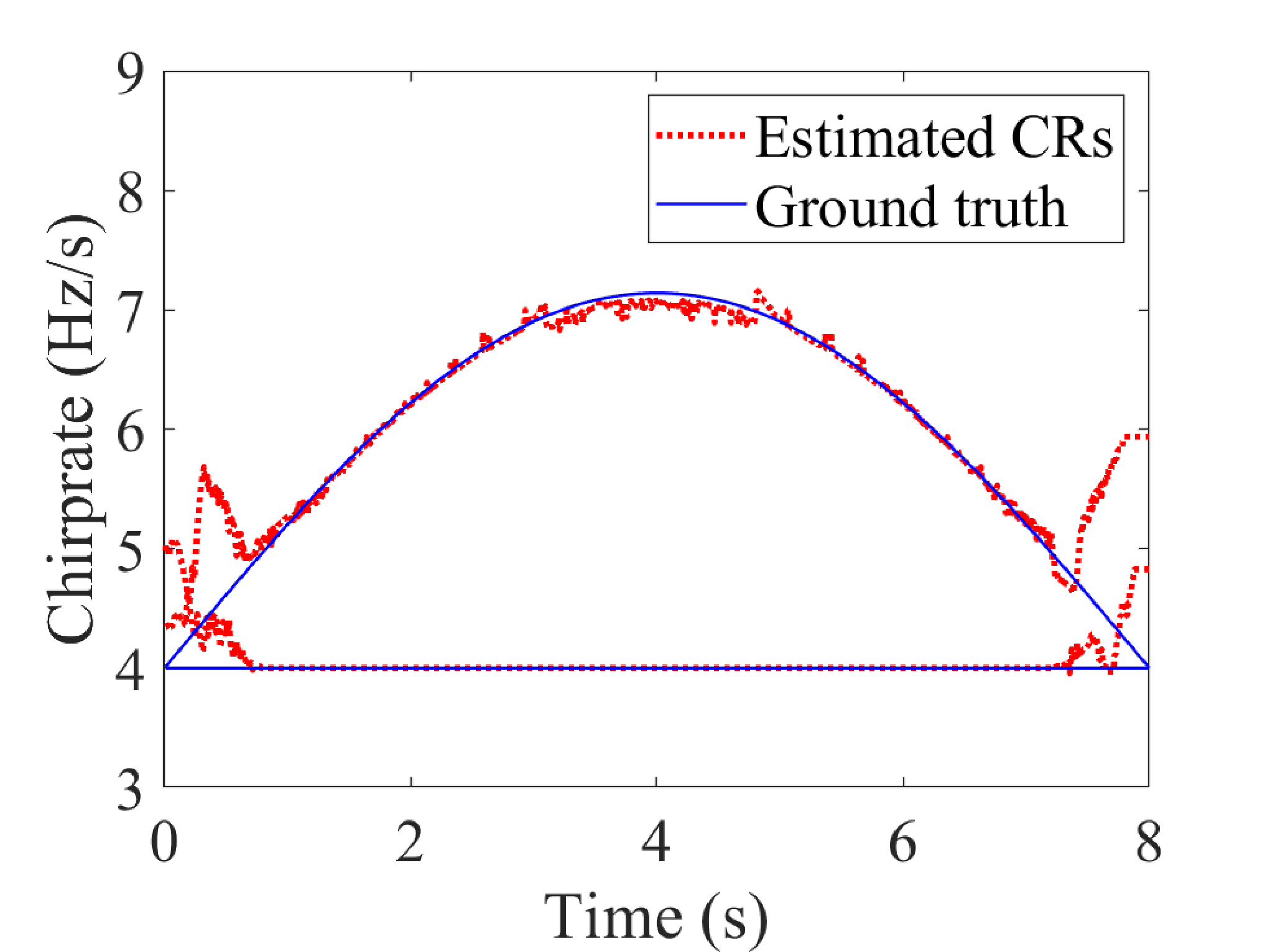}} &
        \resizebox{1.8in}{1.1in}{\includegraphics{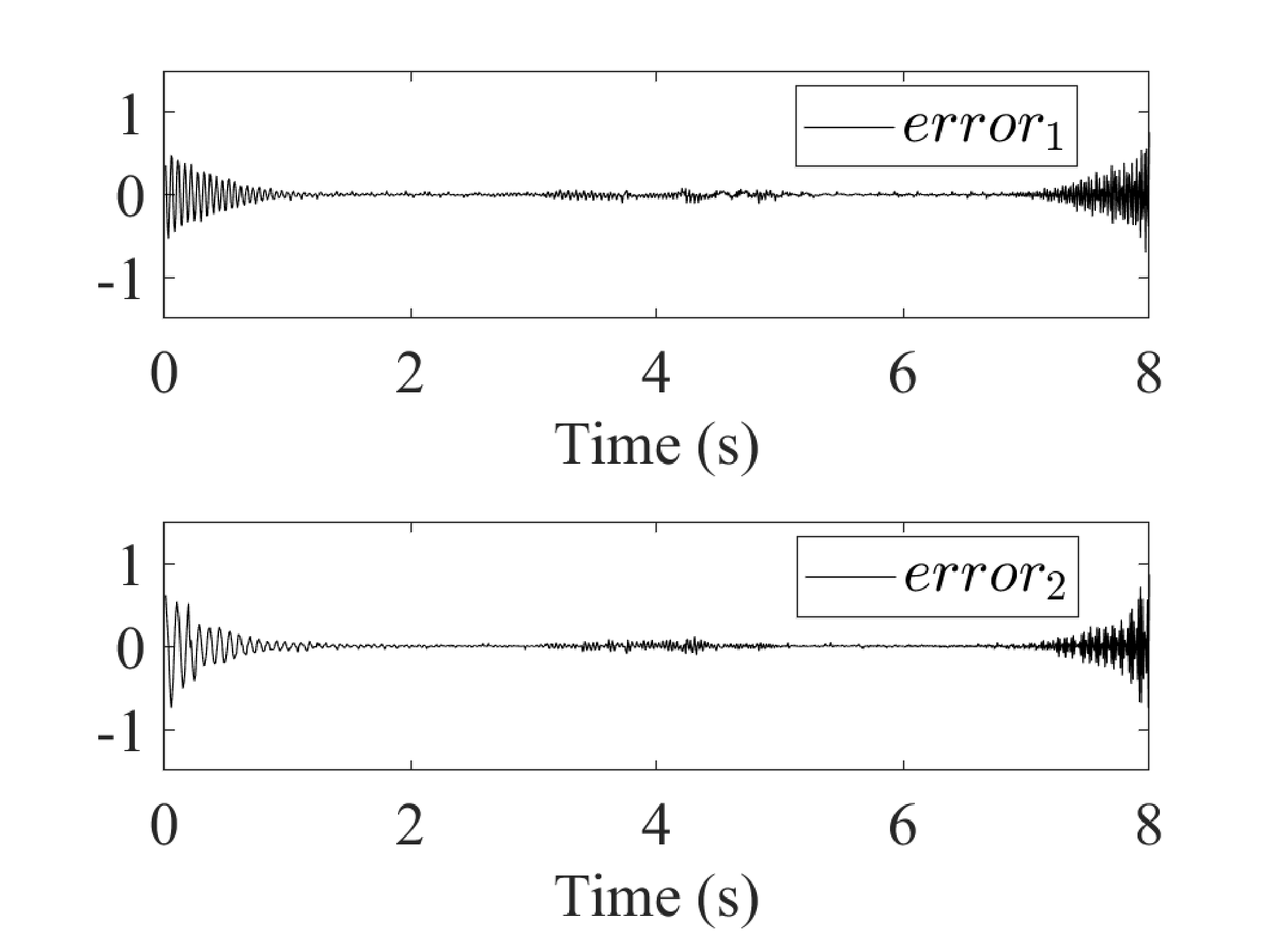}} \\
    \end{tabular}
	\caption{\small IFs and chirprates estimations, and real part errors of mode retrieval by SXWLCT.  
      First row by \(\mathcal{S}^1_z(t, \xi, \gamma)\), Second row by \(\mathcal{S}^5_z(t, \xi, \gamma)\),  Third row by \(\mathcal{S}^2_z(t, \xi, \gamma)\) and Fourth row by \(\mathcal{S}^6_z(t, \xi, \gamma)\).} 
    \label{figure:SXWLCT of $Z$}
\end{figure}

For the signal \( z(t) \), the SXWLCT methods also demonstrate their capability to separate signal modes and accurately extract the IFs, as shown in Fig.~\ref{figure:SXWLCT of $Z$}.
However, slight deviations occur in the estimated chirprate near the intersection points of the sinusoidal components. Despite this, the SXWLCTs achieve good mode retrieval.
Notably, when using $\mathcal{S}^1_x(t, \xi, \gamma)$ and $\mathcal{S}^5_x(t, \xi, \gamma)$ to recover modes, the recovered high-chirprate components exhibit smaller errors compared to the other modes.

\begin{figure}[H]
    \centering
    \begin{tabular}{cc}
        \resizebox{3.20in}{2in}{\includegraphics{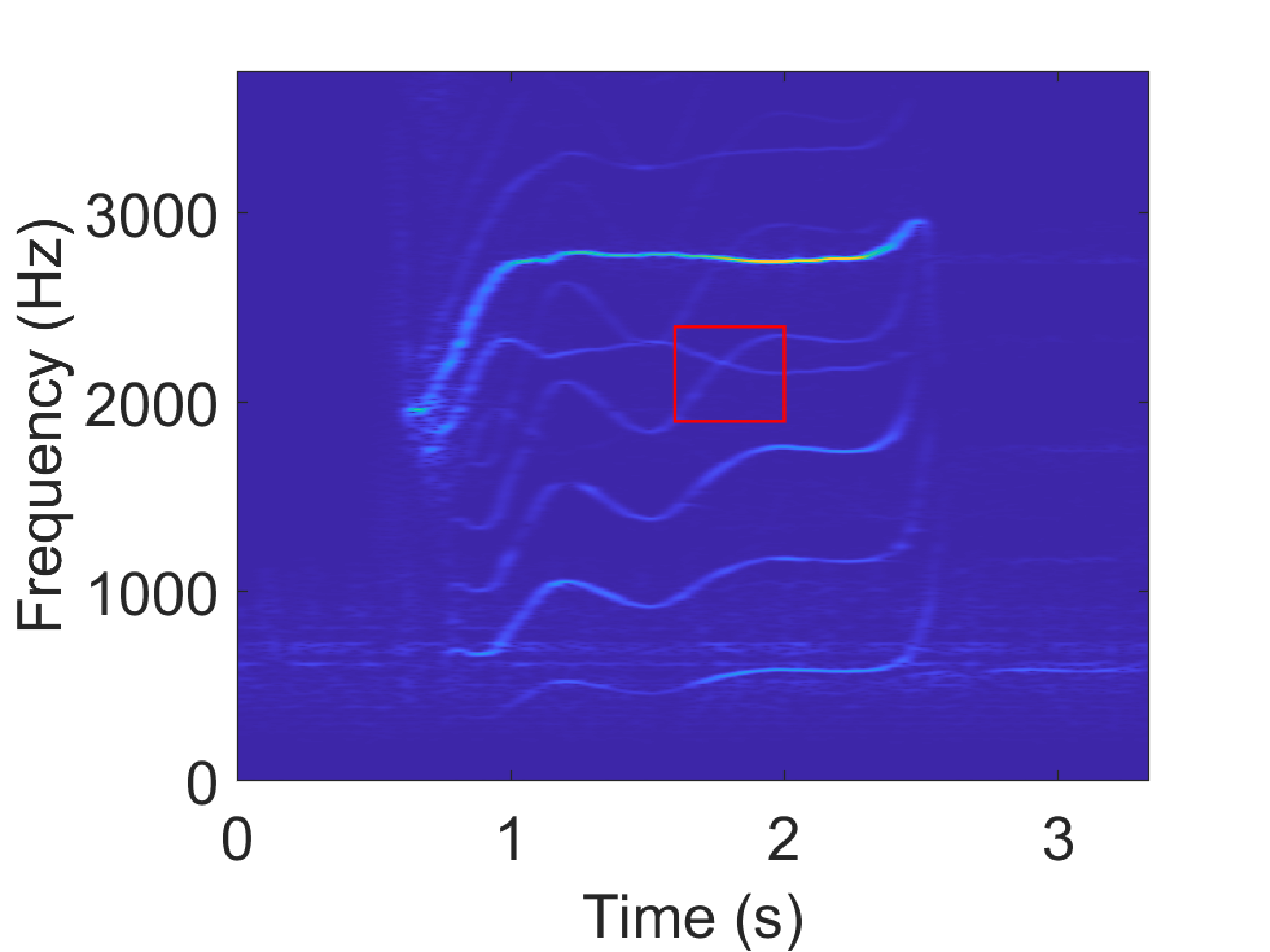}} &
        \resizebox{3.20in}{2in}{\includegraphics{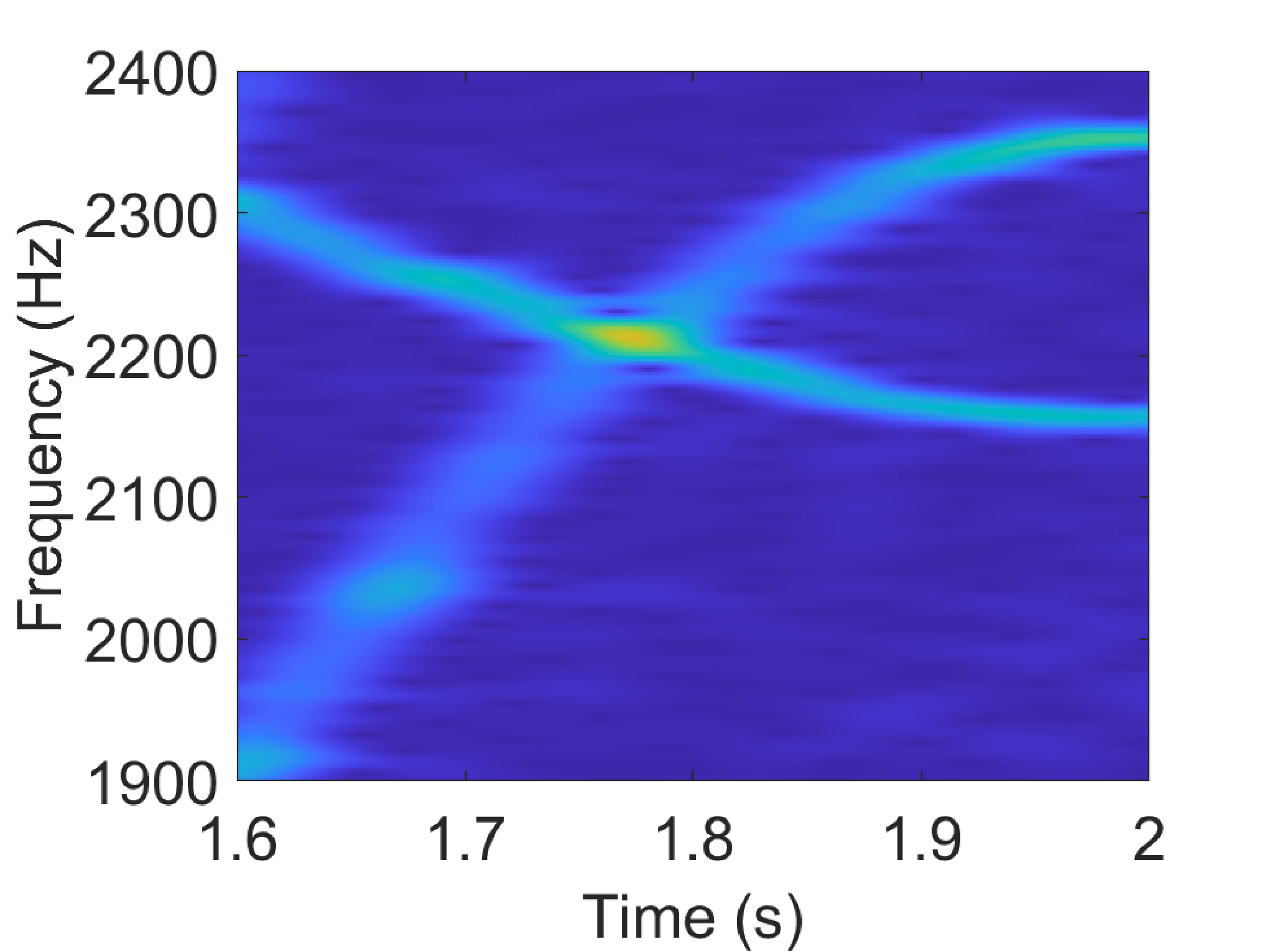}} \\
        (a) & (b) \\
    \end{tabular}
   \caption{\small (a) STFT of killer whale acoustic data; (b) Enlarged part with crossover IFs.}
    \label{figure:STFT of example4}
\end{figure}

At the end of this section, we apply the SWLCTs to a real-world signal to demonstrate its effectiveness in time-frequency analysis. 
The acoustic data used in this experiment are from killer whale group communications \cite{miller2004call,zhang2022local}, which provide valuable insights into marine animal behavior. 
Such signals are challenging to analyze due to the diversity of vocal repertoires and strong background noise in the underwater environment. 
The original data were downsampled to a rate of 7500 Hz.
Fig.~\ref{figure:STFT of example4} shows two crossing IF curves between 1900-2400 Hz during the 1.6-2 s interval,  which is highlighted by a red rectangle in Panel(a) of STFT with an enlarged part presented in Panel(b).

\begin{figure}[H]
    \centering
    \begin{tabular}{cc}
        \resizebox{2.0in}{1.5in}{\includegraphics{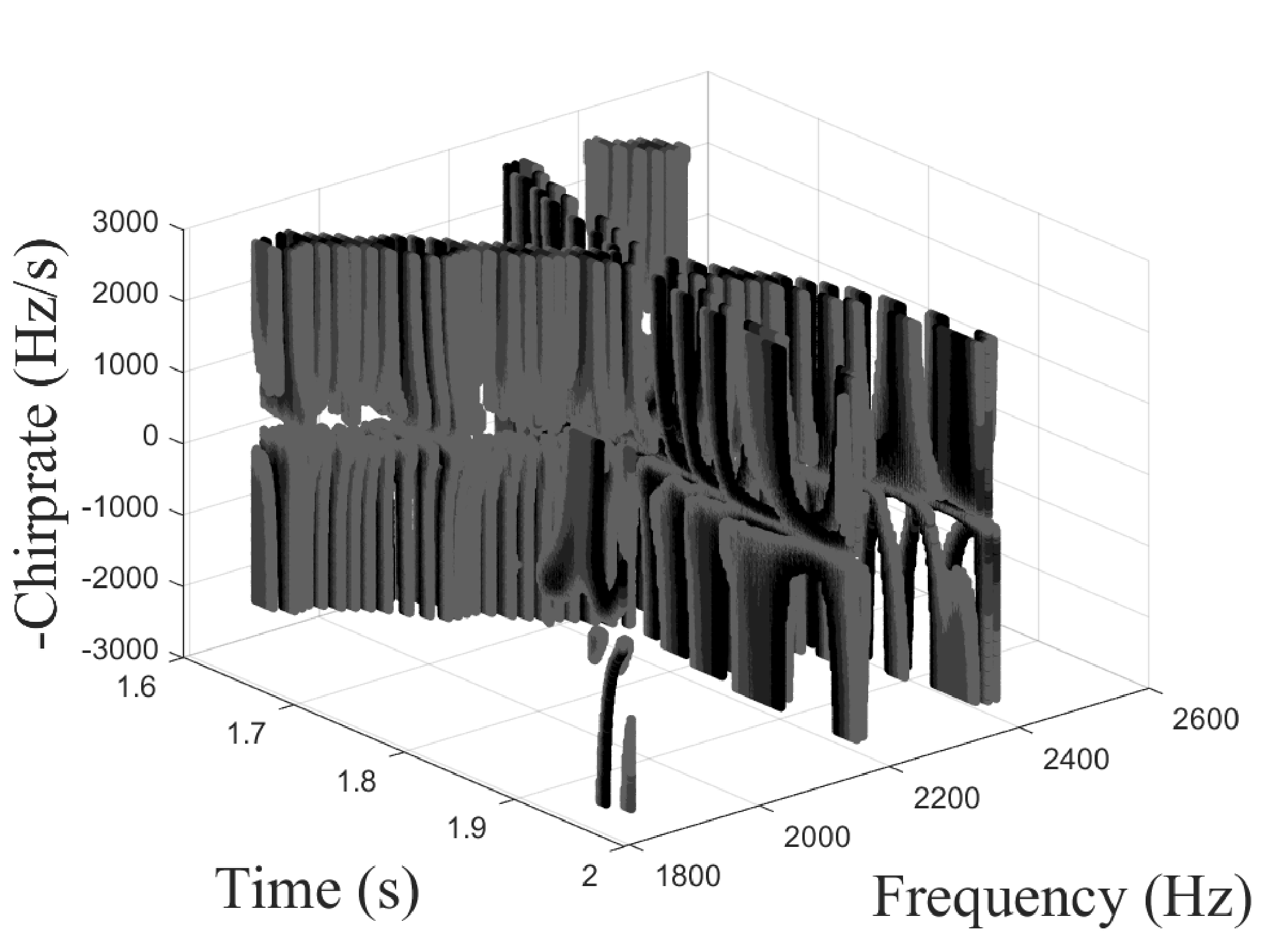}} &
        \resizebox{2.0in}{1.5in}{\includegraphics{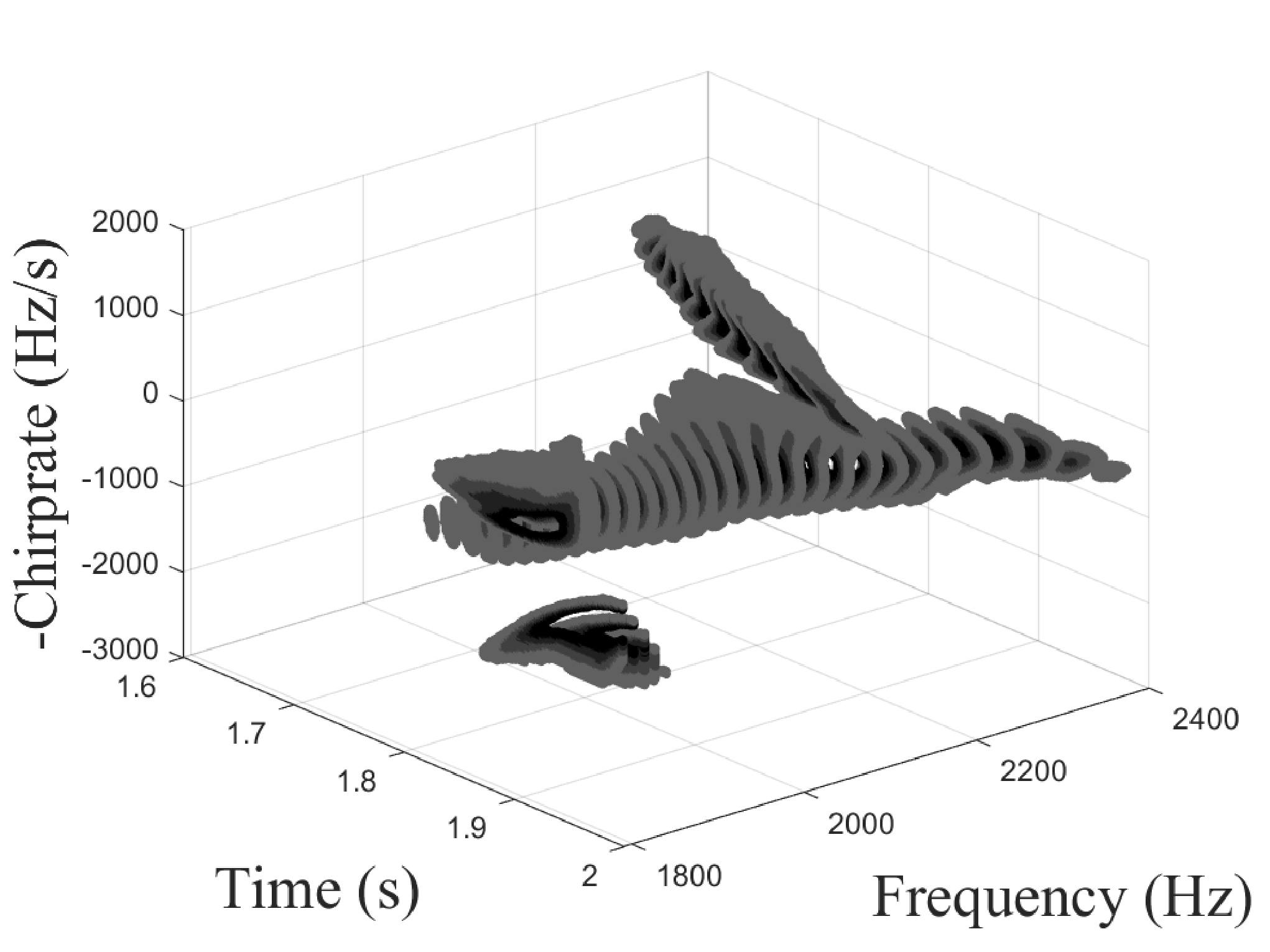}} \\
        (a) & (b) \\
    \end{tabular}
    \caption{\small  (a) WLCT with \(n=2\); (b) XWLCT with \(n=2\).}
    \label{figure:example4_SXWLCT_3D}
\end{figure}

To enable an intuitive assessment, we provide three-dimensional representations of both the WLCT and its X-ray transform, using \(n = 6\) as a representative example (see Fig.~\ref{figure:example4_SXWLCT_3D}).
The analysis, conducted over the time interval of 1.6--2 s and the frequency range of 1900--2400 Hz, clearly illustrates the enhanced energy concentration of the XWLCTs along the chirprate dimension. 
This concentrated representation consequently leads to sharper time-frequency ridges, which are subsequently extracted using the SXWLCTs within the same region.

\begin{figure}[H]
    \centering
    \begin{tabular}{ccc}
        \resizebox{1.8in}{1.1in}{\includegraphics{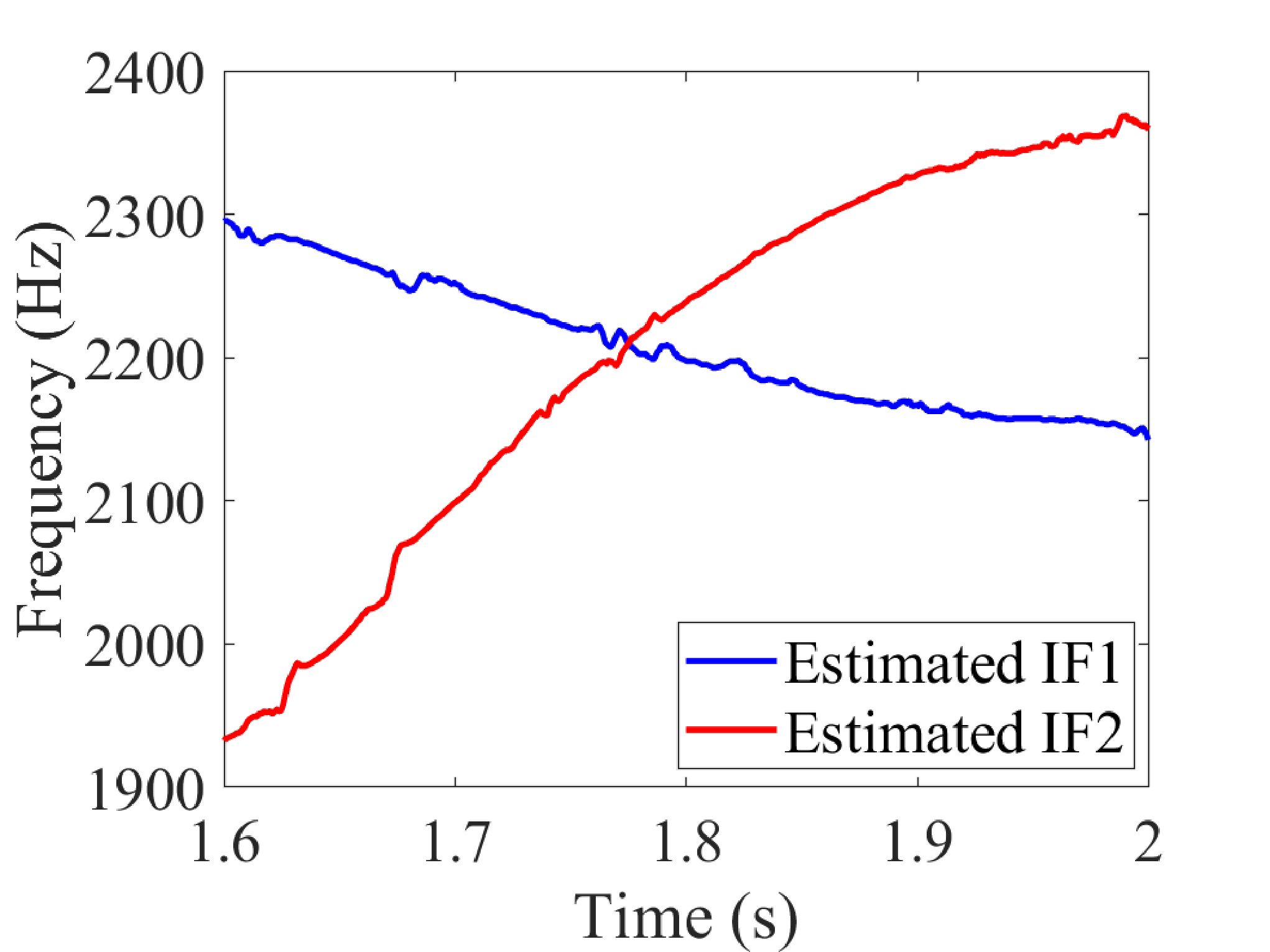}} & 
        \resizebox{1.8in}{1.1in}{\includegraphics{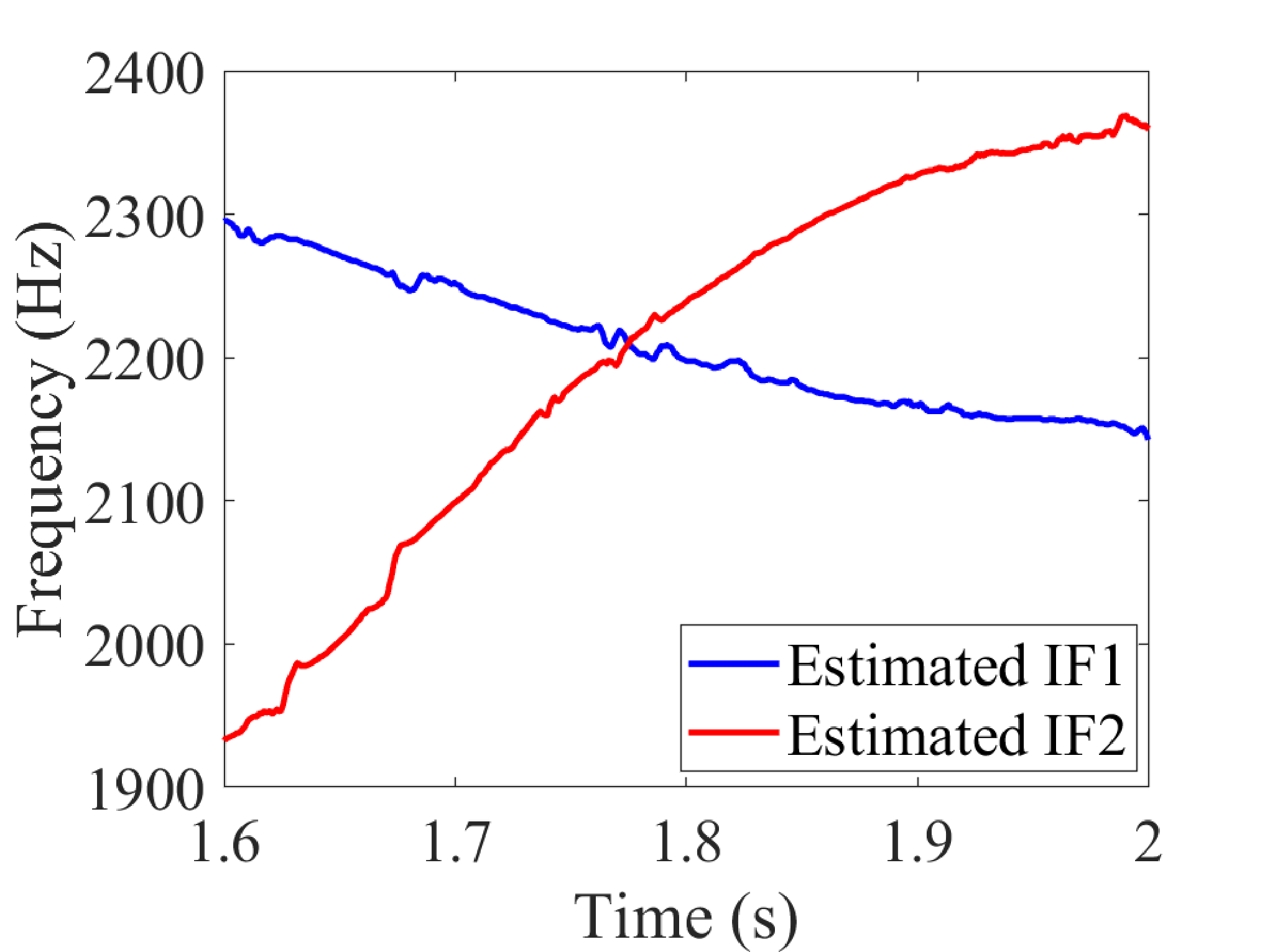}} &
        \resizebox{1.8in}{1.1in}{\includegraphics{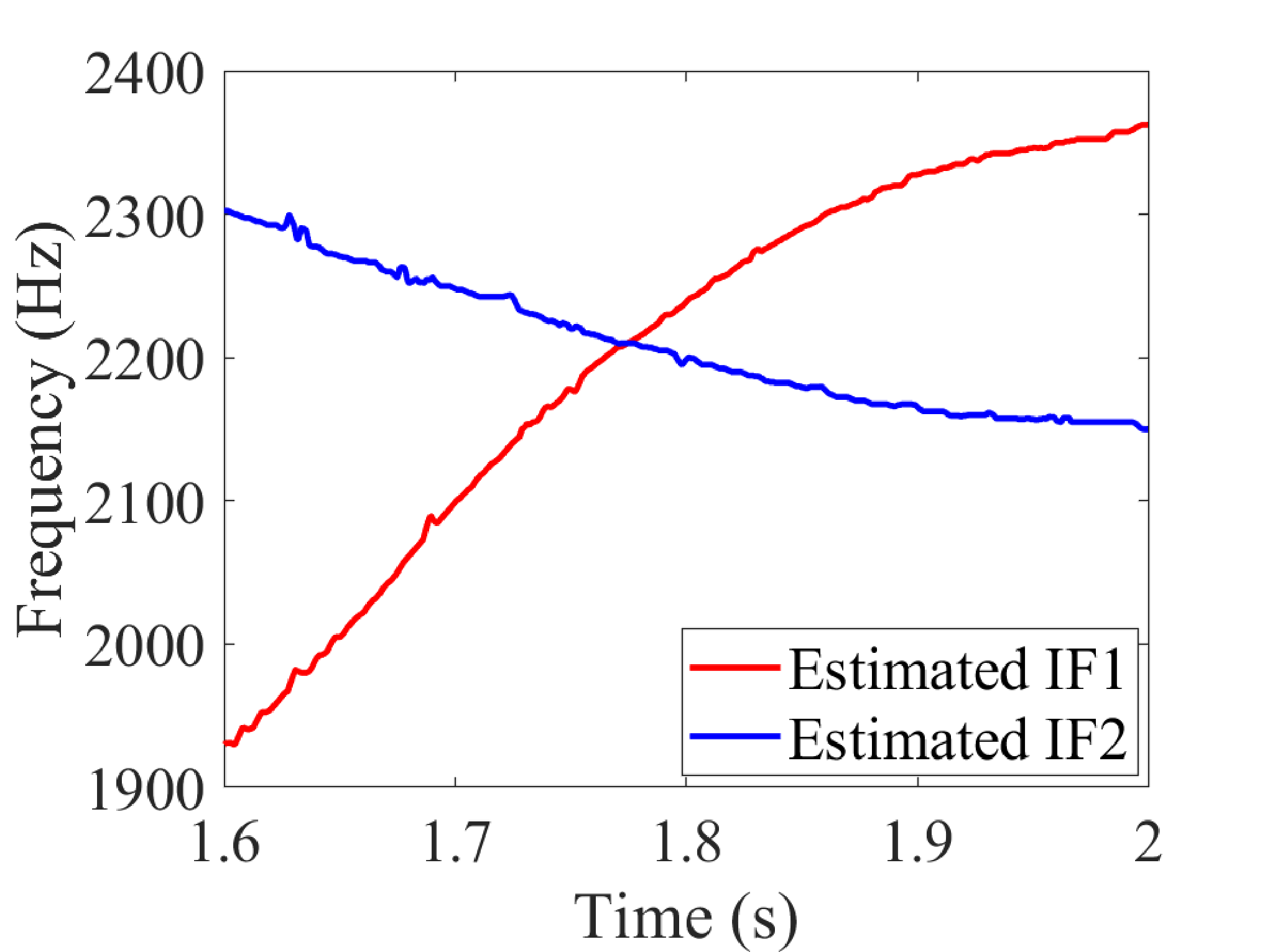}} \\
        \resizebox{1.8in}{1.1in}{\includegraphics{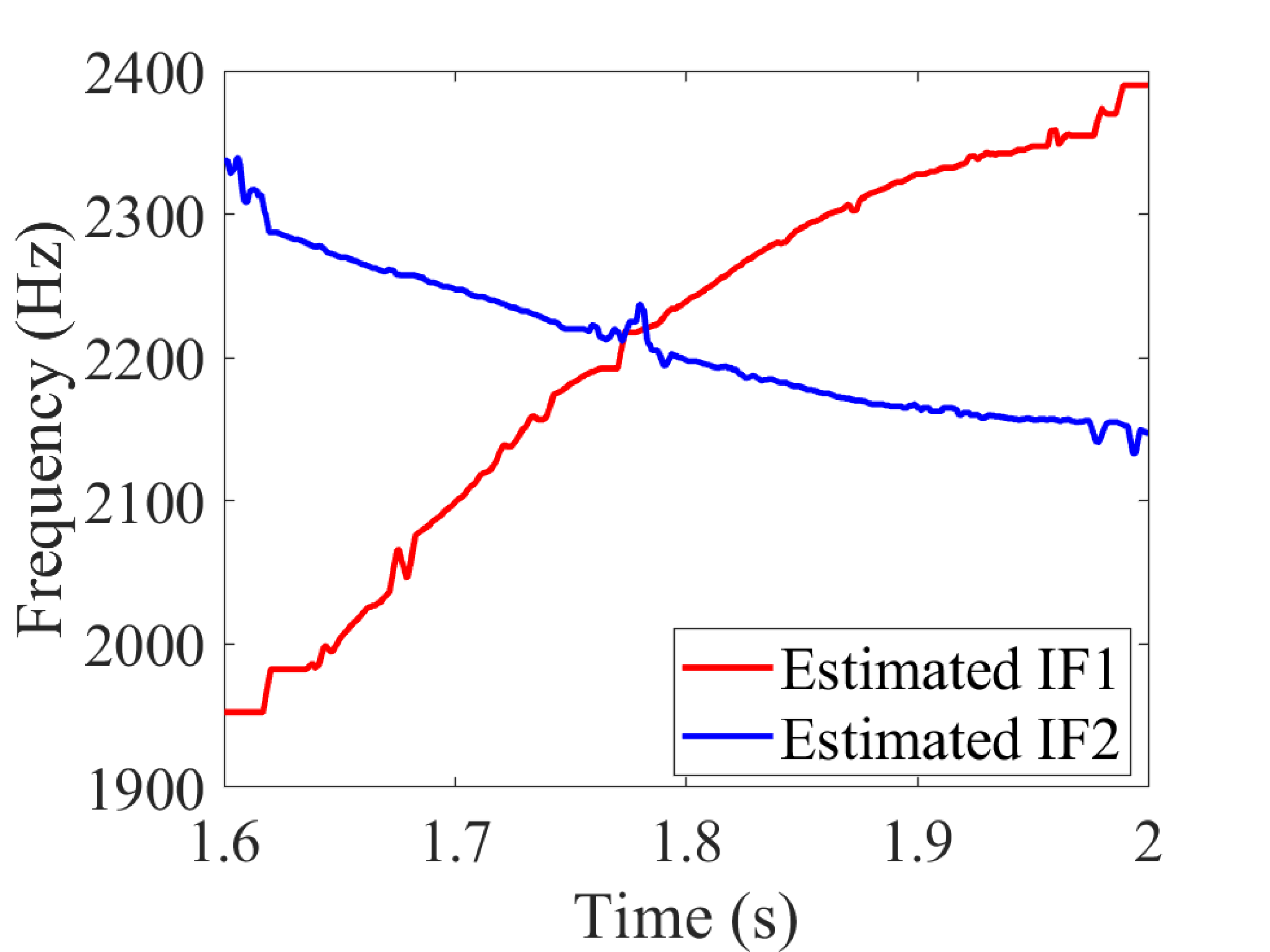}} & 
        \resizebox{1.8in}{1.1in}{\includegraphics{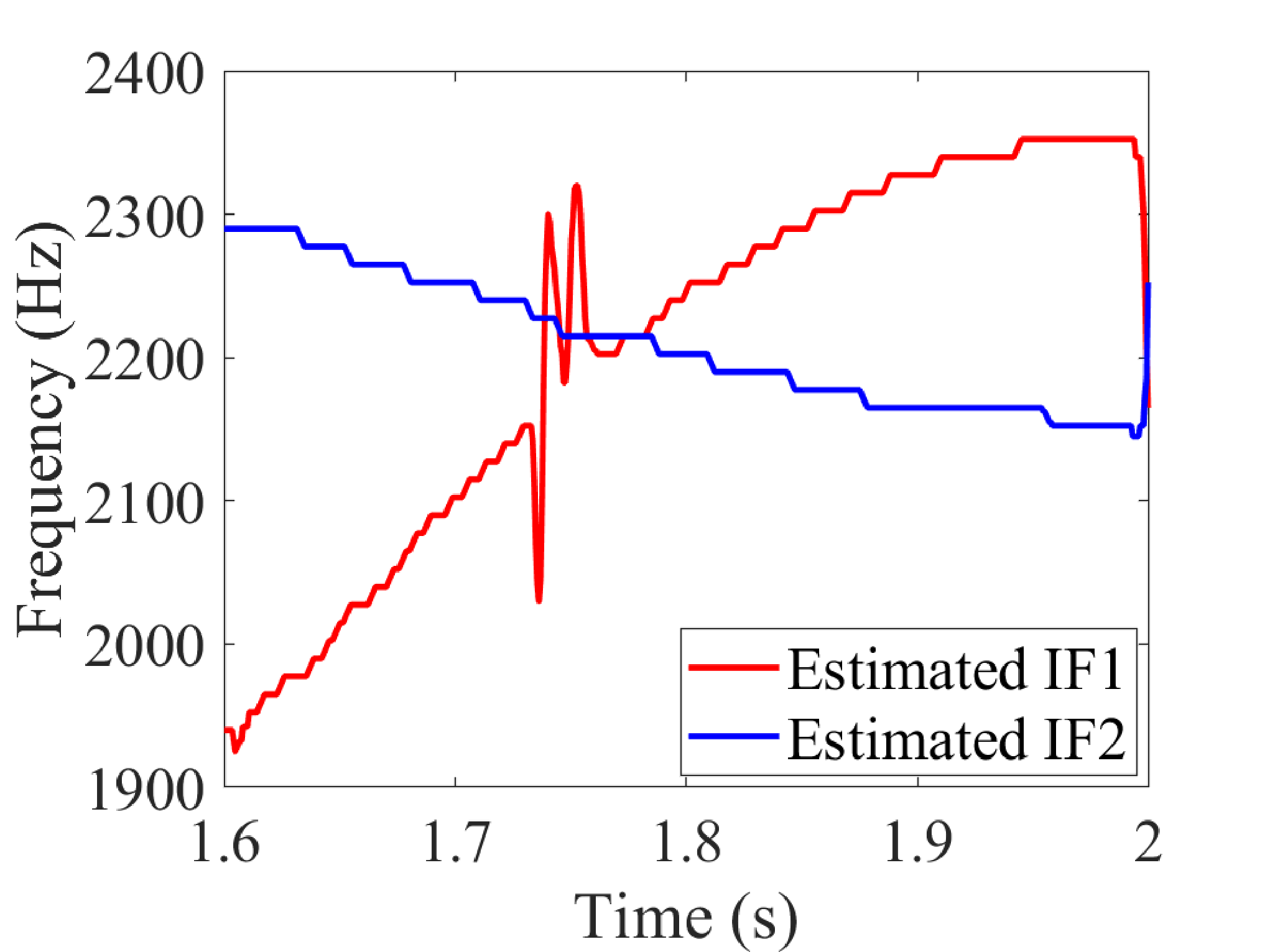}} &
        \resizebox{1.8in}{1.1in}{\includegraphics{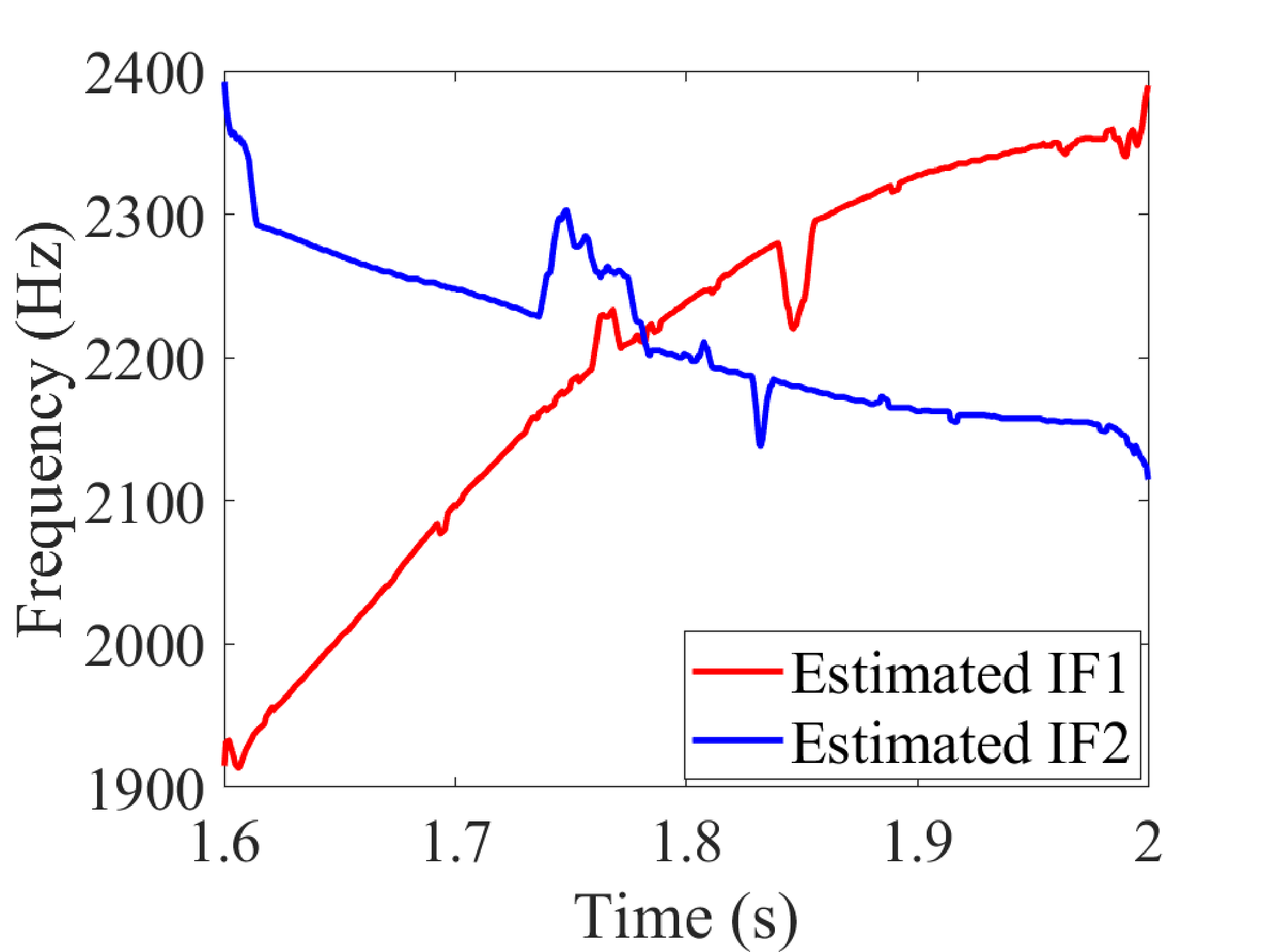}} \\
    \end{tabular}
    \caption{\small IF extraction results of SXWLCTs, TET and MESCT. First row (left to right): SXWLCT $n=1$, SXWLCT $n=5$, SXWLCT $n=2$; Second row (left to right): SXWLCT $n=6$, TET, MESCT (squeezing five times).}
    \label{figure:example4_SXWLCTs_IFs}
\end{figure}

For a comparative analysis, the TET and MESCT methods are also employed to extract the IF ridges.
As evidenced by Fig.~\ref{figure:example4_SXWLCTs_IFs}, the proposed SXWLCTs provide accurate  IF estimations, whereas significant deviations are observed in both TET and MESCT, especially near intersection points.

Overall, the WLCTs demonstrate substantial potential for signal separation in three-dimensional space, establishing a more versatile framework than the CT. 
The X-ray transform further enhances this framework by accelerating  decay along the chirprate dimension, yielding a highly concentrated energy distribution. 
Whereas many existing synchrosqueezing techniques based on CT rely on specific reassignment operators, the proposed WLCT and XWLCT frameworks offer a more adaptable platform for implementing and enhancing such operators. 
The results presented in this section confirm that integrating our framework with these advanced operators holds substantial promise and paves the way for more powerful time-frequency analysis tools.

\section{Algorithm implementation}
In implementing the discrete WLCTs, the core computational step involves constructing the discretized functions \(\cL^{M_{\lambda_l,n}}(g)(\lambda_l,\tau_k)\), while subsequent processing stages maintain close structural parallels with standard discrete WFT implementations.

Suppose the input signal $x(t)$ is discretized uniformly at points
\begin{align}
t_m &= t_0 + m\Delta t, \quad m = 0, 1, \cdots, N-1,
\end{align}
where $\Delta t$ is the time-domain sample step, and $N$ denotes the total number of sampling points. 
The chirprate variable $\lambda$ is discretized into $N_c$ bins, where \(N_c\) is the  number of  chirprate bins (one may set \( N_c= 2 \left\lfloor \frac{N}{2}  \right\rfloor\)). 

The values of $\lambda$ are determined by the choice of \(n\) and the discretization method, as detailed below.
For \(n=2,6\), we adopt uniform sampling. First, we assume that the range of the variable $\lambda$ is given by
\(-R_0 \leq \lambda \leq R_0\), 
where \(R_0\) is a tunable parameter that can be adjusted based on the signal characteristics.
Then, the discrete values of $\lambda$ are defined as
\begin{align}
\label{lambda_2}	\lambda_\ell &= R_0 + (1-\ell) \Delta \lambda, \quad \ell =  1,2, \dots, N_c,
\end{align}
where the sampling interval \(\Delta \lambda\) is given by
\begin{align*}
	\Delta \lambda &= {2R_0}/{(N_c-1)}.
\end{align*}
As for \(n = 1\) or \(5\), we define the discrete chirprates using a dyadic scaling structure. The positive chirprates are given by \(\lambda_\ell^+ = a_0 \cdot 2^{j \Delta a}\), where  \(a_0\) is a fundamental   scaling factor and \(\Delta a\) is a tunable resolution parameter.
The corresponding negative chirprates are defined in a symmetric manner. The complete discrete chirprate set is then:
\begin{align}
    \label{lambda_1}
    \lambda_\ell &= \begin{cases}
        -a_0 \cdot 2^{\left(\left\lfloor N_c/2 \right\rfloor + 1 - \ell\right) \Delta a}, & \ell = 1, 2, \dots, \left\lfloor N_c/2 \right\rfloor \\
        a_0 \cdot 2^{\left(\ell - \left\lfloor N_c/2 \right\rfloor - 1\right) \Delta a}, & \ell = \left\lfloor N_c/2 \right\rfloor + 2, \dots, N_c
    \end{cases}
\end{align}
where the scaling factor \(a_0\) defaults to \(\Delta t\) but can be adapted to  the signal  characteristics.

Based on  \eqref{LCT_gaussion_bneq0}, substituting variables \eqref{lambda_2} or \eqref{lambda_1} into \eqref{LCT_gaussion_bneq0} allows us to obtain the discretized representations of \(C_n(\lambda_\ell)\) and \(P_n(\lambda_\ell)\), respectively. Here, \(C_n\) and \(P_n\) are defined in \eqref{eq:parameter_P} with the parameter matrix \(M_{\gl_l, n}\).
Additionally, the time variable for the discrete window 
functions is defined as 
\begin{align}
 \label{discrete_window_time} \tau_k = (k-\lfloor \frac{N}{2}\rfloor) \Delta t, \quad k = 1, 2, \dots, N, 
\end{align}
where \(\Delta t\) is the time step size.
Thus, we can obtain the discrete representation of \(\cL^{M_{\gl, n}}(g)(\lambda_\ell, \tau_k)\)  as
\begin{align}
	\label{discretized_window}
\cL^{M_{\gl_l, n}}(g)(\lambda_\ell, \tau_k) = C_n(\lambda_\ell) e^{-P_n(\lambda_\ell) {\tau_k}^2}.
\end{align}
In fact, for the function  $ \tau^d \cL^{M_{\gl_l, n}}( g)(\tau)$ where $d= 1, 2$, we can discretize it as
\begin{align}
	\label{discretized_tauwindow}
   \tau_k^d \cL^{M_{\gl_l, n}}( g)(\lambda_\ell, \tau_k)= \tau_k^d  C_n(\lambda_\ell) e^{-P_n(\lambda_\ell) (\tau_k)^2}.
\end{align}

When dealing with the discretization of the frequency variable \(\eta\), we have
\begin{equation}
    \label{discretized_eta}
    \eta_j := \begin{cases}
        j \Delta \eta, & \text{for } 0 \leq j \leq \left\lfloor \frac{N}{2} \right\rfloor, \\
        (j - N) \Delta \eta, & \text{for } \left\lfloor \frac{N}{2} \right\rfloor < j < N,
    \end{cases}
\end{equation}
where \(\Delta \eta := \frac{1}{N \Delta t}\). 
 For the case when \(0 \leq j \leq \left\lfloor \frac{N}{2} \right\rfloor\), the product \(\eta_j \tau_k\) can be expressed as
\[
\eta_j \tau_k = (k-\left\lfloor \frac{N}{2} \right\rfloor) \Delta t \cdot j \Delta \eta = \frac{(k-\lfloor \frac{N}{2} \rfloor) j}{N}.
\]	
We define the discrete transform \( T_x^{g, n}(\lambda_l, \eta_j, t_m) \) as follows:
\begin{align}
    \label{FFTWLCT}
    T_x^{g, n}(\lambda_l, \eta_j, t_m) &= \sum_{k=1}^{N} x\left(t_m + \tau_k\right) \cL^{M_{\gl, n}}(g)(\lambda_\ell, \tau_k) e^{-2\pi i \eta_j \tau_k} \Delta t  \nonumber \\
                        & =  \sum_{k=1}^{N} x\left( t_m+ \tau_k   \right) \cL^{M_{\gl, n}}(g)(\lambda_\ell, \tau_k) e^{-2\pi i \frac{(k-\lfloor \frac{N}{2} \rfloor) j}{N}} \Delta t.
\end{align}
This transform can be efficiently implemented using the fast Fourier transform (FFT).
The overall computational complexity is \( O(N_c N \log N) \), comparable to that of the CT implementation.
As WLCTs produce three-dimensional time-frequency-chirprate representations, processing long signals poses significant challenges in terms of memory requirements and computational load. 
This can be mitigated by reducing the number of chirprate bins and incorporating a hop size parameter (analogous to the method used in the windowed Fourier transform) to accelerate computation and reduce memory  consumption.

Substituting \eqref{discretized_window} with \eqref{discretized_tauwindow},  we obtain the discrete expression for \( T_x^{\tau^d g, n}(\lambda_l, \eta_j, t_m) \),
\begin{align}
    \label{FFTTWLCT}
    T_x^{\tau^d g, n}(\lambda_l, \eta_j, t_m) =  \sum_{k=1}^{N} x\left( t_m+ \tau_k   \right) \tau_k^d \cL^{M_{\gl, n}}(g)(\lambda_\ell, \tau_k) e^{-2\pi i \frac{(k-\lfloor \frac{N}{2} \rfloor) j}{N}} \Delta t.
\end{align}
Then, we can define the discrete IF reference function and the chirprate reference function as follows:
\begin{align} \label{def_discreteLambda}
	\Lambda_x^{g, n}(\lambda_l, \eta_j, t_m)={\rm Re} \Big( iP_n(\lambda_l)+\frac{\left(T_x^{g, n}(\lambda_l, \eta_j, t_m)\right)^2}{{2\pi i  \big(  \left(T_x^{\tau g, n}(\lambda_l, \eta_j, t_m)\right)^2- T_x^{g, n}(\lambda_l, \eta_j, t_m) T_x^{\tau^2g, n}(\lambda_l, \eta_j, t_m) \big) }}\Big),
\end{align}
\begin{align}\label{def_discreteOmega}
	\Omega_x^{g, n}(\lambda_l, \eta_j, t_m)=\eta_j-{\rm Re} \Big(\frac{T_x^{g, n}(\lambda_l, \eta_j, t_m)T_f^{\tau g,n}(\lambda_l, \eta_j, t_m)}{{2\pi i  \big(  \left(T_x^{\tau g, n}(\lambda_l, \eta_j, t_m)\right)^2- T_x^{g, n}(\lambda_l, \eta_j, t_m) T_x^{\tau^2g, n}(\lambda_l, \eta_j, t_m) \big) }}\Big).
\end{align}

Finally, we summarize the above steps in Algorithm \ref{alg:WLCT}.

\begin{algorithm}[H]
    \caption{Implementation Procedure for WLCT and Calculations of Reassignment Operators}
    \label{alg:WLCT}
    \begin{algorithmic}[1]
        \STATE \textbf{Input:} $x(t_m)$, $n$, $R_0$, $N_c$, $a_0$, $\Delta a$, $N$
        \STATE Discretize the chirprate $\lambda_l$ using \eqref{lambda_2} or \eqref{lambda_1} based on $n$.
        \STATE Discretize the frequency $\eta_j$ and window time $\tau_k$ using \eqref{discretized_eta} and \eqref{discrete_window_time}.
        \FOR{$l \gets 1$ to $N_c$}
            \STATE Precompute the discrete windows: $\mathcal{L}^{M_{\lambda_l, n}}(g)(\lambda_\ell, \tau_k)$, $\tau_k \mathcal{L}^{M_{\lambda_l, n}}(g)(\lambda_\ell, \tau_k)$, and $\tau_k^2 \mathcal{L}^{M_{\lambda_l, n}}(g)(\lambda_\ell, \tau_k)$, based on  \eqref{discretized_window} and \eqref{discretized_tauwindow}.
            \FOR{$m \gets 1$ to $N$}
                \STATE Compute $T_x^{g, n}(\lambda_l, \eta_j, t_m)$ via FFT (see \eqref{FFTWLCT}).
                \STATE Compute $T_x^{\tau g, n}(\lambda_l, \eta_j, t_m)$ and $T_x^{\tau^2 g, n}(\lambda_l, \eta_j, t_m)$ via FFT (see \eqref{FFTTWLCT}).
                \STATE Calculate the reassignment operators $\Lambda_x^{g, n}(\lambda_l, \eta_j, t_m)$ and $\Omega_x^{g, n}(\lambda_l, \eta_j, t_m)$ using \eqref{def_discreteLambda} and \eqref{def_discreteOmega}.
            \ENDFOR
        \ENDFOR
        \STATE \textbf{Output:}
        \STATE \quad  The WLCT coefficients $T_x^{g,n}(\lambda_l, \eta_j, t_m)$
        \STATE \quad  The reassignment operators: $\Lambda_x^{g, n}(\lambda_l, \eta_j, t_m)$ and $\Omega_x^{g, n}(\lambda_l, \eta_j, t_m)$
    \end{algorithmic}
\end{algorithm}

 Besides,  for multicomponent signals whose  instantaneous chirprates are all  positive (e.g., signal \(y(t)\) in \eqref{example2} and \(z(t)\) in \eqref{example3}), the computational complexity  can be  reduced by restricting  the analysis to the positive chirprate domain. 
This can be achieved by redefining the chirprate grid with  a bin size of \(\left\lfloor \frac{N_c}{2} \right\rfloor\).
 Specifically, the  chirprate discretization in \eqref{lambda_2} is replaced with
\begin{align*}
    \lambda_\ell &= -\ell  \Delta \lambda, \quad \ell = 1, \dots, \left\lfloor \frac{N_c}{2} \right\rfloor, 
\end{align*}
while the  discretization in \eqref{lambda_1} becomes
\begin{align*}
    \lambda_\ell &= a_0  2^{\ell \Delta a}, \quad \ell = 1, 2, \dots, \left\lfloor \frac{N_c}{2} \right\rfloor.
\end{align*}

For signal \(x(t)\) (given by \eqref{example1}) with signal length \(N=512\), chirprate bin size \(N_c=513\), and frequency bin size \(N_\eta=256\), the required memory is approximately 1.0 GB.
In contrast, for signal \(z(t)\) (given by \eqref{example3}) with \(N=1024\), \(N_\eta=512\), and \(N_c=512\), the memory requirement is about 4.0 GB.
Then,  the corresponding runtimes for implementing the Algorithm \ref{alg:WLCT} are provided in Table \ref{tab:runtime_comparison}, respectively, measured on a standard laptop (Intel Core(TM) i7-14650HX CPU @ 2.20 GHz with 16 GB RAM).

\begin{table}[H]
    \centering
    \caption{Runtime (in seconds) of Algorithm \ref{alg:WLCT} for signals \(x(t)\) and \(z(t)\) with different \(n\)}
    \begin{tabular}{lcccc}
        \toprule
        Signal & \(n=1\) & \(n=2\) & \(n=5\) & \(n=6\) \\
        \midrule
        \(x(t)\) & 62.02  & 56.67  & 66.22  & 64.08 \\
        \(z(t)\) & 210.70 & 199.67 & 203.80 & 216.88 \\
        \bottomrule
    \end{tabular}
    \label{tab:runtime_comparison}
\end{table} 
    
Following this, we consider the reassignment algorithm. To further enhance the precision of the signal frequency and chirprate estimation,  the frequency step size  can be  reset  to $\Delta \xi$ and the chirprate step size to $\Delta \gamma$.
\begin{align*}
    \gamma_p &= -R_0 + (p-1) \Delta \gamma, \quad p = 1, 2, \cdots, N_1, \\
    \xi_q &=  (q-1) \Delta \xi, \quad q = 1, 2, \cdots, N_2,
\end{align*}
where
\begin{align*}
    N_1 = \frac{2R_0}{\Delta \gamma} + 1, \quad
    N_2 = \left\lfloor \frac{N}{2} \right\rfloor \frac{\Delta \eta}{\Delta \xi}.
\end{align*}
Generally, \(\Delta \gamma\) and \(\Delta \xi\) are smaller than \(\Delta \lambda\) and \(\Delta \eta\).
Therefore, the estimated ridges are more precise when this resetting is applied.

Then, we obtain
\begin{align}
    \label{synchrosqueezed WLCT transform}
    S^n_x(\gamma_p,\xi_q, t_m) = \sum_{(j, \ell) \in O_{p, q, m}} T_x^{g, n}(\lambda_\ell, \eta_j, t_m),
\end{align}
where the set $O_{p, q, m}$ is defined as
\begin{equation*}
O_{p, q, m} := \left\{ (j, \ell): 
\begin{array}{l}
|\Lambda_x^{g, n}(\lambda_\ell, \eta_j, t_m) - \gamma_p| \leq \frac{1}{2} \Delta\gamma, \; 
|\Omega_x^{g, n}(\lambda_\ell, \eta_j, t_m)- \xi_q| \leq \frac{1}{2} \Delta\xi, \\
|T_x^{g, n}(\lambda_\ell, \eta_j, t_m)| > \epsilon, \\
\left| \left( T_x^{\tau g, n} (\lambda_\ell, \eta_j, t_m)\right)^2 - T_x^{g, n}(\lambda_\ell, \eta_j, t_m) T_x^{\tau^2 g, n}(\lambda_\ell, \eta_j, t_m) \right| > \epsilon
\end{array}
\right\},
\end{equation*}
with $\epsilon$ being a predefined threshold value.

Afterwards, we consider the implementation of the XWLCTs as described
in  Definition~\ref{defineXWLCT}. 
{We find it is quite flexible for the choice of $h(v)$. }  
In this paper, we define
\begin{equation}\label{def_hvp}
h(v_p) = \frac{1}{\sqrt{2\pi}} e^{-\frac{v_p^2}{2}},
\end{equation}
where the discrete points \( v_p \) are given by:
\begin{align*}
v_p = (p - N_0) \cdot \Delta v, \quad \text{for } p = 1, 2, \cdots, 2N_0 + 1.
\end{align*}
Here, \( \Delta v \) denotes the discretization step size for the window function, with a default value of \( \Delta v = 2 \Delta t \), and \( N_0 \) is set to 64 by default.

The X-ray transform is given by
\begin{align}\label{implementation_X-ray transform}
\mathcal{T}_x^{g, n} (\lambda_l, \eta_j, t_m) =\sum_{p} \left| T_x^{g, n} (\lambda_l, \eta_j + \Lambda_l v_p, t_m + v_p) \right|  h(v_p)  \Delta v, 
\end{align}
where
\begin{align*}
\Lambda_l &=
\begin{cases}
-\lambda_l, & \hbox{for $n=2$ and 6}, \\
\frac{1}{\lambda_l}, & \hbox{for $n=1$ and 5}.
\end{cases}
\end{align*}
The detailed implementation steps are presented in Algorithm \ref{alg:XWLCT}.

\begin{algorithm}[H]
    \caption{Implementation Procedure for XWLCT}
    \label{alg:XWLCT} 
    \begin{algorithmic}[1]
        \STATE \textbf{Input:} Discrete time points $t_m$;  $n$, $\Lambda_l$; $\Delta v$, $N_0$, $\eta_j$, $\Delta t$,$\Delta \eta$,   $N$, $N_c$ $N_f$.
        \STATE Construct the discrete window function $h(v_p)$ according to \eqref{def_hvp}.
        \FOR{$l \gets 1$ to $N_c$} 
            \FOR{$m \gets 1$ to $N$} 
                \FOR{$j \gets 1$ to $N_f$} 
                    \FOR{$p \gets 1$ to $2N_0+1 $} 
                        \IF{$0 < \eta_j + \Lambda_l v_p < N_f \Delta \eta$ \AND $0 < t_m + v_p < N \Delta t$}
                            \STATE $\mathcal{T}_x^{g, n} (\lambda_l, \eta_j, t_m) = \mathcal{T}_x^{g, n} (\lambda_l, \eta_j, t_m) +\left| T_x^{g, n} (\lambda_l, \eta_j + \Lambda_l v_p, t_m + v_p) \right|  h(v_p)  \Delta v$.
                        \ENDIF
                    \ENDFOR
                \ENDFOR
            \ENDFOR
        \ENDFOR
        \STATE \textbf{Output:} The XWLCT coefficients $\mathcal{T}_x^{g,n}(\lambda_l, \eta_j, t_m)$.
    \end{algorithmic}
\end{algorithm}

For the SXWLCTs $\mathcal{S}^n_x(\gamma_p,\xi_ q, t_m)$, simply replace  \( T_x^{g, n} (\lambda_l, \eta_j, t_m) \) in \eqref{synchrosqueezed WLCT transform} with \( \mathcal{T}_x^{g, n} (\lambda_l, \eta_j, t_m) \) to obtain the desired modification. A time-frequency-chirprate representation with concentrated energy is obtained after applying the synchrosqueezed transform. 
The codes are available at: 

\n https://github.com/Lishuixin065/SXWLCT.


\section{Conclusion}

In this paper, we introduce a novel windowed linear canonical transform (WLCT) to address multicomponent signals with overlapping instantaneous frequency (IF) curves. 
By employing the WLCT, we extend the traditional two-dimensional time-frequency plane into a three-dimensional framework, thereby offering richer representations for signal analysis. 
Our analysis and numerical experiments demonstrate that this approach holds significant potential for three-dimensional signal separation, effectively distinguishing and handling complex signal components. 
Furthermore, through the application of the X-ray transform and the synchrosqueezed transform, we achieve a more precise time-frequency-chirprate representation, providing accurate IF and chirprate  estimations.
These advancements further enhance the precision and methodology of signal processing.






\section*{Acknowledgments} 
This work was partially supported by the National Key Research and Development Program of China  under Grants 2022YFA1005703  and the National Natural Science Foundation of China under Grants U21A20455 and 12571109.

\bibliographystyle{elsarticle-num-names} 

\bibliography{IEEEabrv,myref820}

\end{document}